\documentclass[12pt,a4paper,twoside,sort&compress]{book}
\usepackage{epsfig}
\usepackage{amsfonts}
\usepackage{amssymb}
\usepackage{amsmath}
\usepackage{setspace}
\usepackage{slashed}
\usepackage{lscape}
\usepackage{pdfpages}
\RequirePackage[a-1b]{pdfx}
\RequirePackage[latin1,utf8]{inputenc}
\usepackage{csquotes}
\usepackage{amsthm,bm}
\usepackage{latexsym}
\usepackage{psfrag}
\usepackage{graphicx}
\usepackage{rotating}
\usepackage{graphics}
\usepackage{subfigure}
\usepackage{mathenv}
\usepackage{amsthm}
\usepackage{xcolor}
\usepackage[thinlines]{easytable}
\usepackage{multirow}
\usepackage{booktabs}
\usepackage{soul}
\usepackage{lineno,hyperref}
\usepackage{lipsum}
\usepackage{mathtools}
\usepackage{fancyhdr}
\usepackage{subfloat}

\usepackage{color}
\usepackage{float}
\newcommand{\ba}{\begin{eqnarray}}
\newcommand{\ea}{\end{eqnarray}}
\newcommand{\baa}{\begin{align}}
\newcommand{\eaa}{\end{align}}

\newcommand{\nn}{\nonumber\\}

\definecolor{sinopia}{rgb}{0.8,0.25,0.04}

\definecolor{greenopia}{rgb}{0.3,0.65,0.14}

\newcommand{\mdot}{\hspace{-1.2mm}\cdot\hspace{-1.2mm}}
\DeclareMathOperator*{\SumInt}{%
	\mathchoice%
	{\ooalign{$\displaystyle\sum$\cr\hidewidth$\displaystyle\int$\hidewidth\cr}}
	{\ooalign{\raisebox{.14\height}{\scalebox{.7}{$\textstyle\sum$}}\cr\hidewidth$\textstyle\int$\hidewidth\cr}}
	{\ooalign{\raisebox{.2\height}{\scalebox{.6}{$\scriptstyle\sum$}}\cr$\scriptstyle\int$\cr}}
	{\ooalign{\raisebox{.2\height}{\scalebox{.6}{$\scriptstyle\sum$}}\cr$\scriptstyle\int$\cr}}
}

\newcommand{\ignore}[1]{}
\usepackage[latin1]{inputenc}
\usepackage{cite}
\usepackage{subfigure}
\usepackage[small,bf]{caption}

\usepackage[Conny]{fncychap} 
\ChNameVar{\LARGE\sffamily\bfseries}
\ChNumVar{\Huge} \ChTitleVar{\Huge\sffamily\bfseries}
\ChRuleWidth{0.0pt} \ChNameUpperCase \ChTitleUpperCase 
\usepackage{fancyhdr}

\makeatletter
\newsavebox{\@brx}
\newcommand{\llangle}[1][]{\savebox{\@brx}{\(\m@th{#1\langle}\)}%
	\mathopen{\copy\@brx\kern-0.5\wd\@brx\usebox{\@brx}}}
\newcommand{\rrangle}[1][]{\savebox{\@brx}{\(\m@th{#1\rangle}\)}%
	\mathclose{\copy\@brx\kern-0.5\wd\@brx\usebox{\@brx}}}
\makeatother

   {\setlength\paperheight {297mm}%
   \setlength\paperwidth  {210mm}}
\catcode`@=11

\catcode`@=11
\begin{document}

    \pagestyle{fancy}                       
    \fancyfoot{}                            
    \renewcommand{\chaptermark}[1]{         
    \markboth{\chaptername\ \thechapter.\ #1}{}} %
    \renewcommand{\sectionmark}[1]{         
    \markright{\thesection.\ #1}}         %
    \fancyhead[LE,RO]{\bfseries\thepage}    
   \fancyhead[RE]{\bfseries\leftmark}      
   \fancyhead[LO]{\bfseries\rightmark}     
   \renewcommand{\headrulewidth}{0.3pt}    
    \makeatletter
    \def\cleardoublepage{\clearpage\if@twoside \ifodd\c@page\else%
        \hbox{}%
        \thispagestyle{empty}
        \newpage%
        \if@twocolumn\hbox{}\newpage\fi\fi\fi}
    \makeatother

\includepdf[page={1-5}]{sumit_8}


\thispagestyle{empty}
\frontmatter
\clearpage
\phantomsection
\addcontentsline{toc}{chapter}{Abstract}

\begin{center}
{\Huge{\textbf{Abstract}}}
\end{center}
In the present thesis, we have studied and explored some of the critical features of the deconfined state of matter dubbed Quark-Gluon Plasma (QGP). Big Bang theory argues that this state of matter occurred during the universe's initial evolution. Thus, knowledge of this phase of matter is of significant importance for comprehending the evolution of the universe. For this purpose, the Ultra-Relativistic Heavy Ion Collisions (URHIC) program is designed to gain access to the bulk feature of the extreme quantum chromodynamics (QCD) matter created in the experiments, namely Relativistic Heavy Ion Collider (RHIC) at Brookhaven National Laboratory (BNL) and the Large Hadron Collider (LHC) at the European Organization for Nuclear Research (CERN). The different properties of the extreme matter that have been studied in this thesis are next-to-leading order (NLO) dispersion properties for soft-moving quarks, screening masses of mesons, drag and diffusion properties of heavy quarks $\mathrm{(HQs)}$, the specific shear viscosity of QGP and energy loss of $\mathrm{HQs}$ propagating in QGP background. The theoretical techniques which are utilized in order to study these properties are mainly hard thermal loop (HTL) effective theory, which is required when one is interested in studying the soft scale ($\sim gT$) physics of the underlined theory QCD at finite temperatures, having $g$ as the strength of the interaction and the other technique which are utilized here is the nonperturbative resummation approach dubbed as Gribov quantization. This nonperturbative resummation deals with the magnetic scale of the theory through the mass parameter having $g^{2}T$ order.   
\\
We have studied and calculated the quark self-energy at NLO and the corresponding dispersion relations for the soft-moving quarks in the real-time formalism (RTF) using an HTL effective theory. The four-point vertex diagram has been calculated apart from the usual three-point vertex diagram, with the effective quark, gluon propagators, and effective vertices taken into account. Since NLO dispersion relations depend on the estimation of NLO quark self-energy, NLO quark self-energy is expressed in terms of two-quarks-one-gluon and two-quarks-two-gluons HTL effective vertex functions, which are further expressed in terms of solid angle integrals. The solid angles and the momentum integrals in quark self-energy have been computed numerically and plotted as a ratio of quark momentum and quark energy. Utilizing that, we showed the NLO correction to quark damping rate and quark energy for both quark modes. The obtained results for NLO dispersion relations show the converging earlier results in the zero momentum limit. Since HTL resummation deals with the electric scale of the theory, we have incorporated the magnetic scale through the nonperturbative Gribov resummation approach in order to study the mesonic correlation lengths. We have calculated the mesonic screening masses both for quenched QCD and for (2+1) flavor QCD cases following the analogies with the non-relativistic quantum chromodynamics (NRQCD) effective theory. The obtained results show that the Gribov quantization improves the infrared dynamics of the theory, i.e., obtained results of mesonic screening masses improve the earlier perturbative results in the low-temperature domain and are well suited with the recent lattice measurements. Since HQs are considered a suitable probe for QGP, thus we explored the heavy quark $\mathrm{(HQ)}$ dynamics using the Gribov-Zwanziger $\mathrm{(GZ)}$ approach. We have estimated the drag and diffusion coefficient of $\mathrm{HQs}$ moving in the QGP background. The interaction of $\mathrm{HQs}$ with the medium constituents, which are encoded in the matrix elements, is calculated through the $\mathrm{GZ}$ propagator for the mediator gluons to take into account the nonperturbative outcomes pertinent to the phenomenologically accessible temperature regime in QGP laboratory experiments. The obtained drag coefficient has been utilized to estimate the momentum and temperature variation of HQ energy loss propagating in the QGP background. Also, using the transverse diffusion coefficient, the temperature variation of the specific shear viscosity of the background medium (QGP) has been obtained. We reported a higher energy loss of propagating HQs compared to perturbative estimates. The shear viscosity to entropy density ratio is observed to comply with the anti-de Sitter/conformal field theory (AdS/CFT) estimation over a more significant temperature regime compared to the earlier perturbative expectation.
       


\cleardoublepage 
\thispagestyle{empty}

\phantomsection\addcontentsline{toc}{chapter}{Acknowledgments}
\pagestyle{fancy}
\fancyhead[LO,RE]{{\bf Acknowledgments}}
\fancyhead[RO,LE]{{\bf \thepage}}

\begin{center}
{\Huge{\textbf{\emph{Acknowledgments}}}}
\end{center}
First of all, I am grateful to my academic supervisors, Binoy Krishna Patra and Najmul Haque, for their guidance, expertise, and endless support I received during my Ph.D. tenure. Their valuable feedback and insights have helped me to push the boundaries of my research. I am grateful for their dedication, patience, and willingness to go the extra mile in order to ensure a little success I achieved during my research. Moreover, I am indebted to them for their invaluable academic guidance and for the essential life lessons that shaped me professionally. I want to give a special thanks to Najmul Haque for the nonacademic help during my visits to NISER.  
\\
I am grateful to my collaborator, Arghya Mukherjee, for working with me on the heavy quark dynamics project. A fruitful discussion via email regarding my research queries with Abdessamad Abada, Margaret Carrington, Mikko Laine, and Michael Strickland is highly appreciated. Discussions with Robert Pisarski, Michael Strickland, Rishi Sharma, Santosh Kumar Das, and Raktim Abir during conferences were beneficial and helped me to stay motivated and enthusiastic about my research. I want to acknowledge the Student Research Committee (SRC) members, including Dibakar RayChowdhury, Tapas Kumar Mandal, and Praveen Srivastava, for keeping an eye on my progress and the suggestions I received during the research.           
\\
I want to thank my colleagues from the high energy physics research group, namely, Mujjeb Hasan, Shublakshmi Rath, Salman Ahamad Khan, Debarshi Dey, Pushpa Panday, Abhishek Tiwari, Sarthak Sathpathy, Prakash Gochiyat, Gopal Yadav, Hemant Rathi, Jitender pal at IIT Roorkee and Sadaf Madni, Manas Debnath, Sourav Dey, Jaffery at NISER, Odisha for their academic discussion and nonacademic help received during the tenure. The time at IIT Roorkee and outside the campus was excellent with the presence of different people I met and thanks for their insights and help: Shyam Sundar Sharma, Lekhika Malhotra, Nitish Bhardwaj, Vikas Rastogi, Prathul Nath, Rama Krishna, Vikas Kumar, Amit Singh, Purnima, Ankur, Himani Verma, Shikha Rathi, Brij Mohan, Pranjul, Vishwajeet, Gargi Tyagi, Nisha Chahal, Anshika. I want to extend my thanks to all the members of Research Scholar Room 1, including  Aalok Kumar Sahoo, Deepak Patel, Nitish Ghosh, Ayushi Agarwal and all other batchmates from the 2019 July batch. 
\\
I am genuinely grateful for the funding that the Indian government's Ministry of Education (MoE) gave me for my PhD. I want to acknowledge the financial support I received from NISER during my stay at the NISER Institute. Last but not least, I would like to thank my mother and my grandfather, who played a remarkable role in my success, for their continuous support, love, and affection all these years.

\cleardoublepage
\thispagestyle{empty}
\phantomsection
\addcontentsline{toc}{chapter}{Table of Contents}
\pagestyle{fancy}
\fancyhead[LO,RE]{{\bf Contents}}
\fancyhead[RO,LE]{{\bf \thepage}}
\tableofcontents

\clearpage
\thispagestyle{empty} 
\phantomsection
\addcontentsline{toc}{chapter}{List of Publications}

\pagestyle{fancy}
\fancyhead[LO,RE]{{\bf Publications}}
\fancyhead[RO,LE]{{\bf \thepage}}

\begin{center}
{\Large\textbf{List of Publications}}
\vspace{0.6cm}
\end{center}
\textbf{A. Refereed Journal Publications (included in the 
thesis)}
\begin{enumerate}
\item
\textbf{Sumit}, Najmul Haque and Binoy Krishna Patra,
\textquotedblleft{\it NLO quark self-energy and dispersion relation using the hard thermal loop resummation}\textquotedblright, 
JHEP {\bf 05},(2023) 171.\\
\url{https://link.springer.com/article/10.1007/JHEP05(2023)171}
\item
\textbf{Sumit}, Najmul Haque and Binoy Krishna Patra,
\textquotedblleft {\it QCD mesonic screening masses using Gribov quantization}\textquotedblright, 
Phys.Lett.B {\bf 845}, (2023) 138143.\\
\url{https://doi.org/10.1016/j.physletb.2023.138143}
\item
\textbf{Sumit}, Arghya Mukherjee, Najmul Haque and Binoy Krishna Patra,
\textquotedblleft {\it Heavy quark dynamics via Gribov-Zwanziger approach}\textquotedblright, 
Phys.Rev.D 109 (2024) 11, 114043. \\
\url{https://doi.org/10.1103/PhysRevD.109.114043}

\end{enumerate}
\vspace{0.6cm}
\textbf{B. Proceedings in National and International Conferences :}

\begin{enumerate}
\vspace{0.5cm}	
\item 
Santosh K. Das et al.
,\textquotedblleft {\it Dynamics of Hot QCD Matter -- Current Status and Developments} \textquotedblright, Int.J.Mod.Phys. E 31 (2022).\url{https://www.worldscientific.com/doi/10.1142/S0218301322500975}

\end{enumerate}
\vspace{0.6cm}

\textbf{Participation in Conferences/Schools:}
\vspace{0.6cm}
\begin{enumerate}



\item Attended and gave an oral presentation in 
{\bf Hot QCD Matter} conference organized by IIT Goa and Goa University from 12-14 May 2022.


\item Attended {\bf{XQCD Ph.D. school 2022}} held at University of Stavanger and presented a poster at the {\bf{``International Conference on QCD in Extreme Conditions (XQCD 2022),"}} which was held from July 27 to July 29 2022 at the Norwegian University of Science and Technology (NTNU) Norway. 


\item Attended and presented a poster at {\bf{``8th International Conference on Physics and Astrophysics of Quark Gluon Plasma (ICPAQGP 2023)"}} held at Blue Lily Resorts, Puri, Odisha, India from 6-10 February, 2023. 

\item Delivered a talk at {\bf ``8th Shivalik HEPCATS Meeting"} held at NIT Jalandhar, India on May 12, 2023.

\item Attended and presented a poster at {\bf ``2nd Workshop on Dynamics of QCD Matter"} held at NISER, Odisha, India on 7-9 October 2023.

\item Attended and delivered a talk in online mode at {\bf ``India-JINR workshop on elementary particle and nuclear physics, and condensed matter research"} held at JINR, Dubna, Russia on 16-19 October 2023.

\item Attended and delivered a talk at {\bf ``4th Heavy Flavor Meet (HFM 2023)"} held at IIT Goa, India on 1-4 November 2023.

\item Delivered a talk at {\bf ``9th Shivalik HEPCATS Meeting"} held at IISER Mohali, Mohali, India on 28th January, 2024.

\item Attended {\bf{``School for Advanced Topics in Particle Physics (SATPP): Selected Topics in Effective Field Theories''}} held at ICTS, Bengaluru, India from 8-19 April 2024.

\item Attended and gave an oral presentation in 
{\bf 2nd Hot QCD Matter Conference 2024} organized by IIT Mandi from 1-3 July 2024.

\end{enumerate} 

\clearpage
\phantomsection
\addcontentsline{toc}{chapter}{List of Figures}
\pagestyle{fancy}
\fancyhead[LO,RE]{{\bf List of Figures}}
\fancyhead[RO,LE]{{\bf \thepage}}
\listoffigures
\cleardoublepage
\phantomsection
\mainmatter \pagestyle{fancy} \setcounter{chapter}{0}
\setcounter{section}{0} \setcounter{subsection}{0}
\setcounter{tocdepth}{3}


\chapter{Introduction}\label{Chapter_1}
\allowdisplaybreaks
\pagestyle{fancy}
\fancyhead[LE]{\emph{\leftmark}}
\fancyhead[LO]{\emph{\rightmark}}
\rhead{\thepage} 

The quest to know how the universe operates is one of the main things that intrigues humankind. This quest leads to the four forces, i.e., electromagnetic, weak, strong, and gravitational, through which fundamental interactions that exist in nature can be described. Here, we will focus mainly on the nature of strong forces, which are relevant to studying strongly coupled systems. A short overview of the background of QCD, which is the quantum field theory (QFT) of strong interactions, is presented here.\\
Roughly, the history of QCD starts in $20^{th}$ century with the discovery of strongly interacting subatomic particles having a longer lifetime, namely proton and neutron, prior to second world war~\cite{FRS:1919nrm, Chadwick:1932ma, Chadwick:1932wcf}. The other lighter particles like mesons, pions, and kaons were first discovered in 1947 through cosmic rays events~\cite{Lattes:1947mw, Rochester:1947mi}. However, at that time, these and all other recently discovered particles were regarded as elementary. The leptons, which came to light around the same time period, were thought to be elementary, and this is still true today. \\
As the particle accelerator facilities developed with time, the number of strongly interacting particles (including the discoveries of $\Lambda$~\cite{Hopper:1950vs} and $\Delta$~\cite{Anderson:1952ah} and their other variants, as well as charged and uncharged variants of these particles) increased drastically. The idea that all particles are fundamental and indivisible was changed to a new paradigm. How to organize them in a well-mannered fashion was a challenging task, and the first proposed method relied on the representation theory of $SU(3)$ group matrix in 1961~\cite{Neeman:1961jhl, Gell-Mann:1961omu}. This organization of particles was the earlier generalization of $SU(2)$ gauge group based on isospin symmetry, explaining new features of the observed particles. However, it turned out to need to be completed. Nevertheless, later on, after a few years, a final proposal was made~\cite{Gell-Mann:1964ewy, Zweig:1964ruk} which claims that these recently discovered particles were not fundamental at all but are composite, made through an $SU(3)$ triplet of fundamental particles named as $\textit{quarks}$~\cite{Joyce:1939}. Although the same symmetry group had been used before, the consideration of quarks as the fundamental degree of freedom of strong interaction was a fundamentally important step. Many more hadronic resonances have been discovered since then, but their characteristics are now well described utilizing the underlined theory of QCD.  \\
There are just three sorts of quarks in the quark model at first: u(p), d(own), and s(trange) quarks. Although the first two were sufficient to explain the structure of protons and neutrons, the third one was required to understand the features of kaons, which come out to be the first strange particles. The studies of weak interaction properties led to the idea of the fourth quark~\cite{Glashow:1970gm}, namely c(harm) quark, which was discovered experimentally~\cite{E598:1974sol, SLAC-SP-017:1974ind} in the same decade as it was proposed so. The other two quarks' predictions were also theoretically made during the development of weak sector interactions studies~\cite{Kobayashi:1973fv}. Thus, we got the three families of quarks consisting of each of 2, a model that is reliable to date. The experimental verification~\cite{E288:1977xhf} of b(ottom) quark was also made soon after its theoretical proposal. However, the ultimate and heaviest t(op) quark was observed experimentally in 1995~\cite{D0:1995jca, CDF:1995wbb}. Some of the features of quarks are shown in Table~\ref{table:1}. The 'Other' column lists various quantum numbers that vanish unless specifically mentioned.
 \\
\begin{table}[h!]
\centering
\begin{tabular}{ |p{3cm}||p{3cm}|p{3cm}|p{3cm}|  }
 \hline
 \hline
Flavour & Mass & Charge & Other\\
 \hline
u(p) & $2.16^{+0.49}_{-0.26}$ MeV & 2/3 & isospin 1/2 \\
\hline
d(own)  & $4.67^{+0.48}_{-0.17}$ MeV & -1/3 & isospin -1/2 \\
\hline
s(trange)  & $93^{+11}_{-5}$ MeV & -1/3 & strangeness -1 \\
\hline
c(harm)  & $1.27^{+0.02}_{-0.02}$ GeV & 2/3 & charmness 1 \\
\hline
b(ottom)  & $4.18^{+0.03}_{-0.02}$ GeV & -1/3 & bottomness -1 \\
\hline
t(op)  & $172.9^{+0.4}_{-0.4}$ GeV & 2/3 & topness 1 \\
\hline
\end{tabular}
\caption{Different features of quarks~\cite{ParticleDataGroup:2018ovx}}
\label{table:1}
\end{table}
With the quark model resounding success, on the theoretical front, the discovery of $\textit{asymtotic freedom}$ of non-abelian Yang-Mills theory~\cite{Gross:1973id, Politzer:1973fx} in 1973 finally made it possible a QFT of quarks and gluons as fundamental degrees of freedom -now utilizing $SU(3)$ as a local gauge symmetry-this will be discussed in detail further in this Chapter. This QFT is referred to as QCD due to its similarities with the analogous theory of electromagnetic interaction, Quantum Electrodynamics (QED). The other remarkable prior experimental result was the deep inelastic scattering performed at the Stanford Linear Accelerator Center (SLAC) in 1969, demonstrating that the nucleons indeed have substructures~\cite{Bloom:1969kc}, as well as the 3-jet events, confirming the existence of gluons, performed at Deutsches Elektronen-Synchrotron~\cite{Ellis:2014rma} in 1979. Lastly, the deconfined phase of matter - QGP, in which the truly fundamental degrees of freedom are quarks and gluons, was finally discovered in 2000~\cite{Heinz:2000bk}. For the more recent development, one can look at review~\cite{ParticleDataGroup:2018ovx}.     \\
This Chapter will cover the fundamental features of QFT of strong interactions, i.e., QCD, its fundamental properties, and its applications to understanding the deconfined state of matter, i.e., QGP.  
\section{Quantum Chromodynamics (QCD)}
After a brief history of the modern theory of strong interactions QCD~\cite{Griffiths:2008zz, Peskin:1995ev} in which the fundamental partonic degree of freedom are quarks and gluons, let us focus on the mathematical language of the theory. The Lagrangian of the theory is written as 
\begin{align}
\mathcal{L}_{\mathrm{QCD}}=-\frac{1}{4} F_a^{\mu \nu} F_{\mu \nu}^a+\sum_\alpha \bar{\Psi}_\alpha (i \slashed{D}-m_{\alpha}) \Psi_\alpha
\label{QCD_Lagrangian}
\end{align}
The non-abelian gauge field strength tensor $F_{\mu \nu}^{a}$ is given by
\begin{align}
F_{\mu \nu}^a=\partial_\mu \mathcal{A}_\nu^a-\partial_\nu \mathcal{A}_\mu^a + g_{s} f_{a b c} \mathcal{A}_{\mu}^{b} \mathcal{A}_{\nu}^{c},
\label{Gluon_Field_Strength}
\end{align}
where $\mathcal{A}_{\mu}^{a}$ is the non-abelian gauge field with color index a (runs from $1,2, \cdots 8$), $g_{s} = \sqrt{4\pi \alpha_{s}}$ is the strong running coupling strength and $f_{a b c}$ are completely anti-symmetric structure constant. In eq.~\eqref{QCD_Lagrangian}, the index $\alpha = u, d, s, c, b, t$ is the flavor index and $D_{\mu}$ is the covariant derivative defined as $D_\mu=\partial_\mu-i g_{s} \mathcal{A}_\mu$ where $\mathcal{A}_{\mu} = \mathcal{A}_{\mu}^{a} T^{a}$. Here $T^{a}$'s are generators of $SU(3)$ group following the lie algebra as $[T^{a}, T^{b}] = i f^{a b c } T^{c}$. The fermionic field $\Psi_{\alpha}$ can be represented as  
\begin{align}
\Psi_\alpha=\left(\begin{array}{c}
\Psi^{\text {red }} \\
\Psi^{\text {blue }} \\
\Psi^{\text {green }}
\end{array}\right)_\alpha
\end{align}
Here, we have red, blue, and green as the three-color quantum number corresponds to a quark of mass $m_\alpha$. The running coupling term in eq.~\eqref{Gluon_Field_Strength} signifies that $F_{\mu \nu}^{a} F^{\mu \nu}_{a}$ consists of terms like $g_{s} \partial_\nu \mathcal{A}_\mu^a f^{a b c} \mathcal{A}^{\mu b} \mathcal{A}^{\nu c}$ and $g^2 f^{a b c} f^{a j k} \mathcal{A}_\mu^b \mathcal{A}_\nu^c \mathcal{A}^{\mu j} \mathcal{A}^{\nu k}$. These terms contribute to $3$ and $4$-point gluonic vertices in Feynman graphs, respectively. This self-interaction feature of QCD distinct it from the QED, where theory has no self-interactions. This feature of self-interaction between gluon leads to the remarkable property of theory, namely \textit{aymptotic freedom}, which we are going to discuss in the following subsection.
\subsection{Key Properties}
\begin{figure}[t]
\centering
\includegraphics[scale=0.7,keepaspectratio]{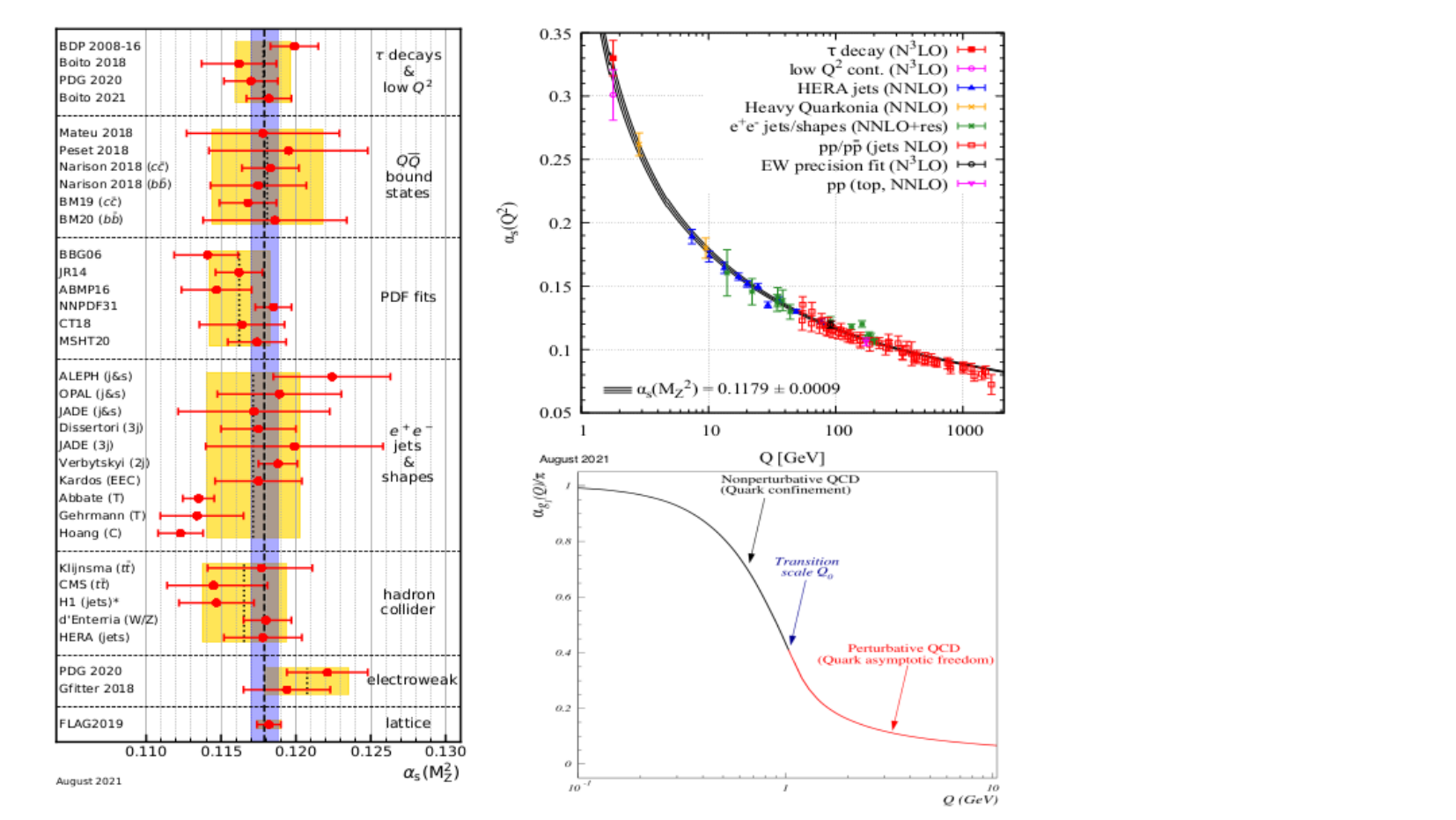}
\caption{\textbf{(Left)} The average value of $\alpha_{s} (M_{z}^{2})$ and its uncertainty for $7$ different sub-fields~\cite{ParticleDataGroup:2022pth}. \textbf{(Top)} measurements of $\alpha_{s} (Q^{2})$ as a function of the energy scale $Q$. (Bottom) There are two zones for running coupling, perturbative domain, and non-perturbative domain, each having unique features~\cite{Deur:2014qfa}.}
\label{alpha_s_measurement}
\end{figure}

In QED, the electric charges are screened due to vacuum fluctuations, while in QCD, due to the self-interaction property of gluons, the color charges get anti-screened. This leads to the fact that as we start increasing the energy scale $(Q)$ to probe the matter, the QCD coupling strength starts decreasing. This unique phenonemenon is known as \textit{aymptotic freedom}~\cite{Gross:1973id,Politzer:1973fx,tHooft:1985mkt}. The significance of this property on the qualitative nature of QCD was first pointed out by Collins and Perry shortly after the discovery of \textit{aymptotic freedom}~\cite{Cabibbo:1975ig, Collins:1974ky}. The one-loop running coupling of QCD is given by~\cite{Gross:1973id, Politzer:1973fx}
\begin{align}
\alpha_s\left(Q^2\right)=\frac{g^2\left(Q^2\right)}{4 \pi}=\frac{4 \pi}{\left(11-\frac{2}{3} N_f\right) \ln \left(\frac{Q^2}{\Lambda_{Q C D}^2}\right)},
\end{align}
where $N_f$ and $Q^2$ represent the number of quark flavors and the four-momentum transfers, respectively. $\Lambda_{Q C D}$ is the typical QCD scale parameter having value roughly around $\sim 0.2 GeV$. Because of the asymptotic property of QCD, perturbative approaches are reliable for studying various features of the theory. In figure~\ref{alpha_s_measurement}, the average value of running coupling estimation is provided and calculated via different sub-field approaches. The precise determination of running coupling and masses of different quarks is essential to studying the different observables of the theory. \\  
The other important property of the QCD is that this theory possesses a property known as \textit{color-confinment}~\cite{Mandelstam:1974pi, Greensite:2011zz}. The fundamental degrees of freedom of QCD, i.e., quarks, are not detected freely as an individual entity in experiments but instead appear in combined forms in bound states like \textit{hadrons}. In principle, one may try to demonstrate this from the QCD lagrangian, but this is a non-trivial task to be performed, and in fact, it has no complete analytic proof till now. Nevertheless, the field theory studies on lattices, i.e., lattice field theory, which is the first principle QCD approach~\cite{Wilson:1974sk}, suggest that in the systems that are based on QCD lagrangians do indeed exhibit confinement and are very successful in explaining these bound states, hadrons properties like mass spectra~\cite{PACS-CS:2007thi}. Apart from lattice studies, the effective models, which are based on the underlying symmetries of QCD, are reliable for understanding the phenomenon of confinement and low energy dynamics of the theory.  
\section{QCD phase diagram}
\begin{figure}[tbh]
\centering
\includegraphics[scale=0.5,keepaspectratio]{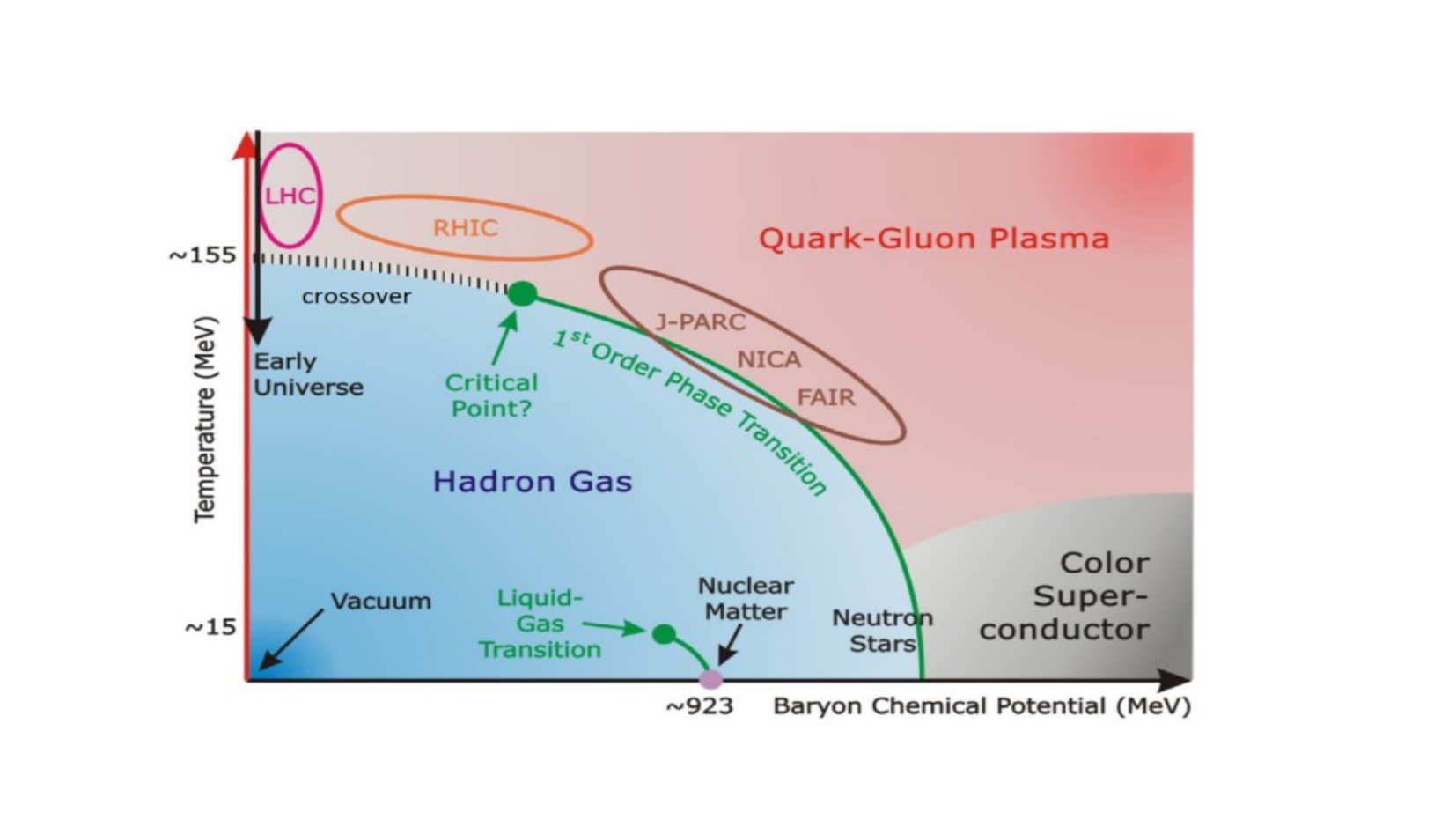}
\caption{Various phases of strongly interacting matter~\cite{CRC-TR 211}}
\label{qcd_phase_diagram}
\end{figure}

In Figure~\ref{qcd_phase_diagram}, the various phases of strongly interacting matter are shown, i.e., quarks and gluon system at different temperatures and baryon chemical potential, which is a measurement of the asymmetry between quarks and anti-quarks. The phase diagram shown in Figure~~\ref{qcd_phase_diagram} can be divided roughly into three different regions, namely the hadronic phase, the QGP phase, and the color-superconducting phase. \\
In the QGP phase, quarks and gluons are in the deconfined state, and chiral symmetry is restored~\cite{Bhattacharya:2014ara}. This phase consists of the partons, which move approximately freely in the medium due to the screening of strong color forces and are believed to have existed in the early cosmos a few microseconds after the Big Bang. On the other hand, the hadronic phase corresponds to the vacuum state of QCD, and within this phase, the partons are confined inside the hadrons. The color-superconducting phase~\cite{Alford:2001dt, Alford:2007xm, Satz:2009hr} arises at very high densities, low temperatures scenario, and it is expected that partons might stick together strongly in this phase. It is anticipated within this phase that after quark pair formation, quarks will operate like a fluid and result in the features of color conduction without any hindrances. \\
For $\mu_{B} = 0$ region of the phase diagram is well understood by Lattice QCD (LQCD) calculations, and thermodynamics of QCD like pressure, energy density, entropy density, and trace-anomaly has been studied from the first principle calculations. It has been found that there is a rise in pressure and energy density around $T = 160$ MeV, which indicates that there is a change in the effective degree of freedom from hadrons to quarks and gluons. This phenomenon is widely known as $\textit{confinemnt-deconfinement}$ transition. With further refined studies of LQCD, it was concluded that this is not a strict phase transition but a rapid $\textit{cross-over}$~\cite{Aoki:2006we}. For finite $\mu_{B}$, LQCD suffers from infamous sign problems and other systematic uncertainties. \\
A number of ways have been opted out to resolve the issue. Some of them include the Taylor series expansion~\cite{Allton:2002zi,Allton:2005gk,Gavai:2008zr,MILC:2008reg,Kaczmarek:2011zz} around $\mu_{B} = 0$ and analytic continuation~\cite{deForcrand:2002hgr,DElia:2002tig} from imaginary to real $\mu_{B}$ are promising. From these studies, it was concluded that the phase transition from hadronic state to QGP is a $\textit{cross-over}$ around the region ${\mu_{B}}/{T} \gtrsim 2$~\cite{Ratti:2018ksb}. Recently, a new Taylor series resummation has been proposed at finite chemical potential~\cite{Mitra:2022vtf}, and the QCD equation of state has been studied using this new approach~\cite{Mitra:2023csk}. However, the QCD phase diagram remains to be explored widely from LQCD techniques. Apart from LQCD, several effective models have been utilized to understand the nature of QCD phase transition, and these models suggest that there is a first-order phase transition at high baryon densities having a critical point at $(T_{c}, \mu_{c})$. The first order transition is not a chiral phase transition. The approximate chiral symmetry does not give an order parameter for this
transition. If $\mu_{B} > \mu_{c}$ then crossover becomes first order chiral phase transition~\cite{Asakawa:1989bq,Barducci:1989wi,Wilczek:1992sf,Berges:1998rc} which ends at another critical point $(T_{F},\mu_{F})$. If these deconfined and chiral phase transitions do not coincide, then there is a possibility of some exotic phases like quarkyonic phase~\cite{Cleymans:1986cq, Kouno:1988bi} which are chiral symmetric but confined phase~\cite{McLerran:2007qj, Hidaka:2008yy} may exist. \\
The existence of this final critical point $\mu_{F}$ has been suggested theoretically~\cite{Halasz:1998qr,Barducci:1993bh,Barducci:1990sv,Kiriyama:2000yp} and later on confirmed through lattice simulations~\cite{Fodor:2004nz}. Although with effective models, this point is known to have some uncertainty, locating it precisely in the phase diagram is still challenging. The Beam Energy Scan (BES) program at RHIC has been developed to find its location precisely by creating QGP with different $\mu_{B}$ and $T$ by varying the colliding energy ($\sqrt{s}$) of heavy nuclei. 
\section{Heavy-ion collisions (HIC): Overview}
\begin{figure}[tbh]
\centering
\includegraphics[scale=0.55,keepaspectratio]{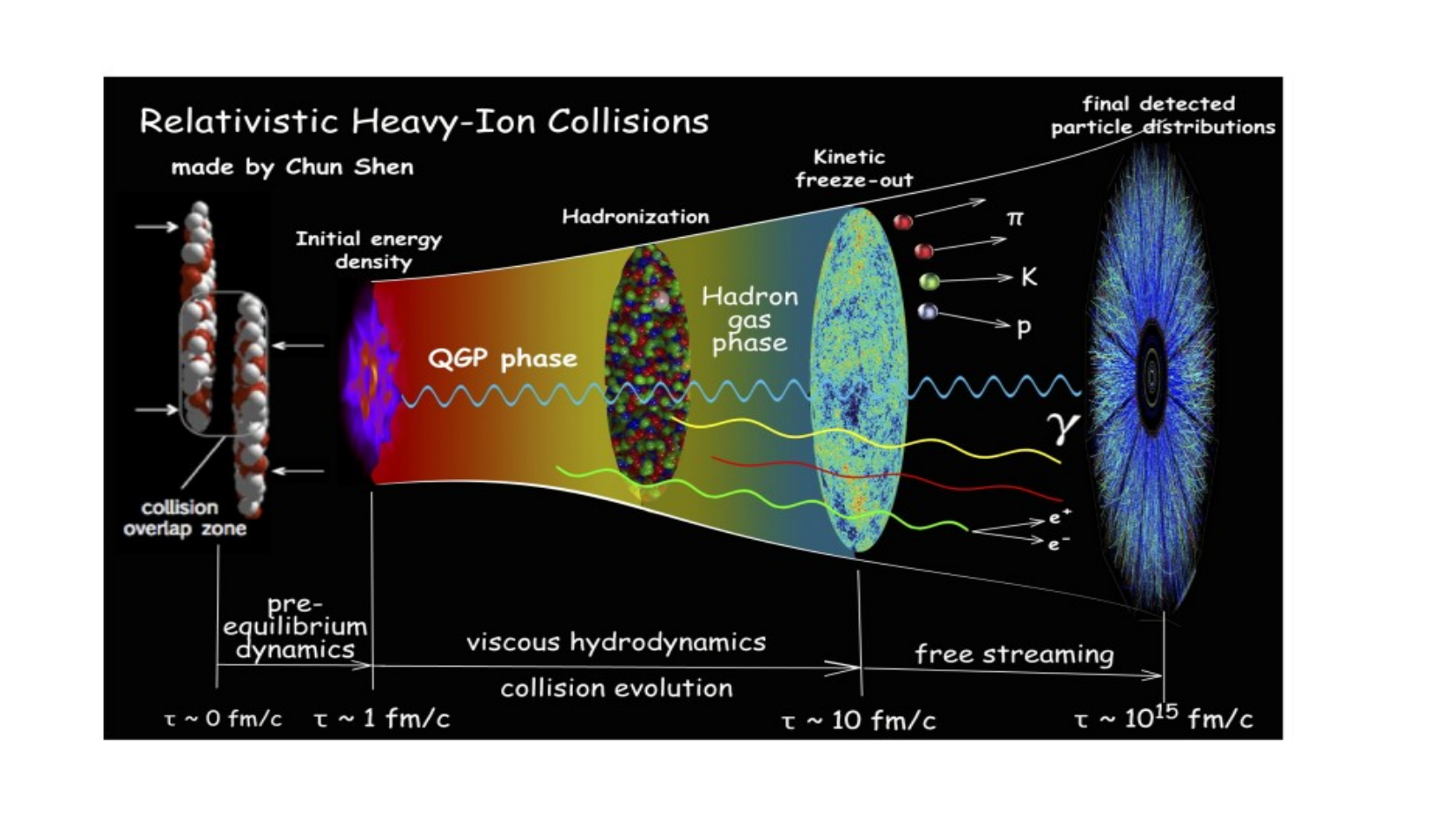}
\caption{Typical diagram for HIC system~\cite{Shen:2014vra}.}
\label{HIC_in_picture}
\end{figure}
\begin{figure}[tbh]
\centering
\includegraphics[scale=0.6,keepaspectratio]{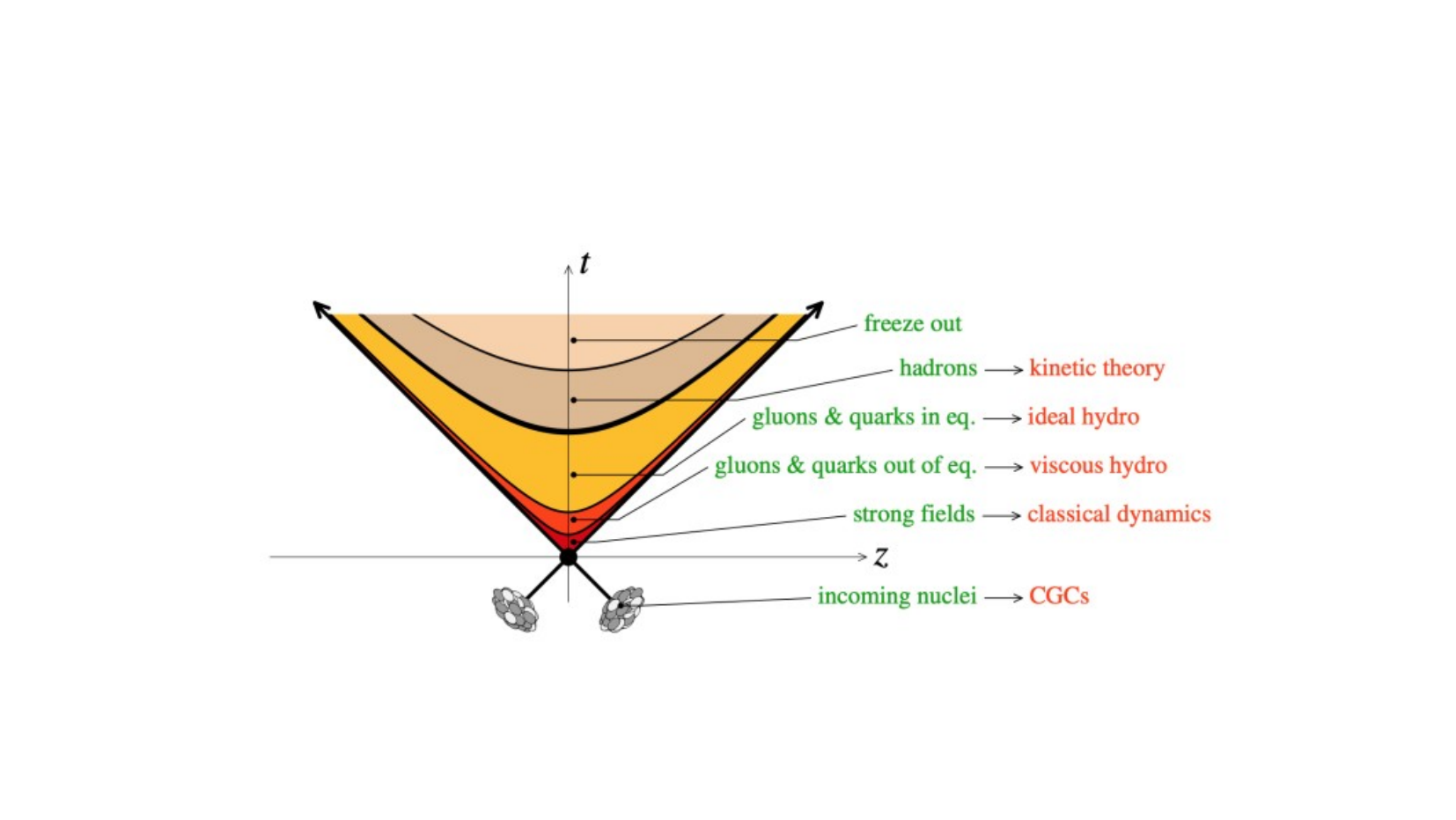}
\caption{Schmetic diagram for the space-time evolution of HIC system~\cite{Gelis:2011xw}}
\label{space_time_evo._of_HIC}
\end{figure}

The QCD phase diagram was covered in the previous section, and the phase in which we are interested in this thesis is the deconfined phase of quarks and gluons, i.e., the QGP phase. In order to study this phase, the HIC experiments are designed in such a way that heavy and stable ions are accelerated with very high energies so that they collide relativistically and create similar situations as the early universe scenario in the laboratory. A typical diagram for the HIC system is shown in Figure~\ref{HIC_in_picture}. The experimental facilities of these HIC systems started with the Bevelac facility in Lawrence Berkley National Laboratory (LBNL)~\cite{Nagamiya:1984vk} where first-time collectivity phenomena were observed~\cite{Gustafsson:1984ka}. In further facilities like Alternating Gradient Synchrotron (AGS) at BNL and Super-Proton Synchrotron (SPS) at ``Conseil Européen pour la Recherche Nucléeaire" (CERN), the energies of the incoming ions have been increased. Later on, the experimental program, namely the RHIC, started in $2000$ in which the center of mass energy of the incoming nucleons is around $\sqrt{s} = 7.7-200$ GeV, and for the first time, the QGP signals were detected~\cite{BRAHMS:2004adc, STAR:2005gfr, PHENIX:2004vcz}. At this energy significant coherence was observed in the created system, which could not be achieved through simple one-to-one nucleon-nucleon collisions~\cite{Muller:2006ee, Gyulassy:2004zy}, and the energy density also lies according to LQCD prediction, which confirms the QGP formation in this HIC systems~\cite {Karsch:2001cy}. At the present day, the maximum energy achieved in HIC experiments is at LHC, which is $\sqrt{s} = 5.02$ TeV in Pb-Pb collision and $\sqrt{s} = 5.44$ TeV in Xe-Xe collision. The other future experiments which are designed to study the high baryon density of QCD phase diagram are Compressed Baryonic Matter (CBM)~\cite{Wilczek:2010ae} at Facility for Antiproton and Ion Research (FAIR) at Gesellschaft für Schwerionenforschung (GSI) and Nuclotron-based Ion Collider fAcility (NICA) at Joint Institute for Nuclear Research (JINR), which will be useful to study the matter in the core of neutron star. \\
As discussed earlier, due to asymptotic freedom at large energies, the strength of the strong interaction is reduced. Consequently, in ultra-relativistic heavy-ion collisions (URHIC), most quarks and the Lorentz-contracted nuclei pass-through each other, taking on a pancake-like shape that releases a substantial amount of energy at the collision's center. Depending on this energy density, the formed medium either consists of a gas of interacting hadrons or a deconfined medium of quarks and gluons. The different stages of the fireball created after these collisions are shown in Figure~\ref{space_time_evo._of_HIC}.  
\subsection{Pre-equilibrium phase}
At RHIC, at time $\tau < 0$ (before collision time), two beams of Lorentz contracted gold nuclei are accelerated with a speed close to the speed of light in the opposite directions and introduced to collide at the particular positions; let us say at $z\pm t$. The collision occurs to happen at $t= z = 0$, and the interacting matter creates a large amount of energy in the central region near $z \sim 0$. In this stage ($0<\tau<\tau_{0}$), the system is out-of-equilibrium, and the evolutions in the transverse plane, i.e., directions that are transverse to the original beam direction, can be considered to be static. As a result, the system's evolution dynamics favor the longitudinal direction expansion of the system. This initial stage can be more accurately described by a dense ensemble of gluon walls using an effective theory known as Colar Class Condensate (CGC)~\cite{Gelis:2010nm, Busza:2018rrf}. The matter produced after this stage needs around $10^{-23}$ seconds to attain the thermal equilibrium state and form the QGP state.    
\subsection{Thermalized QGP phase}
After a time $\tau_{0}$, the system consisting of a deconfined medium of quarks and gluons, i.e., QGP, quickly expands and cools down because of thermal pressure gradients. After the locally equilibrated QGP formation, the subsequent hadronization takes place continuously at the edge of the fireball during the time period $\tau_{0} < \tau < \tau_{f}$ after a critical temperature $T_{c}$ which is estimated to be $160$ MeV using LQCD techniques. The expansion of the system during this stage is well described by the effective theory of hydrodynamics~\cite{Gale:2013da, Strickland:2014pga, Jeon:2015dfa, Jaiswal:2016hex}, which relies mainly on the conservation laws of energy, momentum and other conserved charges of the system. In section~\ref{QGP_methods}, the theoretical approaches needed to investigate this stage of HIC will be covered. It takes roughly a time period of around $10$ fm/c for the QGP state to hadronize, and thus, the system gradually becomes ready for the ultimate freeze-out state, which has a freeze-out time $\tau_{f}$.     
 
\subsection{Freeze-out}
After a time of $\tau \sim 20 > \tau_{f}$ fm/c, the hadron gas undergoes chemical freeze-out, where the inelastic collisions come to an end~\cite{Braun-Munzinger:2003pwq}. Thus, the particle number becomes conserved, but elastic collisions between hadrons occur after chemical freeze-out, and local equilibrium can still be conserved. Kinetic freeze-out happens when the particle's mean free path surpasses the system's size over a period of time. At this point, the hadrons no longer interact in any way, and their momenta are constant. The hydrodynamic description can not be applied directly since the system is out of equilibrium now. Also, the anisotropic hydrodynamics~\cite{Strickland:2014pga, Alqahtani:2017jwl, Alqahtani:2017mhy} has been developed, which considers the large space-momentum anisotropies into consideration. The hadrons coming out after kinetic freeze-out freely move until they reach the detector. The particle yield is well described by the thermal statistical models~\cite{Cleymans:1992zc, Cleymans:1998yb, Becattini:2000jw, Cleymans:2005xv}.     
\section{QGP Probes}
After a brief overview of HIC, whose main aim is to produce and detect the novel state of matter having a fundamental degree of freedom as quarks and gluons, let us discuss the probes of this matter. Due to the collective phenomena of the plasma, the collection of distinct data of the various transitory phases that constitute QGP is required to describe the deconfined nature of matter. Because of the confining traits of the QCD vacuum, the direct signals of QGP can not be detected. Thus, it is mainly studied through indirect probes. The deconfined QCD matter properties are extracted by the final state particles coming out of the fireball and reaching the detector. Some of the essential probes used to access the QGP properties are discussed below.   
\subsection{Electromagnetic probes}
Electromagnetic (EM) probes, i.e., photons and dileptons, are quite helpful probes in characterizing the hot and dense matter created in HIC experiments, see for example~\cite{Stankus:2005eq, Rapp:2009yu, Rapp:2011is, David:2019wpt, Salabura:2020tou, Geurts:2022xmk} for reviews. The primary cause of this is that they mostly interact with the medium particles through electromagnetic (EM) interactions rather than (directly) interacting with them. On the basis of the EM and strong couplings, it can be inferred that the photons and dileptons mean free path is greater than the extent of the generated firball~\cite{Feinberg:1976ua, Kapusta:1991qp}. Thus, photons and dileptons can navigate in the medium with minimal disturbance and convey data from their place of generation to the detector~\cite{Chatterjee:2009rs, Rapp:2014hha}. Small EM coupling also entails the rare production of photons and dileptons in contrast to strongly interacting particles like pions. For instance, in contrast, to decay into pions, the suppression of state $\rho(770)$ vector meson decay into electron-positron duo or muon-antimuon duo is $\sim 10^{-5}$. See~\cite{ParticleDataGroup:2020ssz} for the corresponding branching ratio. \\
Another key feature of these probes is that they are produced at all stages of the collision phenomena. As a result, they can be utilized to gather data from every stage of the fireball that is generated, ranging from the first hard scattering process through the pre-equilibrium phase and QGP to the hadronic gas phase. However, this information over a particular phase is very complicated with the space-time structure of the created medium, which makes it very challenging to obtain the properties of a specific phase, like the equilibrated QGP phase or hadronic phase. These probes, mainly dileptons, are also helpful for learning information regarding various features of the QCD phase diagram, like the precise position of first-order phase transitions and proposed critical endpoints. The light mesons can theoretically describe the thermal dilepton and thermal photon spectra because they contain the same quantum number as the photon. Thus, the light meson can be directly transformed into a real or virtual photon, which, subsequently, can decay into a lepton-antilepton duo, i.e., dilepton. So, light vector mesons like $\rho(770), \omega(782), \phi(1020)$ serve as a bridge between strongly interacting regimes and emitted EM particles. In actuality, light vector mesons with the substantial contribution dominance from vector meson $\rho(770)$, also called vector meson dominance (VMD), can be used to explain the emitted thermal electromagnetic spectrum. See the most recent review~\cite{Geurts:2022xmk} for further information.
\subsection{Jet quenching}
As discussed earlier, experimentally, due to color confinement, the direct detection of high-momentum quarks and gluons is not possible. As a result, they develop and radiate into a parton beam. At the last stage of the shower, these partons hadronize and generate final state hadrons. The final hadron transverse spectrum consists of knowledge about the changes of the parent parton created in the earliest hard scattering process. Although each hadron, on average, consists of some fraction of momentum from the parent parton. However, in order to get a more precise picture, jets are introduced. Jets are objects that are based on clusters of final-state particles built from a quark or a gluon, which are the state-of-the-art approach that could retain a larger proportion of momentum from the mother parton momentum in comparison to individual hadrons.  
\\
Jet quenching, also referred to as attenuation of jets, is a crucial experimental demonstration of parton energy loss in QCD matter and strong experimental proof for that in nucleus-nucleus collision at RHIC~\cite{PHENIX:2004vcz, STAR:2005gfr} and at LHC~\cite{ALICE:2010yje, ALICE:2011gpa} has been found. Apart from the medium-induced effects, jet observables such as jet momentum spectra~\cite{CMS:2016uxf, ATLAS:2018gwx} and dijet angular distributions~\cite{ATLAS:2010isq, CMS:2011iwn} are pretty sensitive to study the initial parton distribution of heavy nuclei. Apart from this, the nuclear modification factor plays a crucial role in studying the parton distribution function utilizing jets in smaller systems such as proton-ion collisions and is very important in HIC data interpretation~\cite{PHENIX:2001hpc, STAR:2002svs}. For some review, look at Ref.~\cite{Majumder:2010qh, Mehtar-Tani:2013pia, Blaizot:2015lma, Apolinario:2022vzg}.
\subsection{Heavy quarkonia suppression}
In the pre-equilibrium stage, strong scattering between partons results in the production of heavy quarks and anti-quarks, which eventually form the bound states known as heavy quarkonia. Charmonium $(J/\Psi)$ refers to the bound states of charm quarks $(c\bar{c})$, while bottomonium $(\Upsilon)$ refers to the bound states of beauty quarks $(b\bar{b})$~\cite{E598:1974sol,SLAC-SP-017:1974ind,E288:1977xhf}. The color charges in the QGP get screened due to the presence of the medium partons in the plasma. This process is widely known as Debye screening. At high temperatures, due to debye screening, the interaction strength between $Q\bar{Q}, $ weakens, leading to the dissociation of the bound state and suppressing its production. The suppression of heavy quarkonia is considered an important probe and acts as a thermometer for QGP and was first proposed by Matsui and Satz in 1986~\cite{Matsui:1986dk}. Experimentally, the signals are obtained through the quarkonium decaying to lepton pairs~\cite{McLerran:1984ay}, and theoretically, quarkonium spectral functions are studied through various potential models~\cite{Shuryak:2004tx, Wong:2004zr} and have been extracted through LQCD~\cite{Satz:2006kba}. 
\subsection{Anisotropic flow}
In HIC experiments, because of the non-centrality, the shape of the produced fireball deforms and looks like an almond (as seen in Fig.~\ref{HIC_non_central}). It occurs due to the pressure gradients developing in opposite directions, resulting in the preferential expansion of produced matter in the longitudinal direction. As a result, the distribution of momentum, number density, or energy with respect to the direction of the beam becomes anisotropic. One of the most significant experimental evidence supporting collective flow in HIC is azimuthal anisotropy in particle generation~\cite{Ollitrault:1992bk, Huovinen:2006jp, Heinz:2009xj, Teaney:2009qa, Bhalerao:2005mm, Mohanty:2011fp}. Fourier coefficients of the azimuthal distribution of the particle yield are given by~\cite{Soudi:2023epi}:
\vspace{-.5cm}
\begin{figure}[tbh]
\centering
\includegraphics[scale=0.5,keepaspectratio]{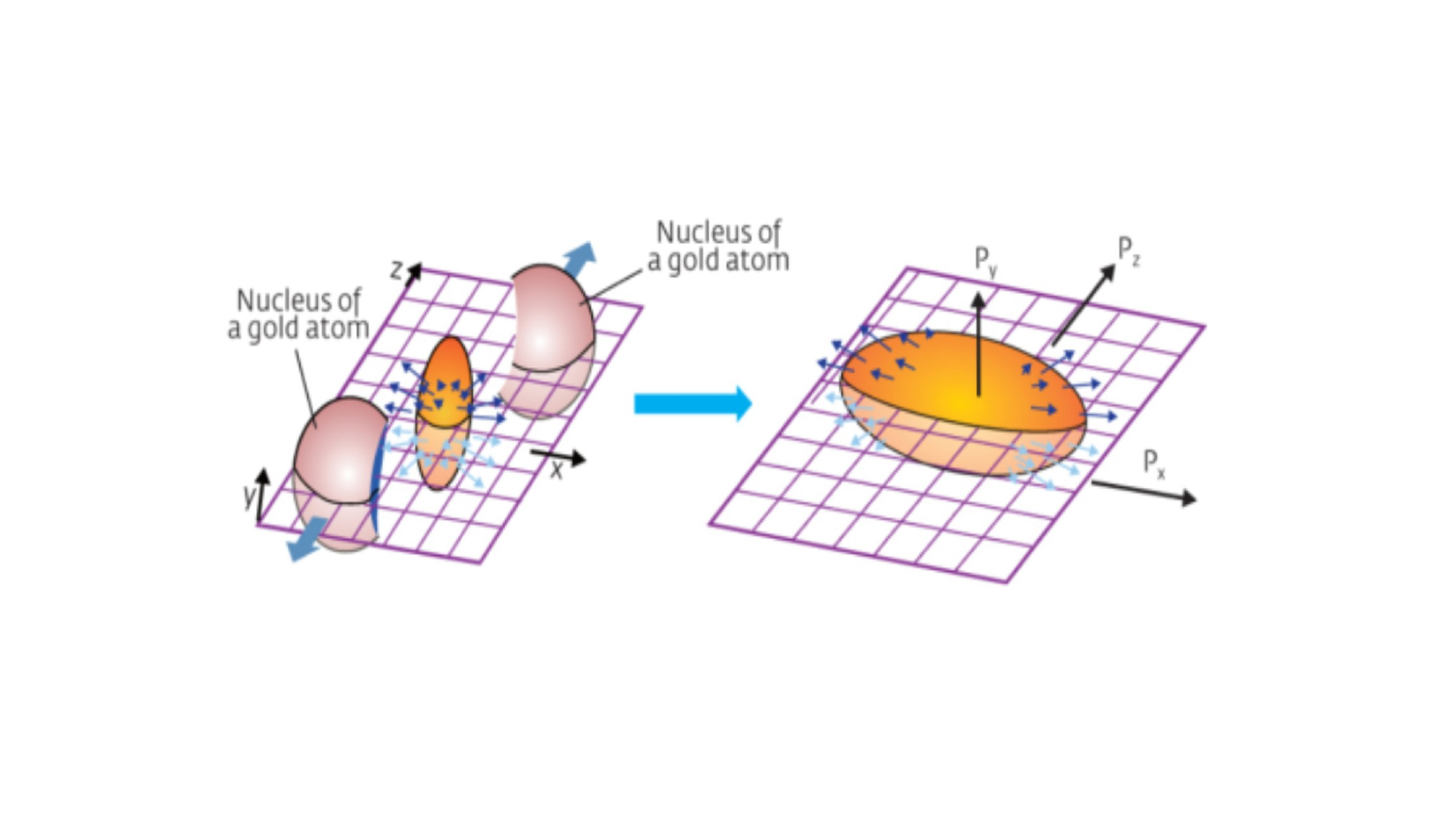}
\vspace{-1.5cm}
\caption{Schematic picture of a non-central HIC}
\label{HIC_non_central}
\end{figure}
\begin{align}
E\frac{d N}{d \phi} \propto 1+\sum_{n=1}^{\infty} 2 v_n \cos \left[n\left(\phi-\Psi_n\right)\right]
\end{align}
Here, $v_n$ are the flow coefficients for the $n^{\text {th}}$ harmonic with respect to the flow plane angle $\Psi_n$. For central collision, $v_{n} = 0$, and only radial flow exists. The direct flow and elliptic flow correspond to $v_{n} = 1$ and $v_{n} = 2$, respectively. A non-zero value of elliptic flow as measured by RHIC experiment shows early thermalization and formation of QGP~\cite{STAR:2004jwm}. The hydrodynamic behavior of QGP, where pressure gradients induce the matter to flow and disperse the momentum anisotropically~\cite{Huovinen:2001cy, ALICE:2019zfl}, can account for the large elliptic flow observation.

\subsection{Strangeness enhancement}
Strangness enhancement is one of the essential probes of QGP in HIC~\cite{Rafelski:1982pu, Rafelski:1982ii, Koch:1986ud, Letessier:1993hi}. In the initial stage of HIC, the strangeness is almost zero since the valence quarks are made up of light up and down quarks. The light partons interact with themselves to produce strange quark pairs in the later stages of HIC. These strange quark pairs become quasi-free and interact with other quark anti-quark pairs to form strange particles. Threshold energy, i.e., 2$M_s$, where $M_s$ is the mass of the strange quark, is required in the initial stage interaction to make the strange pair of quarks. Also, the gluon-gluon interaction dominates the light quark anti-quark interaction for forming the strange quark pairs. In the hadronic phase, the strange particle production is highly suppressed because of the large mass of strange quarks. Thus, the production of strange particles will provide signals for the QGP formation. Kaon production is mainly responsible for strangeness because of their lowest rest mass among strange mesons. The yield of the ratio of strange to non-strange hadrons like $K/\pi, \phi/\omega,$ and $\bar{\Lambda}/\Lambda $ acts as an experimental observable to get an idea about strangeness enhancement~\cite{Singh:1992sp}.   
\section{Methods of studying QGP}\label{QGP_methods}
In order to study QGP or, in general, to understand the strongly interacting system, the compelling framework is perturbation theory, which is very flexible in the way that it can be applied to study real-time observables as well as non-zero densities. Perturbation theory mainly relies on the weak coupling expansion procedure, which depends on expanding the functional integrals to define the physical observables in powers of running coupling. However, a number of different other methods are also available, like computational ones, which are non-perturbative. We will briefly discuss both approaches, namely perturbative and non-perturbative approaches, to study strongly interacting matter.
\subsection{Perturbative approaches}
The perturbative analysis requires the resummation of Feynmann graphs of all loops in order to obtain a valid result in terms of a strong coupling constant of $g$. Due to poor converging results of observables, additional resummations are often required. The problems are mainly connected to the soft collective excitations, which have the momentum scales as $gT, g^{2}T$. Such contribution often leads to infrared (IR) issues in the bare perturbation theory. These soft collective modes require an appropriate determination of the IR degree of freedom (DOF) followed by some effective description for them. Indeed, it is a widely recognized fact in plasma physics that colllective excitation occurs at soft, i.e., $gT$ scale. The collective effects are well organized in a mathematically consistent, up-to-date, and gauge-invariant description in the HTL effective theory~\cite{Andersen:1999fw, Andersen:1999sf, Haque:2014rua}. The idea was originally given by Brateen and Pisarski~\cite{Braaten:1989mz,Braaten:1991gm}, by Frenkel and Taylor~\cite{Frenkel:1989br,Frenkel:1991ts} and Taylor and Wong~\cite{Taylor:1990ia,Blaizot:2001nr}. In Ref.~\cite{Blaizot:2001nr}, HTL's relationship to the kinetic theory of the underlying hard modes is well examined. Apart from HTL effective theory, there are other effective approaches as well, which are based on dimensional reduction~\cite{Mogliacci:2013mca, Andersen:2012wr, Ipp:2006ij, Vuorinen:2002ue, Vuorinen:2003fs}. This technique is quite successful in describing the thermodynamic quantities. We will discuss more about the HTL resummation in the following {\bf Chapter 2} and also look at some reviews on the HTL and related topics in~\cite{Blaizot:2003tw, Kraemmer:2003gd, Andersen:2004fp}.  \\
There are primarily two comprehensive formalisms for perturbative thermal field theory: ``imaginary-time'' formalism (ITF) and RTF. Both formalisms are roughly equivalent in terms of features like IR sensitivity, and which formalism is more suitable to use relies on the observable under consideration. The RTF is applicable even in systems that are out of thermal equilibrium and quite helpful in studying quantities like particle production rates and spectral functions. In comparison, ITF is constructed on the assumption of thermal equilibrium from scratch. This formalism is very useful for the estimation of bulk thermodynamic time-independent observables. For a more recent review, please look at~\cite{Ghiglieri:2020dpq,Haque:2024gva}.  
\subsection{Nonperturbative approaches}
Since perturbative methods are useful only for the high-temperature regime, however, the temperatures obtained at RHIC and LHC experiments are slightly near to critical temperature $T_{c}$. Thus, for the lab-made QGP, near the phase transition, the non-perturbative effects must be considered for a more reliable picture. Many non-perturbative approaches in the literature are quite useful in order to deal with the high coupling scenario. LQCD is among them, which is a first principle numerical approach based on QCD lagrangian~\cite{Wilson:1974sk, Kogut:1982ds}. This theory is formulated on grids (lattices) on space-time points. When the size of the considered lattice is taken to be very large, and the spacing between the lattices is reduced to zero, then the QCD continuum is achieved. LQCD is quite successful in probing the QCD matter near $T_{c}$ where a phase transition occurs from hadronic matter to deconfined QGP phase. QCD thermodynamics and other relevant quantities at finite temperature and zero chemical potential have been calculated very accurately~\cite{Borsanyi:2011sw, Borsanyi:2012cr, Bazavov:2013dta, Bazavov:2013uja, Bernard:2004je, Bazavov:2009zn, HotQCD:2012fhj, Ding:2010ga, Aarts:2002cc, Karsch:2001uw, Aarts:2005hg} using LQCD. \\ 
Apart from LQCD, there are various other non-perturbative models based on the QCD symmetries and the inputs taken from LQCD. Some of them includes the effective models like color singlet model~\cite{Islam:2012kv}, Nambu -Jona Lasinio (NJL) model~\cite{Nambu:1961tp,Nambu:1961fr}, its Polykov loop extended version (PNJL) model~\cite{Fukushima:2003fw,Ratti:2005jh,Islam:2014sea}, chiral perturbation theory~\cite{Leutwyler:1993iq}, quasi-particle model~\cite{Peshier:1995ty}, Linear Sigma model (LSM)~\cite{Petropoulos:1998gt}, Hadron resonance gas (HRG)~\cite{Huovinen:2009yb} and  functional renormalization group method~\cite{Wetterich:1992yh}. Also, the Gribov-Zwanziger prescription is an effective non-perturbative resummation approach that improves the QCD thermodynamics~\cite{Fukushima:2013xsa}, and we will use this approach in {\bf Chapter 4} and {\bf Chapter 5}. More details about this will be provided in {\bf Chapter 2}.   
\section{Outline of the thesis}
This thesis aims to explore some of the properties of deconfined matter using the resummed perturbation theory, namely HTL effective theory, and using the non-perturbative resummation approach, namely Gribov resummation. The properties of deconfined matter that have been studied are dispersion laws for soft moving quarks ({\bf chapter} \ref{Chapter_3}), screening masses of mesons ({\bf chapter} \ref{Chapter_4}) and drag, diffusion coefficients, energy loss of $\mathrm{HQs}$, shear viscosity to entropy density ratio of QGP ({\bf chapter} \ref{Chapter_5}). The thesis has been organized as follows: After this brief introduction, in {\bf chapter} \ref{Chapter_2}, we will review the two formalisms of finite temperature field theory, namely RTF and ITF, and give a brief overview of the HTL effective theory. Later, we will discuss the QCD quantization using the Gribov procedure and will provide relevant details about the Gribov approach to QCD.    
\par
In {\bf Chapter} \ref{Chapter_3}, we studied the quark self-energy at NLO and the associated NLO dispersion relations using the HTL resummation in RTF formalism. In NLO, we have replaced all the propagators and vertices with the HTL-effective ones in the usual quark self-energy diagram. Additionally, a four-point vertex diagram also contributes to the quark NLO self-energy. We calculate the usual quark self-energy diagram and the four-point vertex diagram separately. Using those, we express the NLO quark self-energy in terms of the three- and four-point HTL-effective vertex integrals. Using the Feynman parametrization, we express the integrals containing the three- and four-point HTL effective vertex functions in terms of the solid angles. After completing the solid angle integrals, we numerically calculate the momentum integrals in the quark NLO self-energy expression and show the final results of the said observable as a function of the ratio of quark momentum and quark energy. Utilizing the NLO quark self-energy, we plot the NLO correction to dispersion laws. \par
In {\bf Chapter} \ref{Chapter_4}, we have calculated mesonic correlation lengths of various mesonic observables using the non-perturbative Gribov resummation, both for quenched QCD and $(2 + 1)$ ﬂavor QCD. This study follows the analogies with the NRQCD effective theory, a well-known theory for studying heavy quarkonia at zero temperature. The non-perturbative resummation used improves the earlier perturbative results in the low-temperature domain and is well suited to recent lattice measurements. \par
In {\bf Chapter} \ref{Chapter_5}, we investigate the momentum-dependent drag and diffusion coefficient of $\mathrm{HQs}$ moving in the QGP background. The leading order scattering amplitudes required for this purpose have been obtained using the $\mathrm{GZ}$ propagator for the mediator gluons to consider the non-perturbative estimations relevant to the phenomenologically accessible temperature regime. The $\mathrm{HQ}$ drag and diffusion coefficients so obtained have been implemented to estimate the temperature and momentum variation of the energy loss of the HQ as well as the temperature variation of the shear viscosity to entropy density ration $(\eta/s)$ of the background medium. Our results suggest a higher energy loss of the propagating HQ compared to the perturbative estimates. In contrast, the $\eta/s$ is observed to comply with the AdS/CFT estimation over a significantly more comprehensive temperature range compared to the perturbative expectation.	\par
Finally, we summarize this thesis in {\bf Chapter} \ref{Chapter_6} and give some possible extensions of the work provided in this thesis.			

\chapter{Methodology}\label{Chapter_2}
\allowdisplaybreaks
\pagestyle{fancy}
\fancyhead[LE]{\emph{\leftmark}}
\fancyhead[LO]{\emph{\rightmark}}
\rhead{\thepage} 

In this chapter, we will discuss the methodology used in chapters $3$, $4$, and $5$. We will provide a general overview of the formalism used in order to study the thermal field theory, namely RTF and ITF. After these formalisms, we will discuss the HTL effective theory in detail, which has been utilized in Chapter $3$ calculations. Later on, we will discuss the Gribov ambiguity in QCD and QCD quantization using the Gribov procedure in detail for chapters $4$ and $5$ completeness.  
\section{Thermal quantum field theory}
\begin{figure}[tbh]
\hspace{-2cm}
\includegraphics[scale=.7,keepaspectratio]{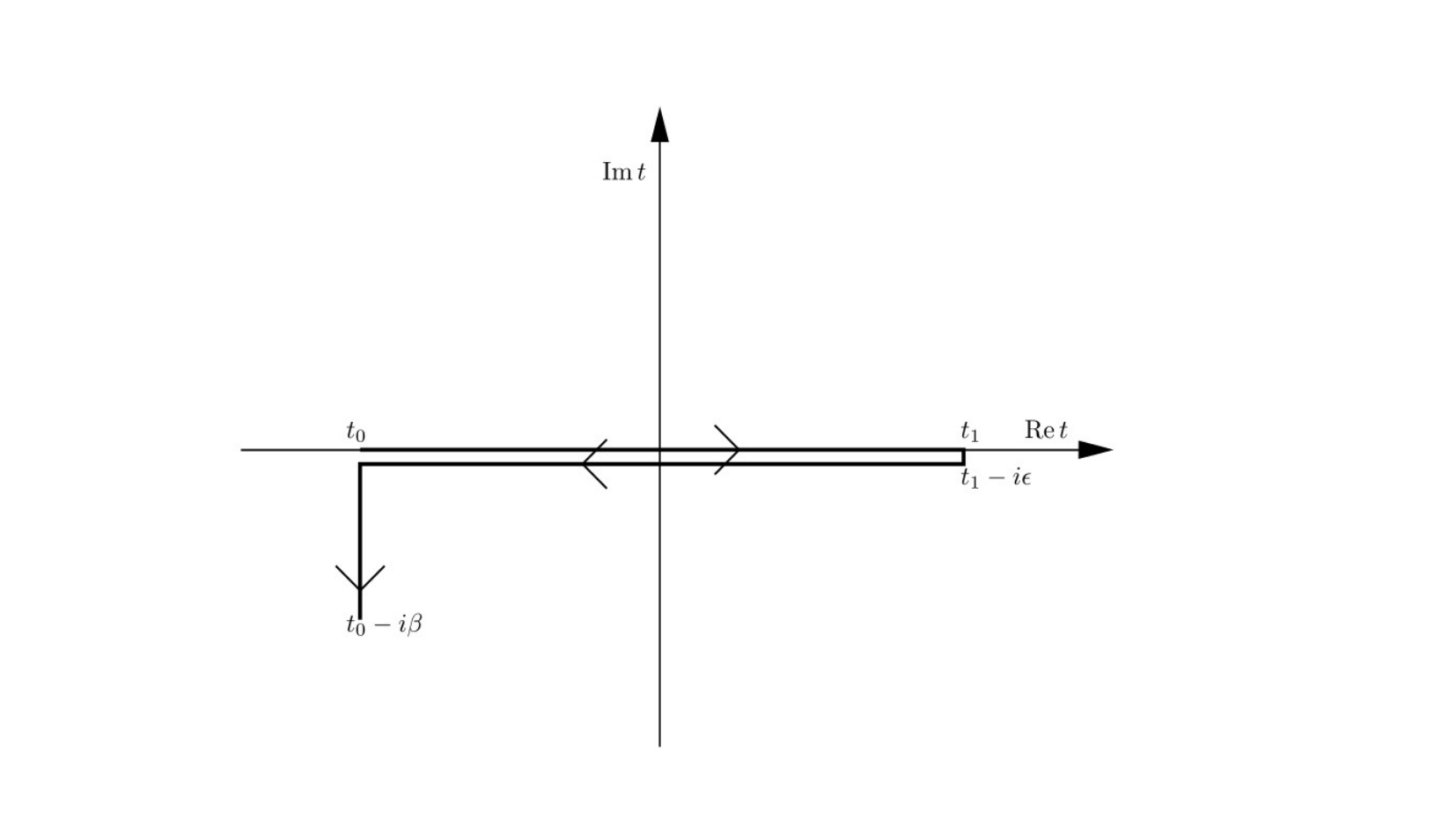}
	\caption{The SK contour on the complex time plane.}
	\label{fig_Keldysh_Contour}
\end{figure}

At $T=0$ field theory, one primarily focus on the $S$ matrix elements which are connected through the reduction formula known as Lehmann-Symanzik-Zimmermann (LSZ) reduction formula (see e.g.~\cite{Peskin:1995ev}) to vacuum expectation values of a collection of time-ordered operators $T[\mathfrak{\hat{O}}]$. By acting on the vaccum states $|\Omega\rangle$ these operators $\mathfrak{\hat{O}}$ create asymptotic states that represent the particles which are infinitely far in the future or past. Statistical field theory differs from the $T = 0 $ field theory mainly in two ways. Firstly, one needs to take into consideration the statistical fluctuations coming via the macroscopic detailing of the state of the system. Let us consider a medium which is described by the states $|i\rangle$  at initial time $t_{0}$ then the ensemble average consisting of both statistical and quantum-mechanical fluctuation is given by   
\begin{align}
\left\langle\hat{\mathfrak{O}}\left(t_0\right)\right\rangle=\operatorname{Tr} \hat{\rho}\left(t_0\right) \hat{\mathfrak{O}}\left(t_0\right), \quad \hat{\rho}\left(t_0\right) \equiv \sum_i p_i\left(t_0\right)|i\rangle\langle i|,
\end{align}
where $\hat{\rho}$ describes the density operator given in terms of probabilities. At the later time say $t_{1}$, the density operator evolves through the time-translation operator, and the expectation value becomes
$\hat{\rho}\left(t_1\right)$
\begin{align}
\left\langle\hat{\mathfrak{O}}\left(t_1\right)\right\rangle=\operatorname{Tr} \hat{\rho}\left(t_1\right) \hat{\mathfrak{O}}\left(t_1\right)=\operatorname{Tr} \mathcal{U}\left(t_1, t_0\right) \hat{\rho}\left(t_0\right) \mathcal{U}\left(t_0, t_1\right) \hat{\mathfrak{O}}\left(t_1\right) .
\label{density_evolution}
\end{align}
In terms of field bases (the field is a bosonic field here, which can be considered as a single component of gauge field $A_{\mu}$), one can write Eq.~\eqref{density_evolution} 
\begin{align}
\left\langle\hat{\mathfrak{O}}\left(t_1\right)\right\rangle=\sum_{i, j, k, l}\left\langle\Phi_i\left|\mathcal{U}\left(t_1, t_0\right)\right| \Phi_j\right\rangle \rho_{j k}\left\langle\Phi_k\left|\mathcal{U}\left(t_0, t_1\right)\right| \Phi_l\right\rangle \mathfrak{O}_{l i},
\label{field_expectation}
\end{align}
where $\rho_{j k} \equiv\left\langle\Phi_j|\hat{\rho}| \Phi_k\right\rangle$ and similar expression can be written for the operator $\mathfrak{O}_{l i}$. Now, the matrix elements of the evolution operator $\mathcal{U}\left(t_1, t_0\right)$ can be represented utilizing the path integral formulation as
\begin{align}
\left\langle\Phi_i\left|\mathcal{U}\left(t_1, t_0\right)\right| \Phi_j\right\rangle=\left\langle\Phi_i\left|e^{-i \hat{\mathcal{H}}\left(t_1-t_0\right)}\right| \Phi_j\right\rangle=\int_{\Phi_1\left(t_0\right)=\Phi_j}^{\Phi_1\left(t_1\right)=\Phi_i} \mathcal{D} \Phi_1(t) e^{i S\left(\Phi_1\right)},
\end{align}
Similarly, using the unitarity property, we get
\begin{align}
\left\langle\Phi_k\left|\mathcal{U}\left(t_0, t_1\right)\right| \Phi_l\right\rangle=\left\langle\Phi_k\left|e^{-i \hat{\mathcal{H}}\left(t_0-t_1\right)}\right| \Phi_l\right\rangle=\int_{\Phi_2\left(t_0\right)=\Phi_k}^{\Phi_2\left(t_1\right)=\Phi_l} \mathcal{D} \Phi_2(t) e^{-i S\left(\Phi_2\right)}
\end{align}
Here, $\hat{\mathcal{H}}$ is the Hamiltonian of the system. Thus Eq.~\eqref{field_expectation} becomes
\begin{align}
\left\langle\hat{\mathfrak{O}}\left(t_1\right)\right\rangle=\sum_{i, j, k, l} \int_{\Phi_1\left(t_0\right)=\Phi_j}^{\Phi_1\left(t_1\right)=\Phi_i} \mathcal{D} \Phi_1(t) \int_{\Phi_2\left(t_0\right)=\Phi_k}^{\Phi_2\left(t_1\right)=\Phi_l} \mathcal{D} \Phi_2(t) e^{i S\left(\Phi_1\right)-i S\left(\Phi_2\right)} \rho_{j k} \mathfrak{O}_{l i} .
\label{final_operator_form}
\end{align}
The indices $ijkl$ refer to the field configuration of the corresponding states $|i\rangle$, while the labelling $1$ and $2$ denote the time evolution of ket $(1)$ and bra $(2)$. The unitarity of the evolution operator accounts for the minus sign in the exponent of the action in Eq.~\eqref{final_operator_form}, which is also referred to as the ``doubling of the degree of freedom". The density operator in the grand canonical ensemble has the following structure.
\begin{align}
\hat{\rho}_{\mathrm{eq}}=\frac{1}{Z} e^{-\beta\left(\hat{\mathcal{H}}-\mu_i \hat{N}_i\right)}, \quad Z=\operatorname{Tr} e^{-\beta\left(\hat{\mathcal{H}}-\mu_i \hat{N}_i\right)},
\label{density_operator_equilibrium}
\end{align}
where the number operator and chemical potential are denoted by $\hat{N}_i$ and $\mu_i$, respectively. It can be seen that the equilibrium form of the density matrix is similar to the evolution operator with a time argument of $-i\beta$ from Eq.~\eqref{density_operator_equilibrium}. Consequently, the path integral form of the density matrix can be expressed as 
\begin{align}
\left(\rho_{\mathrm{eq}}\right)_{j k} \equiv\left\langle\Phi_j\left|\hat{\rho}_{e q}\right| \Phi_k\right\rangle=\frac{1}{Z} \int_{\Phi_E\left(t_0\right)=\Phi_k}^{\Phi_E\left(t_0-i \beta\right)= \pm \Phi_j} \mathcal{D} \Phi_E e^{-S_E\left(\Phi_E\right)},
\end{align}
where $S_E$ is the Euclidean action, $S_E=\int_0^\beta d \tau L_E$. The field at $t=t_0-i \beta$ is equivalent to $\pm \Phi_j$, with the upper sign corresponds to a periodic boundary condition for bosonic fields and the lower one referring to an anti-periodic boundary condition for fermionic fields. \\ 
In equilibrium, the density operator commutes with the Hamiltonian $\left[\hat{\rho}_{e q}, \hat{\mathcal{H}}\right]=0$, and is thus time-translation invariant. While for the operators which are separated in time, such as $\hat{\mathfrak{O}}=\hat{\mathfrak{O}}_i\left(t_1\right) \hat{\mathfrak{O}}_j\left(t_2\right)$ with $t_1<t_2$, the action of the first operator $\mathfrak{O}_i\left(t_1\right)$ on the density operator creates a non-equilibrium state which is characterized by a new density matrix $\hat{\rho}\left(t_1\right)=\hat{\rho}_{\text {eq }} \hat{\mathfrak{O}}_i\left(t_1\right)$.
The new density operator does not commute with the Hamiltonian $\hat{\mathcal{H}}$, and thus, the integral over real branches is no longer trivial. The contour consisting of two real branches and one imaginary one is known as Schwinger-Keldysh (SK) contour~\cite{Schwinger:1960qe, Keldysh:1964ud} shown in Figure~\ref{fig_Keldysh_Contour} (see~\cite{Bellac:2011kqa} for more details).
\\
Another distinction with $T=0$ QFT is that an LSZ reduction in the distant asymptotics from the space region where the interactions occur is not possible due to the random interaction that occurs in a thermal medium. Thus, the $S$ matrix elements are not interesting observables that were very useful at $T=0$ perturbation theory. In a thermal medium, operator ordering plays a crucial role, and most observables depend on the forward or backward Wightman function, describing physical correlations in the medium or describing causation in the medium through retarded and advanced functions. For bosons, the Wightman functions read
\begin{align}
& \mathcal{D}^{>}\left(t_1, t_0\right)=\left\langle\Phi\left(t_1\right) \Phi\left(t_0\right)\right\rangle, \\
& \mathcal{D}^{<}\left(t_1, t_0\right)=\left\langle\Phi\left(t_0\right) \Phi\left(t_1\right)\right\rangle,
\end{align}
Here, the spatial coordinates, color, and Lorentz indices are omitted for simplicity. The retarded and advanced correlators are given by
\begin{align}
& \mathcal{D}^R\left(t_1, t_0\right)=\Theta\left(t_1-t_0\right) \rho_B\left(t_1, t_0\right), \\
& \mathcal{D}^A\left(t_1, t_0\right)=-\Theta\left(t_0-t_1\right) \rho_B\left(t_1, t_0\right),
\end{align}
which are written in terms of the spectral function
\begin{align}
\rho_B\left(t_1, t_0\right)=\left\langle\left[\Phi\left(t_1\right), \Phi\left(t_0\right)\right]\right\rangle .
\end{align}
The corresponding expressions for the fermionic field can be written as 
\begin{align}
& \mathcal{S}^{>}\left(t_1, t_0\right)=\left\langle\Psi\left(t_1\right) \bar{\Psi}\left(t_0\right)\right\rangle, \\
& \mathcal{S}^{<}\left(t_1, t_0\right)=-\left\langle\bar{\Psi}\left(t_0\right) \Psi\left(t_1\right)\right\rangle, \\
& \rho_F\left(t_1, t_0\right)=\left\langle\left\{\Psi\left(t_1\right), \bar{\Psi}\left(t_0\right)\right\}\right\rangle, \\
& \mathcal{S}^R\left(t_1, t_0\right)=\Theta\left(t_1-t_0\right) \rho_F\left(t_1, t_0\right), \\
& \mathcal{S}^A\left(t_1, t_0\right)=-\Theta\left(t_0-t_1\right) \rho_F\left(t_1, t_0\right) .
\end{align}
The different correlation functions can be determined 
by knowing the density matrix $\rho_{B}$ and $\rho_{F}$
\begin{align}
& \rho_B\left(t_1, t_0\right)=\mathcal{D}^R\left(t_1, t_0\right)-\mathcal{D}^A\left(t_1, t_0\right)=\mathcal{D}^{>}\left(t_1, t_0\right)-\mathcal{D}^{<}\left(t_1, t_0\right), \\
& \rho_F\left(t_1, t_0\right)=\mathcal{S}^R\left(t_1, t_0\right)-\mathcal{S}^A\left(t_1, t_0\right)=\mathcal{S}^{>}\left(t_1, t_0\right)-\mathcal{S}^{<}\left(t_1, t_0\right) .
\end{align}
Thus the Wightman function measures correlations while retarded function measures causation. In thermal equilibrium, these correlation functions are related by the fluctuation-dissipation theorem known as Kubo-Martin-Schwinger (KMS) relation~\cite{Kubo:1957mj, Martin:1959jp}. Also, the Wightman functions $\mathcal{D}^{>}(t)$ and $\mathcal{D}^{<}(t)$ are strictly analytic inside the range $-\beta<\operatorname{Im}(t)<0$ and $0<\operatorname{Im}(t)<\beta$, (see~\cite{Bellac:2011kqa, Laine:2016hma} as well) respectively and can be seen through the spectral representation of Wightman functions. 
\begin{align}
\mathcal{D}^{>}\left(t_1, t_0\right)=\frac{1}{Z} \sum_{m, n} e^{-\beta E_n} e^{-i E_n\left(t_1-t_0\right)} e^{i E_m\left(t_0-t_1\right)}|\langle n|\hat{\Phi}(0)| m\rangle|^2 .
\end{align}
It is evident that the sum is convergent if the exponent in the above equation controls the convergence of the Wightman function. Consequently, for the range $-\beta<\operatorname{Im}(t)<0$, the resulting function is analytic. The cyclicity attribute of trace can be used to get the necessary KMS relation, which is
\begin{align}
& \mathcal{D}^{>}(t)=\mathcal{D}^{<}(t+i \beta), \\
& \mathcal{S}^{>}(t)=-e^{-\beta \mu} \mathcal{S}^{<}(t+i \beta),
\end{align}
where we have omitted the chemical potential of the boson case. In momentum space, the above relations take the following form,
\begin{align}
& \mathcal{D}^{>}(\omega) \equiv \int d t e^{i \omega t} \mathcal{D}^{>}(t)=e^{\beta \omega} \mathcal{D}^{<}(\omega), \\
& \mathcal{S}^{>}(\omega) \equiv \int d t e^{i \omega t} \mathcal{S}^{>}(t)=-e^{\beta(\omega-\mu)} \mathcal{S}^{<}(\omega),
\end{align}
In terms of Wightman and spectral functions, the required relations are given by
\begin{align}
n_B(\omega) \rho_B(\omega) & =\mathcal{D}^{<}(\omega),  
\label{Wightman_right}
\\
\left(1+n_B(\omega)\right) \rho_B(\omega) & =\mathcal{D}^{>}(\omega), 
\label{Wighmtman_left}
\\
n_F(\omega-\mu) \rho_F(\omega) & =-\mathcal{S}^{<}(\omega),
\end{align}
where $n_B(\omega)=\left(e^{\beta \omega}-1\right)^{-1}$ and $n_F(\omega)=\left(e^{\beta \omega}+1\right)^{-1}$ are the Bose-Einstein (BE) and Fermi-Dirac (FD) distributions, respectively.
\section{Real-time formalism}
SK contour, as mentioned earlier, is the ``traditional" basis for the well-known ``doubling of degrees of freedom". Two bases are commonly used in this formalism. One is a $1 / 2$ basis having ``1" and ``2" fields as shown in Figure~\ref{fig_Keldysh_Contour}. The other one is ``$r/a$" basis, which is more suitable because the vertices and matrix structure of the propagator becomes more straightforward in this basis compared to ``$1 / 2$" basis. Also, in this basis, the causal connection of amplitudes is more explicit.
\subsection{The $1 / 2$ basis}
The SK integral in Eq.~\eqref{final_operator_form} can be expanded perturbatively, and for that, the generating function is generally introduced by generalizing the field doubled path integral as mentioned in Eq.~\eqref{final_operator_form} as
\begin{align}
Z\left[\mathcal{J}_1, \mathcal{J}_2\right]=\int \mathcal{D} \Phi_E e^{-S_E\left(\Phi_E\right)} \int \mathcal{D} \Phi_1 \mathcal{D} \Phi_2 e^{i S\left(\Phi_1\right)-i S\left(\Phi_2\right)-\int d^4 x\left(\mathcal{J}_1(x) \Phi_1(x)-\mathcal{J}_2(x) \Phi_2(x)\right)} .
\label{Partition_function_one_two_basis}
\end{align}
For more detailed analysis for gauge fields and fermionic fields, follow~\cite{Bellac:2011kqa}. The construction of the perturbative expansion involves disentangling the action's free part from $S_{I}$, the interaction part, which is quadratic in fields.    
\begin{align}
\mathbf{\mathcal{D}}_{i j}=\left.\frac{\delta}{\delta \mathcal{J}_i} \frac{\delta}{\delta \mathcal{J}_j} Z\left[\mathcal{J}_1, \mathcal{J}_2\right]\right|_{J=0},
\label{propagator_matrix_form}
\end{align}
In the above Eq.~\eqref{propagator_matrix_form}, both the propagators and vertices are matrices, and the diagonal component of the propagator is the time- and anti-time-ordered Feynman propagators,
\begin{align}
& \mathcal{D}^F\left(t_1, t_0\right)=\Theta\left(t_1-t_0\right)\left\langle\Phi\left(t_1\right) \Phi\left(t_0\right)\right\rangle+\Theta\left(t_0-t_1\right)\left\langle\Phi\left(t_0\right) \Phi\left(t_1\right)\right\rangle, \\
& \mathcal{D}^{\bar{F}}\left(t_1, t_0\right)=\Theta\left(t_0-t_1\right)\left\langle\Phi\left(t_1\right) \Phi\left(t_0\right)\right\rangle+\Theta\left(t_1-t_0\right)\left\langle\Phi\left(t_0\right) \Phi\left(t_1\right)\right\rangle,
\end{align}
Meanwhile, the off-diagonal terms are forward and backward Wightman functions. Thus, the propagator in a $1/2$ basis becomes
\begin{align}
\mathbf{\mathcal{D}}=\left(\begin{array}{cc}
\left\langle\Phi_1 \Phi_1\right\rangle & \left\langle\Phi_1 \Phi_2\right\rangle \\
\left\langle\Phi_2 \Phi_1\right\rangle & \left\langle\Phi_2 \Phi_2\right\rangle
\end{array}\right)=\left(\begin{array}{cc}
\mathcal{D}^F & \mathcal{D}^{<} \\
\mathcal{D}^{>} & \mathcal{D}^{\bar{F}}
\end{array}\right) .
\end{align}
The time- and anti-time-ordered Feynman propagators can be written in terms of spectral function by using the definition of (anti)-time-ordering, Eqs.~\eqref{Wighmtman_left},~\eqref{Wightman_right} and the relations of Wightman functions, retarded correlator with the spectral function as
\begin{align}
& \mathcal{D}^F(\omega, k)=\frac{1}{2}\left[\mathcal{D}_R(\omega, k)+\mathcal{D}_A(\omega, k)\right]+\left(\frac{1}{2}+n_{\mathrm{B}}(\omega)\right) \rho(\omega, k), \\
& \mathcal{D}^{\bar{F}}(\omega, k)=-\frac{1}{2}\left[\mathcal{D}_R(\omega, k)+\mathcal{D}_A(\omega, k)\right]+\left(\frac{1}{2}+n_{\mathrm{B}}(\omega)\right) \rho(\omega, k) .
\end{align}
\begin{figure}[tbh]
\hspace{3cm}
\includegraphics[scale=1.7,keepaspectratio]{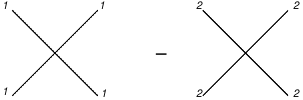}
	\caption{Two vertices within the $1/2$ basis. Minus sign in the vertex with type 2 fields comes because of different signs of action in Eq.~\eqref{Partition_function_one_two_basis}.}
\label{fig_1_2_basis}
\end{figure}
The vertices in this basis have the usual zero temperature field theory apart from the modification that the fields come with indices $1$ and $2$, and vertices with index $2$ come with an opposite sign, as seen in Figure~\ref{fig_1_2_basis}.
\subsection{The $r / a$ basis}
\begin{figure}[tbh]
\hspace{3cm}
\includegraphics[scale=1.7,keepaspectratio]{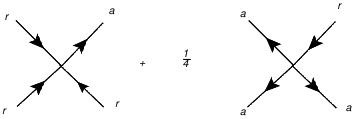}
	\caption{Graphical description of the vertices appearing in the example mentioned above in the r /a basis.}
\label{fig_r_a_basis}
\end{figure}

In this $r/a$  basis, instead of using $1$ and $2$ fields, one can define fields with the combinations of $1$ and $2$ fields as~\cite{Keldysh:1964ud,Chou:1984es}
\begin{align}
\Phi_r \equiv \frac{1}{2}\left(\Phi_1+\Phi_2\right) \quad \Phi_a \equiv \Phi_1-\Phi_2,
\end{align}
In this $r/a$ basis, the propagator matrix becomes
\begin{align}
\mathbf{\mathcal{D}}=\left(\begin{array}{cc}
\left\langle\Phi_r \Phi_r\right\rangle & \left\langle\Phi_r \Phi_a\right\rangle \\
\left\langle\Phi_a \Phi_r\right\rangle & \left\langle\Phi_a \Phi_a\right\rangle
\end{array}\right)=\left(\begin{array}{cc}
\mathcal{D}^{r r} & \mathcal{D}^R \\
\mathcal{D}^A & 0
\end{array}\right) \quad \text { in } r / a \text { basis, }
\end{align}
and the symmetric $rr$-propagator reads as
\begin{align}
\mathcal{D}^{r r}=\frac{1}{2}\left(\mathcal{D}^{>}+\mathcal{D}^{<}\right)
\end{align}
The ``22" component, which consists of twice $a$ fields, is zero to all orders, in and out of equilibrium because of the $\Theta$-functions inclusion in the definitions of different correlation functions. The vertices within this basis have an odd number of $a$ indices which is because in the $1/2$ basis, the corresponding vertices come with opposite signs. For the interacting part $S_I$ of the action, we get
\begin{align}
S_I\left(\Phi_1\right)-S_I\left(\Phi_2\right)=S_I\left(\Phi_r+\frac{1}{2} \Phi_a\right)-S_I\left(\Phi_r-\frac{1}{2} \Phi_a\right) .
\end{align}
If there is an even number of $a$ fields or there is no occurrence of $\Phi_{a}$, in that case, the two contributions in the interacting part of action cancel each other exactly. These results will be utilized in chapter~\ref{Chapter_3} in the calculation of NLO quark self-energy where real-time formalism has been used. Let us consider an example with $\Phi^{4}$ interaction, the full action consisting of $1$ and $2$ field is proportional to
\begin{align}
S_I\left(\Phi_1\right)-S_I\left(\Phi_2\right) \propto \frac{1}{4 !}\left(\Phi_1^4-\Phi_2^4\right)=\frac{1}{2^2} \frac{1}{3 !} \Phi_a^3 \Phi_r+\frac{1}{3 !} \Phi_r^3 \Phi_a .
\end{align}
The normalization factor of $\Phi_{a}$ and $\Phi_{r}$ are chosen in such a way that the correct symmetry factor will be reproduced. The symmetric function measures the correlation between two fields, which is either due to quantum fluctuation or statistical fluctuation. The diagrammatic picture of the propagator in $r/a$ basis is shown in Figure~\ref{fig_r_a_basis}.
\vspace{-1.5cm}
\section{Imaginary time formalism}
In this formalism, Green's functions and, thus, the fields become periodic or anti-periodic functions of imaginary time direction. For a generic $n$-point Green's function $\mathrm{\mathcal{G}}$ for a bosonic field, one gets
\begin{align}
\mathrm{\mathcal{G}}_{\text {bosonic }}\left(t_i\right)=\mathrm{\mathcal{G}}_{\text {bosonic }}\left(t_i-i \beta\right) \quad \text { (bosons), }
\end{align}
For the fermionic field, due to the anti-commuting nature of the field, the relation becomes 
\begin{align}
\mathrm{\mathcal{G}}_{\text {fermionic }}\left(t_i\right)=-\mathrm{\mathcal{G}}_{\text {fermionic }}\left(t_i-i \beta\right) \quad \text { (fermions) },
\end{align}
Therefore, in an imaginary-time direction, Green's function for a fermionic field  turns into an anti-periodic function. All of the bosonic (fermionic) fields are uniquely defined in the time range $0 \leq \tau \leq \beta$ in the imaginary time $\tau=-it$. The other values of $\tau$ can be obtained by applying the periodicity (anti-periodicity) of the fields. The fields themselves have to be (anti-)periodic functions of imaginary time because of the (anti-)periodicity of Green's functions. Consequently, the Fourier integral related to the time direction becomes a discrete Fourier sum following a Fourier decomposition of fields. The permitted frequencies conjugate to the imaginary-time direction for bosonic fields are $\mathcal{P}_0=\omega_n=2 \pi n T$, where $n \in \mathbb{Z}$. The zero-component of the Euclidean (imaginary-time) four-momentum, $\mathcal{P}=\left(\omega_n, \mathbf{p}\right)$, is denoted by $\mathcal{P}_0=-i p_0$. In contrast, the permitted frequencies for fermionic fields are $\mathcal{P}_0=\omega_n=(2 n+1) \pi T$, where $n \in \mathbb{Z}$. The underlined frequencies are known as Matsubara frequencies, and the mode expansion becomes
\begin{align}
\Phi(\tau, \mathbf{x})=\SumInt_p \Phi\left(\omega_n, \mathbf{p}\right) e^{-i\left(\omega_n \tau-\mathbf{p} \cdot \mathbf{x}\right)} \quad \text { (bosonic field), }
\end{align}
where
\begin{align}
\SumInt_\mathcal{P} \equiv T \sum_{\mathcal{P}_0=2 \pi n T} \int \frac{d^3 p}{(2 \pi)^3},
\end{align}
and
\begin{align}
\Phi(\tau, \mathbf{x})=\SumInt_{\{P\}} \Phi\left(\omega_n, \mathbf{p}\right) e^{-i\left(\omega_n \tau-\mathbf{p} \cdot \mathbf{x}\right)} \quad \text { (fermionic field) },
\end{align}
where
\begin{align}
\SumInt_{\{\mathcal{P}\}} \equiv T \sum_{\mathcal{P}_0=(2 n+1) \pi T} \int \frac{d^3 p}{(2 \pi)^3} .
\end{align}
The so-called Euclidean action $S_{\mathrm{E}}[\Phi]=\int_0^\beta d \tau \int d^3 x \mathcal{L}_{\mathrm{E}}[\Phi]$ is used in this formalism to express the path integrals that define different physical quantities, where $\mathcal{L}_{\mathrm{E}}$ reads as $\mathcal{L}_{\mathrm{E}}=-\mathcal{L}_{\mathrm{M}}(t \rightarrow-i \tau)$. 
The contribution to the partition function, for a bosonic field, for instance, takes the form (see~\cite{Laine:2016hma} for the cases of fermionic and gauge fields)
\begin{align}
Z=\int_{\Phi(\tau=0)}^{\Phi(\tau=\beta)} \mathcal{D} \Phi e^{-S_E[\Phi]} .
\end{align}
Also, this formalism has an advantage over RTF in the sense that the Euclidean integrals have naturally more convergence properties compared to Minkowski ones. Thus Euclidean integrals can be evaluated using the lattice techniques. 
\subsection{Imaginary-time Feynman rules}
In ITF, it is convenient to switch from Minkowski metric to Euclidean one, and for that, one can replace
$g_{\mu v} \rightarrow \delta_{\mu v}$. The anti-commutation properties of gamma matrices become 
$\left\{\gamma_\mu^{\mathrm{E}}, \gamma_v^{\mathrm{E}}\right\}=-2 \delta_{\mu \nu}$ with $\gamma_0^{\mathrm{E}} \equiv i \gamma_0$. 
Here are the 'E' labels for Euclidean space. Thus, the free quark propagator in ITF takes the following form 
\begin{align}
\mathcal{S}_0^{i j}=-\delta^{i j} \frac{\slashed{\mathcal{P}}-m}{\omega_n^2+\mathbf{p}^2+m^2},
\end{align}
with $i$ and $j$ are color indices in fundamental representation. The free gluon propagator in a general covariant gauge reads as
\begin{align}
\left(\mathcal{G}_0\right)_{\mu \nu}^{a b}=\frac{\delta^{a b}}{\mathcal{P}^2}\left[\delta_{\mu \nu}-(1-\xi) \frac{\mathcal{P}_\mu \mathcal{P}_\nu}{\mathcal{P}^2}\right],
\end{align}
with $\mathcal{P}^2=\omega_n^2+\mathbf{p}^2$, and $a$ and $b$ adjoint color indices.
Apart from the changes, i.e., transforming to imaginary time and using the Euclidean-space gamma matrices, the QCD vertex functions remain the same in the imaginary time formalism (see appendix~\cite{Ipp:2003qt} for details). 
\subsection{One-loop gluon polarization tensor}
\begin{figure}[tbh]
\includegraphics[scale=1.3,keepaspectratio]{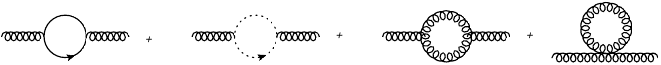}
	\caption{The 1-loop gluon polarization graphs. In these graphs, wavy lines represent gluons, solid lines with arrows denotes quarks and anti-quarks, and dotted lines with arrows denotes the ghost field.}
\label{fig_gluon_self_energy}
\end{figure}

In this subsection, we will discuss the application of ITF for the 1-loop correction to the gluon propagator, also called as gluon polarization tensor $\Pi_{\mu\nu}$. In the Feynman gauge $\zeta = 1$, the inverse of the gluon propagator is given by
\begin{align}
(\mathcal{G}^{-1})^{ab}{_{\mu\nu}} = \mathcal{P}^{2}\delta^{ab}\delta_{\mu\nu} + \Pi^{ab}{_{\mu\nu}}(\mathcal{P}),
\end{align}
The gluon polarization tensor $\Pi^{ab}{_{\mu\nu}}(\mathcal{P})$ can be calculated via diagrams shown in Figure~\ref{fig_gluon_self_energy} with the appropriate Feynman rules. The full computation is shown in review~\cite{Laine:2016hma, Mustafa:2022got}, and we will quote the final result for the gluon polarization tensor in the infrared limit, i.e., setting first $p_{0} = 0$ and then letting $p \rightarrow 0$.  
\begin{align}
\Pi_{\mu \nu}^{a b}\left(p_0=0, p \rightarrow 0\right)=4 g^2 \delta^{a b} n_\mu n_\nu\left(C_A \mathcal{I}_1^0-2 T_F N_f \widetilde{\mathcal{I}}_1^0\right),
\end{align}
Here, $n_{\mu} = \delta_{\mu 0}$ refers the rest frame of the heat bath, $T_{F} = 1/2$ is the Dynkin index of the generator in the fundamental representation, and the integral $\mathcal{I}_1^0$,  $\widetilde{\mathcal{I}}_1^0$ are defined as 
\begin{align}
\mathcal{I}_1^0 \equiv \SumInt_{\mathcal{Q}} = \frac{1}{\mathcal{Q}^{2}}, \quad \quad \widetilde{\mathcal{I}}_1^0  \equiv \SumInt_{\{\mathcal{Q}\}} = \frac{1}{\mathcal{Q}^{2}},
\end{align}
and the final result for the resummed (full) gluon propagator reads as~\cite{Laine:2016hma} 
\begin{align}
\left\langle\tilde{\mathcal{A}}_\mu^a(\mathcal{K}) \tilde{\mathcal{A}}_\nu^b(\mathcal{Q})\right\rangle \stackrel{\mathcal{K} \approx 0}{\approx} \quad \frac{\delta^{a b} \delta_{\mu \nu} \delta(\mathcal{K}+\mathcal{Q})}{\mathcal{K}^2+\delta_{\mu 0} \delta_{\nu 0} m_{\mathrm{E}}^2},
\end{align}
where
\begin{align}
m_D^2=\frac{g^2}{3 N_f} \sum_f\left\{\left(C_A+T_F N_f\right) T^2+\frac{3 \mu_f^2}{\pi^2} T_F N_f\right\} 
\end{align}
$m_{D}$ is a debye mass parameter whose existence corresponds to the fact that the color-electric field $\mathcal{A}_{0}$ gets screened in a thermal plasma, i.e., zeroth Matsubara modes of temporal (electrostatic) gluon field acquires a thermal mass at one-loop order. The definition of a $m_{D}$ parameter becomes ambiguous at higher orders. One way out is to define debye mass as a ``matching coefficient" in the context of effective field theory. This has been done indeed in the Ref.~\cite{Ghisoiu:2015uza}. However, if one wants to define $m_{D}$ as a physical excitation of the system, the result becomes nonperturbative at NLO~\cite{Rebhan:1993az}, and a lattice approach is needed for a appropriate definition~\cite{Arnold:1995bh}. 
\\ 
On the other hand, the magnetostatic fields ($p_0=0$ component of $\mathcal{A}_i$) stay unscreened at this order $gT$ and obtain a nonperturbative screening mass of order $g^{2}T$. It can be understood through the fact connected with the gauge transformation properties of the fields. On breaking the Lorentz invariance due to the presence of a thermal medium, the four dimensional gauge invariance is broken to a three-dimensional one. In this process, the electrostatic field becomes an adjoint scalar field by obtaining a nonzero thermal mass. At the same time, the magnetostatic fields continue to transform as three-dimensional gauge fields and, therefore, must remains massless to all orders in perturbation theory.
\vspace{-1cm}  
\section{High temperature limit}
Now we will discuss the implications of different energy scales present in thermal QCD, i.e., the scale $\pi T$, $g T$ associated with the nonzero Matsubara mode of different fields, and magnetic scale $g^{2}T$ related with the screening of static gluons, i.e., $n=0$ Matsubara modes of these fields. 
\subsection{Breakdown of loop expansion}
Given that ITF is frequently used to comprehend the bulk behaviour of thermodynamic quantities in the $\pi T \gtrsim \mu$ regime, or at high temperatures. It turns out that only a few orders in perturbative expansion have well-defined naive perturbative expansions of physical quantities. A precise way of checking at which particular order of perturbative expansion the terms become non-analytic in $\alpha_{s}$ is to consider the contribution of pressure coming from the non-interacting static gluons. The pressure contribution takes the following form 
\begin{align}
{P}_{\text {gluons }} \sim \int d^3 p \,\, p \, n_B\left(E_p\right)
\end{align}
Here, $n_B$ denotes the BE distribution function, and $E_p$ corresponds to the dispersion relation of the (electrostatic or magnetostatic) gluons. Through the inspection of the hard, soft, and ultrasoft scale, the following pattern emerges:
\begin{align}
& P_{\text {gluons }}^{p \sim \pi T} \sim T^4 n_B(\pi T) \sim T^4+\mathcal{O}\left(g^2\right), \\
& P_{\text {gluons }}^{p \sim g T} \sim(g T)^4 n_B(g T) \sim g^3 T^4+\mathcal{O}\left(g^4\right), \\
& P_{\text {gluons }}^{p \sim g^2 T} \sim\left(g^2 T\right)^4 n_B\left(g^2 T\right) \sim g^6 T^4, 
\label{pressure_magnetostatic}
\end{align}
where it has been considered that $n_B(E) \sim T / E$ if $E \ll T$. The expansion parameters of the three terms, which comprise three distinct scales, are $g^2 n_B(\pi T) \sim g^2, g^2 n_B(g T) \sim g$, and $g^2 n_B\left(g^2 T\right) \sim 1$. These observations suggest that the contribution of magnetostatic gluons to the pressure is essentially nonperturbative. For this reason, $\mathcal{O}\left(g^n\right)$ is absent from Eq.~\eqref{pressure_magnetostatic}. This is the well-known Linde problem~\cite{Linde:1980ts} in the literature, and it illustrates the full breakdown of loop expansion at the order $g^{2}T$. 
\\ 
It should be emphasised that nonperturbative contribution does not always enter the weak coupling expansion of a physical quantity in the same order; this can vary depending on the quantity in question. For going to higher loop order perturbative calculation, some resummation is a must to cure the unphysical IR divergences coming from the problematic field modes. Among these resummations techniques~\cite{Kajantie:1997tt, Braaten:1995cm, Braaten:1995jr, Andersen:1999fw, Andersen:1999sf, Andersen:1999va, Blaizot:1999ip, Blaizot:1999ap, Blaizot:2000fc, Blaizot:2001vr}, dimensionally reduced effective theories, which takes usage of the scale hierarchies present in the system, and Hard Thermal Loop perturbation theory (HTLpt), which uses the HTL effective action into consideration are most important ones.  
\subsection{Dimensional reduction}
This approach is based on the finding that in the weak coupling limit, a scale hierarchy exists between the three energy scales, which contributes to the QCD bulk thermodynamics. If $g \ll 1$, then 
\begin{align}
m_{\text {mag }} \sim g^2 T \ll m_{\text {elec }} \sim g T \ll m_{\text {hard }} \sim \pi T,
\end{align}
Here, $m_{\text {mag }}, \, m_{\text {elec }}$ and $m_{\text {hard }}$ corresponds to magnetostatic, electrostatic screening and nonzero Matsubara frequency. Electrostatic screening at leading order can be found out through the IR limit of one-loop self-energy of $A_{0}$ field, whereas magnetostatic screening appears nonperturbatively. Out of these three scales, two are associated with the $n=0$ field mode. Therefore, one of them can be integrated out from the system, the largest one, i.e., $m_{\text {hard }}$ for the construction of three-dimensional effective theory. Such an effective description holds true for the long-range static field modes and in the limit of high temperature. For the starting works on dimensionally reduced effective field theories, see the Ref.~\cite{Kajantie:1997tt, Ginsparg:1980ef, Appelquist:1981vg, Kajantie:1995dw}.
\\
The most general Lagrangian can be constructed for effective description by respecting the three-dimensional gauge invariance and other underlying symmetry of the theory, ordering the operators in terms of their dimensionality, and truncating the result at a particular order. One obtains through this process the Electrostatic QCD (EQCD) lagrangian as~\cite{Braaten:1995cm, Braaten:1995jr} 
\begin{align}
\mathcal{L}_{\mathrm{E}} = \, &\frac{1}{2} \operatorname{Tr} F_{i j}^2+\operatorname{Tr}\left[D_i, \mathcal{A}_0\right]^2+m_{\mathrm{E}}^2 \operatorname{Tr} \mathcal{A}_0^2+\lambda_{\mathrm{E}}^{(1)}\left(\operatorname{Tr} \mathcal{A}_0^2\right)^2+\lambda_{\mathrm{E}}^{(2)} \operatorname{Tr} \mathcal{A}_0^4 \nn
+ & i \lambda_{\mathrm{E}}^{(3)} \operatorname{Tr} \mathcal{A}_0^3 +\cdots,
\label{EQCD_Lagrangian_Form}
\end{align}
Here the fields $\mathcal{A}_i \equiv \mathcal{A}_i^a T^a, \mathcal{A}_0 \equiv \mathcal{A}_0^a T^a$ are now in three dimensions and field strength tensor $F_{ij}$ and covariant derivative are denoted as
\begin{align}
F_{i j}^a & =\partial_i \mathcal{A}_j^a-\partial_j \mathcal{A}_i^a+g_{\mathrm{E}} f^{a b c} \mathcal{A}_i^b \mathcal{A}_j^c, \\
D_i & =\partial_i-i g_{\mathrm{E}} \mathcal{A}_i,
\end{align}
If one integrates out the temporal fields as well from the Eq.~\eqref{EQCD_Lagrangian_Form}, then one obtains the effective theory known as Magnetostatic QCD (MQCD)
\begin{align}
\mathcal{L}_{\mathrm{M}}=\frac{1}{2} \operatorname{Tr} F_{i j}^2+\cdots,
\end{align}
where we have
\begin{align}
F_{i j}^a=\partial_i \mathcal{A}_j^a-\partial_j \mathcal{A}_i^a+g_{\mathrm{M}} f^{a b c} \mathcal{A}_i^b \mathcal{A}_j^c .
\end{align}
For MQCD, the only dimensionful scale appearing in the theory is $g_M^2=g^2 T+\mathcal{O}\left(g^3\right)$. 
\\
The matching of a set of physical quantities—generally, different Green's functions in EQCD and MQCD to the full theory—determines the parameters of the effective theories. The idea behind this matching is that these effective theories reproduce the original long-range physics. The primary feature of these computations is that they may be carried out in a rigorous loop expansion in the full theory, that is, without requiring any resummation and by controlling the IR and UV divergences with this method. The EQCD and MQCD parameters namely $g_{E}^{2}, m_{\mathrm{E}}^2, \lambda_{\mathrm{E}}^{(1)}, \lambda_{\mathrm{E}}^{(2)} \lambda_{\mathrm{E}}^{(3)}$ and $g_{M}^{2}$ have been obtained to high order in perturbation theory, for details see Ref.~\cite{Ghisoiu:2015uza, Kajantie:1997tt, Laine:2005ai, Hart:2000ha} 
\section{HTL resummation}
Another approach based on HTL effective theory is used to perform a high-temperature resummation. This description contains resummed gluon, quark propagators as well as resummed vertices, which are required in order to preserve the gauge invariance. All these can be collectively described via a efficient HTL effective action, which we will describe further.
\subsection{Fermionic HTL in r/a formalism}
\begin{figure}[tbh]
\hspace{2cm}
\includegraphics[scale=1.5,keepaspectratio]{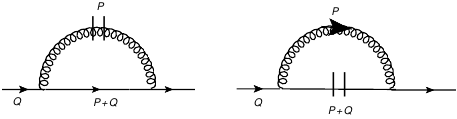}
	\caption{The two different r/a assignments for the retarded fermion self-energy. Here, parallel lines show a symmetric propagator.}
\label{fig_fermionic_HTL}
\end{figure}

For the quark-gluon vertex, only $rra$ and $aaa$ assignments are possible. However, the $aaa$ vertex can not contribute since there is neither an $aa$ propagator nor an $rrr$ vertex. Thus, there are two assignments for the retarded self-energy $\Sigma^R(\mathcal{P})$ in $r/a$ basis shown in Figure~\ref{fig_fermionic_HTL}. Utilizing the Feynman rules of $r/a$ basis yields
\begin{align}
-i \Sigma^R(\mathcal{Q})=(-i g)^2 C_F \int \frac{d^4 \mathcal{P}}{(2 \pi)^4} \gamma^\mu\left[\mathcal{S}^R(\mathcal{P}+\mathcal{Q}) \mathcal{G}_{\mu \nu}^{r r}(\mathcal{P})+\mathcal{S}_{r r}(\mathcal{P}+\mathcal{Q}) \mathcal{G}_{\mu \nu}^A(\mathcal{P})\right] \gamma^\nu,
\end{align} 
The HTL amplitudes are gauge invariant quantities, and using the field theory properties this has been indeed shown in Ref.~\cite{Kobes:1990dc}. The fermionic propagator in $r/a$ basis reads as 
\begin{align}
\mathcal{S}_{R, A}(\mathcal{P})=\frac{i \not \mathcal{P}}{\mathcal{P}^2 \mp i \epsilon p^0}, \quad \mathcal{S}_{r r}(\mathcal{P})=-\not \mathcal{P}\left(\frac{1}{2}-n_{\mathrm{F}}\left(\left|p^0\right|\right)\right) 2 \pi \delta\left(\mathcal{P}^2\right) .
\end{align}
while the gluon propagator in the Feynman gauge reads as
\begin{align}
\mathcal{G}_{R, A}^{\mu v}(\mathcal{Q})=\frac{-i g^{\mu \nu}}{\mathcal{Q}^2 \mp i \epsilon q^0}, \quad \mathcal{G}_{r r}^{\mu \nu}(\mathcal{Q})=g^{\mu v}\left(\frac{1}{2}+n_{\mathrm{B}}\left(\left|q^0\right|\right)\right) 2 \pi \delta\left(\mathcal{Q}^2\right)
\end{align}
Here the matrices notation is $\{\gamma^{\mu}, \, \gamma^{\nu}\} = -2 g^{\mu \nu}$
As we are interested in obtaining the HTL contribution, we can make use of the theory's basic notion, which requires obtaining the leading term of the soft external quark interacting with a hard loop separately. By doing an expansion for $\mathcal{Q} \sim gT \ll \mathcal{P} < T$, taking the leading term and doing the $p_{0}$ and $p$ integration, we get
\begin{align}
\Sigma^R(\mathcal{Q})=\frac{m_{\infty}^2}{2} \int \frac{d \Omega_v}{4 \pi} \frac{\slashed v}{v \cdot \mathcal{Q}-i \epsilon},
\label{fermionic_HTL_form}
\end{align}
where $v \equiv \mathcal{P}/p_{0} = (1, \vec{p}/p_{0})$ and $m_{\infty}^2 \equiv g^2 C_F T^2 / 4$ is the asymptotic mass of the quarks. Gauge invariance property suggests that the eikonal propagator which is given by $\frac{\slashed v}{v \cdot \mathcal{Q}}$ is just the first term in the expansion of $\frac{\slashed v}{v \cdot \mathcal{D}}$, with $D$ being the covariant derivative which indeed comes from the explicit computation of higher-point functions. It has been shown in Refs.~\cite{Braaten:1991gm, Taylor:1990ia}  that HTL amplitudes with two external quark lines can be reproduced by adding an extra, effective term to the QCD lagrangian, and this extra term in Minkowskian metric reads as
\begin{align}
\delta \mathcal{L}_f=i \frac{m_{\infty}^2}{2} \bar{\psi} \int \frac{d \Omega_v}{4 \pi} \frac{\slashed v}{v \cdot D} \psi,
\end{align}
All fermionic HTLs with two external soft quark lines and an arbitrary number of soft gluons are produced by this Lagrangian. These retarded amplitudes are proportional to $g^{2}T^{2}$ and are all gauge invariant. A similar procedure with some extra intricacies relating to gauge fixing and the two-point gluon HTL amplitude requires the next order in the expansion for $\mathcal{Q} \ll \mathcal{P}$, leads to the retarded gluonic HTL, which reads out as  
\begin{align}
\Pi_R^{\mu v}(\mathcal{Q})=m_D^2 \int \frac{d \Omega_v}{4 \pi}\left(\delta_0^\mu \delta_0^v+v^\mu v^v \frac{q^0}{v \cdot \mathcal{Q}-i \epsilon}\right),
\label{gluonic_HTL_form}
\end{align}
where $m_D$ is the leading-order Debye mass, which reads $m_D^2=\left(N_c+T_F n_f\right) g^2 T^2 / 3$, with $T_F=1 / 2$. Eq.~\eqref{gluonic_HTL_form} has a similar structure as Eq.~\eqref{fermionic_HTL_form} with the difference in the different numerator structure for the eikonal propagator of gluonic source and the presence of extra term for temporal gluons. The corresponding Lagrangian which generates all $n \geq 2$-point gluonic functions reads as~\cite{Braaten:1991gm}
\begin{align}
\delta \mathcal{L}_g=\frac{m_D^2}{2} \operatorname{Tr} \int \frac{d \Omega_v}{4 \pi} F^{\mu \alpha} \frac{v_\alpha v_\beta}{(v \cdot D)^2} F^\beta{ }_\mu .
\end{align}
\subsection{Collective modes}
For the two-point gluonic HTL, $\Pi^{\mu v}(\mathcal{Q})$ is transverse to $\mathcal{Q}$. However, there also exist two independent functions, $\Pi_L$ and $\Pi_T$, which are respectively longitudinal and transverse with regard to three momenta $q$. They are defined as $\Pi_L(\mathcal{Q})=\left(\mathcal{Q}^2 / q^2\right) \Pi^{00}(\mathcal{Q})$ and $\Pi_T(\mathcal{Q})=\left(\delta^{i j}-\hat{q}^i \hat{q}^i\right) \Pi^{i j}(\mathcal{Q}) / 2$. After doing the angular integrations in Eq.~\eqref{gluonic_HTL_form}, one obtains,  
\begin{align}
\Pi_R^{00}(\mathcal{Q})=m_D^2\left(1-\frac{q^0}{2 q} \ln \frac{q^0+q+i \epsilon}{q^0-q+i \epsilon}\right), \quad \Pi_T^R(\mathcal{Q})=\frac{m_D^2}{2}-\frac{\mathcal{Q}^2}{2 q^2} \Pi_R^{00}(\mathcal{Q}) .
\end{align}
Similarly, the resummed retarded gluon propagator in coulomb gauge is defined as $\mathcal{G}_R^{00}(\mathcal{Q}) \equiv \mathcal{G}_L^R(\mathcal{Q})$ and $\mathcal{G}_R^{i j}(\mathcal{Q}) \equiv\left(\delta^{i j}-\hat{q}^i \hat{q}^j\right) \mathcal{G}_T^R(\mathcal{Q})$, where
\begin{align}
\mathcal{G}_R^{00}(\mathcal{Q}) & =\frac{i}{q^2+m_D^2\left(1-\frac{q^0}{2 q} \ln \frac{q^0+q+i \epsilon}{q^0-q+i \epsilon}\right)}, 
\label{retarded_longiudinal_HTL}
\\
\mathcal{G}_R^{i j}(\mathcal{Q}) & =\left.\frac{-i\left(\delta^{i j}-\hat{q}^i \hat{q}^j\right)}{\mathcal{Q}^2+\frac{m_D^2}{2}\left(\frac{q_0^2}{q^2}-\left(\frac{q_0^2}{q^2}-1\right) \frac{q^0}{2 q} \ln \frac{q^0+q}{q^0-q}\right)}\right|_{q^0=q^0+i \epsilon} .
\label{retarded_transverse_HTL}
\end{align}
For the time-like region, i.e., $q_{0}> q$, the longitudinal and transverse functions exhibits plasmon poles, i.e., collective effects at order $gT$. In the limit of three-momentum approaching zero, the longitudinal and transverse modes merge, and the plasmon dispersion relation reduces to plasma-frequency $\omega_L(q=0)=\omega_T(q=0)= \pm m_D / \sqrt{3}$. On the other hand, for the asymptotic large momenta $q \gg m_D$, the longitudinal mode approaches the light cone with exponentially vanishing residue~\cite{Pisarski:1989cs}. In contrast, the transverse modes persists with unitary residue and asymptotic mass $M_{\infty}$, i.e. $\omega_T\left(q \gg m_D\right)=\sqrt{q^2+M_{\infty}^2}$, $M_{\infty}=m_D / \sqrt{2}$~\cite{Kalashnikov:1979cy, Weldon:1982aq}.
\\
For the space-like region, i.e., $q_{0}<q$, the logarithms in Eqs.~\eqref{retarded_longiudinal_HTL} and \eqref{retarded_transverse_HTL} clearly acquire an imaginary part which induces a nonvanishing spectral function at $\mathcal{Q}^2>0$. This is known as Landau damping. Another feature in this region is related to the Debye screening: in the static limit of the above HTL propagator, one finds
\begin{align}
\mathcal{G}_L^R(0, q)=\frac{i}{q^2+m_D^2}, \quad \mathcal{G}_T^R(0, q)=\frac{-i}{q^2} .
\end{align}
This implies that the static chromoelectric fields are screened, i.e., at distances larger than the Debye radius $r_{D}= 1/m_{D}$, they vanish exponentially. At the same time, static chromomagnetic fields are not screened in HTL effective theory. At larger distances, the nonperturbative physics arises at scale $g^{2}T$ for the screening of these fields. Also, the dynamics of chromomagnetic fields at this scale can be described using an effective Hamiltonian derived in Refs.~\cite{Bodeker:1998hm, Bodeker:2000da}. 
\\
For the propagators in Eqs.~\eqref{retarded_longiudinal_HTL} and \eqref{retarded_transverse_HTL}, the plasmons have zero width, also known as gluon damping rate. In other words, the position of the plasmon pole is determined by the real part of the self-energy, which is of the order of $gT$, while the plasmon width is of the order of $g^{2}T$. This implies that the real part of the plasmon arises from the one-loop HTL limit, i.e., with hard momenta running through it. The width of the plasmon requires soft momenta through the loop and, thus, a consistent HTL resummation with resummed propagator and vertices. The gluon damping rate calculation and the proof of its gauge invariance was the first milestone in the HTL approach~\cite{Braaten:1989kk, Braaten:1990it}.    
\\
For the collective modes of fermions, the retarded self-energy $\Sigma^R$ given in Eq.~\eqref{fermionic_HTL_form} can be decomposed in modes having the positive or negative chirality to helicity ratio. One finds that, 
\begin{align}
\mathcal{S}_R(\mathcal{Q})=h_q^{+} \mathcal{S}_R^{+}(\mathcal{Q})+h_q^{-} \mathcal{S}_R^{-}(\mathcal{Q}),
\end{align}
where, 
where $h_{\boldsymbol{q}}^{ \pm} \equiv\left(\gamma^0 \mp \gamma^i \hat{q}^i\right) / 2$ and
\begin{align}
\mathcal{S}_R^{ \pm}(\mathcal{Q})=\frac{i}{q^0 \mp\left(q+\Sigma_R^{ \pm}\left(q^0 / q\right)\right)}=\frac{i}{q^0 \mp\left[q+\frac{m_{\infty}^2}{2 q}\left(1-\frac{q^0 \mp q}{2 q} \ln \left(\frac{q^0+q}{q^0-q}\right)\right)\right]},
\end{align}
Here, $q_{0}\equiv q_{0}+ i \epsilon$. For positive (negative) frequencies, the massless bare theory only has a positive (negative) chirality-to-helicity mode, with $\omega^{+}(q)=q\left(\omega^{-}(q)=-q\right)$. While in HTL theory, both modes develop time-like poles and at vanishing momenta $\omega^{+}(q=0)=\omega^{-}(q=0)=m_{\infty}^2 / \sqrt{2}$ is the fermionic counterpart of the plasma frequency as discussed earlier. On the other hand, for aymptotic momenta $\omega^{+}\left(q \gg m_{\infty}^2\right)=q+m_{\infty}^2 /(2 q)$ clearly develops an asymtotic mass $m_{\infty}^{2}$ while $\omega^{-}\left(q \gg m_{\infty}^2\right)=q$, has an exponentially vanishing residue~\cite{Pisarski:1989wb}.
\\ 
At intermediate momenta, the mode having a negative chirality-to-helicity ratio, also known as plasmino mode, shows non-monotonic behavior shown in Figure~\ref{fig_Dispersion_Relation_quarks}. This mode can be understood as a collective excitation where the positive frequency fermion mixes with the negative frequency anti-fermion. These time-like modes, whose pole position is of the order of $gT$, are long-lived, and their widths are of the order of $g^{2}T$, similar to gluonic excitations. The quark damping rate, which requires a consistent HTL-resummed calculation, has been calculated for vanishing momenta in Ref.~\cite{Braaten:1992gd} (see also~\cite{Nakkagawa:1992ew} for gauge invariance discussion). We will present the generalized calculations of the quark damping rate for the soft momenta in Chapter 3. In the space-like domain, the Landau damping also manifests itself for soft quarks, which physically corresponds to the scattering of soft, virtual quarks with the hard constituents of the medium. 
\subsection{HTL perturbation theory}
The compact representation of the HTL effective Lagrangian is~\cite{Braaten:1991gm}
\begin{align}
\mathcal{L}=\mathcal{L}_{\mathrm{QCD}}+\mathcal{L}_{\mathrm{HTL}}
\end{align}
where $\mathcal{L}_{\mathrm{QCD}}$ is the standard QCD Lagrangian in vacuum. The effective Lagrangian's HTL contribution can be expressed as follows:
\begin{align}
\mathcal{L}_{\mathrm{HTL}}=-\frac{1}{2} m_D^2 \operatorname{Tr}\left(G_{\mu \alpha}\left\langle\frac{\mathcal{Y}^\alpha \mathcal{Y}^\beta}{(\mathcal{Y} \cdot D)^2}\right\rangle_{\mathcal{Y}} G^\mu{ }_\beta\right)+i m_q^2 \bar{\psi} \gamma^\mu\left\langle\frac{\mathcal{Y}_\mu}{\mathcal{Y} \cdot D}\right\rangle_y \psi,
\label{HTL_effective_lagrangian}
\end{align}
The gluon field strength tensor is denoted by $G^{\mu\nu}$ as discribed earlier, $D$ stands for the covariant derivative in appropriate representation, $\mathcal{Y}^\mu=(1, \hat{\mathbf{y}})$ is a light-like vector and $\langle\cdots\rangle$ is average over all possible directions of $\hat{\mathbf{y}}$. This effective Lagrangian is gauge invariant and can produce all HTL $n$-point functions like three and four gluon vertex, and all of these functions satisfy the Ward-Takahashi identities by construction~\cite{Braaten:1991gm}.  
\\
A well-known approach for studying QCD thermodynamics, which resolves the IR issues via reorganization of thermal perturbation theory and the HTL resummation, is known as HTL perturbation theory (HTLpt)~\cite{Andersen:1999fw, Andersen:1999sf, Andersen:1999va, Andersen:2002ey, Andersen:2003zk, Andersen:2009tw, Andersen:2009tc, Andersen:2010ct, Andersen:2010wu, Andersen:2011sf, Andersen:2011ug, Mogliacci:2013mca, Haque:2013sja, Haque:2013qta, Haque:2012my, Haque:2014rua, Andersen:2015eoa}. This framework follows the systematic analytic reorganization of perturbation theory and relies on the HTL effective lagrangian. This approach is gauge-invariant and can be applied to studying the static as well as dynamic quantities of the system. This method is a extension of the screened perturbation theory that was previously employed to examine scalar field theories~\cite{Karsch:1997gj, Chiku:1998kd, Andersen:2000yj, Andersen:2001ez, Andersen:2008bz}.  The Lagrangian density in HTLpt is expressed as follows: 
\begin{align}
\mathcal{L}=\left[\mathcal{L}_{\mathrm{QCD}}+(1-\delta) \mathcal{L}_{\mathrm{HTL}}\right]_{g \rightarrow \sqrt{\delta} g}+\Delta \mathcal{L}_{\mathrm{HTL}}
\label{HTLpt_lagrangian}
\end{align}
where $\mathcal{L}_{\text {HTL }}$ represents the HTL contribution to the HTL effective Lagrangian given in Eq.~\eqref{HTL_effective_lagrangian} and $\Delta \mathcal{L}_{\text {HTL }}$ holds the counterterms required for renormalization. The first term in Eq.~\eqref{HTL_effective_lagrangian} is the usual QCD Lagrangian reads as
\begin{align}
\mathcal{L}_{\mathrm{QCD}}=-\frac{1}{2} \operatorname{Tr}\left(G_{\mu \nu} G^{\mu v}\right)+\mathcal{L}_{\text {gf }}+\mathcal{L}_{\text {ghost }}+\Delta \mathcal{L}_{\mathrm{QCD}}
\end{align}
Here, the gluon field strength tensor which is denoted by $G_{\mu \nu}=\partial_\mu \mathcal{A}_\nu-\partial_\nu \mathcal{A}_\mu-i g\left[\mathcal{A}_\mu, \mathcal{A}_\nu\right]$ and $\mathcal{A}_\mu$ is the gluon field described in $N_c \times N_c$ matrix in the $SU\left(N_c\right)$ algebra. The ghost term $\mathcal{L}_{\text {ghost }}$ depends on the choice of the gauge-fixing term $\mathcal{L}_{\text {gf }}$ which one choose while the last term, $\Delta \mathcal{L}_{\mathrm{QCD}}$, contains the vacuum counterterms required for $T=0$ renormalization.
\\
The parameter $\delta$ shown in Eq.~\eqref{HTLpt_lagrangian} functions as an expansion parameter in HTLpt. For $\delta = 1$, there will be no thermal contribution to the $T=0$ QCD Lagrangian. To proceed, one generally does the Taylor expansion of the generating functional around $\delta=0$ and then $\delta^{0}$ terms correspond to freely moving HTL quasiparticles. At the same time, higher order in $\delta$ consists of higher and higher order quasiparticle interactions. In principle, if one does an expansion of results of physical quantities to all orders in $\delta$, then there would be no dependence on the mass parameter $m_{D}$ and $m_{q}$ shown in Eq.~\eqref{HTL_effective_lagrangian}. However, in practice, for any finite order in expansion, one chooses some variational prescription to fix these parameters~\cite{Haque:2013sja, Haque:2014rua}.
\section{Gribov quantization}
Now, shifting from the discussion of effective theories, let us come back to QCD and the QCD quantization procedure. The standard procedure of quantizing Yang-Mills (YM) theory is the Faddev-Popov (FP) quantization~\cite{Faddeev:1967fc}. However, it has been shown that this approach needs to be completed in the sense that the FP approach requires an ideal gauge fixing choice. However, the advantage of this approach is the introduction of new modes, known as ghost modes, which violates the spin-statistics theorem~\cite{Streater:1989vi} and therefore leaves YM theory to be invariant under Becchi Rouet Stora and Tyutin (BRST) transformation~\cite{Becchi:1974xu, Becchi:1974md}.  
\vspace{-.5cm}
\subsection{Brief introduction}
The gauge theory constraint, namely that the configuration $\mathcal{A}$ and the local gauge transformation of this gauge field denoted as ${}^\mathcal{U} \mathcal{A}$, must be identified physically, which leads to a restrictive spectrum in the underlined theory. In the Gribov picture, this constraint changes the QCD spectrum and explains the gluon's absence from the physical spectrum.  
\\ 
In 1978, Gribov demonstrated that within non-abelian gauge group, namely $SU(2)$ and $SU(3)$, the local gauge group condition is stronger in comparison to an abelian gauge group and has more restrictions. For the abelian gauge theories, the gauge is uniquely fixed through the condition $\partial \cdot \mathcal{A}=0$. However, for non-abelian, it has been shown by Gribov that there can exist different configurations $\mathcal{A} \neq \mathcal{A}^{\prime}$ following the gauge constraint $\partial \cdot \mathcal{A}^{\prime}=\partial \cdot \mathcal{A}=0$. These configurations are connected via a 'large' gauge transformation $\mathcal{A}^{\prime}={ }^\mathcal{U} \mathcal{A}$. These are famously known as Gribov copies, and thus, one must identify these Gribov copies physically in a non-abelian gauge theory. The physical configuration, which is devoid of Gribov copies, is also known as a fundamental modular region (FMR). It has been demonstrated in Ref.~\cite{Singer:1978dk} that this situation can not be avoided within non-abelian gauge theory, and physical configuration space is topologically non-trivial. 
\\
The approximate calculations made by Gribov showed that the gluon propagator did not have a physical pole and thus excluded from the physical spectrum because of non-abelian gauge-invariance constraint consequence~\cite{Gribov:1977wm}. It is important to point out that the Gribov form of the gluon propagator $k^{2}/(k^{4}+\gamma_{G}^{4})$ also violates reflection positivity due to the unphysical poles at $k^{2} = \pm i \gamma_{G}^{2}$, which is a crucial postulate of any QFT. 
The semi-classical solution of Gribov to Gribov copies has been generalized at the quantum level by implementing a cut-off at the Gribov horizon through a local and renormalizable action known as the $\mathrm{GZ}$ action. While the gluon confinement via unphysical singularities in the gluon propagator is successfully described by the GZ action, the perturbation series resulting from this action does not produce precise quantitative results.
\vspace{-.8cm}
\subsection{The FP quantization}
Let us recall the FP approach of quantizing YM action at the perturbative level. One starts with the generating functional defined as
\begin{align}
Z(\mathcal{J})=\int[\mathrm{d} \mathcal{A}] \mathrm{e}^{-S_{\mathrm{YM}}+\int \mathrm{d} x \mathcal{J}_\mu^a \mathcal{A}_\mu^a} .
\end{align}
where $S_{\mathrm{YM}}$ is given by
\begin{align}
S_{\mathrm{YM}}=\int \mathrm{d}^d x \frac{1}{4} F_{\mu \nu}^a F_{\mu \nu}^a .
\end{align}
Taking the quadratic part of the YM action and doing a Gaussian integration leads to the following generating functional
\begin{align}
Z(\mathcal{J})_{\mathrm{quadr}}=(\operatorname{det} \mathcal{A})^{-1 / 2} \int[\mathrm{d} \mathcal{A}] \mathrm{e}^{-\frac{1}{2} \int \mathrm{d} x \mathrm{d} y \mathcal{J}_\nu^a(x) \mathcal{A}_{\mu \nu}(x, y)^{-1} \mathcal{J}_\mu^a(y)},
\label{generating_functional_1}
\end{align}
Here, $\mathcal{A}_{\mu \nu}(x, y)=\delta(x-y)\left(\partial^2 \delta_{\mu \nu}-\partial_\mu \partial_v\right)$. One finds that the generating functional in Eq.~\eqref{generating_functional_1} is not well-defined since the matrix $\mathcal{A}_{\mu \nu}(x, y)$ inverse does not exists. Indeed, the matrix $\mathcal{A}_{\mu \nu}(x, y)$ has vectors having zero eigenvalues, e.g., the vector $\mathrm{Y}_{\mu} = \partial_{\mu} \chi(x)$. 
\begin{align}
\int dy \mathcal{A}_{\mu\nu}(x,y) \mathrm{Y}_{\nu}(y) = \int dy [\delta(x-y) (\partial^{2}\delta_{\mu\nu} - \partial_{\mu}\partial_{\nu}) ] \partial_{\nu} \chi(y) = 0. 
\end{align}
Let us consider a gauge transformation $\mathcal{A}_{\mu}^{\prime} = \mathcal{U}\mathcal{A}_{\mu}\mathcal{U}^{\dagger}- \frac{i}{g} (\partial_{\mu}\mathcal{U}) \mathcal{U}^{\dagger} $ of $\mathcal{A}_{\mu} =0$ in order to comprehand the origin of these zero modes, where we take $\mathcal{U}=\exp \left(\operatorname{ig} X^a \chi^a\right)$,
\begin{align}
\mathcal{A}_\mu^{\prime}=-\frac{\mathrm{i}}{\mathrm{g}}\left(\partial_\mu \mathcal{U}\right) \mathcal{U}^{\dagger}=X_a \partial_\mu \chi^a,
\end{align}
or we have $\mathcal{A}_\mu^{a \prime}=\partial_\mu \chi^a$. This shows that zero modes of $\mathrm{Y}_{\mu}$ are gauge transformations of $\mathcal{A}_{\mu} =0$. Since we are doing an integration over the whole space of all gauge fields $\mathcal{A}_{\mu}$, we are also doing an integration over equivalent gauge fields, which give rise to zero modes, i.e., we are considering too many configurations into account. Because of the gauge invariance property of YM action, one can divide the configuration space $\mathcal{A}_{\mu}(x)$ into gauge orbits of different classes, and two points of one equivalency class are connected by the gauge transformation. By doing the gauge fixing, one obtains the following gauge-fixed action : 
\begin{align}
S=S_{\mathrm{YM}}+\underbrace{\int \mathrm{d} x\left(\bar{c}^a \partial_\mu D_\mu^{a b} c^b-\frac{1}{2 \alpha}\left(\partial_\mu \mathcal{A}_\mu^a\right)^2\right)}_{S_{\mathrm{gf}}} .
\label{gauge_fixed_lagrangian}
\end{align}
Here, $\alpha$ denotes the width of the Gaussian distribution having values $\alpha = 0$ and $1$ for the Landau and Feynman gauge, respectively. $c$ and $\bar{c}$ are Grassmann variables, also known as FP ghosts, which arise automatically during gauge fixing conditions. Thus, after gauge-fixing, one obtains the well-defined generating functional as
\begin{align}
Z(\mathcal{J})=\int[\mathrm{d} \mathcal{A}][\mathrm{d} c][\mathrm{d} \bar{c}] \exp \left[-S+\int \mathrm{d} x \mathcal{J}_\mu^a \mathcal{A}_\mu^a\right],
\end{align}
where the action $S$ is mentioned in Eq.~\eqref{gauge_fixed_lagrangian}. 
\subsection{The Gribov ambiguity}
\begin{figure}[tbh]
\hspace{-2cm}
\includegraphics[scale=.7,keepaspectratio]{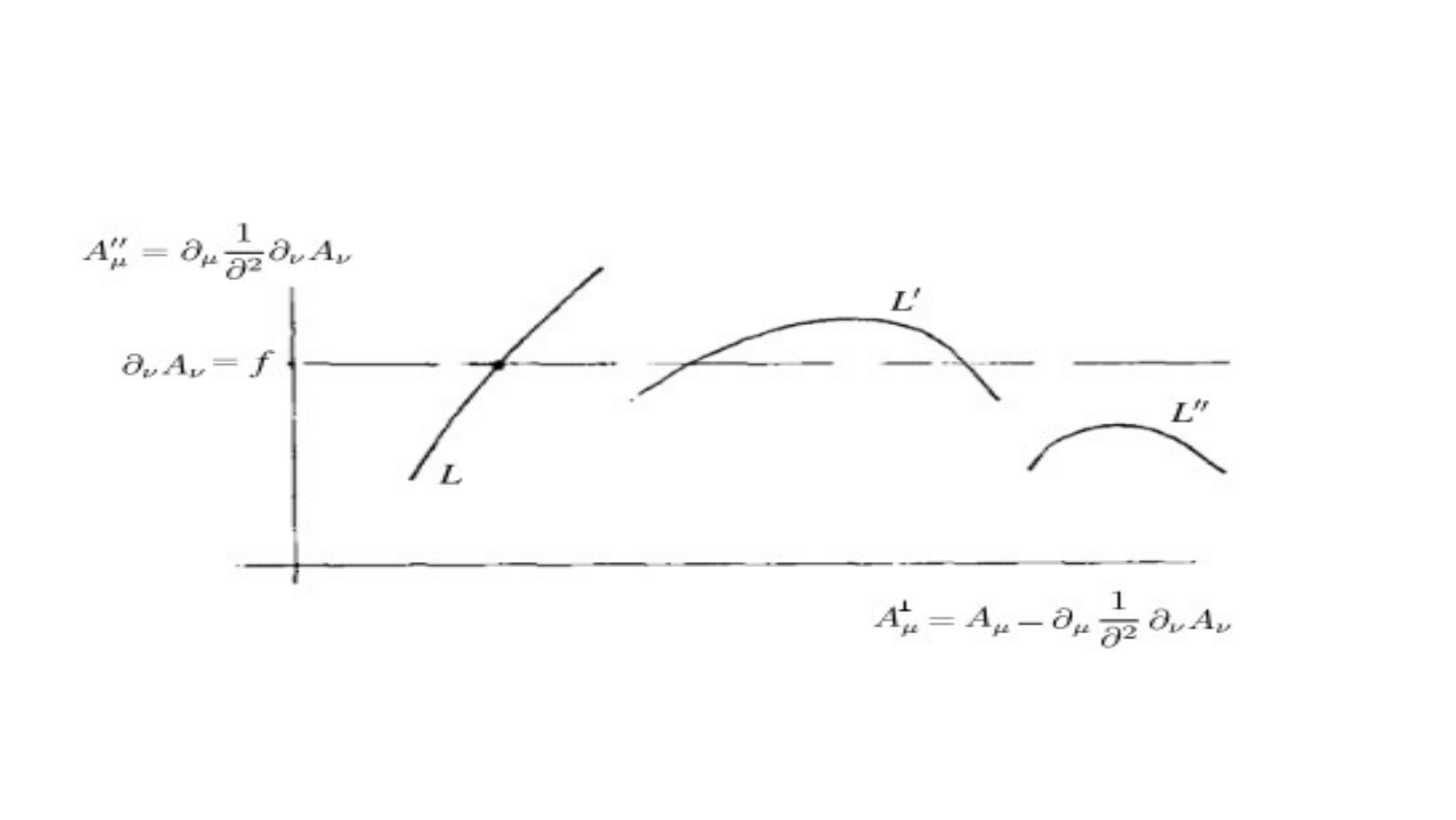}
\vspace{-2cm}
	\caption{The three different possibilities for a gauge orbit with respect to a gauge condition. Figure from Ref.~\cite{Gribov:1977wm}}
	\label{fig_Gribov}
\end{figure}

The gauge condition that has been utilized during the FP method of gauge fixing is given by
\begin{align}
\mathfrak{F}^{a}(\mathcal{A}_{\mu}(x)) = \partial_{\mu}\mathcal{A}^{\mu a} (x) - \mathcal{B}^{a} (x)
\end{align}
where $\mathcal{B}^{a} (x)$ is any arbitrary scalar field. Let us discuss that this gauge condition proposed by FP is not ideal. It was shown by Gribov in 1977 in Ref.~\cite{Gribov:1977wm}, where he considers three different possibilities of a gauge orbit intersection with the gauge condition shown in Figure~\ref{fig_Gribov}. The gauge orbit can have an intersection with the gauge condition only once $(L)$, more than once $(L^{\prime})$, or it can have no intersection at all $(L^{\prime \prime})$. For quantification, let us consider two equivalent gauge fields $\mathcal{A}_{\mu}$ and $\mathcal{A}_{\mu}^{\prime}$ (known as Gribov copies), which are related via a gauge transformation. If both these fields obey the Landau gauge conditions, then we have      
\begin{align}
\begin{array}{r}
\mathcal{A}_\mu^{\prime}=\mathcal{U}\mathcal{A}_\mu \mathcal{U}^{\dagger}-\frac{\mathrm{i}}{g}\left(\partial_\mu \mathcal{U}\right) \mathcal{U}^{\dagger}, \quad \partial_\mu \mathcal{A}_\mu=0 \quad \& \quad \partial_\mu \mathcal{A}_\mu^{\prime}=0, \\
\Downarrow \\
\partial_\mu \mathcal{U} \mathcal{A}_\mu \mathcal{U}^{\dagger}+\mathcal{U} \mathcal{A}_\mu \partial_\mu \mathcal{U}^{\dagger}-\frac{\mathrm{i}}{g}\left(\partial_\mu^2 \mathcal{U}\right) \mathcal{U}^{\dagger}-\frac{\mathrm{i}}{g}\left(\partial_\mu \mathcal{U}\right)\left(\partial_\mu \mathcal{U}^{\dagger}\right)=0 .
\label{Landau_gauge_condition}
\end{array}
\end{align}
For an infinitesimal transformation, $\mathcal{U}=1+\alpha, \mathcal{U}^{\dagger}=1-\alpha$, with $\alpha=\alpha^a X^a$, 
the Eq.~\eqref{Landau_gauge_condition} can be expanded to first order, 
\begin{align}
-\partial_\mu\left(\partial_\mu \alpha+\mathrm{ig}\left[\alpha, \mathcal{A}_\mu\right]\right)=0,
\end{align}
which can be further written in terms of FP operator $\partial_\mu D_\mu $ as
\begin{align}
-\partial_\mu D_\mu  \alpha = 0
\end{align}
Thus, the FP operator $\partial_{\mu} D_{\mu} $ contains zero mode, which is not sensitive to the gauge constraint. In other words, we can say that the existence of infinitesimal Gribov copies is associated with the zero eigenvalues of the FP operator. Also, the FP operator is hermitian i.e. $\partial_{\mu}D_{\mu} \alpha = D_{\mu}\partial_{\mu} \alpha$. The hermiticity of the FP operator implies that it has real eigenvalues. However, the perturbations around the zero modes can produce negative eigenvalues. Thus, the positive definite condition of the FP determinant used during the quantization procedure can not be used. Thus, the introduction of ghost fields becomes meaningless since the positive definite property of the FP determinant is utilized for the introduction of ghost fields.     
\\
Let us consider small values of $\mathcal{A}_{\mu}$ then we have $\partial_{\mu}^{2}\alpha =0$. But for the eigenvalue equation 
\begin{align}
-\partial_\mu^2 \psi=\varepsilon \psi,
\end{align}
has only positive eigenvalues $\varepsilon = p^{2}>0$ (apart from zero eigenvalues). However, for larger $\mathcal{A}_{\mu}$, we can have negative eigenvalues for the FP operator. This implies that the gauge condition is not ideal. For the zero-mode construction of the FP operator, look at Ref.~\cite{Sobreiro:2004us, Henyey:1978qd,vanBaal:1991zw}. Also, note that the issue of Gribov copies is not present in the case of QED. It can be proved as follows. The gauge transformation in QED is given by
\begin{align}
\mathcal{A}_\mu^{\prime}=\mathcal{A}_\mu-\partial_\mu \chi,
\end{align}
Thus, for the Landau gauge condition, $\partial_{\mu}\mathcal{A}_{\mu} =0$, the condition for $\mathcal{A}_{\mu}^{\prime}$ to be a copy of $\mathcal{A}_{\mu}$ becomes
\begin{align}
\partial_\mu \mathcal{A}_\mu^{\prime}=0 \Rightarrow \partial_\mu^2 \chi=0,
\label{plane_wave}
\end{align}
The above Eq.~\eqref{plane_wave} does not have any solutions other than plane waves. Since plane wave has an oscillating nature, they can not be used for the construction of gauge copy $\mathcal{A}_{\mu}^{\prime}$.  
\subsection{Gribov region: possible solution to Gribov ambiguity}
\begin{figure}[tbh]
\hspace{-2cm}
\includegraphics[scale=.7,keepaspectratio]{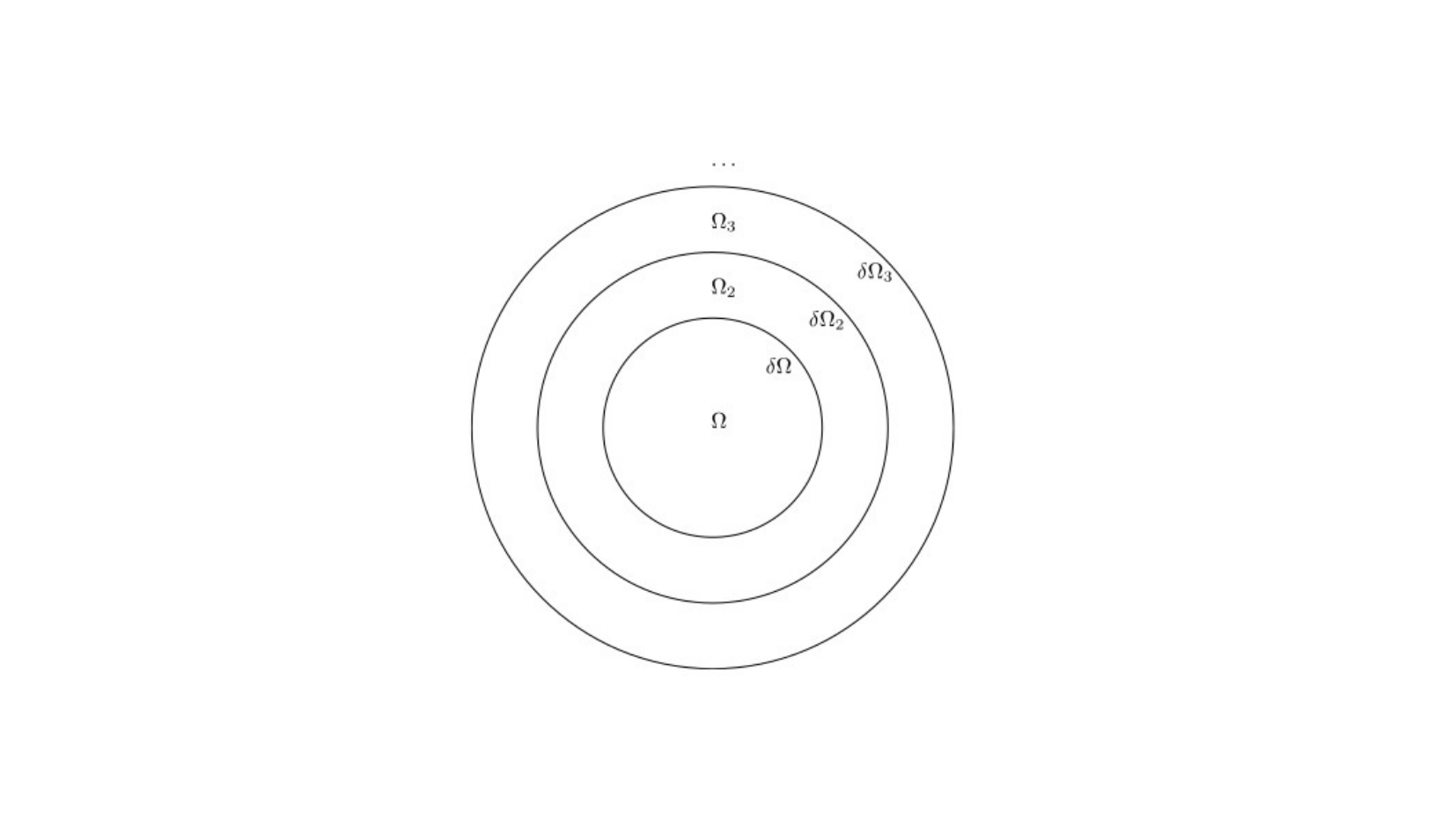}
\vspace{-2cm}
	\caption{The various regions in the hyperspace $\partial \mathcal{A}=0$. Figure from Ref.~\cite{Vandersickel:2012tz}}
	\label{fig_hyperspace}
\end{figure}

As discussed, FP quantization is not complete and thus requires improvisation to gauge fixing. Gribov proposed that one can restrict the integration region to the so-called Gribov region $\Omega$, which is defined in the following manner
\begin{align}
\Omega \equiv\left\{\mathcal{A}_\mu^a, \partial_\mu \mathcal{A}_\mu^a=0, \mathcal{M}^{a b}>0\right\},
\end{align}
where we have FP operator $\mathcal{M}^{a b}$ is defined as 
\begin{align}
\mathcal{M}^{a b}(x, y)=-\partial_\mu D_\mu^{a b} \delta(x-y)=\left(-\partial_\mu^2 \delta^{a b}+\partial_\mu f_{a b c} \mathcal{A}_\mu^c\right) \delta(x-y) .
\end{align}
For this Gribov region, the FP operator is positively definite, and thus, the problem of the positively definite issue of the FP determinant is resolved. The border of the Gribov region is known as the Gribov horizon $\delta\Omega$, and after crossing this horizon, the eigenvalues become negative. Similarly, other horizons can also be defined, as shown in Figure~\ref{fig_hyperspace}, in a very simplified version of the actually complicated space of gauge fields.  
\vspace{-.2cm}
\\
To conclude here, for the correct quantization of YM theory, one should really be constrained to FMR in order to have a completely correct gauge fixing choice where only one gauge configuration is chosen per orbit. For practical calculation, one restricts oneself to the Gribov region. An action known as $\mathrm{GZ}$ action has been developed, which automatically constraints to Gribov region~\cite{Zwanziger:1992qr, Zwanziger:1989mf}. Also, note that there can still be Gribov copies present in the Gribov horizon. However, it has been conjectured that the essential configurations reside on the common boundary region of $\delta \Lambda \bigcap \delta \Omega$ of the Gribov region $\Omega$ and FMR region $\Lambda$~\cite{Zwanziger:1993dh, Greensite:2004ke, Greensite:2004ur}. Thus, the extra Gribov copies will not play a significant role inside the Gribov region, and one can analyze different observable within the Gribov region.  
\subsection{The no-pole condition and Gribov gluon propagator}
In order to constrain the integration region to the first Gribov region, Gribov introduced a factor $\mathcal{V}(\Omega)$ in the generating functional mentioned in Eq.~\eqref{generating_functional_1}. Thus we have, 
\begin{align}
Z(J) & =\int_{\Omega}[\mathrm{d} \mathcal{A}] \exp \left[-S_{\mathrm{YM}}\right] \\
& =\int[\mathrm{d}\mathcal{A}][\mathrm{d} c][\mathrm{d} \bar{c}] \mathcal{V}(\Omega) \delta(\partial \mathcal{A}) \exp \left[-S_{\mathrm{YM}}-\int \mathrm{d} x \bar{c}^a(x) \partial_\mu D_\mu^{a b} c^b(x)\right].
\label{Partition_function_with_V}
\end{align}
Now, in order to calculate this factor $V(\Omega)$, one sees that there is a close connection between the ghost sector of the theory and the FP determinant. This factor has been determined from the poles of the ghost propagator, and the result for the condition that the FP operator has no zero mode reads as
\begin{align}
\mathcal{V}(\Omega)=\Theta(1-\sigma(0, \mathcal{A})),
\end{align}
$\sigma(0, \mathcal{A})$ is the function that appears in the ghost dressing function, and we have
\begin{align}
\mathcal{V}(\Omega)=\int_{-\mathrm{i} \infty+\varepsilon}^{+\mathrm{i} \infty+\varepsilon} \frac{\mathrm{d} \beta}{2 \pi \mathrm{i} \beta} \mathrm{e}^{\beta(1-\sigma(0, \mathcal{A}))},
\end{align}
This function can be inserted in the partition function mentioned in Eq.~\eqref{Partition_function_with_V}, and from there, a free gluon propagator can be derived. We will quote the final results of the computation, and details can be found in Ref.~\cite{Vandersickel:2012tz}. The Gribov gluon propagator in the Landau gauge reads as
\begin{align}
\left\langle \mathcal{A}_\mu^a(k) \mathcal{A}_\nu^b(p)\right\rangle=\delta(k+p) \delta^{a b} \frac{k^2}{k^4+\gamma_{G}^4} P_{\mu \nu}(k),
\end{align}
where $P_{\mu \nu}(k) = (\delta_{\mu \nu} - \frac{k_{\mu} k_{v}}{k^{2}}) $ is the projection tensor, and $\gamma_{G}$ is the Gribov mass parameter which acts as an infrared regulator in the theory. The analytic results for this Gribov mass parameter take the following form at finite temperature as~\cite{Fukushima:2013xsa}
\begin{align}
\gamma_{G} =
\left\{ \begin{array}{rcl}
\mu \exp \left(\frac{5}{12}-\frac{32 \pi^2}{3 N_{\mathrm{c}} g^2}\right) & \mbox{for}
& (T \rightarrow 0), \\ \frac{d-1}{d} \cdot \frac{N_{\mathrm{c}}}{4 \sqrt{2} \pi} g^2 T & \mbox{for} &  (T \rightarrow \infty) 
\end{array}\right.
\end{align}
It is visible here that the magnetic scale emerges at high temperatures, which incorporates the nonperturbative phenomena in the theory, and $\mu$ is some renormalization scale that can be fixed accordingly for a given temperature $T$. Similarly, the ghost propagator reads as~\cite{Vandersickel:2012tz}
\begin{align}
\left\langle \bar{c}_{a}(p) c_b(k)\right\rangle=\delta(k+p) \delta^{a b} \mathfrak{g}(k^{2}),
\end{align}
where 
\begin{align}
\mathfrak{g}(k^{2}) = \frac{1}{k^{4}}\frac{128 \pi^{2}\gamma^{2}}{Ng^{2}}
\end{align}
The above ghost propagator result shows the enhancement of ghost contribution in the $d=4$ dimension. Similar behavior is also observed in $d=2,3$ dimensions. Thus, these results show that limiting the integration to the Gribov region leads to an extra pole in the ghost propagator, which shows that the near boundary region to the Gribov region has important consequences for the ghost propagator. Also, note that a similar form of gluon and ghost propagator can also be obtained from the GZ action~\cite{Vandersickel:2012tz}. 
\subsection{Reflection positivity}
It has been shown in Ref.~\cite{Zwanziger:1990by} that if the gluon propagator $\mathcal{D}(k)$ vanishes at $k=0$, then the reflection positivity would be violated. Reflection positivity and other postulates of QFT satisfy the Kallen Lehmann representation, which is given by
\begin{align}
\mathcal{D}(k)=\int_0^{\infty} \mathrm{d} m^2 \frac{\rho\left(m^2\right)}{k^2+m^2},
\end{align}
Here $\rho\left(\mathrm{m}^2\right) \geq 0$ is a positive weight spectral function. Thus, the gluon propagator in the position i.e. x$-$space becomes 
\begin{align}
\widetilde{\mathcal{D}}(x)=\int_0^{\infty} \mathrm{d} m^2 \rho\left(m^2\right) \widetilde{\mathcal{D}}_m(x),
\end{align}
is positive for all $x, \widetilde{\mathcal{D}}(x)>0$, because the free propagator of a particle of mass $m$ is positive for all $x$. If the propagator at zero momentum vanishes, then
\begin{align}
\mathcal{D}(0)=\int \mathrm{d}^d x \widetilde{\mathcal{D}}(x)=0,
\end{align}
Furthermore, if $\widetilde{\mathcal{D}}(x)$ is positive (or zero) for all $x$, then the gluon propagator vanishes identically, which is false. Thus, if the gluon propagator vanishes at $k=0$, then the reflection positivity is violated. In fact, it is maximally violated in the sense that gluon correlator $\widetilde{\mathcal{D}}(x)$ is positive and negative in equal measure. For the Gribov propagator having form as $\mathcal{D}(k)=k^{2}/\left[\left(k^2\right)^2+m^4\right]$ maximally violated positivity because $\mathcal{D}(0)=0$, and the Kallen-Lehmann representation is violated by having the imaginary poles $k^2= \pm \mathrm{im}^2$.
\\
Lattice investigation also shows that $\mathcal{D}(0)$ vanishes in Euclidean dimesion $d=2$, whearas in $d=3,4$ dimesion $\mathcal{D}(0)>0$. Thus, reflection positivity is maximally violated in the $2$ dimension but not maximally violated for the $3$ and $4$ dimension. Lattice results show the indications of a violation of reflection positivity in $3$ dimension~\cite{Cucchieri:2004mf} and in $4$ dimension both for quenched and unquenched cases~\cite{Bowman:2007du}. According to the confinement scenario, the violation of reflection positivity further indicates that the gluon is not a physical particle.  



\chapter{Quark self-energy and corresponding dispersion relations at NLO using HTL resummation}\label{Chapter_3}
\allowdisplaybreaks
\pagestyle{fancy}
\fancyhead[LE]{\emph{\leftmark}}
\fancyhead[LO]{\emph{\rightmark}}
\rhead{\thepage} 
In this chapter, we have studied and calculated the NLO quark self-energy alongside the NLO dispersion laws using the HTL resummation. This chapter is based on our work:
{\em NLO quark self-energy and dispersion relation using the hard thermal loop resummation}, {\bf Sumit}, Najmul Haque and Binoy Krishna Patra, {\bf \color{blue} JHEP 05 (2023) 171}~\cite{Sumit:2022bor}.

\section{Introduction}
At finite temperatures, the conventional perturbative loop expansion in QCD confronts a number of troubles. One of the main concerns is that the physical quantities, such as dispersion laws, start to give gauge-dependent results. In Ref.~\cite{Kalashnikov:1979cy}, the authors have calculated the QCD polarisation tensor at finite temperature and chemical potential in one-loop order, and then the gluon dispersion properties to leading order (LO) in the QCD coupling $g$. Also, the authors of Ref.~\cite{Kalashnikov:1979cy} have shown that the dispersion laws are gauge-independent, but the damping rates for gluons at one-loop-order are not in the large-wavelength limit. One of the important outcomes of reference, as mentioned above, tells us that the Debye screening does occur in the LO in the one-loop calculation in the chromoelectric mode. However, chromomagnetic screening is absent in the LO. In LO, the absence of chromomagnetic field screening is also studied in Refs.~\cite{Linde:1978px, Linde:1980ts,Gross:1980br}. In Ref.~\cite{Klimov:1981ka}, the massless spectrum of the elementary quark excitations at the LO in $g$ is studied, and the full quasiparticle spectra to leading order for the whole momentum range are given in Ref.~\cite{Klimov:1982bv} at the high-temperature limit. The quasiparticle spectra were also studied in Ref.~\cite{Weldon:1982aq} for gluons and~\cite{Weldon:1982bn} for quarks.\\	
The authors of Ref.~\cite{Kobes:1987bi} have examined the problem of the gauge dependency of the gluon damping rates, which have been estimated up to one-loop order, in particular at zero momentum, in various gauges and schemes, with differing results. Subsequently, it was determined~\cite{Pisarski:1988vd} that higher-order diagrams can contribute to lower orders in powers of the QCD coupling and that the LO finding is incomplete. In other words, in powers of the QCD coupling $g$, the conventional loop expansion is no longer applicable. The problem is solved in a series of papers by Braaten and Pisarski. They created an organized theory for an effective perturbative expansion known as HTL resummation, which sums the higher-order terms into effective vertices and effective propagators~\cite{Braaten:1989mz, Braaten:1989kk, Frenkel:1989br, Bellac:2011kqa}. Ref.~\cite{Braaten:1990it} showed that the transverse component of the gluon damping rate $\gamma_{t}(0)$ at vanishing momentum was finite, positive, and gauge-independent using the effective HTL propagators and vertices.\\
Once developed, the important thing to check out is the reliability of the HTL-summed perturbative calculations in gauge theories at high temperatures. If so, it can be considered an important candidate for describing the QGP properties. One of the exciting questions is the IR sensitivity of the massless gauge theories, which worsens at finite temperatures due to the Bose-Einstein distribution BE function. At finite temperature, physical quantities are more infrared sensitive~\cite{Burgess:1991wc,Rebhan:1992ca,Rebhan:1993az} because the BE distribution function acts like $1/k$ for extremely vanishing gluon energies $k$.\\	
Using the HTL resummed propagators and vertices, many studies have been performed in perturbative QCD to acknowledge the thermodynamic attributes of plasma. For example, the pressure and quark number susceptibilities have been studied using the thermodynamic potential up to two- and three-loop order~\cite{Haque:2014rua, Haque:2013sja, Haque:2013qta, Haque:2012my, Andersen:2011sf, Andersen:2010wu, Jiang:2010jz}. In addition, the electric and magnetic features of the plasma were investigated in Ref.~\cite{Liu:2011if}. 
Massless quarks and gluons are found to acquire the thermal masses of order $gT$, $m_q$, and $m_g$, respectively, using the HTL summation~\cite{Kalashnikov:1979cy, Klimov:1981ka, Klimov:1982bv, Weldon:1982aq, Weldon:1982bn}. This indicates that the IR region is 'okay' for the lowest order $gT$ in effective perturbation. Nevertheless, as we have already said, the static chromomagnetic field screens at the NLO, which is known as magnetic screening~\cite{Linde:1978px, Linde:1980ts, Gross:1980br} rather than the LO. Therefore, we need to go beyond the LO calculations to illustrate the IR sector of the HTL perturbative expansion. Recently, many other calculations have been performed at NLO using thermal field theory to probe the hot and dense QCD matter. For example, the transport coefficient at NLO, namely the ratio of shear viscosity to entropy density $\eta/s$ and the ratio of the quenching parameter to temperature $\hat{q}/T^{3} $ have been obtained in Ref.~{\cite{Muller:2021wri}}, HTL lagrangian has been derived at NLO for the photon in Ref.~\cite{Carignano:2019ofj}, two loops HTL’s have been derived for any general model recently in Ref.~\cite{Ekstedt:2023anj}, explicit results for the gluon self-energy in semi-QGP at NLO have been obtained in Ref.~\cite{Wang:2022dcw}. Also, the complete calculation of soft photon self-energy at NLO in QED is presented in Ref.~\cite{Gorda:2022fci}. Using that cold and dense electron gas pressure at N$^{3}$LO has been obtained in Ref.~\cite{Gorda:2022zyc}.\\	
To probe the IR sector of HTL perturbative expansion, the first calculation performed using NLO HTL-dressed perturbative expansion is the non-moving gluon damping rate~\cite{Braaten:1990it}. In a similar line, the non-moving quark damping rate has been calculated~\cite{Kobes:1992ys, Braaten:1992gd} in imaginary time formalism~\cite{Bellac:2011kqa, Mustafa:2022got, Kapusta:2006pm}. The quark damping rate of a non-moving quark was recently calculated in Ref.~\cite{Carrington:2006gb} using real-time formalism. After the study of non-moving gluon and quark damping rates in 1992, later, by using this formalism,  the damping rates of slow-moving longitudinal~\cite{Abada:1998ue, Abada:1997vm} and transverse gluons~\cite{Abada:2004dr} in the one-loop order HTL, quarks damping rate~\cite{Abada:2007opj, Abada:2005na, Abada:2000hh}, electrons~\cite{Abada:2007zz} and photons \cite{Abada:2011cc} damping rate in QED, and also quasiparticles energy up to NLO in scalar QED~\cite{Abada:2005jq} have been studied.\\
After the rigorous studies of boson and fermion damping rates, people also studied the quasiparticle energies up to NLO. That would come from studying the real part of the HTL effective one-loop self-energies. The NLO quasiparticle energy calculation is much non-trivial than the computation of the damping rate. It was started by determining the plasma frequency $\omega_g(0)$ for pure gluon case up to NLO in the long-wavelength limit in Ref.~\cite{Schulz:1993gf}. A gauge-invariant plasma frequency in next-to-leading order for pure-gluon plasma is established with $N_{f}=0$ in imaginary time formalism as
\begin{equation}
\omega_{g}(0)=\frac{\sqrt{c_{A}}}{3} g T\left(1-0.09 \sqrt{c_{A}} g+\ldots\right), 
\label{omega_nlo}
\end{equation}
with Casimir number $c_A=N_c$ and $N_c$ represents the number of colors in QCD, $g$ is the QCD coupling constant, and $T$ is the temperature. The first term of Eq.~\eqref{omega_nlo} represents the leading order part, whereas the second represents the NLO contribution. Later, fermion mass up to NLO in QCD at high temperature (and QED) was calculated in Refs.~\cite{Carrington:2006xj, Carrington:2006gb, Carrington:2008dw}. The NLO quark mass for two-flavour QCD is obtained in Ref.~\cite{Carrington:2008dw} in the real-time formalism~\cite{Bellac:2011kqa,Landsman:1986uw,Chou:1984es} as
\begin{equation}
m_q^{nlo}=\frac{g T}{\sqrt{6}}\left(1+\frac{1.87}{4 \pi} g+\ldots\right).
\label{m_q_nlo}
\end{equation}
The NLO contributions to the plasma frequency and quark mass are from the one-loop diagrams involving soft momenta. Indeed, using the general power-counting arguments done in Ref.~\cite{Mirza:2013ula}, it has been shown that, except for the photon self-energy, NLO corrections come only from the soft one-loop resummed graphs with effective vertices and propagators. For the former case, two-loop graphs with hard internal momentum also contribute. In work \cite{Carrington:2006gb} using the power-counting arguments, it has been shown that in imaginary-time formalism, one would not get the correct information about the number of terms that give subleading contributions. In Ref.~\cite{Abada:2014bma}, the authors have set up a framework to calculate the NLO dispersion relation for a slow-moving quark. The authors used the setup to show how one can proceed to evaluate the different terms in the NLO part of quark self-energy. In the present work, we calculate all sixteen terms of NLO quark self-energy to complete the NLO dispersion relations, i.e., energy and damping rates, for quarks moving slowly within HTL approximation in real-time formalism. This has been done by utilizing the closed-time-path (CTP) approach of thermal field theory~\cite{Martin:1959jp, Keldysh:1964ud}. The benefit of avoiding the analytic continuation from Matsubara frequencies to real energies. For the latter case,  it is non-trivial to obtain the analytical responses of physical observables.
Nevertheless, one issue is that, because of the doubling of degrees of freedom, each $N$-point function gets a tensor structure, in which we have to work with $2N$ components. Thus, there will be enough rise in graphs containing three and four-point vertex integrals. Also, this computation will not be easy even if one sets the quark momentum to be zero from the starting point, as done in Ref.~\cite{Chou:1984es}, where they replace the momentum contractions of effective vertices by effective self-energy differences using Ward identities.\\	
This chapter is sketched in this way. In section~\ref{NLO}, we discuss the lowest-order dispersion relations for quarks and the expressions of the effective quark and gluon propagators. Then, in section~\ref{section_3}, we present the NLO formalism for slow-moving quarks. In this section, we present the expression for NLO dispersion laws from which one can extract the NLO quark energies and damping rates of order $g^{2}T$. These quantities directly correspond with the HTL effective NLO quark self-energy $\Sigma^{(1)}$. This contribution $\Sigma^{(1)}$ to the NLO quark self-energy is calculated further in this section, and we present a detailed analytic result of $\Sigma^{(1)}$ in terms of the three- and four-point HTL effective vertex integrals. In section~\ref{sec4}, we evaluate the different terms of NLO quark self-energy and describe how these terms have been evaluated numerically. In section~\ref{sec:result}, we have plotted the NLO quark self-energy w.r.t. the ratio of momentum and energy. Then, we plotted the NLO correction to dispersion relations using the above quantity. We encapsulate the chapter in the last section~\ref{sec:summary}.

 
\section{Lowest order dispersion relations}\label{NLO} 
\begin{figure}[t]
	\centering
\includegraphics[width=7cm,height=9cm,keepaspectratio]{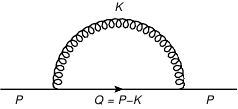}
	\caption{Feynman graph for the quark self-energy in leading order}
	\label{fig1}
\end{figure}
\begin{figure}[tbh]
\centering
\includegraphics[scale=.7,keepaspectratio]{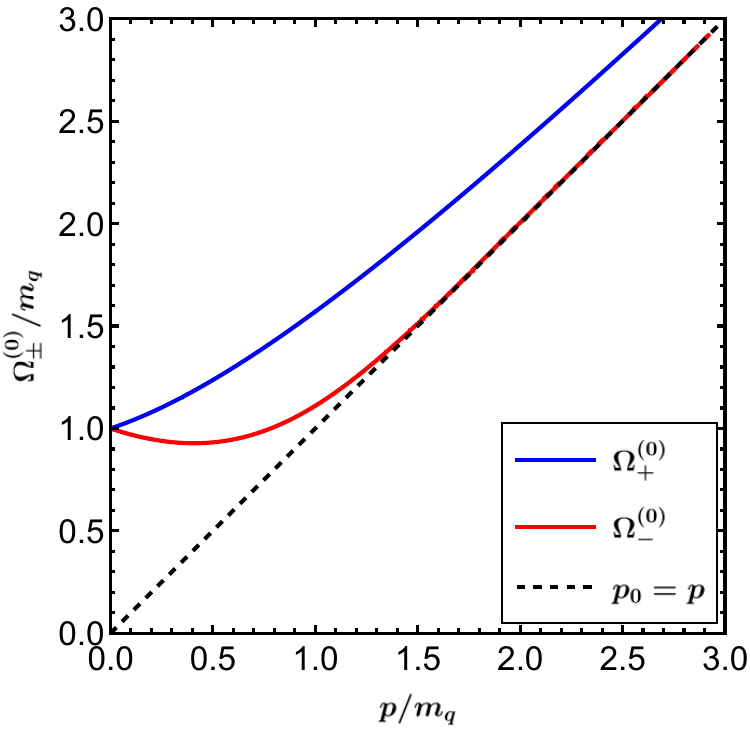}  
\caption{Lowest-order dispersion laws for quark excitations}
\label{fig_Dispersion_Relation_quarks}
\end{figure}

A one-loop effective quark propagator can be written as
\begin{equation}
S(P)=\frac{1}{\slashed{P} - \Sigma(P)},\label{q_prop}
\end{equation}
Where $P = (p_0, \vec{p})\equiv (p_0, \hat{p}\,p)$ is the four-momentum of external quark and $\Sigma{(P)}$ is the one-loop HTL quark self-energy shown in Figure~\ref{fig1}.
The zeros of the denominator of the propagator in Eq.~\eqref{q_prop} give the dispersion  relations as
\begin{equation}\label{q_disp}
\text{det} [\slashed{P} - \Sigma(P)]=0,
\end{equation}
The self-energy $\Sigma(P)$ can be decomposed into helicity eigenstates as
\ba
\Sigma(P) &=& \gamma_{+p}\  \Sigma_{-}(P)+\gamma_{-p}\  \Sigma_{+}(P) .\label{q_self_energy}
\ea 
Here, $\gamma_{\pm p} \equiv\left(\gamma^{0} \mp \vec{\gamma} \cdot \hat{p}\right) / 2 $ , with $\hat{p} = \vec{p}/p $ and $\gamma^{\mu}$ are the Dirac matrices. In the lowest order, the quark self-energy $\Sigma_{{\pm}} (P)$ in Eq.~\eqref{q_self_energy} is calculated as
\begin{equation}\label{q_self_lo}
\Sigma_{\pm}(\omega, p)= \frac{m_{q}^{2}}{p}\left[\pm 1+\frac{1}{2}\left(1 \mp \frac{\omega}{p}\right) \ln \frac{\omega+p}{\omega-p}\right] ,
\end{equation}
\noindent where $m_{q}^2= \frac{C_F}{8} g^2T^2 $ represents the square of the thermal mass of a quark in leading order at zero chemical potential with $ C_{F} = (N^{2}_{c}-1)/2N_{c}$. Eq.~\eqref{q_disp} can be summarized in the following two dispersion relations: 
\begin{equation}\label{disp_two}
p_0 \mp p - \Sigma_{\pm}(P) = 0 .
\end{equation}
The solution of the dispersion relations in the lowest order is $p_0=\Omega^{(0)}_\pm$, and they are displayed in Figure~\ref{fig_Dispersion_Relation_quarks}. 
The $\Omega^{(0)}_+(p)$ mode represents the propagation of an ordinary quark with a momentum-dependent thermal mass, and the ratio of chirality to helicity for this mode is $+1$. On the other hand, the $\Omega_-^{(0)}(p)$ mode represents the propagation of a quark mode for which the chirality to helicity ratio is $-1$. This mode is called the plasmino mode, and it is absent at zero temperature but appears as a consequence of the thermal medium due to the broken Lorentz invariance~\cite{Mustafa:2002pb}. At large momenta, two modes go to the light cone very quickly. Whereas the soft portion is effectively constricted. For soft external momenta $(p/m_{q} < 1)$, the solution of the dispersion relations can be expanded as
\begin{equation}\label{disp_sol}
\Omega^{(0)}_{\pm}(p)=m_{q}\bigg[1 \pm \frac{1}{3}\frac{p}{m_q}+\frac{1}{3} \frac{p^2}{m_q^2} \mp \frac{16}{135} \frac{p^3}{m_q^3}+\frac{1}{54} \frac{p^{4}}{m_q^4} \pm \frac{32}{2835} \frac{p^{5}}{m_q^5}-\frac{139}{12150} \frac{p^{6}}{m_q^6} \pm \ldots\bigg].
\end{equation}
Using the HTL self-energies $\Sigma_\text{HTL}$ defined in Eq.~\eqref{q_self_lo}, one can write the one-loop effective quark propagator, which can also be decomposed into the helicity eigenstates as
\begin{align}
\Delta(P) = & \, \gamma_{+p} \Delta_{-}(P)+\gamma_{-p} \Delta_{+}(P) ; \nn
\Delta_{\pm}^{-1}(P)  = & \, p_{0} \pm p-\frac{m_{q}^{2}}{p}\left[\mp 1+\frac{1}{2 p} m_{q}^{2}\left(p \pm p_{0}\right) \ln \frac{p_{0}+p}{p_{0}-p}\right] .
\label{q_prop_htl}
\end{align}
Since the quark damping rate comes from the negative of the imaginary part of self-energy, there is no quark damping at the lowest order, and it starts to contribute from NLO.\\
In addition to the leading order quark self-energy, we also require an HTL-dressed gluon propagator to calculate the NLO contribution of the quark self-energy. In covariant gauge, a one-loop HTL resummed gluon propagator is 
\begin{equation}
D_{\mu \nu}(K)=\frac{\xi K_\mu K_\nu}{K^4}+D_{T}(K) A_{\mu \nu}+D_{L}(K) B_{\mu \nu},
\end{equation}
Here, $A_{\mu \nu}$ and $B_{\mu \nu}$ are the transverse and longitudinal projection operators, respectively, and can be expressed as
\begin{equation}
A_{\mu \nu}=g_{\mu \nu}-\frac{K_{\mu} K_{\nu}}{K^{2}}-B_{\mu\nu} ; \quad B_{\mu \nu}=-\frac{K^2}{k^{2}}\left(u_{\mu}-\frac{k_{0} K_{\mu}}{K^{2}}\right)\left(u_{\nu}-\frac{k_{0} K_{\nu}}{K^{2}}\right),
\end{equation}
Where $u^\mu$ is the four-velocity of the heat bath and in the plasma rest-frame, $u^\mu = (1, \vec{0}) $. The quantities $D_{L, T}$ are the longitudinal and transverse HTL effective gluon propagators, respectively, and given by
\begin{align}
D_{L}^{-1}(K)= & \, K^{2}+2 m_{g}^{2} \frac{K^{2}}{k^{2}}\left(1-\frac{k_{0}}{2 k} \ln \frac{k_{0}+k}{k_{0}-k}\right); \nn
D_{T}^{-1}(K) = & \, K^{2}-m_{g}^{2}\left[1+\frac{K^{2}}{k^{2}}\left(1-\frac{k_{0}}{2 k} \ln \frac{k_{0}+k}{k_{0}-k}\right)\right].\label{glu_prop}
\end{align}
In the above Eq.~\eqref{glu_prop}, $ m_g = \frac{1}{6}\sqrt{(N_c + s_{F})}\ gT $ is the gluon thermal mass to LO. Here, $s_{F} = N_{f}/2$, and  $N_{f}$ represents the number of quark flavors. 
\section{NLO formalism}\label{section_3}
For on shell quarks, we write the (complex) quark energy $p_0 \equiv \Omega(p)$ 
as
\begin{equation}\label{disp_decom}
\begin{aligned}
\Omega(p)= \Omega^{(0)}(p) + \Omega^{(1)}(p) + \cdots \hspace{2mm} .
\end{aligned}
\end{equation}
A similar kind of decomposition can also be done for self-energy $\Sigma$, as well
\begin{equation}\label{sigma_nlo}
\begin{aligned}
\Sigma(P)= \Sigma_\text{HTL}(P)+ \Sigma^{(1)}(P)+\cdots ,
\end{aligned}
\end{equation}
where $\Sigma_\text{HTL}$ is the LO quark self-energy having $gT$ order, whereas $\Sigma^{(1)}$ the NLO contribution of quark self-energy, having $g^{2}T$ order. Similarly, the first term $\Omega^{(0)}(p)$ and second term $\Omega^{(1)}(p)$ in Eq.~\eqref{disp_decom} are of the same order as terms on r.h.s. of Eq.~\eqref{sigma_nlo}. Thus, Eq.~\eqref{disp_two} will take the form as
\begin{equation}
\Omega^{(0)}_{\pm}(p)+\Omega_{\pm}^{(1)}(p) +\cdots= \pm p + \left.\Sigma_{\mathrm{HTL} \pm}\left(\omega, p\right)\right|_{\omega\rightarrow\Omega_{\pm}(p)}+\left.\Sigma_{\pm}^{(1)}\left(\omega, p\right)\right|_{\omega\rightarrow\Omega_{\pm}(p)}+\ldots \hspace{2mm} . 
\end{equation}
\begin{figure}[t]
\centering	\includegraphics[width=8cm,height=8cm,keepaspectratio]{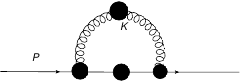}
\caption{Feynman graph for the NLO HTL resummed quark self-energy $ \Sigma_{1}^{(1)} $. The black blobs indicate HTL effective quantities. All momenta are soft.}
\label{fig3}
\end{figure}
Since we are interested in slow-moving quarks, we can take $p \sim gT$, and we get
\begin{equation}\label{disp_nlo0}
\Omega_{\pm}^{(1)}(p)=\frac{\Sigma_{\pm}^{(1)}\left(\Omega^{(0)}_{\pm}(p), p\right)}{1-\left.\partial_{\omega} \Sigma_{\mathrm{HTL} \pm}(\omega, p)\right|_{\omega=\Omega^{(0)}_{\pm}(p)}} .
\end{equation}
Here $\partial_{\omega}$ represents variation w.r.t $\omega$. Real values of above Eq.~\eqref{disp_nlo0} give us the NLO corrections to the momentum-dependent quark energies, whereas the NLO contribution to the quark damping rate comes from the negative of the imaginary part of Eq.~\eqref{disp_nlo0}. Also, by using the expression of $\Sigma_\text{HTL}$ mentioned in Eq.~\eqref{q_self_lo}, the NLO dispersion relations will take the final form as
\begin{equation}\label{disp_nlo}
\Omega_{\pm}^{(1)}(p)=\frac{{\Omega^{(0)}_{\pm}}^{2}(p)-p^{2}}{2 m_{q}^{2}} \Sigma_{\pm}^{(1)}\left(\Omega^{(0)}_{\pm}(p), p\right) . 
\end{equation}
We need to determine the NLO quark self-energy to evaluate the above equation. For that, one needs to consider two one-loop graphs with effective vertices, shown \footnote{The Feynman graphs are drawn using jaxodraw software~\cite{Binosi:2003yf}.} in Figure~\ref{fig3} and Figure \ref{fig4}. The graph in Figure~\ref{fig3} can be written as
\begin{equation}
\begin{aligned}
\Sigma_{1}^{(1)}(P)=-i g^{2} C_{F} \int \frac{d^{4} K}{(2 \pi)^{4}} \Gamma^{\mu}(P, Q) \Delta(Q) \Gamma^{\nu}(Q, P) D_{\mu \nu}(K),
\end{aligned}
\end{equation}
where $Q = P - K$. Similarly, from the vertex graph in Figure~\ref{fig4}, we can write
\begin{equation}
\begin{aligned}
\Sigma_{2}^{(1)}(P)=\frac{-i g^{2} C_{F}}{2} \int \frac{d^{4} K}{(2 \pi)^{4}} \Gamma^{\mu \nu}(P, K) D_{\mu \nu}(K).
\end{aligned}
\end{equation}
\begin{figure}[t]
	\centering
	\includegraphics[scale=1.9,keepaspectratio]{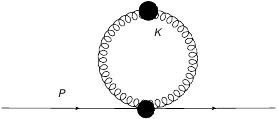}
	\caption{Feynman diagram for the NLO quark self-energy $ \Sigma_{2}^{(1)}$. The black blobs represent HTL effective quantities. All momenta are soft.}
	\label{fig4}
\end{figure}
We have three possible summation indices for the above equations: Lorentz (explicit), RTF, and Dirac. Firstly, we start with the Keldysh indices on the ``r/a'' basis, utilizing the thermal field theory's CTP formulation. For the fermion,  the retarded (R), advanced (A), and symmetric (S) propagators can be defined as 
\begin{align}
\Delta^{\mathrm{R}}(K) \equiv& \, \Delta^{\mathrm{ra}}(K)=\Delta\left(k_{0}+i \varepsilon, \vec{k}\right); \quad \quad
\Delta^{\mathrm{A}}(K)  \equiv   \Delta^{\mathrm{ar}}(K)=\Delta\left(k_{0}-i \varepsilon, \vec{k}\right);\nn
\Delta^{\mathrm{S}}(K) \equiv& \, \Delta^{\mathrm{rr}}(K)=N_{F}\left(k_{0}\right) \operatorname{sign}\left(k_{0}\right)\left[\Delta^{\mathrm{R}}(K)-\Delta^{\mathrm{A}}(K)\right].\label{RAS_F}
\end{align}
Similarly, for boson
\begin{align}
D_{\mu\nu}^{\mathrm{R}}(K) \equiv& \, D_{\mu\nu}^{\mathrm{ra}}(K)=D_{\mu\nu}\left(k_{0}+i \varepsilon, \vec{k}\right); \quad \quad
D_{\mu\nu}^{\mathrm{A}}(K)  \equiv  D_{\mu\nu}^{\mathrm{ar}}(K)=D_{\mu\nu}\left(k_{0}-i \varepsilon, \vec{k}\right);\nn
D_{\mu\nu}^{\mathrm{S}}(K) \equiv& \, D_{\mu\nu}^{\mathrm{rr}}(K)=N_{B}\left(k_{0}\right) \operatorname{sign}\left(k_{0}\right)\left[D_{\mu\nu}^{\mathrm{R}}(K)-D_{\mu\nu}^{\mathrm{A}}(K)\right],\label{RAS_B}
\end{align}
where $N_{F}$ in Eq.~\eqref{RAS_F} and $N_{B}$ in Eq.~\eqref{RAS_B} are defined as follows
\begin{equation}\label{dist_func}
N_{B,F}(k_{0}) = 1\pm 2n_{B,F}(k_{0});   \quad  n_{B,F}(k_{0})= \frac{1}{e^{|k_{0}|/T} \mp 1}.
\end{equation}
The modulus value in the argument of the BE and FD distribution function is required to avoid the blow-up of the function. Thus, for the two components of $\Sigma^{(1)}_{1}$, we have the following explicit expressions.
\begin{align}
\Sigma_{(1) \pm}^{(1)}(P) = & \, \frac{-i g^{2} C_{F}}{2} \int \frac{d^{4} K}{(2 \pi)^{4}} \operatorname{tr} \gamma_{\pm p}\bigg[\Gamma_{\operatorname{arr}}^{\mu}(P, Q) \Delta^{\mathrm{R}}(Q) \Gamma_{\operatorname{arr}}^{\nu}(Q, P) D_{\mu \nu}^{\mathrm{S}}(K)\nn
+ & \, \Gamma_{\mathrm{arr}}^{\mu}(P, Q) \Delta^{\mathrm{S}}(Q) \Gamma_{\operatorname{rar}}^{\nu}(Q, P) D_{\mu \nu}^{\mathrm{A}}(K) +\Gamma_{\mathrm{arr}}^{\mu}(P, Q) \Delta^{\mathrm{R}}(Q) \Gamma_{\text {aar }}^{\nu}(Q, P) D_{\mu \nu}^{\mathrm{A}}(K) \nn
+ & \, \Gamma_{\operatorname{arr}}^{\mu}(P, Q) \Delta^{\mathrm{R}}(Q) \Gamma_{\operatorname{arr}}^{\nu}(Q, P) D_{\mu \nu}^{\mathrm{R}}(K) +\Gamma_{\operatorname{ara}}^{\mu}(P, Q) \Delta^{\mathrm{A}}(Q) \Gamma_{\mathrm{rar}}^{\nu}(Q, P) D_{\mu \nu}^{\mathrm{A}}(K)\bigg] ,
\label{sigma_1}
\end{align}
and for the two components of $\Sigma^{(1)}_{2}$, we get
\begin{align}
\Sigma_{(2) \pm}^{(1)}(P) = & \, \frac{-i g^{2} C_{F}}{4} \int \frac{d^{4} K}{(2 \pi)^{4}} \operatorname{tr} \gamma_{\pm p}\left[\Gamma_{\operatorname{arrr}}^{\mu \nu}(P, K) D_{\mu \nu}^{\mathrm{S}}(K) + \Gamma_{\operatorname{aarr}}^{\mu \nu}(P, K) D_{\mu \nu}^{\mathrm{R}}(K) \right. \nn
+ & \, \left. \Gamma_{\operatorname{arar}}^{\mu \nu}(P, K) D_{\mu \nu}^{\mathrm{A}}(K)\right]. \label{sigma_2}
\end{align}
Eq.~\eqref{sigma_1} and Eq.~\eqref{sigma_2} are the results of the Mathematica program developed in Ref.~\cite{Carrington:2006gb}, which can take care of all the real-time-field indices of a given Feynman diagram. We rederive all the three and four-point HTL-effective vertex integrals using their corresponding Feynman graphs. Thus, for the two-quarks-one-gluon effective vertices, we will get
\begin{align}
\Gamma_{\mathrm{arr}}^{\mu}(P, K) = & \, \gamma^{\mu}+I_{--}^{\mu}(P, K); \quad  \Gamma_{\mathrm{rar}}^{\mu}(P, K) =\gamma^{\mu}+I_{+-}^{\mu}(P, K); \nn
\Gamma_{\mathrm{aar}}^{\mu}(P, K) = & \,\Gamma_{\mathrm{ara}}^{\mu}(P, K)=0,
\label{vert_func_3}
\end{align}
and two-quarks-two-gluons effective vertices give 
\begin{equation}\label{vert_func_4}
\Gamma_{\mathrm{arrr}}^{\mu \nu}(P, K)=I_{--}^{\mu \nu}(P, K) ; \quad \Gamma_{\mathrm{aarr}}^{\mu \nu}(P, K)=\Gamma_{\mathrm{arar}}^{\mu \nu}(P, K)=0.
\end{equation}
In Eq.~\eqref{vert_func_3} and Eq.~\eqref{vert_func_4}, the four vectors $(I$'s) are HTL contribution which are given by
\begin{align}
I_{\eta_{1} \eta_{2}}^{\mu}(P, Q) = & \, m_{q}^{2} \int \frac{d \Omega_{s}}{4 \pi} \frac{S^{\mu} \slashed{{S}}}{\left(P S+i \eta_{1} \varepsilon\right)\left(Q S+i \eta_{2} \varepsilon\right)} ;\label{solid_ang_int1} \\ 
I_{\eta_{1} \eta_{2}}^{\mu \nu}(P, K) = \, & m_{q}^{2} \int \frac{d \Omega_{s}}{4 \pi} \frac{-S^{\mu} S^{\nu} \slashed{{S}}}{\left(P S+i \eta_{1} \varepsilon\right)\left(P S+i \eta_{2} \varepsilon\right)} 
\left[\frac{1}{(P+K) S+i \eta_{1} \varepsilon} \right. \nn
+ & \, \left. \frac{1}{(P-K) S+i \eta_{2} \varepsilon}\right] .
\label{solid_ang_int2}
\end{align}
Here $S\equiv (1,\hat{s}$) and the indices $\eta_{1} ,\eta_{2} = \pm 1$. Also, the quantity $PS $ and $QS $ are the dot product of four vectors defined as 
$PS= p_{0}-(\vec{p}\!\cdot\!\hat{s})$.
Using the above results, one can rewrite $\Sigma^{(1)}_{(1)\pm}$ from Eq.~\eqref{sigma_1} as
\begin{align}
\Sigma_{(1) \pm}^{(1)}(P) = & \, \frac{-i g^{2} C_{F}}{2} \int \frac{d^{4} K}{(2 \pi)^{4}} \operatorname{tr} \gamma_{\pm p}\left[\left\{\gamma^{\mu}+I_{--}^{\mu}(P, Q)\right\} \Delta^{\mathrm{R}}(Q)\left\{\gamma^{\nu}+I_{--}^{\nu}(Q, P)\right\}\right.\nn
\times & \,  D_{\mu \nu}^{\mathrm{S}}(K) +\left.\left\{\gamma^{\mu}+I_{--}^{\mu}(P, Q)\right\} \Delta^{\mathrm{S}}(Q)\left\{\gamma^{\nu}+I_{+-}^{\nu}(Q, P)\right\} D_{\mu \nu}^{\mathrm{A}}(K)\right].\label{Sigma1pm}
\end{align}
Similarly, $\Sigma^{(1)}_{(2)\pm}$ can be rewritten from Eq.~\eqref{sigma_2} as
\ba
\hspace{-10mm}\Sigma_{(2) \pm}^{(1)}(P)&=&\frac{-i g^{2} C_{F}}{4} \int \frac{d^{4} K}{(2 \pi)^{4}} \operatorname{tr} \gamma_{\pm p} I_{--}^{\mu \nu}(P, K) D_{\mu \nu}^{\mathrm{S}}(K).\label{Sigma2pm}
\ea
Eq.~\eqref{Sigma1pm} can be divided into three kinds of terms as
\begin{align}
\Sigma_{(1)\pm}^{(1)}(P) = & \, \frac{-i g^{2} C_{F}}{2} \int \frac{d^{4} K}{(2 \pi)^{4}}\left[F_{\pm ; 0}^{\mathrm{SR}}(P, K)+F_{\pm ; 0}^{\mathrm{AS}}(P, K)+2 F_{\pm ;--}^{\mathrm{SR}}(P, K)+F_{\pm ;--}^{\mathrm{AS}}(P, K)\right.\nn
+ & \, \left. F_{\pm ;-+}^{\mathrm{AS}}(P, K)+F_{\pm ;--;--}^{\mathrm{SR}}(P, K)+F_{\pm ;--;+-}^{\mathrm{AS}}(P, K)
\right].\label{Sigma1pm2}
\end{align}
The first two terms within the square bracket in Eq.~\eqref{Sigma1pm2} are due to the bare part of the vertex and can be written in a general form as
\begin{align} 
F_{\epsilon_{p} ; 0}^{\mathrm{XY}}(P, K) \equiv & \, \operatorname{tr}\left(\gamma_{\epsilon_{p}} \gamma^{\mu} \gamma_{\epsilon_{q}} \gamma^{\nu}\right) D_{\mu \nu}^{\mathrm{X}}(K) \Delta_{-\epsilon_{q}}^{\mathrm{Y}}(Q)\nn
= & \, -2\left(1-\hat{p}_{\epsilon}\mdot\hat{k} \,\hat{q}_{\epsilon}\mdot\hat{k}\right) D_{T}^{\mathrm{X}}(K) \Delta_{-\epsilon_{q}}^{\mathrm{Y}}(Q) \nn
- & \, \left[k_{0}^{2}\left(1-\hat{p}_{\epsilon}\mdot\hat{q}_{\epsilon}+2 \hat{p}_{\epsilon}\mdot\hat{k} \,\hat{q}_{\epsilon}\mdot\hat{k}\right)- 2k_{0} k\left(\hat{p}_{\epsilon}\mdot\hat{k}+\hat{q}_{\epsilon}\mdot\hat{k}\right)+k^{2}\left(1+\hat{p}_{\epsilon}\mdot\hat{q}_{\epsilon}\right)\right] \nn \times & \, \tilde{D}_{L}^{\mathrm{X}}(K) \Delta_{-\epsilon_{q}}^{\mathrm{Y}}(Q) 
\label{bare_terms}
\end{align}
Here, $\tilde{D}_{L}(K)={D}_{L}(K)/K^{2}$. Also, $\hat{p}_{\epsilon}= \epsilon_{p}\hat{p} $ with $\epsilon_{p} = \pm $ and similarly for $\hat{q}_{\epsilon}$, with summation over $\epsilon_{q}$. The superscripts $X$ and $Y$ can take the RTF indices values $(R, A,$ and $S)$. The third, fourth, and fifth terms within the square bracket in Eq.~\eqref{Sigma1pm2} are the contribution that involves one HTL vertex function and can be written in a general form as
\begin{align}
F_{\epsilon_{p};\eta_{1}\eta_{2}}^{\mathrm{XY}}(P,K)  \equiv & \, \operatorname{tr}\left(\gamma_{\epsilon_{p}} I_{\eta_{1} \eta_{2}}^{\mu} \gamma_{\epsilon_{q}} \gamma^{\nu}\right) D_{\mu \nu}^{\mathrm{X}}(K) \Delta_{-\epsilon_{q}}^{\mathrm{Y}}(Q) \nn
= & \, m_{q}^{2} \int \frac{d \Omega_{s}}{4 \pi}\frac{1}{\left(PS+i \eta_{1} \varepsilon\right)\left(QS+i \eta_{2} \varepsilon\right)}
\times\bigg[D_{T}^{\mathrm{X}}(K) \Delta_{-\epsilon_{q}}^{\mathrm{Y}}(Q)\nn
\times & \, \Big\{1-\hat{p}_{\epsilon}\mdot\hat{q}_{\epsilon}-\hat{p}_{\epsilon}\mdot\hat{s}-\hat{q}_{\epsilon}\mdot\hat{s}+\hat{p}_{\epsilon}\mdot\hat{k} \,\hat{k}\mdot\hat{s}+\hat{q}_{\epsilon}\mdot\hat{k} \,\hat{k}\mdot\hat{s}-(\hat{k}\mdot\hat{s})^{2}+\hat{p}_{\epsilon}.\hat{q}_{\epsilon}\,(\hat{k}\mdot\hat{s})^{2} \nn
+ & \, 2 \hat{p}_{\epsilon}.\hat{s} \,\hat{q}_{\epsilon}.\hat{s} - \hat{p}_{\epsilon}.\hat{k} \,\hat{q}_{\epsilon}\mdot\hat{s}\,\hat{k}\mdot\hat{s}-\hat{q}_{\epsilon}\mdot\hat{k}\,\hat{p}_{\epsilon}\mdot \hat{s}\,\hat{k}\mdot\hat{s}\Big\} \nn
- & \, \Big\{k_{0}^{2}\left(\hat{p}_{\epsilon}\mdot\hat{k} \, \hat{k}\mdot\hat{s}+\hat{q}_{\epsilon}\mdot\hat{k} \,\hat{k}\mdot\hat{s}-(\hat{k}\mdot\hat{s})^{2} +\hat{p}_{\epsilon}\mdot\hat{q}_{\epsilon} \,(\hat{k}\mdot\hat{s})^{2} -\hat{p}_{\epsilon}\mdot\hat{k} \,\hat{q}_{\epsilon}\mdot\hat{s} \,\hat{k}.\hat{s}-\hat{q}_{\epsilon}.\hat{k} \,\hat{p}_{\epsilon}\mdot\hat{s}\, \hat{k}\mdot\hat{s}\right) \nn
+ & \, k^{2}\left(1+\hat{p}_{\epsilon}\mdot\hat{q}_{\epsilon}-\hat{p}_{\epsilon}\mdot\hat{s}-\hat{q}_{\epsilon}\mdot\hat{s}\right) - k_{0} k \left(\hat{p}_{\epsilon}\mdot\hat{k}+\hat{q}_{\epsilon}\mdot\hat{k}+2 \hat{p}_{\epsilon}\mdot\hat{q}_{\epsilon} \,\hat{k}\mdot\hat{s} -\hat{q}_{\epsilon}\mdot\hat{k} \,\hat{p}_{\epsilon}\mdot\hat{s} \right. \nn
- &\,\left. \hat{p}_{\epsilon}\mdot\hat{k} \, \hat{q}_{\epsilon}\mdot\hat{s}-\hat{q}_{\epsilon}\mdot\hat{s} \, \hat{k}\mdot\hat{s}-\hat{p}_{\epsilon}\mdot\hat{s} \,\hat{k}\mdot\hat{s}\right)
\Big\} \tilde{D}_{L}^{\mathrm{X}}(K) \Delta_{-\epsilon_{q}}^{\mathrm{Y}}(Q)\bigg].
\label{htl_terms_1}
\end{align}
Since $\eta_{1},\eta_{2} = \pm 1$ and because of symmetry in $D_{\mu\nu}$, the other contributions with one HTL vertex integral are the same as above when changing $\eta_{1}$ into $\eta_{2}$, namely,
\begin{equation}
\begin{aligned}
\operatorname{tr}\left(\gamma_{\epsilon_{p}} \gamma^{\mu} \gamma_{\epsilon_{q}} I_{\eta_{1} \eta_{2}}^{\nu}\right) D_{\mu \nu}^{\mathrm{X}}(K) \Delta_{-\varepsilon_{q}}^{\mathrm{Y}}(Q)=\operatorname{tr}\left(\gamma_{\epsilon_{p}} I_{\eta_{2} \eta_{1}}^{\mu} \gamma_{\epsilon_{q}} \gamma^{\nu}\right) D_{\mu \nu}^{\mathrm{X}}(K) \Delta_{-\epsilon_{q}}^{\mathrm{Y}}(Q).
\end{aligned}
\end{equation}
The sixth and seventh terms inside the square bracket of Eq.~\eqref{Sigma1pm2} contribute to two HTL vertex functions. It can be written in a general form as
\begin{align}
F_{\epsilon_{p} ; \eta_{1} \eta_{2} ; \eta_{1}^{\prime} \eta_{2}^{\prime}}^{\mathrm{XY}}(P, K) \equiv & \, \operatorname{tr}\left(\gamma_{\epsilon_{p}} I_{\eta_{1} \eta_{2}}^{\mu} \gamma_{\epsilon_{q}} I_{\eta_{1}^{\prime} \eta_{2}^{\prime}}^{\nu}\right) D_{\mu \nu}^{\mathrm{X}}(K) \Delta_{-\epsilon_{q}}^{\mathrm{Y}}(Q)\nn
= & \, m_{q}^{4} \int \frac{d \Omega_{s}}{4 \pi} \frac{1}{\left(P S+i \eta_{1} \varepsilon\right)\left(Q S+i \eta_{2} \varepsilon\right)} \int \frac{d \Omega_{s^{\prime}}}{4 \pi} \frac{1}{\left(P S^{\prime}+i \eta_{2}^{\prime} \varepsilon\right)\left(Q S^{\prime}+i \eta_{1}^{\prime} \varepsilon\right)}\nn
\times & \, \bigg[\Big\{-\hat{s} \mdot\hat{s}^{\prime}-\hat{p}_{\epsilon}\mdot\hat{q}_{\epsilon} \, \hat{s}\mdot\hat{s}^{\prime}+\hat{p}_{\epsilon}\mdot\hat{s} \, \hat{s}\mdot\hat{s}^{\prime}+\hat{q}_{\epsilon}\mdot\hat{s} \, \hat{s}\mdot\hat{s}^{\prime} + \hat{p}_{\epsilon}\mdot\hat{s}^{\prime}\, \hat{s}\mdot\hat{s}^{\prime}+\hat{q}_{\epsilon}\!\cdot\!\hat{s}^{\prime} \, \hat{s}\mdot \hat{s}^{\prime} 
\nn
-& \, \hat{p}_{\epsilon}\mdot\hat{s} \, \hat{q}_{\epsilon}\!\cdot\!\hat{s}^{\prime} \, \hat{s}\!\cdot\! \hat{s}^{\prime}-\hat{q}_{\epsilon}\mdot\hat{s} \, \hat{p}_{\epsilon}\!\cdot\!\hat{s}^{\prime} \, \hat{s}\mdot \hat{s}^{\prime} - \left(\hat{s}\mdot\hat{s}^{\prime}\right)^{2}+\hat{p}_{\epsilon}\mdot\hat{q}_{\epsilon} \, \left(\hat{s} . \hat{s}^{\prime}\right)^ {2} +  \hat{k}\mdot\hat{s}^{\prime}\, \hat{k}\!\cdot\!\hat{s}\nn
+ & \, \hat{p}_{\epsilon}\mdot\hat{q}_{\epsilon}\, \hat{k}\mdot\hat{s}' \, \hat{k}\!\cdot\!\hat{s} -\hat{p}_{\epsilon}\mdot\hat{s} \, \hat{k}\mdot\hat{s}'\, \hat{k}\mdot\hat{s} -\hat{q}_{\epsilon}\mdot\hat{s} \, \hat{k}\mdot\hat{s} \, \hat{k}\mdot\hat{s}^{\prime} -\hat{p}_{\epsilon}\mdot\hat{s}^{\prime}\, \hat{k}\mdot\hat{s} \,\hat{k}.\hat{s}^{\prime} -\hat{q}_{\epsilon}\mdot\hat{s}^{\prime} \,\hat{k}\mdot\hat{s}\, \hat{k}\mdot\hat{s}^{\prime} \nn
+ &\,  \hat{p}_{\epsilon}\mdot\hat{s} \, \hat{q}_{\epsilon}\mdot\hat{s}^{\prime}\, \hat{k}.\hat{s} \,\hat{k}\mdot\hat{s}^{\prime}+\hat{q}_{\epsilon}\mdot\hat{s} \,\hat{p}_{\epsilon}\mdot\hat{s}^{\prime} \,\hat{k}\mdot\hat{s}\, \hat{k}\mdot\hat{s}^{\prime}   + \hat{s}\mdot\hat{s}^{\prime} \, \hat{k}.\hat{s} \,\hat{k}\mdot\hat{s}^{\prime} -\hat{p}_{\epsilon}\mdot\hat{q}_{\epsilon} \, \hat{s}\mdot\hat{s}^{\prime} \,\hat{k}\mdot\hat{s} \,\hat{k}.\hat{s}^{\prime}\Big\} \nn 
\times & \, D_{T}^{\mathrm{X}}(K) \Delta_{-\epsilon_{q}}^{\mathrm{Y}}(Q)\nn
- & \, \Big\{k^{2}\left(1+\hat{p}_{\epsilon}.\hat{q}_{\epsilon}-\hat{p}_{\epsilon}\mdot\hat{s}-\hat{q}_{\epsilon}\mdot\hat{s}-\hat{p}_{\epsilon}\mdot\hat{s}^{\prime}-\hat{q}_{\epsilon}\mdot\hat{s}^{\prime}+\hat{p}_{\epsilon}\mdot\hat{s} \,\hat{q}_{\epsilon}\mdot\hat{s}^{\prime}+\hat{q}_{\epsilon}\mdot\hat{s} \,\hat{p}_{\epsilon}\mdot\hat{s}^{\prime}+\hat{s}.\hat{s}^{\prime} \right.\nn
-& \, \left. \hat{p}_{\epsilon}. \hat{q}_{\epsilon}\,\hat{s}.\hat{s}^{\prime}\right) - k_{0} k\left(\hat{k}\cdot\hat{s} + \hat{k}.\hat{s}^{\prime}\right.-\left.\hat{p}_{\epsilon}.\hat{q}_{\epsilon} \,\hat{k}. \hat{s} + \hat{p}_{\epsilon}.\hat{q}_{\epsilon} \,\hat{k}.\hat{s}^{\prime}-\hat{p}_{\epsilon}.\hat{s} \,\hat{k}\mdot\hat{s}-\hat{p}_{\epsilon}.\hat{s} \,\hat{k}.\hat{s}^{\prime} \right. \nn
-& \, \left. \hat{q}_{\epsilon}.\hat{s}\, \hat{k}\mdot\hat{s}-\hat{q}_{\epsilon}.\hat{s} \,\hat{k}.\hat{s}^{\prime}\right. -\hat{p}_{\epsilon}.\hat{s}^{\prime} \,\hat{k}. \hat{s}-\hat{p}_{\epsilon}.\hat{s}^{\prime} \,\hat{k}\mdot\hat{s}^{\prime}-\hat{q}_{\epsilon}.\hat{s}^{\prime} \,\hat{k}\mdot\hat{s} - \hat{q}_{\epsilon}\!\cdot\!\hat{s}^{\prime} \,\hat{k}.\hat{s}^{\prime} \nn
+ & \,  \hat{p}_{\epsilon}.\hat{s} \, \hat{q}_{\epsilon} .\hat{s}^{\prime} \,\hat{k}.\hat{s}+\hat{p}_{\epsilon}\!\cdot\!\hat{s} \,\hat{q}_{\epsilon}\!\cdot\!\hat{s}^{\prime} \,\hat{k}\!\cdot\!\hat{s}^{\prime}
+\left.\hat{q}_{\epsilon}\!\cdot\!\hat{s} \,\hat{p}_{\epsilon}.\hat{s}^{\prime} \,\hat{k}\mdot\hat{s}+\hat{q}_{\epsilon}.\hat{s} \,\hat{p}_{\epsilon} .\hat{s}^{\prime} \,\hat{k}\mdot\hat{s}^{\prime}+\hat{k}.\hat{s} \,\hat{s}.\hat{s}^{\prime}\right.\nn
+& \, \left. \hat{k}.\hat{s}^{\prime} \,\hat{s}\mdot\hat{s}^{\prime}-\hat{p}_{\epsilon}.\hat{q}_{\epsilon} \,\hat{k}.\hat{s} \,\hat{s}\mdot\hat{s}^{\prime}-\hat{p}_{\epsilon}.\hat{q}_{\epsilon} \,\hat{k}.\hat{s}^{\prime} \,\hat{s}. \hat{s}^{\prime}\right) + k_{0}^{2}\left(\hat{k}\mdot\hat{s}\,\hat{k}\mdot\hat{s}^{\prime}+\hat{p}_{\epsilon}\mdot\hat{q}_{\epsilon}\,\hat{k}\mdot\hat{s}\,\hat{k}\mdot\hat{s}^{\prime}\right.\nn
- & \, \left.\left.\hat{p}_{\epsilon}.\hat{s}\,\hat{k}.\hat{s}\,\hat{k}\mdot\hat{s}^{\prime}-\hat{q}_{\epsilon}\mdot\hat{s}\,\hat{k}\mdot\hat{s}\,\hat{k}\mdot\hat{s}^{\prime}-\hat{p}_{\epsilon}\mdot\hat{s}^{\prime}\, \hat{k}\mdot\hat{s}\,\hat{k}.\hat{s}^{\prime}-\hat{q}_{\epsilon}\!\cdot\!\hat{s}^{\prime}\,\hat{k}\mdot\hat{s}\,\hat{k}\mdot\hat{s}^{\prime}\right.\right.\nn
+ & \, \left.\left.\left.\hat{p}_{\epsilon}\mdot\hat{s} \,\hat{q}_{\epsilon}\mdot\hat{s}^{\prime} \,\hat{k}\mdot \hat{s} \,\hat{k}\mdot\hat{s}^{\prime}+\hat{q}_{\epsilon}\mdot\hat{s} \,\hat{p}_{\epsilon}\mdot\hat{s}^{\prime} \,\hat{k}\mdot\hat{s} \,\hat{k}\mdot\hat{s}^{\prime}+\hat{k}\mdot\hat{s} \,\hat{k}\mdot\hat{s}^{\prime} \,\hat{s}\mdot\hat{s}^{\prime}-\hat{p}_{\epsilon}\mdot\hat{q}_{\epsilon} \,\hat{k}\mdot\hat{s} \,\hat{s}\mdot\hat{s}^{\prime}\,\hat{k}\!\cdot\!\hat{s}^{\prime} \right)\right\}\right. \nn
\times & \, \left. \tilde{D}^{X}_{L}(K)\Delta^{Y}_{-\epsilon_{q}}(Q)\right].
\label{htl_terms_2}
\end{align}
The integrand in $\Sigma^{(1)}_{2}$ contains only one HTL and can be written from Eq.~\eqref{Sigma2pm} as
\ba
\Sigma_{(2)\pm}^{(1)}(P)&=& \frac{-i g^{2} C_{F}}{2} \int \frac{d^{4} K}{(2 \pi)^{4}}\ G_{\pm ;--}^{\mathrm{S}}(P, K),
\ea
where\footnote{A typo in~\cite{Abada:2014bma} is corrected}
\begin{align}
G_{\epsilon_{p} ; \eta_{1} \eta_{2}}^{\mathrm{S}}(P, K) \equiv & \, \frac{1}{2} \operatorname{tr}\left(\gamma_{\epsilon_{p}} I_{\eta_{1} \eta_{2}}^{\mu \nu}\right) \Delta_{\mu \nu}^{\mathrm{X}}(K)\nn
= & \, m_{q}^{2} \int \frac{d \Omega_{s}}{4 \pi} \frac{1}{\left[P S+i \eta_{1} \varepsilon\right]\left[P S+i \eta_{2} \varepsilon\right]}  \left[\frac{1}{(P+K) S+i \eta_{1} \varepsilon}+\frac{1}{(P-K) S+i \eta_{2} \varepsilon}\right] \nn
\times & \, \Bigg[\left(1-\hat{p}_{\epsilon}\mdot\hat{s}\right)\left(1-(\hat{k}\mdot \hat{s})^{2}\right)D_{T}^{\mathrm{X}}(K) 
+\ \Big\{k^{2}-2 k_{0} k\hat{k}.\hat{s}\ +\ k_{0}^{2} \, \,(\hat{k}.\hat{s})^{2}\Big\}\left(1-\hat{p}_{\epsilon}\mdot\hat{s}\right)\nn
\times & \, \tilde{D}_{L}^{\mathrm{X}}(K)\Bigg].\label{sigma2_htl}
\end{align}
Using the above terms, the NLO one-loop HTL-summed quark self-energy can be expressed in compact form as
\begin{align}
\Sigma_{\pm}^{(1)}(P) = & \, -\frac{i g^{2} C_{F}}{2} \int \frac{d^{4} K}{(2 \pi)^{4}}\left[F_{\pm ; 0}^{\mathrm{SR}}(P, K)+F_{\pm ; 0}^{\mathrm{AS}}(P, K)+2 F_{\pm ;--}^{\mathrm{SR}}(P, K)+F_{\pm ;--}^{\mathrm{AS}}(P, K)\right.\nn
+ & \, \left.F_{\pm ;-+}^{\mathrm{AS}}(P, K)+F_{\pm ;--;--}^{\mathrm{SR}}(P, K)+F_{\pm ;--;+-}^{\mathrm{AS}}(P, K)+G_{\pm ;--}^{\mathrm{S}}(P, K)\right].\label{sigma_final}
\end{align}
\section{Evaluation of NLO quark self-energy}\label{sec4}
\begin{figure}
\centering
{\includegraphics[scale=0.8,keepaspectratio]{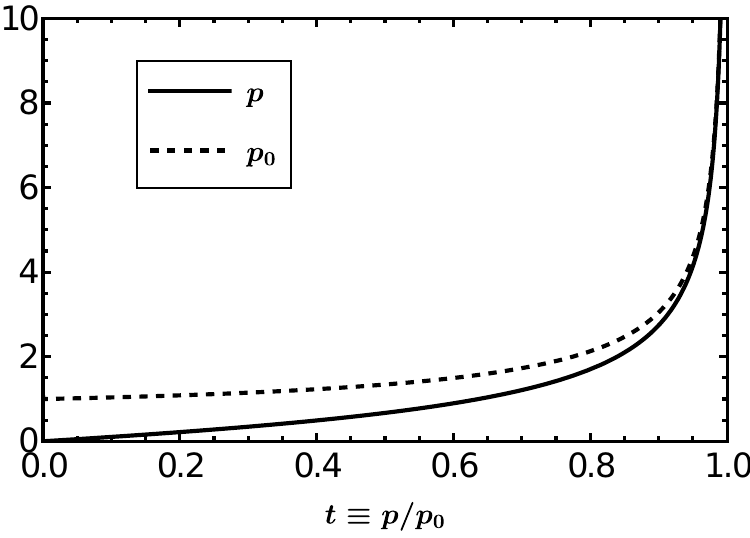}}
\caption{Variation of scaled quark momentum $p$(solid) and mass $p_{0}$ (dashed) w.r.t. $``\text{t}\equiv p/p_{0}"$ variable. For slow-moving quarks $t \lesssim 0.64$.}
	\label{fig5}
\end{figure}

\begin{figure}[t]
\centering
{\includegraphics[scale=0.8,keepaspectratio]{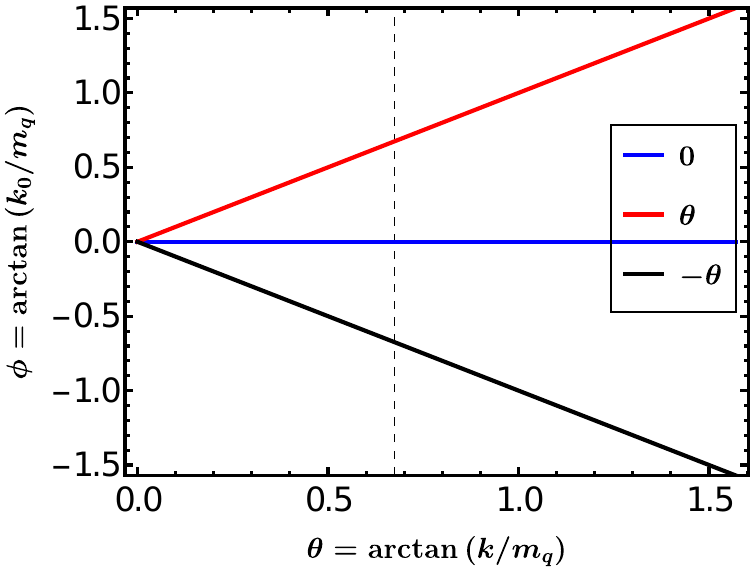}}
\caption{Domains in ($k,k_0$) plane at which the integrand in Eq.~\eqref{Sigma11_final} has sharp jumps. Here, we have used $ t \equiv p/p_{0} = 0.45 $ and $x\equiv \cos\theta = 0.8 $. }
\label{fig6} 
\end{figure}

\begin{figure}[t]
\centering
{\includegraphics[height=9cm, width=12cm]{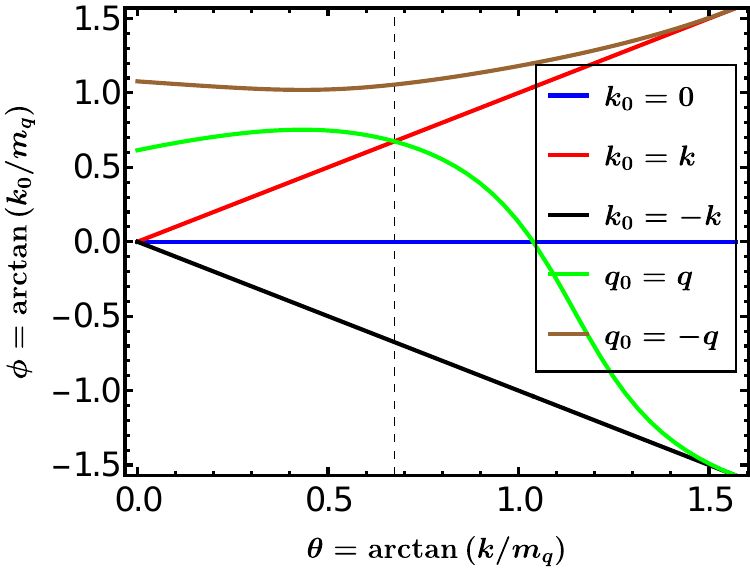}}
\caption{Domains in ($k,k_0$) plane at which the integrand in Eq.~(\eqref{Sigma13}) has divergences. Here we have used $ t \equiv p/p_{0} = 0.45 $ and $x \equiv \cos\theta = 0.8 $.} 
\label{fig7}    
\end{figure}

In this section, we will consider all the terms present in Eq.~\eqref{sigma_final} more elaborately and show how these terms have been evaluated. We can write Eq.~\eqref{sigma_final} as
\begin{align}
\Sigma_{\pm}^{(1)}(P) = & \, \Sigma_{1\pm}^{(1)}(P)+\Sigma_{2\pm}^{(1)}(P)+\Sigma_{3\pm}^{(1)}(P)+\Sigma_{4\pm}^{(1)}(P)+\Sigma_{5\pm}^{(1)}(P)+\Sigma_{6\pm}^{(1)}(P)+\Sigma_{7\pm}^{(1)}(P) \nn
+ & \, \Sigma_{8\pm}^{(1)}(P).\label{Sigma1-8}
\end{align}
We consider quark thermal mass to be $1$, i.e., $m_q$ = $1$. Thus, we will get
\begin{equation}
\frac{m_{g}^{2}}{m_{q}^{2}}=\frac{16 N_{c}\left(N_{c}+s_{F}\right)}{6\left(N_{c}^{2}-1\right)}=\left(3+s_{F}\right),\label{ratio}
\end{equation}
Here, we will fix the number of flavor and color charges. In the above equation, we have considered the value of $N_{c}$ to be $3$, and for the value of $N_{f} = 2$, this ratio in Eq.~\eqref{ratio} is equal to $4$, which we will consider in further computation. Now, we define a variable $t$ as the ratio of $ p/p_0$. The scaled quark momentum can be written in terms of the variable $t$ as 
\begin{equation}\label{t_vari}
p(t) / m_{q}=\sqrt{\frac{t}{1-t}-\frac{1}{2} \ln \left(\frac{1+t}{1-t}\right)} . 
\end{equation}
Since we have considered quark thermal mass $m_q = 1 $. Eq.~\eqref{t_vari} can be derived by using the leading-order quark dispersion relation $ \Delta_{-}^{-1}(p_0,p) = 0 $ 
The variation of the scaled quark momentum $p(t)$ and mass $p_0(t) = \Omega^{(0)}_{-}(p(t))$ is shown in Figure~\ref{fig5}. Also, for a slow-moving quark, one would get $p \lesssim m_q$, which gives us the limit on $t$ variable, i.e., $t \lesssim 0.64$. Beyond this value of the t variable, quarks will be considered fast-moving, which is not an interesting region.
In order to evaluate Eq.~\eqref{sigma_final}, we need retarded transverse $D_{T}^{R}(k,k_0,\varepsilon)$, retarded longitudinal $D_{L}^{R}(k,k_0,\varepsilon)$ gluon propagators which are derived by using Eq.~\eqref{glu_prop} and Eq.~\eqref{RAS_B} (see appendix~\ref{sec:appendix_A_3}) as.
\begin{align}
D_{T}^{R{(-1)}}(k,k_0,\varepsilon) = & \, -\left[\frac{4k_0^{2}}{k^{2}} + \left(k^2 - k_0^{2}\right) \left\{1 - \frac{k_0}{k^3} \ln \frac{(k_0 - k)^2 + \varepsilon^2}{(k_0+k)^2 + \varepsilon^2}\right\}\right]\nn
- & \,  i \left[\frac{2k_{0}}{k^3} \left(k_0^{2} - k^2 \right) \left\{\tan^{-1} \left(\frac{\varepsilon}{k_0 - k}\right) - \tan^{-1} \left(\frac{\varepsilon}{k_0 + k}\right) \right\} - \varepsilon \Theta(k_0) \right],
\end{align}
where $ \Theta(k_0)$ is the step function. Similarly, the longitudinal part of the gluon propagator is
\begin{align}
\tilde{D}_{L}^{R (-1)}(k,k_0,\varepsilon) = & \, \left(k_{0}^{2} -k^{2} \right)^2 \left[1 + \frac{8}{k^2} + \frac{2k_0}{k^3} \ln \frac{(k_0 - k)^2 + \varepsilon^2}{(k_0+k)^2 + \varepsilon^2}\right. \nn
+ & \, \left.i \left[\frac{4k_0}{k^3}  \left\{\tan^{-1} \left(\frac{\varepsilon}{k_0 - k}\right) -  \tan^{-1} \left(\frac{\varepsilon}{k_0 + k}\right) \right\} + \varepsilon \Theta(k_0) \right] \right].
\end{align}
The other quantity required is the retarded quark propagators $\Delta_{\pm}^{R}(q,q_{0},\varepsilon)$ and can be obtained by using Eq.~\eqref{q_prop_htl} and Eq.~\eqref{RAS_F} (For details see ~\ref{sec:appendix_A_3}).
\begin{align}
\Delta_{+}^{R(-1)}(q,q_0,\varepsilon) = & \, q_0 + q + \frac{1}{q} - \frac{q_0+q}{4q^2} \ln \frac{(q_{0}+q)^2+ \varepsilon^2}{(q_{0}-q)^2+ \varepsilon^2}+\frac{\varepsilon}{2 q^2} \left\{\tan^{-1} \left(\frac{\varepsilon}{q_0 + q }\right) \right.\nn
- & \, \left. \tan^{-1} \left(\frac{\varepsilon}{q_0-q}\right)\right\} - i \left[-\varepsilon+  \frac{\varepsilon}{4q^2} \ln  \frac{(q_{0}+q)^2+ \varepsilon^2}{(q_{0}-q)^2+ \varepsilon^2}\right.\nn
+ & \, \left. \frac{q_0+q}{2 q^2} \left\{\tan^{-1} \left( \frac{\varepsilon}{q_0 + q }\right) - \tan^{-1} \left(\frac{\varepsilon}{q_0-q}\right)\right\}\right].
\end{align}
Similarly, the retarded quark propagator for plasmino mode comes out to be  (see appendix~\ref{sec:appendix_A_3})
\begin{align}
\Delta_{-}^{R(-1)}(q,q_0,\varepsilon) = & \, -\left[\frac{1}{q} + q - q_0 - \frac{q_0-q}{4q^2} \ln \frac{(q_{0}+q)^2+ \varepsilon^2}{(q_{0}-q)^2+ \varepsilon^2}+\frac{\varepsilon}{2 q^2}\right. \left\{\tan^{-1} \left(\frac{\varepsilon}{q_0 + q }\right) \right. \nn
- & \, \left.\left. - \tan^{-1} \left(\frac{\varepsilon}{q_0-q}\right)\right\} - i \left[\varepsilon+  \frac{\varepsilon}{4q^2} \ln  \frac{(q_{0}+q)^2+ \varepsilon^2}{(q_{0}-q)^2+ \varepsilon^2}\right.\right.\nn
+ & \, \left.\left. \frac{q_0-q}{2 q^2} \left\{\tan^{-1} \left( \frac{\varepsilon}{q_0 + q }\right) - \tan^{-1} \left(\frac{\varepsilon}{q_0-q}\right)\right\}\right]\right].
\end{align}
Now, let us consider the first term of Eq.~\eqref{sigma_final}, which is 
\begin{equation}
\Sigma_{1\pm}^{(1)}(P)= \frac{-ig^2 C_F}{2} \int \frac{d^4 K}{(2\pi)^4}\Big[F_{\pm ; 0}^{\mathrm{SR}}(P, K)\Big],
\label{Sigma1_1}
\end{equation}
where $F_{\pm ; 0}^{\mathrm{SR}}(P, K) $ is given in Eq.~\eqref{bare_terms}. The inclination between $\vec{p}$ and $\vec{k}$ is $x (=\cos{\theta})$. Also, the angle between $\hat{p}$ and $\hat{q}$ is
\begin{equation}
\hat{p}\cdot\!\hat{q} = \frac{\vec{p}\!\cdot\!\vec{q}}{p q} =   \frac{\vec{p}.(\vec{p}-\vec{k})}{p|\vec{p}-\vec{k}|} = \frac{p^2 - p k x}{p q} = \frac{p}{q} - \frac{k}{q}x,\label{p.q}
\end{equation}
and
\begin{equation}
q = |\vec{q}| =  \sqrt{p^2+k^2-2pkx}.
\end{equation}
Similarly, the dot product of the $\hat{k}$ and $\hat{q}$ is
\begin{equation}
\hat{k}\!\cdot\!\hat{q} = \frac{\vec{k}\!\cdot\!\vec{q}}{k q} =   \frac{\vec{k}.(\vec{p}-\vec{k})}{k|\vec{p}-\vec{k}|} = \frac{pkx-k^2}{kq} = \frac{p}{q}x - \frac{k}{q};\label{k.q}
\end{equation}
%
The first term of Eq.~\eqref{Sigma1-8} can be written using Eq.~\eqref{bare_terms}
\begin{align}
\Sigma_{1\pm}^{(1)}(P) = & \, \frac{-ig^2 C_F}{2} \int \frac{d^4 K}{(2\pi)^4}\left[-2\Big(1-\hat{p}_{\epsilon}.\hat{k} \,\hat{q}_{\epsilon}.\hat{k}\Big) {D}^{S}_{T}(K) \Delta^{R}_{\mp}(Q) - \left[k_{0}^{2}\left(1-\hat{p}_{\epsilon}\mdot\hat{q}_{\epsilon}\right.\right.\right.\nn
+ & \, 2 \left. \hat{p}_{\epsilon}\mdot\hat{k} \,\hat{q}_{\epsilon}\mdot\hat{k}\right) - 2k_{0} k \left.\left.\left(\hat{p}_{\epsilon}\mdot\hat{k}+\hat{q}_{\epsilon}.\hat{k}\right)+k^{2}\left(1+\hat{p}_{\epsilon}\mdot\hat{q}_{\epsilon}\right)\right] \tilde{D}_{L}^{\mathrm{S}}(K) \Delta_{\mp}^{\mathrm{R}}(Q)\right].\label{Sigma11}
\end{align}
Now denoting $\hat{p}_{\epsilon} = \epsilon_{p} \hat{p}$, and using  Eqs.~\eqref{p.q}~\eqref{k.q}, Eq.~\eqref{Sigma11} becomes
\begin{align}
\Sigma _{1\pm}^{(1)}(P) = & \,  \frac{-ig^2 C_F}{2(2\pi)^4}  (2\pi) \int_{-\infty}^\infty dk_0 \int_{0}^\infty k^2 dk  \int_{-1}^1 dx \left[-2\left(1-x\left(\frac{p}{q}x - \frac{k}{q}\right)\right) D^{S}_{T}(K) \Delta^{R}_{\mp}(Q) \right.\nn
- & \, \left.\bigg\{k_{0}^{2}\left(1-\frac{p}{q} + \frac{k}{q}x +2 x\left(\frac{p}{q}x - \frac{k}{q}\right)\right)\mp 2k_{0} k\right.
\left(x+ \frac{px}{q} - \frac{k}{q} \right)\nn
+ & \, k^{2}\left(1+\frac{p}{q} - \frac{k}{q}x\right)\bigg\} \tilde{D}_{L}^{\mathrm{S}}(K) \Delta_{\mp}^{\mathrm{R}}(Q)\bigg]
\label{Sigma11_final}
\end{align}
Similarly, the second term of Eq.~\eqref{Sigma1-8} can be written using Eq.~\eqref{Sigma11_final} as
\begin{align}
\Sigma _{2\pm}^{(1)}(P) = & \,  \frac{-ig^2 C_F}{2(2\pi)^4}  (2\pi) \int_{-\infty}^\infty dk_0 \int_{0}^\infty k^2 dk  \int_{-1}^1 dx \left[-2\left(1-x\left(\frac{p}{q}x - \frac{k}{q}\right)\right)\right.\nn
\times & \, \left.  D^{A}_{T}(K) \Delta^{S}_{\mp}(Q)  \bigg\{k_{0}^{2}\left(1-\frac{p}{q} - \frac{k}{q}x+2 x\left(\frac{px}{q} - \frac{k}{q}\right)\right) \right.  \nn
\mp & \, 2k_{0} k \left(x+\frac{px}{q} - \frac{k}{q}\right) + k^{2}\left(1+\frac{p}{q} - \frac{kx}{q}\right)\bigg\} \,  \left. \tilde{D}_{L}^{\mathrm{A}}(K) \Delta_{\mp}^{\mathrm{S}}(Q)\right]
\label{Sigma12_final}
\end{align}
To evaluate Eq.~\eqref{Sigma11_final} and Eq.~\eqref{Sigma12_final} numerically, we encounter a few issues. One of those issues is that the integrand has a discontinuity because of the terms $\arctan$ in the propagators and the BE distribution function at $k_{0} = 0$. Such discontinuity would cause fatal issues in any integration method. So, the numerical outputs are either unreliable or produce unsatisfactory results. This instability in the results is more prolonged if we consider our tuning parameter $\varepsilon$ too small. So, to make further progress, we have partitioned the integration region into the domains bounded by the lines, which causes discontinuity as
\begin{equation}\label{diverg_pts}
\begin{aligned}
&k_{0}=0 ; \quad k_{0}=\pm k ; 
\quad k=k_{t} \equiv \frac{1}{2} \frac{p_{0}^{2}-p^{2}}{p_{0}-x p}=\frac{1}{2 t} \frac{1-t^{2}}{1-x t} \sqrt{\frac{t}{1-t}-\frac{1}{2} \ln \left(\frac{1+t}{1-t}\right)}.
\end{aligned} 
\end{equation}
By doing the swapping of variables $\theta = \tan^{-1}{k}$ and $\phi = \tan^{-1}{k_0}$, these domains are shown in Figure~\ref{fig6}. The last (vertical) line shown in Figure~\ref{fig6} is discontinuity line $ k = p_{0}-q $.
We evaluate Eqs.~\eqref{Sigma11_final} and Eq.~\eqref{Sigma12_final} in each of the domains as shown in Figure~\ref{fig6} separately or in the domains shown in Eq.~\eqref{diverg_pts} numerically and summed up to get the results. 
Now, the third term of Eq.~\eqref{Sigma1-8} is given by
\begin{equation}
\Sigma _{3\pm}^{(1)}(P) = \frac{-ig^2 C_F}{2} \int \frac{d^4K}{(2\pi)^4}\left[2F_{\pm ; --}^{\mathrm{SR}}(P, K)\right], \label{Sigma13}
\end{equation}
where $F_{\pm ; --}^{\mathrm{SR}}(P, K)$ can be written using Eq.~\eqref{htl_terms_1} as 
\begin{align}
F_{\pm;--}^{\mathrm{SR}}(P,K) = & \, m_{q}^{2} \int \frac{d \Omega_{s}}{4 \pi}\frac{1}{\left(PS-i\varepsilon\right)\left(QS-i\varepsilon\right)}\bigg[\Big\{1-\hat{p}_{\epsilon}\mdot\hat{q}_{\epsilon}-\hat{p}_{\epsilon}\mdot\hat{s}-\hat{q}_{\epsilon}\mdot\hat{s} + \hat{p}_{\epsilon}\mdot\hat{k} \,\hat{k}\mdot\hat{s} \nn
+ & \,\hat{q}_{\epsilon}\mdot\hat{k} \,\hat{k}\mdot\hat{s}-(\hat{k}\mdot\hat{s})^{2}+\hat{p}_{\epsilon}.\hat{q}_{\epsilon}\,(\hat{k}\mdot\hat{s})^{2} + 2 \hat{p}_{\epsilon}.\hat{s} \,\hat{q}_{\epsilon}.\hat{s} - \hat{p}_{\epsilon}.\hat{k} \,\hat{q}_{\epsilon}\mdot\hat{s}\,\hat{k}\mdot\hat{s}-\hat{q}_{\epsilon}\mdot\hat{k}\,\hat{p}_{\epsilon}\mdot \hat{s}\,\hat{k}\mdot\hat{s}\Big\}
\nn
\times & \, D_{T}^{\mathrm{S}}(K) \Delta_{\mp}^{\mathrm{R}}(Q) - \Big\{k_{0}^{2}\left(\hat{p}_{\epsilon}\mdot\hat{k} \, \hat{k}\mdot\hat{s}+\hat{q}_{\epsilon}\mdot\hat{k} \,\hat{k}\mdot\hat{s}-(\hat{k}\mdot\hat{s})^{2}\right. + \hat{p}_{\epsilon}\mdot\hat{q}_{\epsilon} \,(\hat{k}\mdot\hat{s})^{2} \nn
- & \, \left.\hat{p}_{\epsilon}\mdot\hat{k} \,\hat{q}_{\epsilon}\mdot\hat{s} \,\hat{k}.\hat{s}-\hat{q}_{\epsilon}.\hat{k} \,\hat{p}_{\epsilon}\mdot\hat{s}\, \hat{k}\mdot\hat{s}\right) - k_{0} k \left(\hat{p}_{\epsilon}\mdot\hat{k}+\hat{q}_{\epsilon}\mdot\hat{k}+2 \hat{p}_{\epsilon}\mdot\hat{q}_{\epsilon} \,\hat{k}\mdot\hat{s} -\hat{q}_{\epsilon}\mdot\hat{k} \,\hat{p}_{\epsilon}\mdot\hat{s}\right.\nn
- & \, \left. \hat{p}_{\epsilon}\mdot\hat{k} \, \hat{q}_{\epsilon}\mdot\hat{s} - \hat{q}_{\epsilon}\mdot\hat{s} \, \hat{k}\mdot\hat{s}-\hat{p}_{\epsilon}\mdot\hat{s} \,\hat{k}\mdot\hat{s}\right) +k^{2}\left(1+\hat{p}_{\epsilon}\mdot\hat{q}_{\epsilon}-\hat{p}_{\epsilon}\mdot\hat{s}-\hat{q}_{\epsilon}\mdot\hat{s}\right)\Big\}\nn
\times & \, \tilde{D}_{L}^{\mathrm{S}}(K) \Delta_{\mp}^{\mathrm{R}}(Q)\bigg].\label{F_pm_SR_--}
\end{align}
If $F_{1\pm;--}^{SR}(P,K)$ denote the terms in Eq.~\eqref{F_pm_SR_--} without $ \hat{s}$, then
\begin{align}
F_{1\pm;--}^{SR}(P,K)  = & \,  m_{q}^{2} \int \frac{d \Omega_{s}}{4 \pi} \frac{1}{\left(P S-i \varepsilon\right)\left(Q S-i \varepsilon\right)} \bigg[\left(1-\hat{p}_{\epsilon}\mdot\hat{q}_{\epsilon}\right){D}_{T}^{\mathrm{S}}(K) \Delta_{\mp}^{\mathrm{R}}(Q) \nn
- & \,  \left\{k^{2}\left(1+\hat{p}_{\epsilon}\mdot\hat{q}_{\epsilon}\right) \right. \left.- k_{0} k \left(\hat{p}_{\epsilon}\mdot\hat{k}+\hat{q}_{\epsilon}\mdot\hat{k}\right) \right\}\tilde{D}_{L}^{\mathrm{S}}(K) \Delta_{\mp}^{\mathrm{R}}(Q)\bigg].
\label{F1_pm_SR_--}
\end{align}
The solid angle integration in Eq.~\eqref{F1_pm_SR_--} can be computed using Eq.~\eqref{J00_2} as 
\ba
\int \frac{d \Omega_{s}}{4 \pi} \frac{1}{\left(P S-i \varepsilon\right)\left(Q S-i \varepsilon\right)} = V\left(t,k,k_0,x,\varepsilon\right),\label{V_k_k0}
\ea
where
\begin{align}
V\left(t, k, k_0, x, \varepsilon\right) = & \, \int_{0}^{1} d u \frac{1}{Z u^{2}+2 Y u + X} \nn
= & \, \frac{1}{2 \sqrt{\Delta}}\left[\frac { 1 } { 2 } \ln \frac{\left(\left(r_{1}-1\right)^{2}+i_{1}^{2}\right)\left(r_{2}^{2}+i_{2}^{2}\right)}{\left(\left(r_{2}-1\right)^{2}+i_{2}^{2}\right)\left(r_{1}^{2}+i_{1}^{2}\right)} + \ i\left\{\tan^{-1} \left(\frac{r_{1}}{i_{1}}\right)\right.\right. \nn
- & \, \tan^{-1} \left(\frac{r_{1}-1}{i_{1}}\right) \left.\left.-\tan^{-1} \left(\frac{r_{2}}{i_{2}}\right)+\tan^{-1} \left(\frac{r_{2}-1}{i_{2}}\right)\right\}\right]. 
\label{func_V}
\end{align}
Here, $r_{i}=\operatorname{Re} s_{i}$ and $i_{i}=\operatorname{Im} s_{i}$ with
$
s_{1}=\frac{-Y+\sqrt{\Delta}}{Z} ; \quad s_{2}=\frac{-Y-\sqrt{\Delta}}{Z} ; \quad \Delta=Y^{2}-X Z,
$
and
\begin{equation}
\begin{aligned}
&X=\left(p_{0}-i \varepsilon\right)^{2}-p^{2}; \quad
&Y=\vec{k}\!\cdot\!\vec{p}-\left(k_{0}\right)\left(p_{0}-i \varepsilon\right); \quad Z=k_{0}^{2}-k^{2};
\end{aligned}
\end{equation}
Thus, Eq.~\eqref{F1_pm_SR_--} becomes
\begin{align}
F_{1\pm;--}^{SR}(P,K) = & \, V\left(t,k,k_0,x,\varepsilon\right) \bigg[\left(1-\frac{p}{q} + \frac{k}{q}x\right) D^{S}_{T}(K) \Delta^{R}_{\mp}(Q) -\bigg\{k^{2}\bigg(1+\frac{p}{q} - \frac{k}{q}x \bigg) \nn
 \mp & \, k_{0} k \left(x+ \frac{p}{q}x - \frac{k}{q} \right)\bigg\} \tilde{D}_{L}^{\mathrm{S}}(K) \Delta_{\mp}^{\mathrm{R}}(Q) \bigg] .
 \label{F1pm}
\end{align}
With the expression of $F_{1\pm;--}^{SR}(P,K)$ from Eq.~\eqref{F1pm} , the terms of Eq.~\eqref{Sigma13} without s becomes
\begin{align}
\Sigma _{3(1)\pm}^{(1)}(P) = & \, \frac{-ig^2 C_F}{(2\pi)^4}  (2\pi) \int_{-\infty}^\infty dk_0 \int_{0}^\infty k^2 dk \int_{-1}^1 dx \,\, V\left(t,k,k_0,x,\varepsilon\right) \bigg[ \left(1-\frac{p}{q} + \frac{k}{q}x\right) \nn 
\times & \, D^{S}_{T}(K) \Delta^{R}_{\mp}(Q) -\bigg\{k^{2}\bigg(1+ \frac{p}{q} - \frac{k}{q}x \bigg) \mp k_{0} k \left(x+ \frac{p}{q}x - \frac{k}{q} \right)\bigg\}\nn
\times & \, \tilde{D}_{L}^{\mathrm{S}}(K) \Delta_{\mp}^{\mathrm{R}}(Q) \bigg] .
\label{Sigma13_1}
\end{align}
Now consider the terms with $ \hat{s}$ in Eq.~\eqref{F_pm_SR_--}
\begin{align}
F_{2\pm; - -}^{SR}(P,K) = & \, \int \frac{d \Omega_{s}}{4 \pi} \frac{1}{\left(P S-i\varepsilon\right)\left(Q S-i\varepsilon\right)}\bigg[\Big(-\hat{p}_{\epsilon} \mdot \hat{s}-\hat{q}_{\epsilon}.\hat{s} +\ \hat{p}_{\epsilon}.\hat{k} \,\hat{k}.\hat{s}+\hat{q}_{\epsilon}.\hat{k} \,\hat{k}.\hat{s}\Big) 
\nn
\times & \, {D}_{T}^{\mathrm{S}}(K) \Delta_{\mp}^{\mathrm{R}}(Q) - \left\{k_{0}^{2}\left(\hat{p}_{\epsilon}\mdot\hat{k} \, \hat{k}\mdot\hat{s}+\hat{q}_{\epsilon}\mdot\hat{k} \,\hat{k}\mdot\hat{s}\right) - k_{0} k \left(2 \hat{p}_{\epsilon}\mdot\hat{q}_{\epsilon} \,\hat{k}\mdot\hat{s} \right.\right.\nn
- & \, \left.\left. \hat{q}_{\epsilon}\mdot\hat{k} \,\hat{p}_{\epsilon}\mdot\hat{s}-\hat{p}_{\epsilon}\mdot\hat{k} \, \hat{q}_{\epsilon}\mdot\hat{s}\right) -k^{2}\left(\hat{p}_{\epsilon}\mdot\hat{s}+\hat{q}_{\epsilon}\mdot\hat{s}\right)\right\}
 \tilde{D}_{L}^{\mathrm{S}}(K) \Delta_{\mp}^{\mathrm{R}}(Q)\bigg]
 \label{F2_pm_SR_--}
\end{align}
By using Eq.~\eqref{J0i}, we get
\begin{align}
\int \frac{d \Omega_{s}}{4 \pi} \frac{\hat{s}^{i}}{\left(P S-i \varepsilon\right)\left(Q S-i \varepsilon\right)} = & \, 
\int_{0}^{1} d u\left[\frac{r_{0}}{r_{0}^{2}-r^{2}}-\frac{1}{2 r} \ln \frac{r_{0}+r}{r_{0}-r}\right] \frac{r^{i}}{r^{2}} \nn
= & \, \int_{0}^{1} d u \, V_1 \, \hat{r}^{i},
\label{1s_int}
\end{align}
where \begin{equation}
V_1(r_0,r) = \left[\frac{r_{0}}{r_{0}^{2}-r^{2}}-\frac{1}{2 r} \ln \frac{r_{0}+r}{r_{0}-r}\right] \frac{1}{r},
\end{equation}
with
\begin{equation}
r_0 = p_0 - i\varepsilon - k_0 u  ; \quad r = \sqrt{p^2+k^2u^2-2pkux}.
\end{equation}
Now, the required angles in order to solve Eq.~\eqref{F2_pm_SR_--} are
\ba
\hat{p}\!\cdot\!\hat{r} &=& \frac{p}{r} - \frac{k u}{r}x, \quad
\hat{k}\!\cdot\!\hat{r} = \frac{p}{r}x - \frac{k u}{r}, \quad
\hat{q}\!\cdot\!\hat{r} =  \frac{p^2}{qr} - \frac{pkux}{qr}-\frac{pkx}{qr} + \frac{k^2 u}{qr}.
\ea
%
Thus, Eq.~\eqref{F2_pm_SR_--} becomes
\begin{align}
F_{2\pm;--}^{SR}(P,K) = & \, \pm \int_{0}^{1} d u \,V_1(r_0,r)\,\bigg[ \frac{p(p+q)(x^2-1)}{qr}
{D}_{T}^{\mathrm{S}}(K) \Delta_{\mp}^{\mathrm{R}}(Q) +
\bigg\{ k^{2}\bigg(\frac{p}{r}\nn
- & \, \frac{ku}{r}x  + \frac{p^2}{qr} - \frac{pkux}{qr}-\frac{pkx}{qr} 
+ \frac{k^2 u}{qr}\bigg) \mp \frac{k_0k^2p(2u-1)(x^2-1)}{qr} \nn
- & \, k_{0}^{2}\bigg(x+\frac{px}{q} -\frac{k}{q}\bigg)\left(\frac{px}{r}-\frac{ku}{r}\right)  \bigg\}\tilde{D}_{L}^{\mathrm{S}}(K) \Delta_{\mp}^{\mathrm{R}}(Q)\bigg]
\label{F2_PM_SR_--_Sigma13}
\end{align}
So, the $\hat{s}$ contribution in the third term of Eq.~\eqref{Sigma1-8} is 
\begin{equation}\label{Sigma13_2}
\Sigma _{3(2)\pm}^{(1)}(P) = \frac{-ig^2 C_F}{(2\pi)^4}  (2\pi) \int_{-\infty}^\infty dk_0 \int_{0}^\infty k^2 dk  \int_{-1}^1 dx \, F_{2\pm;--}^{SR}(P,K)
\end{equation}
%
Let us consider terms with two $\hat{s}$ in Eq.~\eqref{F_pm_SR_--}
\begin{align}
F_{3\pm;--}^{SR}(P,K) = & \, \int \frac{d \Omega_{s}}{4 \pi} \frac{\hat{s}^{i} \hat{s}^{j}}{\left(P S-i\varepsilon\right)\left(Q S-i\varepsilon\right)}\bigg[\bigg(-\hat{k}_{i}\hat{k}_{j}+\hat{p}.\hat{q}\, \hat{k}_{i}\hat{k}_{j}+2 \hat{p}_{i} \, \hat{q}_{j}-\hat{p}.\hat{k} \, \hat{q}_{i}\hat{k}_{j}\nn
- & \, \hat{q}\!\cdot\!\hat{k} \, \hat{p}_{i} \,\hat{k}_{j}\bigg) \, {D}_{T}^{\mathrm{S}}(K) \Delta_{\mp}^{\mathrm{R}}(Q)\nn
+ & \, \bigg\{k_{0}^{2}\bigg((\hat{k}\mdot\hat{s})^{2} - \hat{p}_{\epsilon}\mdot\hat{q}_{\epsilon} \,(\hat{k}\mdot\hat{s})^{2} + \hat{p}_{\epsilon}\mdot\hat{k} \,\hat{q}_{\epsilon}\mdot\hat{s} \,\hat{k}.\hat{s}+\hat{q}_{\epsilon}\mdot\hat{k} \,\hat{p}_{\epsilon}\mdot\hat{s}\, \hat{k}\mdot\hat{s}\bigg) \nn
- & \, k_{0} k \left(\hat{q}_{\epsilon}\mdot\hat{s} \, \hat{k}\mdot\hat{s} + \hat{p}_{\epsilon}\mdot\hat{s} \,\hat{k}\mdot\hat{s}\right)\bigg\} \, \tilde{D}_{L}^{\mathrm{S}}(K) \Delta_{\mp}^{\mathrm{R}}(Q) \bigg]\label{F3sr2}
\end{align}
From Eq.~\eqref{Jij_int}, we can write
\begin{equation}\label{2s_int}
\begin{aligned}
\int \frac{d \Omega_{s}}{4 \pi} \frac{\hat{s}^{i} \hat{s}^{j}}{\left(P S-i \varepsilon\right)\left(Q S-i \varepsilon\right)} = \int_{0}^{1} d u\left(A \delta^{i j}+B \hat{r}^{i} \hat{r}^{j}\right),
\end{aligned}
\end{equation}
where \begin{equation}\label{A_B_Def}
\begin{aligned}
A &=-\frac{1}{r^{2}}\left(1-\frac{r_{0}}{2 r} \ln \frac{r_{0}+r}{r_{0}-r}\right) ,\quad \quad
B &=\frac{1}{r_{0}^{2}-r^{2}}+\frac{3}{r^{2}}\left(1-\frac{r_{0}}{2 r} \ln \frac{r_{0}+r}{r_{0}-r}\right)
\end{aligned}
\end{equation}
Using Eq.~\eqref{2s_int}, Eq.~\eqref{F3sr2} becomes
\begin{align}
F_{3\pm;--}^{SR}(P,K) = & \, \int_{0}^{1} d u \bigg[\bigg\{ A \left(3 \hat{p}\mdot\hat{q} - 1 -2 x \,\hat{q} \mdot \hat{k}\right) + B \left(\big(\hat{k}\!\cdot\!\hat{r}\big)^2 \,\left(\hat{p}\!\cdot\!\hat{q} -1\right) \right. \nn
+ & \, \left. 2\hat{p}.\hat{r} \, \hat{q}.\hat{r} - x \, \hat{q}.\hat{r} \,\hat{k}.\hat{r} - \hat{q}\!\cdot\!\hat{k} \, \hat{p}\!\cdot\!\hat{r} \, \hat{k}\!\cdot\!\hat{r} \right)\bigg\} {D}_{T}^{\mathrm{S}}(K) \Delta_{\mp}^{\mathrm{R}}(Q) \nn
- & \, \bigg\{A\bigg(k_{0}^{2} \big(\hat{p}\!\cdot\!\hat{q} - 1 -2 x \,\hat{q}\!\cdot\!\hat{k}\big)\pm k_{0}k \big(x+\,\hat{q}.\hat{k}\big) \bigg)+ B\bigg(k_{0}^{2}\bigg(\big(\hat{k}\!\cdot\!\hat{r}\big)^2 \, \nn
\times & \, \left(\hat{p}\!\cdot\!\hat{q} -1\right) -\hat{q}.\hat{k} \,\hat{p}\mdot\hat{r} \, \hat{k}\mdot\hat{r}  - x \, \hat{q}\!\cdot\!\hat{r} \,\hat{k}\!\cdot\!\hat{r}\bigg)\pm k_{0}k \, \hat{k}\!\cdot\!\hat{r} \left(\hat{p}\!\cdot\!\hat{r} + \hat{q}\!\cdot\!\hat{r} \right)\bigg)\bigg\}\nn 
\times & \, \tilde{D}_{L}^{\mathrm{S}}(K) \Delta_{\mp}^{\mathrm{R}}(Q)\bigg] 
\end{align}
Thus, the contribution coming from 2s terms of Eq.~\eqref{Sigma13} is 
\begin{equation}\label{Sigma13_3}
\Sigma _{3(3)\pm}^{(1)}(P) = \frac{-ig^2 C_F}{(2\pi)^4}  (2\pi) \int_{-\infty}^\infty dk_0 \int_{0}^\infty k^2 dk  \int_{-1}^1 dx \, F_{3\pm;--}^{SR}(P,K)
\end{equation}
As we have seen in Eq.~\eqref{Sigma11_final}, there are sudden jumps in the integrand from the gluon propagator, which causes instability. In this term, an additional divergence will come from Eq.~\eqref{func_V} i.e., from $V(t,k,k_0,x,\varepsilon)$. The divergence that comes from Eq.~\eqref{1s_int} and Eq.\eqref{2s_int} is the same as we get from Eq.~\eqref{func_V}. Thus, all the lines of discontinuity are
\begin{align}
k_{0} = & \, 0 ; \quad k_{0}=\pm k ; \quad
k_0 = \, p_0 \pm \sqrt{p^2+k^2-2pkx} ; \nn
k = &  \, k_{t} = \frac{1}{2 t} \frac{1-t^{2}}{1-x t} \sqrt{\frac{t}{1-t}-\frac{1}{2} \ln \left(\frac{1+t}{1-t}\right)}
\label{diverg_pts_1}
\end{align}
These domains are shown in Figure~\ref{fig7}.
We have evaluated  Eqs.~\eqref{Sigma13_1}, \eqref{Sigma13_2}, \eqref{Sigma13_3} in each of the domains of Figure~\ref{fig7} numerically and summed up the results. 
Let us consider the fourth term of Eq.~\eqref{sigma_final}, which is 
\begin{equation}\label{Sigma1_4}
\Sigma _{4\pm}^{(1)}(P) = \frac{-ig^2 C_F}{2} \int \frac{d^4K}{(2\pi)^4}\Big[F_{\pm ; --}^{\mathrm{AS}}(P, K)\Big]
\end{equation}
This term is analogous to the third term of Eq.~\eqref{sigma_final} except for the change of real-time field indices (i.e., $SR \rightarrow AS$) and a factor of $1/2$. Thus, by using Eq.~\eqref{Sigma13_1}, we will get  
\begin{align}
\Sigma _{4(1)\pm}^{(1)}(P)  = & \, \frac{-ig^2 C_F}{2(2\pi)^4}  (2\pi) \int_{-\infty}^\infty dk_0 \int_{0}^\infty k^2 dk \int_{-1}^1 dx \,\,V\left(t,k,k_0,x,\varepsilon\right) \bigg[\left(1-\frac{p}{q} + \frac{k}{q}x\right) \nn 
\times & \, D^{A}_{T}(K) \Delta^{S}_{\mp}(Q) - \bigg\{k^{2}\bigg(1+\frac{p}{q} - \frac{k}{q}x\bigg) \mp k_{0} k \left(x+\frac{p}{q}x - \frac{k}{q}\right)\bigg\}\nn
\times & \, \tilde{D}_{L}^{\mathrm{A}}(K) \Delta_{\mp}^{\mathrm{S}}(Q) \bigg] .
\label{Sigma14_1}
\end{align}
By using Eq.~\eqref{Sigma13_2}, we get
\begin{equation}\label{sigma14_2}
\Sigma _{4(2)\pm}^{(1)}(P) = \frac{-ig^2 C_F}{2(2\pi)^4}  (2\pi) \int_{-\infty}^\infty dk_0 \int_{0}^\infty k^2 dk  \int_{-1}^1 dx \, F_{2\pm;--}^{AS}(P,K)
\end{equation}
where
\begin{align}
F_{2\pm;--}^{AS}(P,K) = & \, \int_{0}^{1} d u \,V_1(r_0,r)\,\bigg[ \pm \frac{p(p+q)(x^2-1)}{qr}
{D}_{T}^{\mathrm{A}}(K) \Delta_{\mp}^{\mathrm{S}}(Q) -\bigg\{\mp k^{2} \nn
\times & \, \bigg(\frac{p}{r}-\frac{ku}{r}x +\frac{p^2}{qr} - \frac{pkux}{qr}-\frac{pkx}{qr} 
+ \frac{k^2 u}{qr}\bigg) - \frac{k_0k^2p(2u-1)(x^2-1)}{qr} \nn 
\pm & \, k_{0}^{2}\bigg(x+\frac{px}{q} -\frac{k}{q}\bigg)\left(\frac{px}{r}-\frac{ku}{r}\right)  \bigg\}\tilde{D}_{L}^{\mathrm{A}}(K) \Delta_{\mp}^{\mathrm{S}}(Q)\bigg]
\end{align}
Similarly, by using Eq.~\eqref{Sigma13_3}, we will get
\begin{equation}
\Sigma _{4(3)\pm}^{(1)}(P) = \frac{-ig^2 C_F}{2(2\pi)^4}  (2\pi) \int_{-\infty}^\infty dk_0 \int_{0}^\infty k^2 dk  \int_{-1}^1 dx \, F_{3\pm;--}^{AS}(P,K)
\end{equation}
where
\begin{align}
F_{3\pm;--}^{AS}(P,K) = & \, \int_{0}^{1} d u \bigg[\bigg\{ A \left(3 \hat{p}\!\cdot\!\hat{q} - 1 -2 x \,\hat{q}\!\cdot\!\hat{k}\right) + B \bigg(\left(\hat{k}\!\cdot\!\hat{r}\right)^2 \left(\hat{p}\!\cdot\!\hat{q} -1\right) \nn
+ & \, 2\hat{p}\!\cdot\!\hat{r} \, \hat{q}\!\cdot\!\hat{r} - x \, \hat{q}\!\cdot\!\hat{r} \,\hat{k}\!\cdot\!\hat{r} - \hat{q}\!\cdot\!\hat{k} \, \hat{p}.\hat{r} \, \hat{k}.\hat{r} \bigg)\bigg\} \, {D}_{T}^{\mathrm{A}}(K) \Delta_{\mp}^{\mathrm{S}}(Q)\nn
- & \, k_0^2\bigg\{A\bigg( \hat{p}\!\cdot\!\hat{q} - 1 -2 x \,\hat{q}\!\cdot\!\hat{k} \pm \frac{k}{k_0} \left(x+\,\hat{q}\!\cdot\!\hat{k}\right) \bigg)+ B\bigg(\left(\hat{k}\!\cdot\!\hat{r}\right)^2 \,\left(\hat{p}\!\cdot\!\hat{q} -1\right) \nn
- & \, \hat{q}\!\cdot\!\hat{k} \,\hat{p}\!\cdot\!\hat{r} \, \hat{k}\!\cdot\!\hat{r}  - x \, \hat{q}.\hat{r} \,\hat{k}\!\cdot\!\hat{r} \pm \frac{k}{k_0} \, \hat{k}\!\cdot\!\hat{r} \left(\hat{p}.\hat{r} + \hat{q}\!\cdot\!\hat{r} \right)\bigg)\bigg\}\tilde{D}_{L}^{\mathrm{A}}(K) \Delta_{\mp}^{\mathrm{S}}(Q)\bigg] 
\end{align}
Let us consider the fifth term of Eq.~\eqref{sigma_final}, which is 
\begin{equation}\label{Sigma1_5}
\Sigma _{5\pm}^{(1)}(P)= \frac{-ig^2 C_F}{2} \int \frac{d^4K}{(2\pi)^4}\Big[F_{\pm ; -+}^{\mathrm{AS}}(P, K)\Big]
\end{equation}
This term is analogous to the fourth term of Eq.~\eqref{sigma_final}) except for the change of $ k_0$. Here, $k_0 \rightarrow k_0 - 2 i \varepsilon $ in the definition of the function $V(t,k,k_{0},x,\varepsilon) , A$ and $B$. 
So, $\Sigma 5_{\pm}^{(1)}(P)$ becomes
\begin{align}
\Sigma _{5(1)\pm}^{(1)}(P) = & \, \frac{-ig^2 C_F}{2(2\pi)^4}  (2\pi) \int_{-\infty}^\infty dk_0 \int_{0}^\infty k^2 dk \int_{-1}^1 dx \,\,V\left(t,k,k_0,x,\varepsilon\right) \bigg[\left\{1-\,\left(\frac{p}{q} - \frac{k}{q}x\right) \right.\nn 
\times & \, \left. D^{A}_{T}(K) \Delta^{S}_{\mp}(Q)\right\} -\bigg\{k^{2}\bigg(1+\left(\frac{p}{q} - \frac{k}{q}x\right)\bigg) \mp k_{0} k \left(x+\left(\frac{p}{q}x - \frac{k}{q}\right)\right)\bigg\}\nn
\times & \, \tilde{D}_{L}^{\mathrm{A}}(K) \Delta_{\mp}^{\mathrm{S}}(Q) \bigg] .\label{Sigma15_1}
\end{align}
Similarly, using Eq.~\eqref{sigma14_2}, one can get
\begin{equation}
\Sigma _{5(2)\pm}^{(1)}(P) = \frac{-ig^2 C_F}{2(2\pi)^4}  (2\pi) \int_{-\infty}^\infty dk_0 \int_{0}^\infty k^2 dk  \int_{-1}^1 dx \, F_{2\pm;-+}^{AS}(P,K)
\end{equation}
where
\begin{align}
F_{2\pm;--}^{AS}(P,K) = & \, \int_{0}^{1} d u \,V_1(r_0,r)\,\bigg[ \pm \frac{p(p+q)(x^2-1)}{qr}
{D}_{T}^{\mathrm{A}}(K) \Delta_{\mp}^{\mathrm{S}}(Q) -\bigg\{\mp k^{2} \nn
\times & \, \bigg(\frac{p}{r}-\frac{ku}{r}x +\frac{p^2}{qr} - \frac{pkux}{qr}-\frac{pkx}{qr} 
+ \frac{k^2 u}{qr}\bigg) - \frac{k_0k^2p(2u-1)(x^2-1)}{qr} \nn 
\pm & \, k_{0}^{2}\bigg(x+\frac{px}{q} -\frac{k}{q}\bigg)\left(\frac{px}{r}-\frac{ku}{r}\right)  \bigg\}\tilde{D}_{L}^{\mathrm{A}}(K) \Delta_{\mp}^{\mathrm{S}}(Q)\bigg]
\end{align}
The contribution of the terms with 2$\hat{s}$ of Eq.~\eqref{Sigma1_5} is given by 
\begin{equation}
\Sigma _{5(3)\pm}^{(1)}(P) = \frac{-ig^2 C_F}{2(2\pi)^4}  (2\pi) \int_{-\infty}^\infty dk_0 \int_{0}^\infty k^2 dk  \int_{-1}^1 dx \, F_{3\pm;-+}^{AS}(P,K),
\end{equation}
where 
\begin{align}
F_{3\pm;--}^{AS}(P,K) = & \, \int_{0}^{1} d u \bigg[\bigg\{ A \left(3 \hat{p}\!\cdot\!\hat{q} - 1 -2 x \,\hat{q}\!\cdot\!\hat{k}\right) + B \bigg(\left(\hat{k}\!\cdot\!\hat{r}\right)^2 \left(\hat{p}\!\cdot\!\hat{q} -1\right) \nn
+ & \, 2\hat{p}\!\cdot\!\hat{r} \, \hat{q}\!\cdot\!\hat{r} - x \, \hat{q}\!\cdot\!\hat{r} \,\hat{k}\!\cdot\!\hat{r} - \hat{q}\!\cdot\!\hat{k} \, \hat{p}.\hat{r} \, \hat{k}.\hat{r} \bigg)\bigg\} \, {D}_{T}^{\mathrm{A}}(K) \Delta_{\mp}^{\mathrm{S}}(Q)\nn
- & \, k_0^2\bigg\{A\bigg( \hat{p}\!\cdot\!\hat{q} - 1 -2 x \,\hat{q}\!\cdot\!\hat{k} \pm \frac{k}{k_0} \left(x+\,\hat{q}\!\cdot\!\hat{k}\right) \bigg)+ B\bigg(\left(\hat{k}\!\cdot\!\hat{r}\right)^2 \,\left(\hat{p}\!\cdot\!\hat{q} -1\right) \nn
- & \, \hat{q}\!\cdot\!\hat{k} \,\hat{p}\!\cdot\!\hat{r} \, \hat{k}\!\cdot\!\hat{r}  - x \, \hat{q}.\hat{r} \,\hat{k}\!\cdot\!\hat{r} \pm \frac{k}{k_0} \, \hat{k}\!\cdot\!\hat{r} \left(\hat{p}.\hat{r} + \hat{q}\!\cdot\!\hat{r} \right)\bigg)\bigg\}\tilde{D}_{L}^{\mathrm{A}}(K) \Delta_{\mp}^{\mathrm{S}}(Q)\bigg] 
\end{align}
Now, the sixth term of Eq.~\eqref{sigma_final} is 
\begin{equation}\label{Sigma1_6}
\Sigma _{6\pm}^{(1)}(P) = \frac{-ig^2 C_F}{2} \int \frac{d^4K}{(2\pi)^4}\left[F_{\pm ; --;--}^{\mathrm{SR}}(P, K)\right]
\end{equation}
The terms involved in $ F_{\pm ; --;--}^{\mathrm{SR}}(P, K) $ are
\begin{align}
F_{\pm ; -- ; --}^{\mathrm{SR}}(P, K) 
= & \, \int \frac{d \Omega_{s}}{4 \pi} \frac{1}{\big(P S-i \varepsilon\big)\left(Q S-i \varepsilon\right)}\times \int \frac{d \Omega_{s^{\prime}}}{4 \pi} \frac{1}{\left(P S^{\prime}-i \varepsilon\right)\left(Q S^{\prime}-i  \varepsilon\right)}\nn
\times & \,\bigg[\bigg\{\bigg(-\hat{s}\mdot\hat{s}^{\prime}-\hat{p}_{\epsilon}\mdot\hat{q}_{\epsilon}\, \hat{s}\mdot\hat{s}^{\prime}+\hat{p}_{\epsilon}\mdot\hat{s} \, \hat{s}\mdot\hat{s}^{\prime}+\hat{q}_{\epsilon}\mdot\hat{s} \, \hat{s}\mdot\hat{s}^{\prime}+\hat{p}_{\epsilon}\mdot\hat{s}^{\prime} \, \hat{s}\mdot\hat{s}^{\prime}+\hat{q}_{\epsilon}\mdot \hat{s}^{\prime} \, \hat{s}\mdot\hat{s}^{\prime} \nn
-& \, \hat{p}_{\epsilon}\mdot\hat{s} \,\hat{q}_{\epsilon}\mdot\hat{s}^{\prime} \, \hat{s}.\hat{s}^{\prime} - \hat{q}_{\epsilon}\mdot \hat{s} \,\hat{p}_{\epsilon}\mdot\hat{s}^{\prime} \, \hat{s}\mdot\hat{s}^{\prime}  -(\hat{s}\mdot\hat{s}^{\prime})^2 + \hat{p}_{\epsilon}\mdot\hat{q}_{\epsilon}\,(\hat{s}\mdot\hat{s}^{\prime})^2+(\hat{k}\mdot\hat{s})\,(\hat{k}\mdot\hat{s}^{\prime})   \nn
+& \, \hat{p}_{\epsilon}\mdot\hat{q}_{\epsilon}\,\hat{k}\mdot\hat{s}\,\hat{k}\mdot\hat{s}^{\prime} - \hat{p}_{\epsilon}\mdot \hat{s} \,\hat{k}\mdot\hat{s}\,\hat{k}\mdot\hat{s}^{\prime} -\hat{q}_{\epsilon}\mdot \hat{s} \,\hat{k}\mdot\hat{s}\,\hat{k}\mdot\hat{s}^{\prime} -\hat{p}_{\epsilon}\mdot\hat{s}^{\prime} \, \hat{k}\mdot\hat{s}\, \hat{k}\mdot\hat{s}^{\prime} -\hat{q}_{\epsilon}\mdot \hat{s}^{\prime} \, \hat{k}\mdot\hat{s} \, \hat{k}\mdot\hat{s}^{\prime}\nn
+& \, \hat{p}_{\epsilon}\mdot\hat{s} \, \hat{q}_{\epsilon}\mdot\hat{s}^{\prime}\, \hat{k}\mdot\hat{s} \, \hat{k}\mdot\hat{s}^{\prime}+\hat{q}_{\epsilon}\mdot\hat{s} \, \hat{p}_{\epsilon} \mdot\hat{s}^{\prime}\, \hat{k}\mdot\hat{s} \,\hat{k}\mdot\hat{s}^{\prime} +\hat{s}\mdot\hat{s}^{\prime} \,\hat{k}\mdot\hat{s} \,\hat{k}\mdot\hat{s}^{\prime} -\hat{p}_{\epsilon}\mdot\hat{q}_{\epsilon} \, \hat{s}\mdot\hat{s}^{\prime} \, \hat{k}\mdot\hat{s} \, \hat{k}\mdot\hat{s}^{\prime}\bigg)\nn
\times & D_{T}^{\mathrm{S}}(K) \Delta_{\mp}^{\mathrm{R}}(Q) \bigg\}\nn  
-& \,\bigg\{k^{2}\bigg(1+\hat{p}_{\epsilon}.\hat{q}_{\epsilon}-\hat{p}_{\epsilon}\mdot\hat{s} -\hat{q}_{\epsilon}\mdot\hat{s}-\hat{p}_{\epsilon}\mdot\hat{s}^{\prime}-\hat{q}_{\epsilon}\mdot\hat{s}^{\prime}+\hat{p}_{\epsilon}\mdot\hat{s} \,\hat{q}_{\epsilon}\mdot\hat{s}^{\prime}+\hat{q}_{\epsilon}\mdot\hat{s} \,\hat{p}_{\epsilon}\mdot\hat{s}^{\prime}\nn
+& \, \hat{s}.\hat{s}^{\prime}-\hat{p}_{\epsilon}. \hat{q}_{\epsilon}\,\hat{s}.\hat{s}^{\prime}\bigg) - k_{0} k\bigg(\hat{k}.\hat{s} + \hat{k}.\hat{s}^{\prime}+\hat{p}_{\epsilon}.\hat{q}_{\epsilon} \,\hat{k}. \hat{s}  +\hat{p}_{\epsilon}.\hat{q}_{\epsilon} \,\hat{k}.\hat{s}^{\prime}-\hat{p}_{\epsilon}.\hat{s} \,\hat{k}\mdot\hat{s}\nn
- & \, \hat{p}_{\epsilon}.\hat{s} \,\hat{k}.\hat{s}^{\prime}-\hat{q}_{\epsilon}.\hat{s}\, \hat{k}\mdot\hat{s}-\hat{q}_{\epsilon}.\hat{s} \,\hat{k}.\hat{s}^{\prime}-\hat{p}_{\epsilon}.\hat{s}^{\prime} \,\hat{k}. \hat{s}-\hat{p}_{\epsilon}.\hat{s}^{\prime} \,\hat{k}\mdot\hat{s}^{\prime}-\hat{q}_{\epsilon}.\hat{s}^{\prime} \,\hat{k}\mdot\hat{s}-\hat{q}_{\epsilon}\!\cdot\!\hat{s}^{\prime} \,\hat{k}.\hat{s}^{\prime}\nn
+& \, \hat{p}_{\epsilon}.\hat{s} \, \hat{q}_{\epsilon} .\hat{s}^{\prime} \,\hat{k}.\hat{s}+\hat{p}_{\epsilon}\!\cdot\!\hat{s} \,\hat{q}_{\epsilon}\!\cdot\!\hat{s}^{\prime} \,\hat{k}\!\cdot\!\hat{s}^{\prime}+\hat{q}_{\epsilon}\!\cdot\!\hat{s} \,\hat{p}_{\epsilon}.\hat{s}^{\prime} \,\hat{k}\mdot\hat{s}+\hat{q}_{\epsilon}.\hat{s} \,\hat{p}_{\epsilon} .\hat{s}^{\prime} \,\hat{k}\mdot\hat{s}^{\prime}+\hat{k}.\hat{s} \,\hat{s}.\hat{s}^{\prime}\nn
+& \, \hat{k}.\hat{s}^{\prime} \,\hat{s}\mdot\hat{s}^{\prime}-\hat{p}_{\epsilon}.\hat{q}_{\epsilon} \,\hat{k}.\hat{s} \,\hat{s}\mdot\hat{s}^{\prime}-\hat{p}_{\epsilon}.\hat{q}_{\epsilon} \,\hat{k}.\hat{s}^{\prime} \,\hat{s}. \hat{s}^{\prime}\bigg) 
+k_{0}^{2}\bigg(\hat{k}\mdot\hat{s}\,\hat{k}\mdot\hat{s}^{\prime}+\hat{p}_{\epsilon}\mdot\hat{q}_{\epsilon}\,\hat{k}\mdot\hat{s}\,\hat{k}\mdot\hat{s}^{\prime} \nn
-& \, \hat{p}_{\epsilon}.\hat{s}\,\hat{k}.\hat{s}\,\hat{k}\mdot\hat{s}^{\prime}-\hat{q}_{\epsilon}\mdot\hat{s}\,\hat{k}\mdot\hat{s}\,\hat{k}\mdot\hat{s}^{\prime}-\hat{p}_{\epsilon}\mdot\hat{s}^{\prime}\, \hat{k}\mdot\hat{s}\,\hat{k}.\hat{s}^{\prime}-\hat{q}_{\epsilon}\!\cdot\!\hat{s}^{\prime}\,\hat{k}\mdot\hat{s}\,\hat{k}\mdot\hat{s}^{\prime}\nn
+& \, \hat{p}_{\epsilon}\mdot\hat{s} \,\hat{q}_{\epsilon}\mdot\hat{s}^{\prime} \,\hat{k}\mdot \hat{s} \,\hat{k}\mdot\hat{s}^{\prime}+\hat{q}_{\epsilon}\mdot\hat{s} \,\hat{p}_{\epsilon}\mdot\hat{s}^{\prime} \,\hat{k}\mdot\hat{s} \,\hat{k}\mdot\hat{s}^{\prime}+\hat{k}\mdot\hat{s} \,\hat{k}\mdot\hat{s}^{\prime} \,\hat{s}\mdot\hat{s}^{\prime}-\hat{p}_{\epsilon}\mdot\hat{q}_{\epsilon} \,\hat{k}\mdot\hat{s} \,\hat{s}\mdot\hat{s}^{\prime}\,\hat{k}\!\cdot\!\hat{s}^{\prime} \bigg)\bigg\}\nn
\times & \, \tilde{D}^{S}_{L}(K)\Delta^{R}_{\mp}(Q)\bigg].
\label{F_pm_SR_--_--}
\end{align}
Let us take the terms of Eq.~\eqref{F_pm_SR_--_--} without $s$.
\begin{align}
F_{1\pm ; -- ; --}^{\mathrm{SR}}(P, K) = & \,
\big(-k^{2}\big)\int \frac{d \Omega_{s}}{4 \pi} \frac{1}{\left(P S-i \varepsilon\right)\left(Q S-i \varepsilon\right)}\int \frac{d \Omega_{s^{\prime}}}{4 \pi} \frac{1}{\left(P S^{\prime}-i \varepsilon\right)\left(Q S^{\prime}-i\varepsilon\right)} \nn
\times & \, \Big(1+\hat{p}.\hat{q}\Big) \tilde{D}^{S}_{L}(K)\Delta^{R}_{\mp}(Q)
\label{F1_SR_--_--}
\end{align}
Using Eq.~\eqref{V_k_k0}, the above Eq.~\eqref{F1_SR_--_--} can be written as 
\begin{align} 
F_{1\pm ; -- ; --}^{\mathrm{SR}}(P, K) = & \,
-k^{2}\left[1+\left(\frac{p}{q}-\frac{kx}{q}\right)\right]\,V(t,k,k_{0},x,\varepsilon)^{2}\,\tilde{D}^{S}_{L}(K)\Delta^{R}_{\mp}(Q)
\end{align}
Let us take the terms of Eq.~\eqref{F_pm_SR_--_--} consisting of $1s$, which can be written as 
\begin{align}
F_{2\pm ; -- ; --}^{\mathrm{SR}}(P, K) = & \,
\bigg[\int \frac{d \Omega_{s}}{4 \pi} \frac{\hat{s}^{i}}{\left(P S-i \varepsilon\right)\left(Q S-i \varepsilon\right)}\int \frac{d \Omega_{s^{\prime}}}{4 \pi} \frac{1}{\left(P S^{\prime}-i \varepsilon\right)\left(Q S^{\prime}-i\varepsilon\right)} \nn
\times & \, \bigg\{\pm k^{2}\bigg(-\hat{p}_{i}-\hat{q}_{i}\bigg) - k_{0}k \bigg(\hat{k}_{i}+\hat{p}\mdot\hat{q}\, \hat{k}_{i}\bigg)\bigg\} \nn 
+ &\int \frac{d \Omega_{s}}{4 \pi} \frac{1}{\left(P S-i \varepsilon\right)\left(Q S-i \varepsilon\right)}\int \frac{d \Omega_{s^{\prime}}}{4 \pi} \frac{\hat{s}^{\prime
i}}{\left(P S^{\prime}-i \varepsilon\right)\left(Q S^{\prime}-i\varepsilon\right)}\nn 
\times & \, \bigg\{\pm k^{2}\bigg(-\hat{p}_{i}-\hat{q}_{i}\bigg) -k_{0}k \bigg(\hat{k}_{i}+\hat{p}\mdot\hat{q}\, \hat{k}_{i}\bigg)\bigg\} \bigg] \bigg(-\tilde{D}^{S}_{L}(K)\Delta_{\mp}^{\mathrm{R}}(Q)\bigg) 
\label{F2_SR_--_--}
\end{align}
Using Eq.~\eqref{1s_int}, Eq.~\eqref{F2_SR_--_--} becomes
\begin{align}
F_{2\pm ; -- ; --}^{\mathrm{SR}}(P, K) = & \, -2V_{1}(r_{0},r) \,V(t,k,k_{0},x,\varepsilon)\,\tilde{D}^{S}_{L}(K)\Delta^{R}_{\mp}(Q)\nn
\times & \, \left[\mp k^{2}\left(\hat{p}\mdot \hat{r} + \hat{q}\mdot\hat{r}\right)
- k_{0}k \,\hat{k}\mdot\hat{r}\left(1+\hat{p}\mdot\hat{q}\right)\right]
\end{align}
Let us consider the terms of Eq.~\eqref{F_pm_SR_--_--} with $2s$, which can be written as 
\begin{align}
F_{3\pm ; -- ; --}^{\mathrm{SR}}(P, K) = & \,
-\int \frac{d \Omega_{s}}{4 \pi} \frac{\hat{s}^{i}}{\left(P S-i \varepsilon\right)\left(Q S-i \varepsilon\right)}\int \frac{d \Omega_{s^{\prime}}}{4 \pi} \frac{\hat{s}^{\prime}_{i}}{\left(P S^{\prime}-i \varepsilon\right)\left(Q S^{\prime}-i\varepsilon\right)} \nn
\times & \, \Big(1+\hat{p}.\hat{q}\Big) D_{T}^{\mathrm{S}}(K) \Delta_{\mp}^{\mathrm{R}}(Q) \nn
+ & \, \int\! \frac{d \Omega_{s}}{4 \pi} \frac{\hat{s}^{i}}{\left(P S-i \varepsilon\right)\left(Q S-i \varepsilon\right)}\int\! \frac{d \Omega_{s^{\prime}}}{4 \pi} \frac{\hat{s}^{\prime {j}}}{\left(P S^{\prime}-i \varepsilon\right)\left(Q S^{\prime}-i\varepsilon\right)} \nn
\times & \, \Big(\hat{k}_{i}\hat{k}_{j} + \hat{p}.\hat{q}\,\hat{k}_{i}\hat{k}_{j}\Big) D_{T}^{\mathrm{S}}(K) \Delta_{\mp}^{\mathrm{R}}(Q)\nn
- & \, \bigg[ \int\! \frac{d \Omega_{s}}{4 \pi} \frac{\hat{s}^{i}}{\left(P S-i \varepsilon\right)\left(Q S-i \varepsilon\right)}\int\! \frac{d \Omega_{s^{\prime}}}{4 \pi} \frac{\hat{s}^{\prime {j}}}{\left(P S^{\prime}-i \varepsilon\right)\left(Q S^{\prime}-i\varepsilon\right)}\nn
\times & \, \bigg\{k^{2}\Big(\hat{p}_{i}\hat{q}_{j} +\hat{q}_{i}\hat{p}_{j}+1- \hat{p}.\hat{q}\Big) + k_{0}^{2}\big(\hat{k}_{i}\hat{k}_{j}\big(1+\hat{p}\mdot\hat{q}\big)\big)\nn
\mp & \, k_{0}k \big(-\hat{p}_{i}\hat{k}_{j}-\hat{q}_{i}\hat{k}_{j}-\hat{k}_{i}\hat{p}_{j}-\hat{k}_{i}\hat{q}_{j} \big)\bigg\} \nn
+ & \, \int\! \frac{d \Omega_{s}}{4 \pi} \frac{\hat{s}^{i}\,\hat{s}^{j}}{\left(P S-i \varepsilon\right)\left(Q S-i \varepsilon\right)}\int\! \frac{d \Omega_{s^{\prime}}}{4 \pi} \frac{1}{\left(P S^{\prime}-i \varepsilon\right)\left(Q S^{\prime}-i\varepsilon\right)} \nn
\times & \, \bigg\{\mp k_{0} k \big(-\hat{p}_{i}\hat{k}_{j}-\hat{q}_{i}\hat{k}_{j}\big)\bigg\}\nn
+ & \, \int\! \frac{d \Omega_{s}}{4 \pi} \frac{1}{\left(P S-i \varepsilon\right)\left(Q S-i \varepsilon\right)}\int\! \frac{d \Omega_{s^{\prime}}}{4 \pi} \frac{\hat{s}^{\prime i}\,\hat{s}^{\prime j}}{\left(P S^{\prime}-i \varepsilon\right)\left(Q S^{\prime}-i\varepsilon\right)} \nn
\times & \, \bigg\{\mp k_{0} k \big(-\hat{p}_{i}\hat{k}_{j}-\hat{q}_{i}\hat{k}_{j}\big)\bigg\}\bigg]
\bigg(\tilde{D}^{S}_{L}(K)\Delta^{R}_{\mp}(Q)\bigg)
\label{F1}
\end{align}
Using Eqs.~\eqref{V_k_k0}~\eqref{1s_int}~\eqref{2s_int}, Eq.~\eqref{F1} becomes
\begin{align}
F_{3\pm ; -- ; --}^{\mathrm{SR}}(P, K)  = & \, \int_{0}^{1} d u \left(V_1 \left(r_0,r \right)\right)^2  D_{T}^{\mathrm{S}}(K) \Delta_{\mp}^{\mathrm{R}}(Q) \left(1 + \hat{p}\mdot\hat{q}\right)\left\{\left(\hat{k}\mdot\hat{r}\right)^2 - 1 \right\}\nn
- & \, \int_{0}^{1} d u\bigg(\tilde{D}^{S}_{L}(K)\Delta^{R}_{\mp}(Q)\bigg) \bigg[\left(V_1 \left(r_0,r \right)\right)^2 \bigg\{k^{2}\bigg(2\hat{p}\mdot\hat{r}\,\hat{q}\mdot\hat{r}-\hat{p}\mdot\hat{q}+1\bigg) \nn
+ & \, k_{0}^{2} \bigg(\hat{k}\mdot\hat{r}\bigg)^{2}\bigg(1+\hat{p}\mdot\hat{q}\bigg)\pm 2k_{0}k \bigg(\hat{k}\mdot\hat{r}\big(\hat{p}\mdot\hat{r}+\hat{q}\mdot\hat{r}\big)\bigg)\bigg\}\nn
\pm & \, 2V(t,k,k_{0},x,\varepsilon)\, k_{0}k \,\bigg\{A\left(x+\hat{q}\mdot\hat{k}\right)+B\, \hat{k}\mdot\hat{r}\,\big(\hat{p}\mdot\hat{r}+\hat{q}\mdot\hat{r}\big) \bigg\}\bigg]
\label{F1_int}
\end{align}
Now, let us consider the terms of Eq.~\eqref{F_pm_SR_--_--} with $3s$, and it can be expressed as 
\begin{align}
F_{4\pm ; -- ; --}^{\mathrm{SR}}(P, K) = & \,
\Bigg[\int \frac{d \Omega_{s}}{4 \pi} \frac{\hat{s}^{i}\hat{s}^{j}}{\left(P S-i \varepsilon\right)\left(Q S-i \varepsilon\right)}\int \frac{d \Omega_{s^{\prime}}}{4 \pi} \frac{\hat{s}^{\prime}_{j}}{\left(P S^{\prime}-i \varepsilon\right)\left(Q S^{\prime}-i\varepsilon\right)} \nn
\times & \, \Big(\hat{p}_{i}+ \hat{q}_{i} \Big) \nn
+ & \,  \int \frac{d \Omega_{s}}{4 \pi} \frac{\hat{s}^{i}\hat{s}^{j}}{\left(P S-i \varepsilon\right)\left(Q S-i \varepsilon\right)}\int \frac{d \Omega_{s^{\prime}}}{4 \pi} \frac{\hat{s}^{\prime k}}{\left(P S^{\prime}-i \varepsilon\right)\left(Q S^{\prime}-i\varepsilon\right)} \nn 
\times & \, \left(-\hat{p}_{i} \hat{k}_{j}\hat{k}_{k} - \hat{q}_{i} \hat{k}_{j}\hat{k}_{k} \right) \nn
+ & \, \int \frac{d \Omega_{s}}{4 \pi} \frac{\hat{s}^{j}}{\left(P S-i \varepsilon\right)\left(Q S-i \varepsilon\right)}\int \frac{d \Omega_{s^{\prime}}}{4 \pi} \frac{\hat{s}^{\prime i}\hat{s}^{\prime}_{j}}{\left(P S^{\prime}-i \varepsilon\right)\left(Q S^{\prime}-i\varepsilon\right)} \Big(\hat{p}_{i} + \hat{q}_{i} \Big) \nn
- & \, \int \frac{d \Omega_{s}}{4 \pi} \frac{\hat{s}^{j}}{\left(P S-i \varepsilon\right)\left(Q S-i \varepsilon\right)}\int \frac{d \Omega_{s^{\prime}}}{4 \pi} \frac{\hat{s}^{\prime i}\hat{s}^{\prime k}}{\left(P S^{\prime}-i \varepsilon\right)\left(Q S^{\prime}-i\varepsilon\right)}\nn
\times & \, \left(\hat{p}_{i} \hat{k}_{j}\hat{k}_{k} + \hat{q}_{i} \hat{k}_{j}\hat{k}_{k} \right)\Bigg] \pm D_{T}^{\mathrm{S}}(K) \Delta_{\mp}^{\mathrm{R}}(Q) \nn 
- & \, \bigg[ \int \frac{d \Omega_{s}}{4 \pi} \frac{\hat{s}^{i}\hat{s}^{j}}{\left(P S-i \varepsilon\right)\left(Q S-i \varepsilon\right)}\int \frac{d \Omega_{s^{\prime}}}{4 \pi} \frac{\hat{s}^{\prime k}}{\left(P S^{\prime}-i \varepsilon\right)\left(Q S^{\prime}-i\varepsilon\right)}\nn
\times & \, \bigg\{-k_{0}k\bigg(\hat{p}_{i}\hat{k}_{j}\hat{q}_{k}+\hat{q}_{i}\hat{k}_{j}\hat{p}_{k}\bigg)\pm k_{0}^{2}\bigg(-\hat{p}_{i}\hat{k}_{j}\hat{k}_{k}-\hat{q}_{i}\hat{k}_{j}\hat{k}_{k}\bigg)\bigg\}\nn
+ & \, \int \frac{d \Omega_{s}}{4 \pi} \frac{\hat{s}^{k}}{\left(P S-i \varepsilon\right)\left(Q S-i \varepsilon\right)}\int \frac{d \Omega_{s^{\prime}}}{4 \pi} \frac{\hat{s}^{\prime i}\,\hat{s}^{\prime j}}{\left(P S^{\prime}-i \varepsilon\right)\left(Q S^{\prime}-i\varepsilon\right)} \nn
\times & \, \bigg\{-k_{0}k\bigg(\hat{q}_{i}\hat{k}_{j}\hat{p}_{k}+\hat{p}_{i}\hat{k}_{j}\hat{q}_{k}\bigg)\pm k_{0}^{2}\bigg(-\hat{p}_{i}\hat{k}_{j}\hat{k}_{k}-\hat{q}_{i}\hat{k}_{j}\hat{k}_{k}\bigg)\bigg\}\nn
+ & \, \int \frac{d \Omega_{s}}{4 \pi} \frac{\hat{s}^{i}\hat{s}^{j}}{\left(P S-i \varepsilon\right)\left(Q S-i \varepsilon\right)}\int \frac{d \Omega_{s^{\prime}}}{4 \pi} \frac{\hat{s}^{\prime}_{j}}{\left(P S^{\prime}-i \varepsilon\right)\left(Q S^{\prime}-i\varepsilon\right)}\nn
\times & \, \bigg\{-k_{0}k \bigg(\hat{k}_{i}-\hat{p}\mdot\hat{q}\,\hat{k}_{i}\bigg)\bigg\}\nn
+ & \, \int \frac{d \Omega_{s}}{4 \pi} \frac{\hat{s}_{j}}{\left(P S-i \varepsilon\right)\left(Q S-i \varepsilon\right)}\int \frac{d \Omega_{s^{\prime}}}{4 \pi} \frac{\hat{s}^{\prime i}\hat{s}^{\prime j}}{\left(P S^{\prime}-i \varepsilon\right)\left(Q S^{\prime}-i\varepsilon\right)}\nn
\times & \, \bigg\{-k_{0}k \bigg(\hat{k}_{i}-\hat{p}\mdot\hat{q}\,\hat{k}_{i}\bigg)\bigg\} \bigg(\tilde{D}^{S}_{L}(K)\Delta^{R}_{\mp}(Q)\bigg)
\label{F2SR_2s}
\end{align}
The above Eq.~\eqref{F2SR_2s} can be simplified using Eqs.~\eqref{1s_int},\eqref{2s_int}. Thus, we will get
\begin{align}
F_{4\pm ; -- ; --}^{\mathrm{SR}}(P, K) = & \, \pm 2 \int_{0}^{1} d u \,  V_1(r_0,r) \left[ A \left(\hat{p}\mdot\hat{r} + \hat{q}\mdot\hat{r} -\hat{p}\mdot\hat{k} \, \hat{r}\mdot\hat{k}-\hat{q}\mdot\hat{k} \, \hat{r}\mdot\hat{k}\right)\right.  \nn
+ & \, \left. B \left(\hat{p}\mdot\hat{r} + \hat{q}\mdot\hat{r} -\hat{p}.\hat{r} \, (\hat{k}.\hat{r})^2 - \hat{q}.\hat{r} \, (\hat{k}.\hat{r})^2 \right)\right]D_{T}^{\mathrm{S}}(K) \, \Delta_{\mp}^{\mathrm{R}}(Q) \nn
+ & \,  2 \int_{0}^{1} d u\ k_0^2 \,  V_1(r_0,r)\,\tilde{D}^{S}_{L}(K)\Delta^{R}_{\mp}(Q) \bigg\{\frac{k}{k_0} \, \hat{k}\mdot \hat{r} \, \big(A+B \big) \big(1-\hat{p}\mdot\hat{q}\big)\nn
+ & \, A \bigg( \frac{k}{k_0} \, \big(\hat{p}\mdot \hat{k} \, \hat{q}\mdot \hat{r} + \hat{q}\mdot \hat{k} \, \hat{p}\mdot \hat{r}\big) \pm  \big(\hat{p}\mdot \hat{k} \, \hat{k}\mdot \hat{r}+\hat{q}\mdot \hat{k} \, \hat{k}\mdot \hat{r}\big) \bigg) \nn 
+ & \,  B \bigg(\frac{k}{ k_0} \, \big(2\hat{p}\mdot \hat{r} \, \hat{q}\mdot \hat{r}\, \hat{k}\mdot \hat{r} \big) \pm   \big(\hat{k}\mdot \hat{r}\big)^{2} \left(\hat{p}\mdot \hat{r} +\hat{q}\mdot \hat{r}\right) \bigg)\! \bigg\}
\end{align}
Let us consider the terms of Eq.~\eqref{F_pm_SR_--_--} with $4s$, and it can be written as
\begin{align}
F_{5\pm ; -- ; --}^{\mathrm{SR}}(P, K) 
= & \, \Bigg[\int \frac{d \Omega_{s}}{4 \pi} \frac{\hat{s}^{i}\hat{s}^{k}}{\left(P S-i \varepsilon\right)\left(Q S-i \varepsilon\right)}\int \frac{d \Omega_{s^{\prime}}}{4 \pi} \frac{\hat{s}^{\prime j} \hat{s}^{\prime}_{k}}{\left(P S^{\prime}-i \varepsilon\right)\left(Q S^{\prime}-i\varepsilon\right)}\nn
\times & \, \Big(-\hat{p}_{i}\hat{q}_{j} - \hat{q}_{i}\hat{p}_{j} \Big) \nn
+ & \, \int \frac{d \Omega_{s}}{4 \pi} \frac{\hat{s}^{i}\hat{s}^{j}}{\left(P S-i \varepsilon\right)\left(Q S-i \varepsilon\right)}\int \frac{d \Omega_{s^{\prime}}}{4 \pi} \frac{\hat{s}^{\prime}_{i} \hat{s}^{\prime}_{j}}{\left(P S^{\prime}-i \varepsilon\right)\left(Q S^{\prime}-i\varepsilon\right)} \nn
\times & \, \Big(\hat{p}.\hat{q} - 1  \Big)\nn
+ & \, \int \frac{d \Omega_{s}}{4 \pi} \frac{\hat{s}^{i}\hat{s}^{k}}{\left(P S-i \varepsilon\right)\left(Q S-i \varepsilon\right)}\int \frac{d \Omega_{s^{\prime}}}{4 \pi} \frac{\hat{s}^{\prime j} \hat{s}^{\prime l}}{\left(P S^{\prime}-i \varepsilon\right)\left(Q S^{\prime}-i\varepsilon\right)} \nn 
\times & \, \left(\hat{p}_{i}\hat{q}_{j} \hat{k}_{k} \hat{k}_{l} + \hat{q}_{i}\hat{p}_{j} \hat{k}_{k} \hat{k}_{l} \right)\nn
+ & \, \int \frac{d \Omega_{s}}{4 \pi} \frac{\hat{s}^{i}\hat{s}^{j}}{\left(P S-i \varepsilon\right)\left(Q S-i \varepsilon\right)}\int \frac{d \Omega_{s^{\prime}}}{4 \pi} \frac{\hat{s}^{\prime}_{i} \hat{s}^{\prime k}}{\left(P S^{\prime}-i \varepsilon\right)\left(Q S^{\prime}-i\varepsilon\right)}\nn
\times & \, \hat{k}_{j} \hat{k}_{k} \Big(1 - \hat{p}.\hat{q} \Big)\Bigg] {D}_{T}^{\mathrm{S}}(K) \Delta_{\mp}^{\mathrm{R}}(Q)\nn
+ & \, \bigg[\int \frac{d \Omega_{s}}{4 \pi} \frac{\hat{s}^{i}\hat{s}^{j}}{\left(P S-i \varepsilon\right)\left(Q S-i \varepsilon\right)}\int \frac{d \Omega_{s^{\prime}}}{4 \pi} \frac{\hat{s}^{\prime k} \hat{s}^{\prime l}}{\left(P S^{\prime}-i \varepsilon\right)\left(Q S^{\prime}-i\varepsilon\right)} \nn
\times & \, \left(\hat{p}_{i}\hat{k}_{j} \hat{q}_{k} \hat{k}_{l} + \hat{q}_{i}\hat{k}_{j} \hat{p}_{k} \hat{k}_{l} \right)\nn
+ & \, \int \frac{d \Omega_{s}}{4 \pi} \frac{\hat{s}^{i}\hat{s}^{j}}{\left(P S-i \varepsilon\right)\left(Q S-i \varepsilon\right)}\int \frac{d \Omega_{s^{\prime}}}{4 \pi} \frac{\hat{s}^{\prime k} \hat{s}^{\prime}_{j}}{\left(P S^{\prime}-i \varepsilon\right)\left(Q S^{\prime}-i\varepsilon\right)} \nn
\times & \, \left(\hat{k}_{i}\hat{k}_{k} - \hat{p}\mdot\hat{q} \hat{k}_{i} \hat{k}_{k} \right)\bigg]
\times \bigg(k_{0}^{2} \, \tilde{D}^{S}_{L}(K)\Delta^{R}_{\mp}(Q)\bigg)
\label{F3SR_4s}
\end{align}
Eq.~\eqref{F3SR_4s} can further be simplified in terms of $A$ and $B$ as defined in Eq.~\eqref{A_B_Def} as
\begin{align}
F_{5\pm ; -- ; --}^{\mathrm{SR}}(P, K)
  = & \,  \int_{0}^{1} du \, \Bigg[2A^2 \left(   x\, \hat{q}\mdot\hat{k} -1 \right) + B^2 \left(\hat{p}\mdot\hat{q} - 2 \hat{p}\mdot\hat{r} \, \hat{q}\mdot\hat{r}  - 1 + 2 \hat{p}\mdot\hat{r} \, \hat{q}\mdot\hat{r} \, \big(\hat{k}\mdot\hat{r}\big)^2 \right. \nn
+ & \, \left. \big(\hat{k}\mdot\hat{r}\big)^2 \left( 1 - \hat{p}.\hat{q}\right) \right) +  AB \left\{-4 \hat{p}.\hat{r} \, \hat{q}.\hat{r} + 2 \left(\hat{p}.\hat{q} - 1 \right) + 2 x \, \hat{q}.\hat{r} \, \hat{k}.\hat{r} \right. \nn
+ & \, \left. 2 \hat{q}.\hat{k} \, \hat{p}.\hat{r} \,\hat{k}\cdot\hat{r} + 2 \left(\hat{k}.\hat{r}\right)^2 \left( 1 - \hat{p}\mdot\hat{q}\right) \right\}\Bigg]  D_{T}^{\mathrm{S}}(K) \Delta_{\mp}^{\mathrm{R}}(Q)\nn
- &  \int_{0}^{1} du \,k_{0}^{2}\,\tilde{D}^{S}_{L}(K)\Delta^{R}_{\mp}(Q) \bigg[\left(1-\hat{p}\mdot\hat{q}\right)\bigg(A^{2} + \left(B^{2}+2AB\right)\big(\hat{k}\mdot\hat{r}\big)^{2}\bigg)\nn
+ & \ 2A^{2} \, \hat{p}\mdot\hat{k}\, \hat{q}\mdot\hat{k} + 2B^{2} \, \hat{p}\mdot\hat{r} \, \hat{q}\mdot\hat{r} \, \big(\hat{k}\mdot\hat{r}\big)^{2} + 2AB \, \left(\hat{p}\mdot\hat{k} \, \hat{q}\mdot\hat{r} \, \hat{k}\mdot\hat{r} + \hat{q}\mdot\hat{k} \, \hat{p}\mdot\hat{r} \, \hat{k}\mdot\hat{r} \right)  \bigg]
\end{align}
Adding all the five individual contributions, Eq.~\eqref{Sigma1_6} becomes
\begin{align}
\Sigma _{6\pm}^{(1)}(P) = & \, \frac{-ig^2 C_F}{2(2\pi)^4}  (2\pi) \int_{-\infty}^\infty dk_0 \int_{0}^\infty k^2 dk  \int_{-1}^1 dx \, \,\bigg[F_{1\pm;--;--}^{SR}(P,K)+ F_{2\pm;--;--}^{SR}(P,K) \nn
+& F_{3\pm;--;--}^{SR}(P,K)+F_{4\pm;--;--}^{SR}(P,K)+F_{5\pm;--;--}^{SR}(P,K)\bigg].
\end{align}
Let us consider the seventh term of Eq.~\eqref{sigma_final}, which is 
\begin{equation}
\Sigma _{7\pm}^{(1)}(P) = \frac{-ig^2 C_F}{2} \int \frac{d^4K}{(2\pi)^4}\Big[F_{\pm ; --;+-}^{\mathrm{AS}}(P, K)\Big]
\label{Sigma1_7}
\end{equation}
The terms involved in $ F_{\pm ; --;+-}^{\mathrm{AS}}(P, K) $ are
\begin{align}
F_{\pm ; -- ; +-}^{\mathrm{AS}}(P, K) 
= & \, \int \frac{d \Omega_{s}}{4 \pi} \frac{1}{\big(P S-i \varepsilon\big)\left(Q S-i \varepsilon\right)}\times \int \frac{d \Omega_{s^{\prime}}}{4 \pi} \frac{1}{\left(P S^{\prime}+i \varepsilon\right)\left(Q S^{\prime}-i  \varepsilon\right)}\nn
\times & \, \bigg[\bigg\{\bigg(-\hat{s}\mdot\hat{s}^{\prime}-\hat{p}_{\epsilon}\mdot\hat{q}_{\epsilon}\, \hat{s}\mdot\hat{s}^{\prime}+\hat{p}_{\epsilon}\mdot\hat{s} \, \hat{s}\mdot\hat{s}^{\prime}+\hat{q}_{\epsilon}\mdot\hat{s} \, \hat{s}\mdot\hat{s}^{\prime}+\hat{p}_{\epsilon}\mdot\hat{s}^{\prime} \, \hat{s}\mdot\hat{s}^{\prime}+\hat{q}_{\epsilon}\mdot \hat{s}^{\prime} \, \hat{s}\mdot\hat{s}^{\prime} \nn
- & \, \hat{p}_{\epsilon}\mdot\hat{s} \,\hat{q}_{\epsilon}\mdot\hat{s}^{\prime} \, \hat{s}.\hat{s}^{\prime} - \hat{q}_{\epsilon}\mdot \hat{s} \,\hat{p}_{\epsilon}\mdot\hat{s}^{\prime} \, \hat{s}\mdot\hat{s}^{\prime} - ( \hat{s}\mdot\hat{s}^{\prime})^2 + \hat{p}_{\epsilon}\mdot\hat{q}_{\epsilon}\,(\hat{s}\mdot\hat{s}^{\prime})^2+(\hat{k}\mdot\hat{s})\,(\hat{k}\mdot\hat{s}^{\prime})   \nn
+ & \, \hat{p}_{\epsilon}\mdot\hat{q}_{\epsilon}\,\hat{k}\mdot\hat{s}\,\hat{k}\mdot\hat{s}^{\prime} - \hat{p}_{\epsilon}\mdot \hat{s} \,\hat{k}\mdot\hat{s}\,\hat{k}\mdot\hat{s}^{\prime} -\hat{q}_{\epsilon}\mdot \hat{s} \,\hat{k}\mdot\hat{s}\,\hat{k}\mdot\hat{s}^{\prime} -\hat{p}_{\epsilon}\mdot\hat{s}^{\prime} \, \hat{k}\mdot\hat{s}\, \hat{k}\mdot\hat{s}^{\prime} \nn
- & \, \hat{q}_{\epsilon}\mdot \hat{s}^{\prime} \, \hat{k}\mdot\hat{s} \, \hat{k}\mdot\hat{s}^{\prime}+\hat{p}_{\epsilon}\mdot\hat{s} \, \hat{q}_{\epsilon}\mdot\hat{s}^{\prime}\, \hat{k}\mdot\hat{s} \, \hat{k}\mdot\hat{s}^{\prime}+\hat{q}_{\epsilon}\mdot\hat{s} \, \hat{p}_{\epsilon} \mdot\hat{s}^{\prime}\, \hat{k}\mdot\hat{s} \,\hat{k}\mdot\hat{s}^{\prime} +\hat{s}\mdot\hat{s}^{\prime} \,\hat{k}\mdot\hat{s} \,\hat{k}\mdot\hat{s}^{\prime}\nn
- & \, \hat{p}_{\epsilon}\mdot\hat{q}_{\epsilon} \, \hat{s}\mdot\hat{s}^{\prime} \, \hat{k}\mdot\hat{s} \, \hat{k}\mdot\hat{s}^{\prime}\bigg) D_{T}^{\mathrm{A}}(K) \Delta_{\mp}^{\mathrm{S}}(Q) \bigg\}\nn  
- & \, \bigg\{k^{2}\bigg(1+\hat{p}_{\epsilon}.\hat{q}_{\epsilon}-\hat{p}_{\epsilon}\mdot\hat{s} -\hat{q}_{\epsilon}\mdot\hat{s}-\hat{p}_{\epsilon}\mdot\hat{s}^{\prime}-\hat{q}_{\epsilon}\mdot\hat{s}^{\prime}+\hat{p}_{\epsilon}\mdot\hat{s} \,\hat{q}_{\epsilon}\mdot\hat{s}^{\prime}+\hat{q}_{\epsilon}\mdot\hat{s} \,\hat{p}_{\epsilon}\mdot\hat{s}^{\prime} \nn
+ & \, \hat{s}.\hat{s}^{\prime}-\hat{p}_{\epsilon}. \hat{q}_{\epsilon}\,\hat{s}.\hat{s}^{\prime}\bigg) - k_{0} k\bigg(\hat{k}.\hat{s} + \hat{k}.\hat{s}^{\prime}+\hat{p}_{\epsilon}.\hat{q}_{\epsilon} \,\hat{k}. \hat{s}  +\hat{p}_{\epsilon}.\hat{q}_{\epsilon} \,\hat{k}.\hat{s}^{\prime} \nn
- & \, \hat{p}_{\epsilon}.\hat{s} \,\hat{k}\mdot\hat{s}-\hat{p}_{\epsilon}.\hat{s} \,\hat{k}.\hat{s}^{\prime}-\hat{q}_{\epsilon}.\hat{s}\, \hat{k}\mdot\hat{s}-\hat{q}_{\epsilon}.\hat{s} \,\hat{k}.\hat{s}^{\prime}-\hat{p}_{\epsilon}.\hat{s}^{\prime} \,\hat{k}. \hat{s} -\hat{p}_{\epsilon}.\hat{s}^{\prime} \,\hat{k}\mdot\hat{s}^{\prime}\nn
- & \, \hat{q}_{\epsilon}.\hat{s}^{\prime} \,\hat{k}\mdot\hat{s}-\hat{q}_{\epsilon}\!\cdot\!\hat{s}^{\prime} \,\hat{k}.\hat{s}^{\prime}+\hat{p}_{\epsilon}.\hat{s} \, \hat{q}_{\epsilon} .\hat{s}^{\prime} \,\hat{k}.\hat{s}+\hat{p}_{\epsilon}\!\cdot\!\hat{s} \,\hat{q}_{\epsilon}\!\cdot\!\hat{s}^{\prime} \,\hat{k}\!\cdot\!\hat{s}^{\prime}+\hat{q}_{\epsilon}\!\cdot\!\hat{s} \,\hat{p}_{\epsilon}.\hat{s}^{\prime} \,\hat{k}\mdot\hat{s} \nn
+ & \, \hat{q}_{\epsilon}.\hat{s} \,\hat{p}_{\epsilon} .\hat{s}^{\prime} \,\hat{k}\mdot\hat{s}^{\prime} + \hat{k}.\hat{s} \,\hat{s}.\hat{s}^{\prime}+\hat{k}.\hat{s}^{\prime} \,\hat{s}\mdot\hat{s}^{\prime}-\hat{p}_{\epsilon}.\hat{q}_{\epsilon} \,\hat{k}.\hat{s} \,\hat{s}\mdot\hat{s}^{\prime}-\hat{p}_{\epsilon}.\hat{q}_{\epsilon} \,\hat{k}.\hat{s}^{\prime} \,\hat{s}. \hat{s}^{\prime}\bigg) \nn
+ & \, k_{0}^{2}\bigg(\hat{k}\mdot\hat{s}\,\hat{k}\mdot\hat{s}^{\prime}+\hat{p}_{\epsilon}\mdot\hat{q}_{\epsilon}\,\hat{k}\mdot\hat{s}\,\hat{k}\mdot\hat{s}^{\prime}-\hat{p}_{\epsilon}.\hat{s}\,\hat{k}.\hat{s}\,\hat{k}\mdot\hat{s}^{\prime}-\hat{q}_{\epsilon}\mdot\hat{s}\,\hat{k}\mdot\hat{s}\,\hat{k}\mdot\hat{s}^{\prime}-\hat{p}_{\epsilon}\mdot\hat{s}^{\prime}\, \hat{k}\mdot\hat{s}\,\hat{k}.\hat{s}^{\prime}\nn
- & \, \hat{q}_{\epsilon}\!\cdot\!\hat{s}^{\prime}\,\hat{k}\mdot\hat{s}\,\hat{k}\mdot\hat{s}^{\prime} + \hat{p}_{\epsilon}\mdot\hat{s} \,\hat{q}_{\epsilon}\mdot\hat{s}^{\prime} \,\hat{k}\mdot \hat{s} \,\hat{k}\mdot\hat{s}^{\prime}+\hat{q}_{\epsilon}\mdot\hat{s} \,\hat{p}_{\epsilon}\mdot\hat{s}^{\prime} \,\hat{k}\mdot\hat{s} \,\hat{k}\mdot\hat{s}^{\prime} \nn
+ & \, \hat{k}\mdot\hat{s} \,\hat{k}\mdot\hat{s}^{\prime} \,\hat{s}\mdot\hat{s}^{\prime}-\hat{p}_{\epsilon}\mdot\hat{q}_{\epsilon} \,\hat{k}\mdot\hat{s} \,\hat{s}\mdot\hat{s}^{\prime}\,\hat{k}\!\cdot\!\hat{s}^{\prime} \bigg)\bigg\} \tilde{D}^{A}_{L}(K)\Delta^{S}_{\mp}(Q)\bigg].
\label{F_pm_AS_--_+-}
\end{align}
Let us take the terms of Eq.~\eqref{F_pm_AS_--_+-} without $s$.
\begin{align}
F_{1\pm ; -- ; +-}^{\mathrm{AS}}(P, K) = & \, 
\int \frac{d \Omega_{s}}{4 \pi} \frac{1}{\left(P S-i \varepsilon\right)\left(Q S-i \varepsilon\right)}\int \frac{d \Omega_{s^{\prime}}}{4 \pi} \frac{1}{\left(P S^{\prime}+i \varepsilon\right)\left(Q S^{\prime}-i\varepsilon\right)} \nn
\times & \, \big(-k^{2}\big) \Big(1+\hat{p}.\hat{q}\Big)\tilde{D}^{A}_{L}(K)\Delta^{S}_{\mp}(Q)
\label{F1_AS_--_+-}
\end{align}
Using Eq.~\eqref{V_k_k0}, the above Eq.~\eqref{F1_AS_--_+-} can be written as
\begin{align}
F_{1\pm ; -- ; +-}^{\mathrm{AS}}(P, K) = & \,
-k^{2}\left[1+\bigg(\frac{p}{q}-\frac{kx}{q}\bigg)\right] V(t,k,k_{0},x,\varepsilon)\,V^{\prime}(t,k,k_{0},x,\varepsilon)\, \nn
\times & \tilde{D}^{A}_{L}(K)\Delta^{S}_{\mp}(Q)
\end{align}
The expression of $ V^{\prime}\left(t,k,k_{0},x,\varepsilon\right)$ is same as $V\left(t,k,k_{0},x,\varepsilon\right)$ except the change of $k_0 \rightarrow k_0 + 2 i \varepsilon$.
Let us take the terms of Eq.~\eqref{F_pm_AS_--_+-} consisting of $1s$, which can be written as 
\begin{align}
F_{2\pm ; -- ; +-}^{\mathrm{AS}}(P, K) = & \,
\bigg[\int \frac{d \Omega_{s}}{4 \pi} \frac{\hat{s}^{i}}{\left(P S-i \varepsilon\right)\left(Q S-i \varepsilon\right)}\int \frac{d \Omega_{s^{\prime}}}{4 \pi} \frac{1}{\left(P S^{\prime}+i \varepsilon\right)\left(Q S^{\prime}-i\varepsilon\right)}\nn
\times & \, \bigg\{\pm k^{2}\bigg(-\hat{p}_{i}-\hat{q}_{i}\bigg)  - k_{0}k \bigg(\hat{k}_{i}+\hat{p}\mdot\hat{q}\, \hat{k}_{i}\bigg)\bigg\} \nn
+ & \, \int \frac{d \Omega_{s}}{4 \pi} \frac{1}{\left(P S-i \varepsilon\right)\left(Q S-i \varepsilon\right)}\int \frac{d \Omega_{s^{\prime}}}{4 \pi} \frac{\hat{s}^{\prime
i}}{\left(P S^{\prime}+i \varepsilon\right)\left(Q S^{\prime}-i\varepsilon\right)}\nn 
\times & \, \bigg\{\pm k^{2}\bigg(-\hat{p}_{i}-\hat{q}_{i}\bigg) -k_{0}k \bigg(\hat{k}_{i}+\hat{p}\mdot\hat{q}\, \hat{k}_{i}\bigg)\bigg\} \bigg] \bigg(-\tilde{D}^{A}_{L}(K)\Delta_{\mp}^{\mathrm{S}}(Q)\bigg) 
\label{F2_AS_--_+-}
\end{align}
Using Eqs.~\eqref{V_k_k0}~\eqref{1s_int}, Eq.~\eqref{F2_AS_--_+-} becomes
\begin{align}
F_{2\pm ; -- ; +-}^{\mathrm{AS}}(P, K) = & \, -\bigg[\bigg\{\bigg(V_{1}(r_{0},r) \,V^{\prime}(t,k,k_{0},x,\varepsilon)+V^{\prime}_{1}(r_{0},r) \,V(t,k,k_{0},x,\varepsilon)\bigg)\,\nn
\times & \, \tilde{D}^{A}_{L}(K)\Delta^{S}_{\mp}(Q)\bigg\}\left\{\mp k^{2} \left(\hat{p}\mdot \hat{r} + \hat{q}\mdot\hat{r}\right) - k_{0}k \,\hat{k}\mdot\hat{r}\left(1+\hat{p}\mdot\hat{q}\right)\right\}\bigg]
\end{align}
The expression of $ V_{1}^{\prime} \left(r_0,r \right)$ is same as $V_{1} \left(r_0,r \right)$ except the change of $k_0 \rightarrow k_0 + 2 i \varepsilon$. Let us consider the terms of Eq.~\eqref{F_pm_AS_--_+-} with $2s$, which can be written as 
\begin{align}
F_{3\pm ; -- ; +-}^{\mathrm{AS}}(P, K) = & \,\bigg\{ 
\int \frac{d \Omega_{s}}{4 \pi} \frac{\hat{s}^{i}}{\left(P S-i \varepsilon\right)\left(Q S-i \varepsilon\right)}\int \frac{d \Omega_{s^{\prime}}}{4 \pi} \frac{\hat{s}^{\prime}_{i}}{\left(P S^{\prime}+i \varepsilon\right)\left(Q S^{\prime}-i\varepsilon\right)} \Big(-1-\hat{p}.\hat{q}\Big) \nn
+ & \, \int\! \frac{d \Omega_{s}}{4 \pi} \frac{\hat{s}^{i}}{\left(P S-i \varepsilon\right)\left(Q S-i \varepsilon\right)}\int\! \frac{d \Omega_{s^{\prime}}}{4 \pi} \frac{\hat{s}^{\prime {j}}}{\left(P S^{\prime}+i \varepsilon\right)\left(Q S^{\prime}-i\varepsilon\right)}\nn
\times & \, \Big(\hat{k}_{i}\hat{k}_{j} + \hat{p}.\hat{q}\,\hat{k}_{i}\hat{k}_{j} \Big) \bigg\} D_{T}^{\mathrm{A}}(K) \Delta_{\mp}^{\mathrm{S}}(Q) \nn
- & \, \bigg[ \int\! \frac{d \Omega_{s}}{4 \pi} \frac{\hat{s}^{i}}{\left(P S-i \varepsilon\right)\left(Q S-i \varepsilon\right)}\int\! \frac{d \Omega_{s^{\prime}}}{4 \pi} \frac{\hat{s}^{\prime {j}}}{\left(P S^{\prime}+i \varepsilon\right)\left(Q S^{\prime}-i\varepsilon\right)}\nn
\times & \, \bigg\{k^{2}\Big(\hat{p}_{i}\hat{q}_{j} +\hat{q}_{i}\hat{p}_{j}+1- \hat{p}.\hat{q}\Big) + k_{0}^{2}\big(\hat{k}_{i}\hat{k}_{j}\big(1+\hat{p}\mdot\hat{q}\big)\big) \nn
\mp & \, k_{0}k \big(-\hat{p}_{i}\hat{k}_{j}-\hat{q}_{i}\hat{k}_{j}-\hat{k}_{i}\hat{p}_{j}-\hat{k}_{i}\hat{q}_{j} \big)\bigg\} \nn
+ & \, \int\! \frac{d \Omega_{s}}{4 \pi} \frac{\hat{s}^{i}\,\hat{s}^{j}}{\left(P S-i \varepsilon\right)\left(Q S-i \varepsilon\right)}\int\! \frac{d \Omega_{s^{\prime}}}{4 \pi} \frac{1}{\left(P S^{\prime}+i \varepsilon\right)\left(Q S^{\prime}-i\varepsilon\right)} \nn
\times & \, \bigg\{\mp k_{0} k \big(-\hat{p}_{i}\hat{k}_{j}-\hat{q}_{i}\hat{k}_{j}\big)\bigg\}\nn
+ & \, \int\! \frac{d \Omega_{s}}{4 \pi} \frac{1}{\left(P S-i \varepsilon\right)\left(Q S-i \varepsilon\right)}\int\! \frac{d \Omega_{s^{\prime}}}{4 \pi} \frac{\hat{s}^{\prime i}\,\hat{s}^{\prime j}}{\left(P S^{\prime}+i \varepsilon\right)\left(Q S^{\prime}-i\varepsilon\right)}  \nn
\times & \, \bigg\{\mp k_{0} k \big(-\hat{p}_{i}\hat{k}_{j}-\hat{q}_{i}\hat{k}_{j}\big)\bigg\}\bigg] \bigg(\tilde{D}^{A}_{L}(K)\Delta^{S}_{\mp}(Q)\bigg)
\label{F3_AS_--_+-}
\end{align}
Using Eqs.~\eqref{V_k_k0}~\eqref{1s_int}~\eqref{2s_int}, Eq.~\eqref{F3_AS_--_+-} becomes
\begin{align}
F_{3\pm ; -- ; +-}^{\mathrm{AS}}(P, K) = & \, \int_{0}^{1} d u \left(V_1 \left(r_0,r \right)\right)\left(V^{\prime}_1 \left(r_0,r \right)\right)  D_{T}^{\mathrm{A}}(K) \Delta_{\mp}^{\mathrm{S}}(Q) \left[\left(1 + \hat{p}\mdot\hat{q}\right)\left\{\big(\hat{k}\mdot\hat{r}\big)^2 - 1 \right\}\right]\nn
- & \, \int_{0}^{1} d u \tilde{D}^{A}_{L}(K)\Delta^{S}_{\mp}(Q) \bigg[ V_1 \left(r_0,r \right) \, V^{\prime}_1 \left(r_0,r \right) \left\{k^{2}\left(2\hat{p}\mdot\hat{r}\,\hat{q}\mdot\hat{r}-\hat{p}\mdot\hat{q}+1\right)\right. \nn
+ & \, \left. k_{0}^{2} \big(\hat{k}\mdot\hat{r}\big)^{2}\left(1+\hat{p}\mdot\hat{q}\right)\pm 2k_{0}k \,\hat{k}\mdot\hat{r}\big(\hat{p}\mdot\hat{r}+\hat{q}\mdot\hat{r}\big) \right\}  V^{\prime}(t,k,k_{0},x,\varepsilon) \, \pm k_{0}k \nn
\times & \, \left\{A\big(x+\hat{q}\mdot\hat{k}\big)+B\hat{k}\mdot\hat{r}\left(\hat{p}\mdot\hat{r}+\hat{q}\mdot\hat{r}\right)\right\} \pm V(t,k,k_{0},x,\varepsilon) k_{0}k \nn 
\times & \, \left\{A^{\prime}\big(x+\hat{q}\mdot\hat{k}\big)+B^{\prime}\hat{k}\mdot\hat{r}\left(\hat{p}\mdot\hat{r}+\hat{q}\mdot\hat{r}\right)\right\}\bigg]\qquad\quad
\label{F3_int_AS_--_+-}
\end{align}
The expression of $ A^{\prime}$ and $B^{\prime}$ are the same as $A$ and $B$ except the change of $k_0 \rightarrow k_0 + 2 i \varepsilon$. Now, let us consider the terms of Eq.~\eqref{F_pm_AS_--_+-} with $3s$, and it can be expressed as 
\begin{align}
F_{4\pm ; -- ; +-}^{\mathrm{AS}}(P, K) = & \,
\Bigg[\int \frac{d \Omega_{s}}{4 \pi} \frac{\hat{s}^{i}\hat{s}^{j}}{\left(P S-i \varepsilon\right)\left(Q S-i \varepsilon\right)}\int \frac{d \Omega_{s^{\prime}}}{4 \pi} \frac{\hat{s}^{\prime}_{j}}{\left(P S^{\prime}+i \varepsilon\right)\left(Q S^{\prime}-i\varepsilon\right)}\Big(\hat{p}_{i}+ \hat{q}_{i} \Big) \nn
+ & \,  \int \frac{d \Omega_{s}}{4 \pi} \frac{\hat{s}^{i}\hat{s}^{j}}{\left(P S-i \varepsilon\right)\left(Q S-i \varepsilon\right)}\int \frac{d \Omega_{s^{\prime}}}{4 \pi} \frac{\hat{s}^{\prime k}}{\left(P S^{\prime}+i \varepsilon\right)\left(Q S^{\prime}-i\varepsilon\right)}\nn
\times & \, \left(-\hat{p}_{i} \hat{k}_{j}\hat{k}_{k} - \hat{q}_{i} \hat{k}_{j}\hat{k}_{k} \right) \nn
+ & \, \int \frac{d \Omega_{s}}{4 \pi} \frac{\hat{s}^{j}}{\left(P S-i \varepsilon\right)\left(Q S-i \varepsilon\right)}\int \frac{d \Omega_{s^{\prime}}}{4 \pi} \frac{\hat{s}^{\prime i}\hat{s}^{\prime}_{j}}{\left(P S^{\prime}+i \varepsilon\right)\left(Q S^{\prime}-i\varepsilon\right)} \Big(\hat{p}_{i} + \hat{q}_{i} \Big) \nn
- & \, \int \frac{d \Omega_{s}}{4 \pi} \frac{\hat{s}^{j}}{\left(P S-i \varepsilon\right)\left(Q S-i \varepsilon\right)}\int \frac{d \Omega_{s^{\prime}}}{4 \pi} \frac{\hat{s}^{\prime i}\hat{s}^{\prime k}}{\left(P S^{\prime}+i \varepsilon\right)\left(Q S^{\prime}-i\varepsilon\right)}\nn
\times & \, \left(\hat{p}_{i} \hat{k}_{j}\hat{k}_{k} + \hat{q}_{i} \hat{k}_{j}\hat{k}_{k} \right)\Bigg]\pm D_{T}^{\mathrm{A}}(K) \Delta_{\mp}^{\mathrm{S}}(Q)\nn
- & \, \bigg[ \int \frac{d \Omega_{s}}{4 \pi} \frac{\hat{s}^{i}\hat{s}^{j}}{\left(P S-i \varepsilon\right)\left(Q S-i \varepsilon\right)}\int \frac{d \Omega_{s^{\prime}}}{4 \pi} \frac{\hat{s}^{\prime k}}{\left(P S^{\prime}+i \varepsilon\right)\left(Q S^{\prime}-i\varepsilon\right)}\nn
\times & \, \bigg\{-k_{0}k\bigg(\hat{p}_{i}\hat{k}_{j}\hat{q}_{k}+\hat{q}_{i}\hat{k}_{j}\hat{p}_{k}\bigg)\pm k_{0}^{2}\bigg(-\hat{p}_{i}\hat{k}_{j}\hat{k}_{k}-\hat{q}_{i}\hat{k}_{j}\hat{k}_{k}\bigg)\bigg\}\nn
+ & \, \int \frac{d \Omega_{s}}{4 \pi} \frac{\hat{s}^{k}}{\left(P S-i \varepsilon\right)\left(Q S-i \varepsilon\right)} \int \frac{d \Omega_{s^{\prime}}}{4 \pi} \frac{\hat{s}^{\prime i}\,\hat{s}^{\prime j}}{\left(P S^{\prime}+i \varepsilon\right)\left(Q S^{\prime}-i\varepsilon\right)} \nn
\times & \, \bigg\{-k_{0}k\bigg(\hat{q}_{i}\hat{k}_{j}\hat{p}_{k}+\hat{p}_{i}\hat{k}_{j}\hat{q}_{k}\bigg)\pm k_{0}^{2}\bigg(-\hat{p}_{i}\hat{k}_{j}\hat{k}_{k}-\hat{q}_{i}\hat{k}_{j}\hat{k}_{k}\bigg)\bigg\}\nn
+ & \, \int \frac{d \Omega_{s}}{4 \pi} \frac{\hat{s}^{i}\hat{s}^{j}}{\left(P S-i \varepsilon\right)\left(Q S-i \varepsilon\right)}\int \frac{d \Omega_{s^{\prime}}}{4 \pi} \frac{\hat{s}^{\prime}_{j}}{\left(P S^{\prime}+i \varepsilon\right)\left(Q S^{\prime}-i\varepsilon\right)} \nn
\times & \, \bigg\{-k_{0}k \bigg(\hat{k}_{i}-\hat{p}\mdot\hat{q}\,\hat{k}_{i}\bigg)\bigg\}\nn
+ & \, \int \frac{d \Omega_{s}}{4 \pi} \frac{\hat{s}_{j}}{\left(P S-i \varepsilon\right)\left(Q S-i \varepsilon\right)}\int \frac{d \Omega_{s^{\prime}}}{4 \pi} \frac{\hat{s}^{\prime i}\hat{s}^{\prime j}}{\left(P S^{\prime}+i \varepsilon\right)\left(Q S^{\prime}-i\varepsilon\right)}\nn
\times & \, \bigg\{-k_{0}k \bigg(\hat{k}_{i}-\hat{p}\mdot\hat{q}\,\hat{k}_{i}\bigg)\bigg\} \bigg] \bigg(\tilde{D}^{A}_{L}(K)\Delta^{S}_{\mp}(Q)\bigg)
\label{F4AS_3s}
\end{align}
The above Eq.~\eqref{F4AS_3s} can be simplified using Eqs.~\eqref{1s_int},\eqref{2s_int}. Thus, we will get
\begin{align}
F_{4\pm ; -- ; +-}^{\mathrm{AS}}(P, K) = & \, \pm \int_{0}^{1} d u \, D_{T}^{\mathrm{A}}(K) \Delta_{\mp}^{\mathrm{S}}(Q)\, \bigg [\left\{V_{1}^{\prime}(r_0,r) A + V_1(r_0,r) A^{\prime}\right\} \nn
\times & \, \left(\hat{p}.\hat{r} + \hat{q}.\hat{r} -\hat{p}.\hat{k} \, \hat{r}.\hat{k} -\hat{q}.\hat{k} \, \hat{r}.\hat{k}\right)+\left\{V_{1}^{\prime}(r_0,r) B +  V_1(r_0,r) B^{\prime}\right\}\nn
\times & \,  \left(\hat{p}\mdot\hat{r} + \hat{q}.\hat{r} -\hat{p}.\hat{r} \, (\hat{k}.\hat{r})^2 - \hat{q}.\hat{r} \, (\hat{k}.\hat{r})^2 \right)\bigg] + \int_{0}^{1} d u \,\bigg[\bigg(V_{1}^{\prime}(r_0,r) A  \nn
+ & \, V_1(r_0,r) A^{\prime}\bigg)  \left(k_{0} k \left(\hat{p}\mdot \hat{k} \, \hat{q}\mdot \hat{r} + \hat{q}\mdot \hat{k} \, \hat{p}\mdot \hat{r} +\hat{k}\mdot \hat{r} \big(1-\hat{p}\mdot\hat{q}\big)\right)  \right.\nn
\pm & \,\left. k_{0}^{2} \big(\hat{p}\mdot \hat{k} \, \hat{k}\mdot \hat{r}+\hat{q}\mdot \hat{k} \, \hat{k}\mdot \hat{r}\big)\right)  + \left(V_{1}^{\prime}(r_0,r) B + V_1(r_0,r) B^{\prime}\right)   \nn
\times & \, \bigg(k_{0} k \left(2\hat{p}\mdot \hat{r} \, \hat{q}\mdot \hat{r}\, \hat{k}\mdot \hat{r} +\hat{k}\mdot \hat{r} \big(1-\hat{p}\mdot\hat{q}\big)\right) \pm k_{0}^{2} \big(\hat{k}\mdot\hat{r}\big)^{2}\big(\hat{p}\mdot \hat{r}+\hat{q}\mdot \hat{r}\big)\bigg)\bigg]\nn
\times & \, \tilde{D}^{A}_{L}(K)\Delta^{S}_{\mp}(Q)
\end{align}
Let us consider the terms of Eq.~\eqref{F_pm_AS_--_+-} with $4s$, and it can be written as
\begin{align}
F_{5\pm ; -- ; +-}^{\mathrm{AS}}(P, K) 
= & \, \Bigg[\int \frac{d \Omega_{s}}{4 \pi} \frac{\hat{s}^{i}\hat{s}^{k}}{\left(P S-i \varepsilon\right)\left(Q S-i \varepsilon\right)}\int \frac{d \Omega_{s^{\prime}}}{4 \pi} \frac{\hat{s}^{\prime j} \hat{s}^{\prime}_{k}}{\left(P S^{\prime}+i \varepsilon\right)\left(Q S^{\prime}-i\varepsilon\right)} \nn
\times & \, \Big(-\hat{p}_{i}\hat{q}_{j} - \hat{q}_{i}\hat{p}_{j} \Big) \nn
+ & \, \int \frac{d \Omega_{s}}{4 \pi} \frac{\hat{s}^{i}\hat{s}^{j}}{\left(P S-i \varepsilon\right)\left(Q S-i \varepsilon\right)}\int \frac{d \Omega_{s^{\prime}}}{4 \pi} \frac{\hat{s}^{\prime}_{i} \hat{s}^{\prime}_{j}}{\left(P S^{\prime}+i \varepsilon\right)\left(Q S^{\prime}-i\varepsilon\right)}\nn 
\times & \, \Big(\hat{p}.\hat{q} - 1  \Big)\nn
+ & \, \int \frac{d \Omega_{s}}{4 \pi} \frac{\hat{s}^{i}\hat{s}^{k}}{\left(P S-i \varepsilon\right)\left(Q S-i \varepsilon\right)}\int \frac{d \Omega_{s^{\prime}}}{4 \pi} \frac{\hat{s}^{\prime j} \hat{s}^{\prime l}}{\left(P S^{\prime}+i \varepsilon\right)\left(Q S^{\prime}-i\varepsilon\right)} \nn 
\times & \, \left(\hat{p}_{i}\hat{q}_{j} \hat{k}_{k} \hat{k}_{l} + \hat{q}_{i}\hat{p}_{j} \hat{k}_{k} \hat{k}_{l} \right)\nn
+ & \, \int \frac{d \Omega_{s}}{4 \pi} \frac{\hat{s}^{i}\hat{s}^{j}}{\left(P S-i \varepsilon\right)\left(Q S-i \varepsilon\right)}\int \frac{d \Omega_{s^{\prime}}}{4 \pi} \frac{\hat{s}^{\prime}_{i} \hat{s}^{\prime k}}{\left(P S^{\prime}+i \varepsilon\right)\left(Q S^{\prime}-i\varepsilon\right)}\nn
\times & \,\hat{k}_{j} \hat{k}_{k} \Big(1 - \hat{p}.\hat{q} \Big)\Bigg] {D}_{T}^{\mathrm{A}}(K) \Delta_{\mp}^{\mathrm{S}}(Q)\nn
+ & \, \bigg[\int \frac{d \Omega_{s}}{4 \pi} \frac{\hat{s}^{i}\hat{s}^{j}}{\left(P S-i \varepsilon\right)\left(Q S-i \varepsilon\right)}\int \frac{d \Omega_{s^{\prime}}}{4 \pi} \frac{\hat{s}^{\prime k} \hat{s}^{\prime l}}{\left(P S^{\prime}+i \varepsilon\right)\left(Q S^{\prime}-i\varepsilon\right)}\nn
\times & \, \left(\hat{p}_{i}\hat{k}_{j} \hat{q}_{k} \hat{k}_{l} + \hat{q}_{i}\hat{k}_{j} \hat{p}_{k} \hat{k}_{l} \right)\nn
+ & \, \int \frac{d \Omega_{s}}{4 \pi} \frac{\hat{s}^{i}\hat{s}^{j}}{\left(P S-i \varepsilon\right)\left(Q S-i \varepsilon\right)}\int \frac{d \Omega_{s^{\prime}}}{4 \pi} \frac{\hat{s}^{\prime k} \hat{s}^{\prime}_{j}}{\left(P S^{\prime}+i \varepsilon\right)\left(Q S^{\prime}-i\varepsilon\right)} \nn
\times & \, \left(\hat{k}_{i}\hat{k}_{k} - \hat{p}\mdot\hat{q} \hat{k}_{i} \hat{k}_{k} \right)\bigg] \bigg(k_{0}^{2} \, \tilde{D}^{A}_{L}(K)\Delta^{S}_{\mp}(Q)\bigg)
\label{F3AS_4s}
\end{align}
Now, the Eq.~\eqref{F3AS_4s} can be rewritten in terms of $A,B,A'$ and $B'$ variables as defined in Eq.~\eqref{A_B_Def} as
\begin{align}
F_{5\pm ; -- ; +-}^{\mathrm{AS}}(P, K)
 = & \,  \int_{0}^{1} du \, \Bigg[2AA^{\prime} \left( x\, \hat{q}\mdot\hat{k} -1\right) + BB^{\prime} \left(\hat{p}\mdot\hat{q} - 2 \hat{p}.\hat{r} \, \hat{q}\mdot\hat{r}  - 1 + 2 \hat{p}\mdot\hat{r} \, \hat{q}\mdot\hat{r} \, \big(\hat{k}\mdot\hat{r}\big)^2
\right. \nn
+ & \, \left.\big(\hat{k}\mdot\hat{r}\big)^2 \left( 1 - \hat{p}\mdot\hat{q}\right) \right) + \big(A^{\prime}B+AB^{\prime}\big)\, \left\{-2 \hat{p}.\hat{r} \, \hat{q}\mdot\hat{r} +  \left(\hat{p}\mdot\hat{q} - 1 \right) \right.  \nn
+ & \, \left. x \, \hat{q}\mdot\hat{r} \, \hat{k}\mdot\hat{r} +  \hat{q}.\hat{k} \, \hat{p}\mdot\hat{r} \,\hat{k}\mdot\hat{r} +  \left(\hat{k}\mdot\hat{r}\right)^2 \left( 1 - \hat{p}\mdot\hat{q}\right) \right\}\Bigg] D_{T}^{\mathrm{A}}(K) \Delta_{\mp}^{\mathrm{S}}(Q) \nn
- & \, \int_{0}^{1} du \,k_{0}^{2}\,\bigg[\left(1-\hat{p}\mdot\hat{q}\right)\bigg(AA^{\prime} + \left(BB^{\prime}+A^{\prime}B+AB^{\prime}\right)\big(\hat{k}\mdot\hat{r}\big)^{2}\bigg) \nn
+ & \, 2AA^{\prime} \, \hat{p}\mdot\hat{k}\, \hat{q}\mdot\hat{k} + 2BB^{\prime} \, \hat{p}\mdot\hat{r} \, \hat{q}\mdot\hat{r} \, \big(\hat{k}\mdot\hat{r}\big)^{2} \big(A^{\prime}B+AB^{\prime}\big) \nn
\times & \left(\hat{p}\mdot\hat{k} \, \hat{q}\mdot\hat{r} \, \hat{k}\mdot\hat{r} + \hat{q}\mdot\hat{k} \, \hat{p}\mdot\hat{r} \, \hat{k}\mdot\hat{r} \right)  \bigg] \, \tilde{D}^{A}_{L}(K)\Delta^{S}_{\mp}(Q) 
\end{align}
Adding all the five individual contributions, Eq.~\eqref{Sigma1_7} becomes
\begin{align}
\Sigma _{7\pm}^{(1)}(P) = & \, \frac{-ig^2 C_F}{2(2\pi)^4}  (2\pi) \int_{-\infty}^\infty dk_0 \int_{0}^\infty k^2 dk  \int_{-1}^1 dx \, \,\bigg[F_{1\pm;--;+-}^{AS}(P,K)+F_{2\pm;--;+-}^{AS}(P,K) \nn
+ & \, F_{3\pm;--;+-}^{AS}(P,K)+F_{4\pm;--;+-}^{AS}(P,K)+F_{5\pm;--;+-}^{AS}(P,K)\bigg].
\end{align}
Eqs.~\eqref{Sigma1_4},~\eqref{Sigma1_5},~\eqref{Sigma1_6} and~\eqref{Sigma1_7} have the same kind of divergences as we seen in Eq.~\eqref{Sigma13}. Thus, we get the same sudden jumps as we mentioned in Eq.~\eqref{diverg_pts_1}. So, these equations are also evaluated numerically in each of the Figure~\ref{fig7} domains and summed up. Now, the eighth term of Eq.~\eqref{sigma_final} is 
\begin{equation}\label{Sigma1_8}
\Sigma _{8\pm}^{(1)}(P) = \frac{-ig^2 C_F}{2} \int \frac{d^4K}{(2\pi)^4}\Big[G_{\pm ; --}^{\mathrm{S}}(P, K)\Big]
\end{equation}
The complete terms of $ G_{\pm ; --}^{\mathrm{S}}(P, K) $ is 
\begin{align}
G_{\pm ; --}^{\mathrm{S}}(P, K)  = & \, \int \frac{d \Omega_{s}}{4 \pi} \frac{1}{\left[P S-i\varepsilon\right]\left[P S-i \varepsilon\right]}  \left[\frac{1}{(P+K) S-i \varepsilon}+\frac{1}{(P-K) S-i \varepsilon}\right] \nn
\times & \, \Bigg[\left\{1-\hat{p}_{\epsilon}.\hat{s}-\left(\hat{k}. \hat{s}\right)^{2}+\hat{p}_{\epsilon}.\hat{s}\,\left(\hat{k}.\hat{s}\right)^{2}\right\} D_{T}^{\mathrm{S}}(K) \nn
+ & \, \ \Big\{k^{2}\left(1-\hat{p}_{\epsilon}.\hat{s}\right)-2 k_{0} k(\hat{k}.\hat{s}-\hat{p}_{\epsilon}.\hat{s}
\,\hat{k}.\hat{s})+k_{0}^{2} \,(1-\hat{p}_{\epsilon}.\hat{s}) \,(\hat{k}.\hat{s})^{2}\Big\}\tilde{D}_{L}^{\mathrm{S}}(K)\Bigg].
\label{Sigma18_tran}
\end{align}
The term of Eq.~\eqref{Sigma18_tran} without any s are 
\begin{align}
G_{1\pm ; - - }^{\mathrm{S}}(P, K) &=& \int \frac{d \Omega_{s}}{4 \pi} \frac{1}{\left[P S\right]\left[P S\right]}\left[\frac{1}{(P+K) S}+\frac{1}{(P-K) S}\right] \bigg\{ {D}_{T}^{\mathrm{S}}(K) + k^{2} \,\tilde{D}_{L}^{\mathrm{S}}(K)\bigg\} 
\label{G1pm}
\end{align}
Using Eq.~\eqref{J_mu_nu_alpha}, Eq.~\eqref{G1pm} can be written as
\begin{equation}
\begin{aligned}
G_{1\pm ; - - }^{\mathrm{S}}(P, K) &=& \left[J^{000}_{--} \left(P,K\right) + J^{000}_{--} \left(P,-K\right) \right] \bigg\{ {D}_{T}^{\mathrm{S}}(K) + k^{2} \,\tilde{D}_{L}^{\mathrm{S}}(K)\bigg\} 
\label{G1pm}
\end{aligned}
\end{equation}
Now, using Eq.~\eqref{J000}, we get
\begin{equation}
\begin{aligned}
J_{--}^{000}(P, K) &=2 \int_{0}^{1} d u_{1} u_{1} \int_{0}^{1} d u_{2} \frac{t_{0}}{\left(t_{0}^{2}-t^{2}\right)^{2}}
\end{aligned}
\end{equation}
where \begin{equation}
\begin{aligned}
t_0 = u_1 u_2 k_0 + p_0 - i \varepsilon , \hspace{5mm}\vec{t} = u_1 u_2 \vec{k} + \vec{p}\ \text{ and } \
 t = \sqrt{p^2+k^2 u_{1}^2 u_{2}^2 + 2 p k u_1 u_2 x}.
\end{aligned}
\end{equation}
Thus, Eq.~\eqref{G1pm} becomes
\begin{equation}
\begin{aligned}
G_{1\pm ; - - }^{\mathrm{S}}(P, K) = 2 \int_{0}^{1} d u_{1} u_{1} \int_{0}^{1} d u_{2} \,\bigg[\frac{t_{0}}{\left(t_{0}^{2}-t^{2}\right)^{2}}+ \frac{t^{\prime}_{0}}{\left(t^{\prime 2}_{0}-t^{\prime 2}\right)^{2}}\bigg]\, \bigg\{ {D}_{T}^{\mathrm{S}}(K) + k^{2} \,\tilde{D}_{L}^{\mathrm{S}}(K)\bigg\} 
\end{aligned}
\end{equation}
The variables $t^{\prime}_{0}$ and $t^{\prime}$ have the same expression as $t_{0}$ and $t$ with negative gluon four-momentum $K$. Now, the required angle in order to solve the other terms of Eq.~\eqref{sigma2_htl} are
\begin{equation}
\hat{p}\cdot\!\hat{t} = \frac{p}{t} + \frac{k u_1 u_2}{t}x \quad \quad
\hat{k}\cdot\!\hat{t} = \frac{p}{t}x + \frac{k u_1 u_2}{t}
\end{equation}
Let us consider the terms of Eq.~\eqref{Sigma1_8} with 1s i.e. $G_{2\pm ; - - }^{\mathrm{S}}(P, K)$. This term can be simplified using Eq.~\eqref{J_mu_nu_alpha} and Eq.~\eqref{J00i} and having the final form as 
\begin{align}
G_{2\pm ; - - }^{\mathrm{S}}(P, K) = & \,  \mp \, 2 \int_{0}^{1} d u_{1} u_{1} \int_{0}^{1} d u_{2} \bigg(\frac{1}{\left(t_{0}^{2}-t^{2}\right)^{2}}+\frac{1}{\left(t_{0}^{\prime 2}-t^{\prime 2}\right)^{2}}\bigg) \nn
\times & \, \left[\hat{p}\mdot\hat{t} \, {D}_{T}^{\mathrm{S}}(K)
+ \bigg( k^{2} \,\hat{p}\mdot\hat{t}\pm 2k_{0}k \, \hat{k}\mdot\hat{t}\bigg)\,\tilde{D}_{L}^{\mathrm{S}}(K)
\right]
\label{G2pmS}
\end{align}
In the similar manner, the term with 2s of Eq.~\eqref{Sigma1_8} can be expressed using Eqs.~\eqref{J_mu_nu_alpha} and ~\eqref{J0ij} as
\begin{align}
G_{3\pm ; - - }^{\mathrm{S}}(P, K) = & \, 2 \int_{0}^{1} d u_{1} u_{1} \int_{0}^{1} d u_{2} \, \bigg[-{D}_{T}^{\mathrm{S}}(K) \bigg\{C(P,K) + C (P,-K) \nn 
+ & \, \left(D(P,K) + D(P,-K)\right)  \big(\hat{k}.\hat{t}\big)^2 \bigg\} \nn
+ & \, \tilde{D}_{L}^{\mathrm{S}}(K)  \bigg\{\left(C(P,K) + C (P,-K)\right)\left( k_{0}^{2} \pm 2k_{0}k \,\hat{p}\mdot\hat{k} \right)\nn
+ & \, \left(D(P,K) + D(P,-K)\right) \left( k_{0}^{2} \,\big(\hat{k}\mdot\hat{t}\big)^{2} \pm 2k_{0}k \,\hat{p}\mdot\hat{t}\,\hat{k}\mdot\hat{t} \right)\bigg\}              
\bigg]\ 
\end{align}
where
\begin{equation}
\begin{aligned}
C (P,K) = \frac{t_{0}}{2 t^{2}\left(t_{0}^{2}-t^{2}\right)}-\frac{1}{4 t^{3}} \ln \frac{t_{0}+t}{t_{0}-t}, \quad \quad
D (P,K )  =\frac{t_{0}\left(5 t^{2}-3 t_{0}^{2}\right)}{2 t^{2}\left(t_{0}^{2}-t^{2}\right)^{2}}+\frac{3}{4 t^{3}} \ln \frac{t_{0}+t}{t_{0}-t}
\end{aligned}
\end{equation}
Let us take the terms of Eq.~\eqref{Sigma1_8} consisting of 3s terms, which take the form after doing some simplification 
\begin{align}
G_{4\pm ; - - }^{\mathrm{S}}(P, K) = & \, \pm 2 \int_{0}^{1} d u_{1} u_{1} \int_{0}^{1} d u_{2} \, \bigg[{D}_{T}^{\mathrm{S}}(K) \Big\{\left(E(P,K) + E (P,-K)\right)\left(\hat{p}\mdot\hat{t} + 2x \hat{k}\mdot\hat{t} \right) \nn
+ & \, \left(F(P,K) + F(P,-K)\right)  \hat{p}\mdot\hat{t} \big(\hat{k}\mdot\hat{t}\big)^2 \Big\}-\tilde{D}_{L}^{\mathrm{S}}(K) \bigg\{\left(E(P,K) + E (P,-K)\right) \nn
\times & \, \left(\hat{p}\mdot\hat{t} + 2x \hat{k}\mdot\hat{t} \right) \left(F(P,K) + F(P,-K)\right)\  \hat{p}\mdot\hat{t} \left(\hat{k}\mdot\hat{t}\right)^2 
\bigg\}\bigg]
\end{align}
with
\begin{align}
E (P,K ) = & \, \frac{1}{2 t^{3}}\left[2+\frac{t_{0}^{2}}{t_{0}^{2}-t^{2}}-\frac{3 t_{0}}{2 t} \ln \frac{t_{0}+t}{t_{0}-t}\right] ,\quad \quad \text{and}  \nn
F(P,K) = & \, \frac{t}{\left(t_{0}^{2}-t^{2}\right)^{2}}-\frac{5}{2 t^{3}}\left[2+\frac{t_{0}^{2}}{t_{0}^{2}-t^{2}}-\frac{3 t_{0}}{2 t} \ln \frac{t_{0}+t}{t_{0}-t}\right].
\end{align}
\begin{figure}
\centering
{\includegraphics[scale=0.8,keepaspectratio]{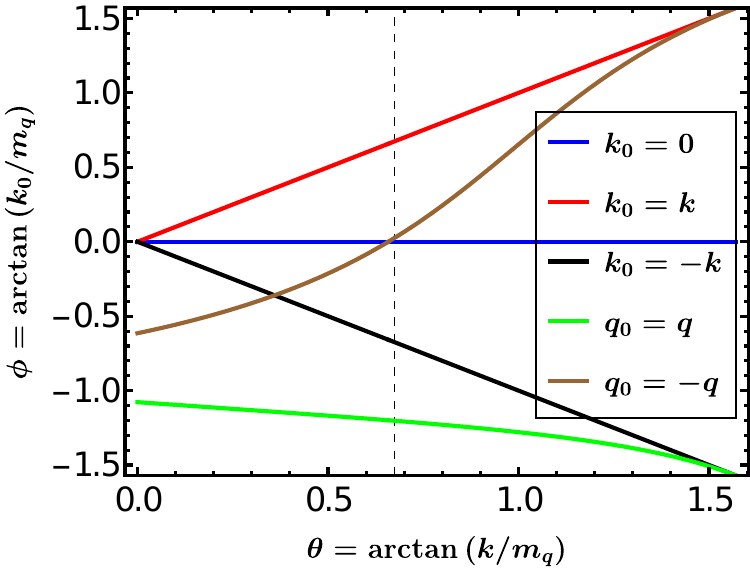}}
\caption{Domains in $(k,k_0)$ plane at which the integrand in Eq.~\eqref{Sigma8_final} has sudden jumps. Here we have used $t\equiv p/p_{0} = 0.45 $ and $x \equiv \cos\theta= 0.8 $} 
\label{fig8}    
\end{figure} 
After adding four-contributions of $\Sigma 8_{\pm}^{(1)}(P)$, Eq.~\eqref{Sigma1_8} becomes
\begin{align}
\Sigma _{8\pm}^{(1)}(P) = & \, \frac{-ig^2 C_F}{(2\pi)^3}  \int\limits_{-\infty}^\infty dk_0 \int\limits_{0}^\infty k^2 dk  \int_{-1}^1 dx \Big[G_{1\pm;--}^{S}(P,K) +G_{2\pm;--}^{S}(P,K) \nn 
+ & \, G_{3\pm;--}^{S}(P,K)+G_{4\pm;--}^{S}(P,K)\Big].\label{Sigma8_final}
\end{align}
Now, to evaluate Eq.~\eqref{Sigma8_final}, we find sudden jumps in the integrand at the following points:
\begin{align}
k_{0} = & \, 0 ; \quad k_{0}=\pm k ; \quad
k_0 = -p_0 \pm \sqrt{p^2+k^2+2pkx} ;  \nn
k = & \, \frac{1}{2 t} \frac{1-t^{2}}{1-x t} \sqrt{\frac{t}{1-t}-\frac{1}{2} \ln \left(\frac{1+t}{1-t}\right)}\label{domain2}
\end{align}
Figure~\ref{fig8} depicts the domains of Eq.~\eqref{domain2}. We have evaluated Eq.~\eqref{Sigma8_final} numerically in each of the individual domains of Figure~\ref{fig8} and added the individual contribution to get the final result.
\section{Results and Discussion}\label{sec:result}
\begin{figure}
\centering
\includegraphics[scale=0.5,keepaspectratio]{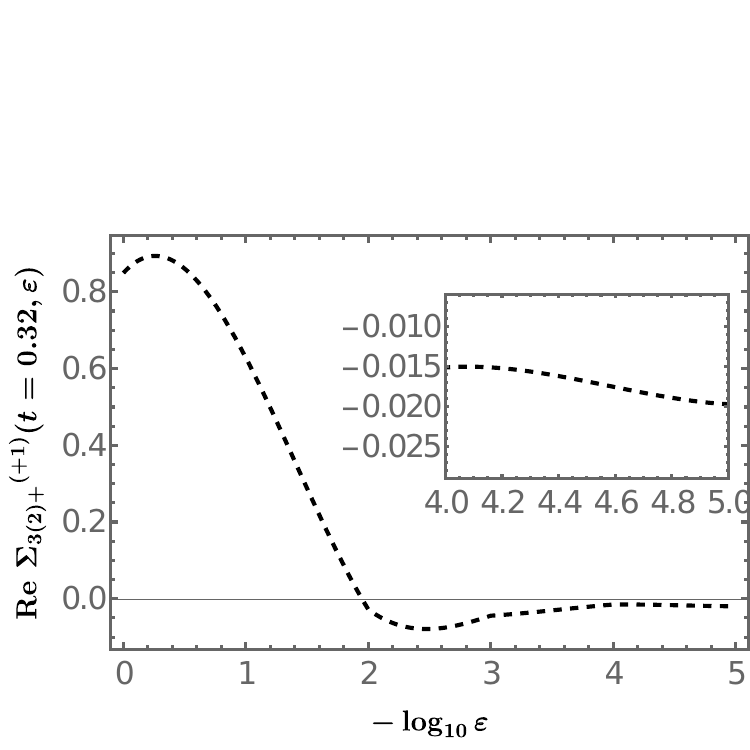}
\qquad
\includegraphics[scale=0.5,keepaspectratio]{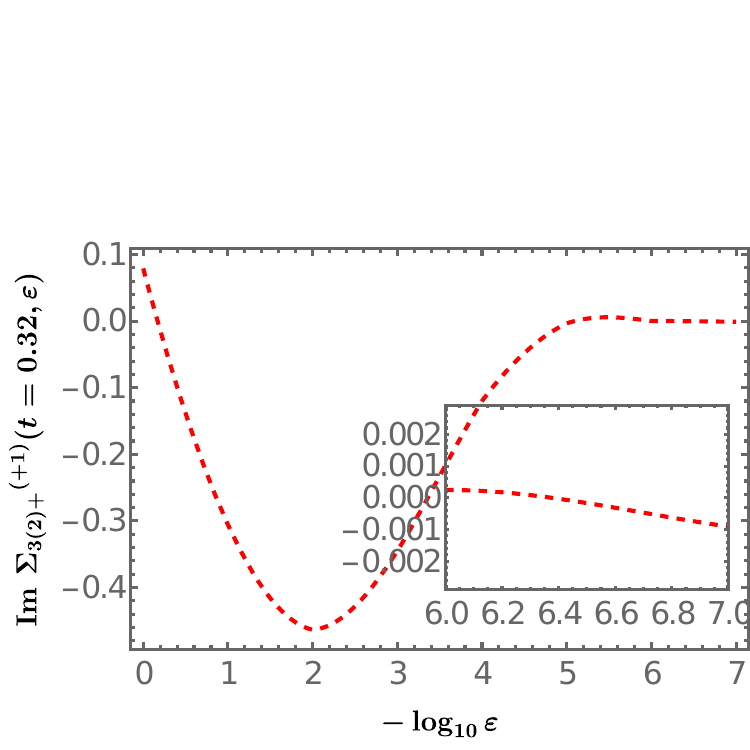}
\caption{The $\varepsilon$ variation of the Real and Imaginary part of the integral $\Sigma_{3(2)\pm}^{(+1)}$ at $t=0.32$ containing only transverse contribution of Eq.~\eqref{F2_PM_SR_--_Sigma13} in multiple of $gm_{q}$.}
\label{fig9}
\end{figure}  
All the terms of NLO quark self-energy in Eq.~\eqref{sigma_final} have a non-trivial dependence on $\varepsilon $. So, in order to be more precise in the integral results, one needs to check the stability for each of the integrals very carefully, which depends non-trivially on the $\varepsilon$ parameter. Here, we have checked the stability for each term by plotting the integrand of that particular integral with $ -\log_{10}\varepsilon$. This is an essential task because different terms have different stability regions, and if one does the integration beyond those regions, then numerical values lose reliability. Even one needs to perform the integrals, which are divided into the different domains as shown in Figures~\ref{fig6}, \ref{fig7}, \ref{fig8}, depending on their stability. For demonstration purposes, we have shown the $\varepsilon$ dependence plot in Figure~\ref{fig9} for the transverse part of the integral mentioned in Eq.~\eqref{F2_PM_SR_--_Sigma13}. For this particular term, on average, we found the stability region around $\sim 10^{-5}$ for the real part and $\sim 10^{-6}$ for the imaginary part, respectively. Then, the integral for this particular term has been done around these stable regions. Similarly, other integrals have been handled in calculating NLO quark self-energy. \\
Note that there are a total of 92 terms (46 for $\Sigma_{+}^{(1)}$ and 46 for $\Sigma_{-}^{(1)}$) for which we needed to check the convergence. All these terms have convergence for different values of $\varepsilon$. For example, as we have mentioned, for a particular term in Eq.~\eqref{F2_PM_SR_--_Sigma13}, the stability region is around $\sim 10^{-5}$ for the real part and $\sim 10^{-6}$ for the imaginary part. In principle, the $\varepsilon$ value should be zero, but our numerical evaluation can not handle that. Thus, because of the finite value of the $\varepsilon$ parameter, numerical errors are introduced in the evaluation of NLO quark self-energy. We have done the numerical estimation of the percentage of error in the following way: We have taken a few values of $\varepsilon$ in the stability region. Then, we have extrapolated the value of that particular $\Sigma_{\pm}^{(1)}$ term to the limit of $\varepsilon$ approaches to $0$. We estimated the error of that particular term from the difference between the considered value and the extrapolated value. The error of 46 terms for $\Sigma_{+}^{(1)}$ is calculated, and we estimated the total error in the evaluation of NLO quark self-energy as $\frac{\delta\Sigma_{+}^{(1)}}{\Sigma_{+}^{(1)}} = \sum_{i} \sqrt{{ \bigg(\frac{\delta \Sigma_{+ i}^{(1)}}{\Sigma_{+ i}^{(1)}}\bigg)}^2}$. We have estimated the maximum error for the measurement of the NLO quark mass due to the finite $\varepsilon$ value being about $11\%$, and for the damping rate, it is about $10\%$.
\begin{figure}
\centering
\includegraphics[scale=0.5,keepaspectratio]{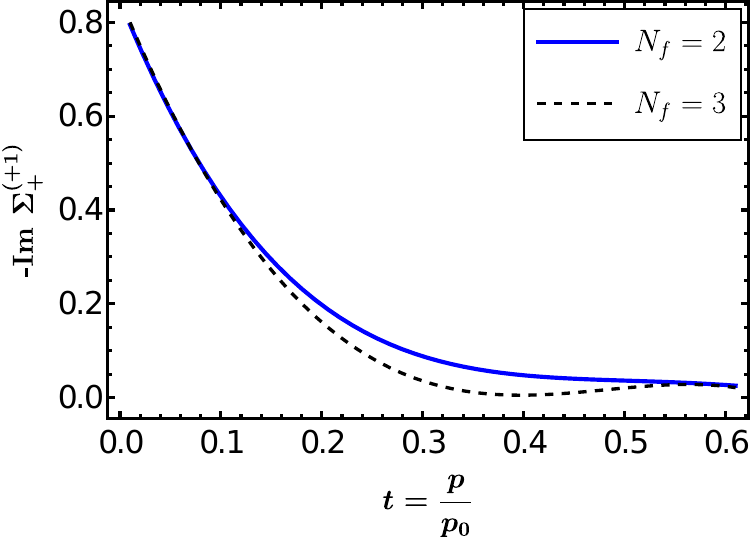}
\qquad
\includegraphics[scale=0.5,keepaspectratio]{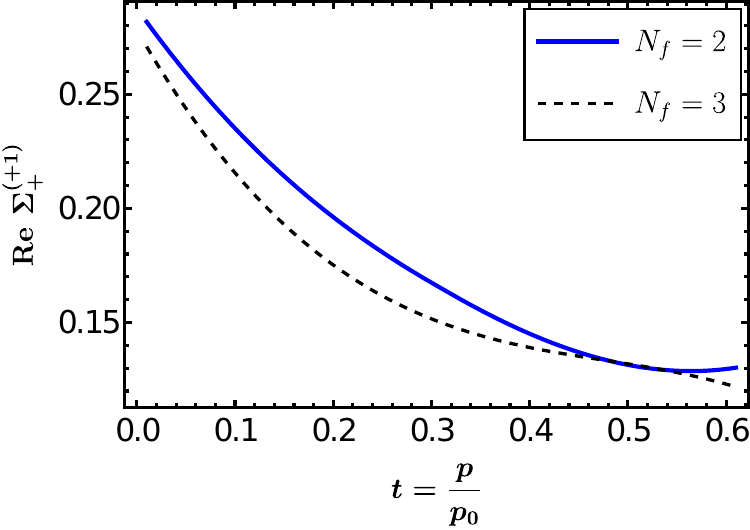}
\caption{The Imaginary part and Real part of $\Sigma_{+}^{(1)}$, scaled with a coefficient of $g m_{q}$, with respect to parameter $t=p/p_{0}$}
\label{fig10}
\end{figure}  
\\
In section~\ref{sec4}, we have given the expressions for each term of Eq.~\eqref{sigma_final} more elaborately and evaluated each term numerically. We add the numerical results of all those terms to get the final result for the expression mentioned in Eq.~\eqref{sigma_final}. The results shown in Figures \ref{fig10}$-$\ref{fig13} are scaled with coefficient $g m_{q}$ on the y-axis and plotted over $p/m_{q}$ on the x-axis. In Figure~\ref{fig10}, we have shown how the imaginary part and real part of $\Sigma_{+}^{(1)}$ (transverse contribution + longitudinal contribution) scaled with a coefficient of $g m_{q}$ varies for two and three flavors. Figure~\ref{fig10}(a) shows the variation of imaginary part of $\Sigma_{+}^{(1)}$ with $p/p_0$. From this plot, one can get the NLO damping rate for quarks with $'+'$ mode, i.e., for ordinary quarks for $N_{f}=2$ and $N_{f}=3$, respectively. Figure~\ref{fig10}(b) shows the variation of real part of $\Sigma_{+}^{(1)}$ with $p/p_0$. We will get the  NLO quark energy from this plot with $`+'$ mode. Using the Eq.~\eqref{disp_nlo}, damping rate and quark energy for soft momentum, $p$ plotted in Figure~\ref{fig10} for '+' quark mode. Figure~\ref{fig11}(a) shows that the damping rate of real quark mode decreases with the increase of soft momentum and then becomes constant, as expected. Similarly, Figure~\ref{fig11}(b) shows how the NLO correction to mass for $'+'$ quark mode behaves.
\begin{figure}
\centering
\includegraphics[scale=0.5,keepaspectratio]{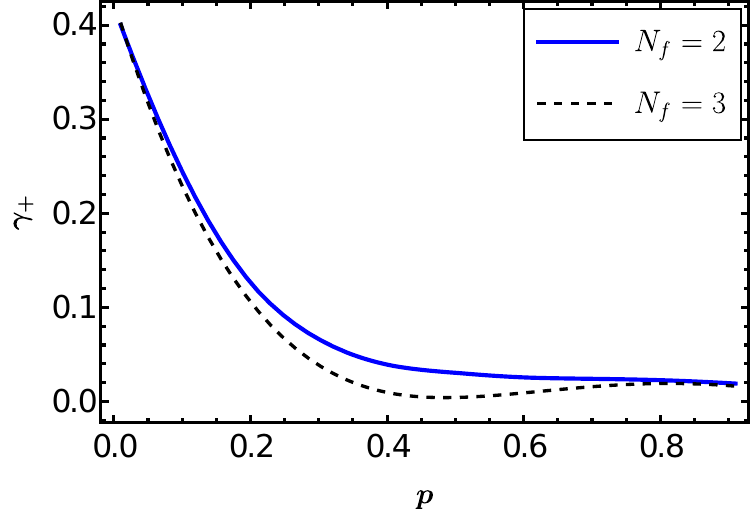} 
\qquad
\includegraphics[scale=0.5,keepaspectratio]{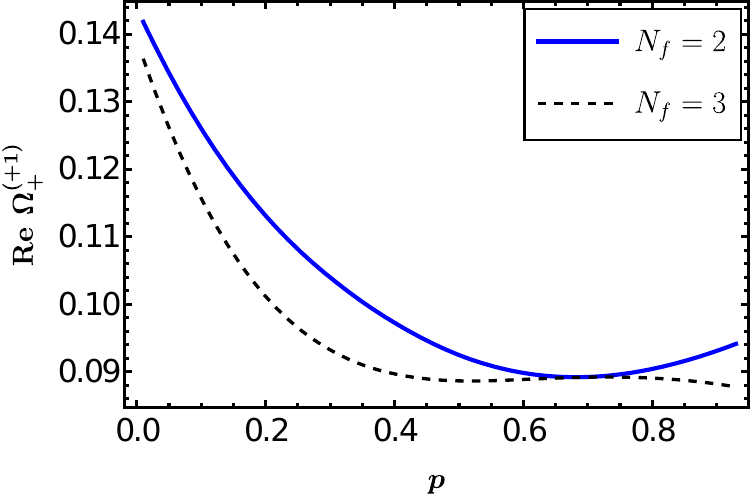}
\caption{Damping rate and quark energy variation with soft momentum $p/m_{q}$ for `+' mode scaled with a coefficient $g m_{q}$.}
\label{fig11}
\end{figure}
Now, Figure~\ref{fig12} shows the behavior of the imaginary part and real part of $\Sigma_{-}^{(1)}$ (transverse contribution + longitudinal contribution) scaled with a coefficient of $g m_{q}$. 
\begin{figure}
\centering
\includegraphics[scale=0.5,keepaspectratio]{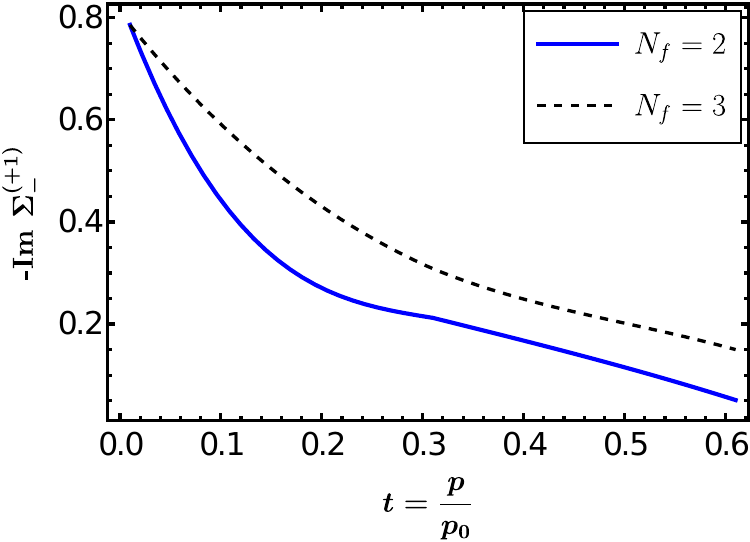} 
\qquad
\includegraphics[scale=0.5,keepaspectratio]{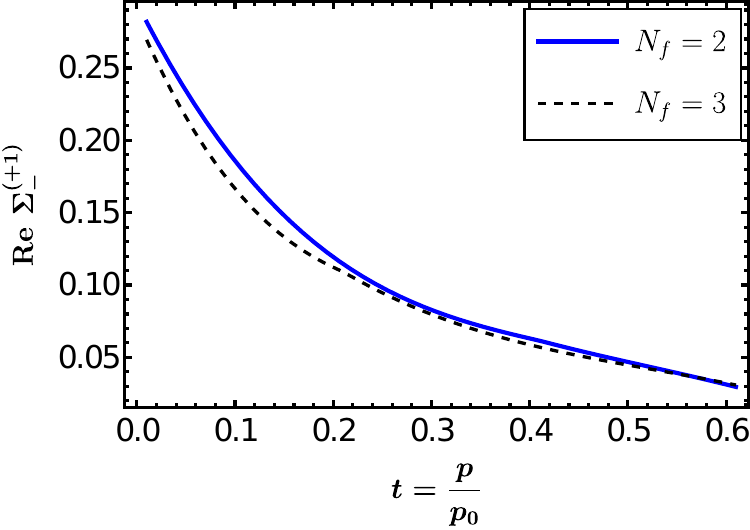}
\caption{The Imaginary part and Real part of $\Sigma_{-}^{(1)}$, scaled with a coefficient of $g m_{q}$, with respect to parameter $t=p/p_{0}$}
\label{fig12}
\end{figure}  
Figure~\ref{fig12}(a) shows the variation of imaginary part of $\Sigma_{-}^{(1)}$ with $p/p_0$. One can get the NLO damping rate for plasmino mode from this plot. Figure~\ref{fig12}(b) shows the variation of real part of $\Sigma_{-}^{(1)}$ with $p/p_0$. One can get the NLO quark mass from this plot with $`-'$ mode. Using the expressions from Eq.~\eqref{disp_nlo}, we have plotted the damping rate and quark energy w.r.t. soft momentum $p$ in Figure~\ref{fig13} for this plasmino mode. Figure~\ref{fig13}(a) shows the behavior of the damping rate for plasmino mode. Figure~\ref{fig13}(b) shows the variation of NLO mass for $`-'$ mode w.r.t soft momentum $p$.
In the limit of zero momentum, one can see that the damping rate and correction to NLO mass approach the same value for both quark modes. The other significant outcome of the above results shows that we can handle the instabilities that arise from the gluon propagator's transverse and longitudinal components, respectively.\\
\begin{figure}
\centering
\includegraphics[scale=0.5,keepaspectratio]{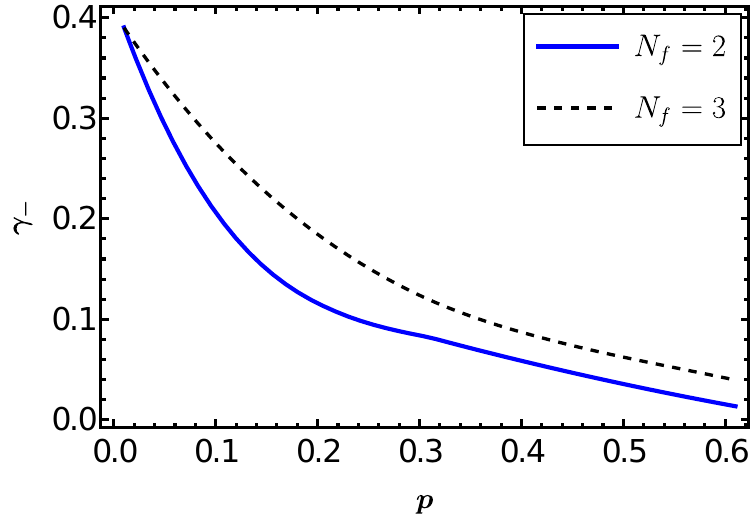} 
\qquad
\includegraphics[scale=0.5,keepaspectratio]{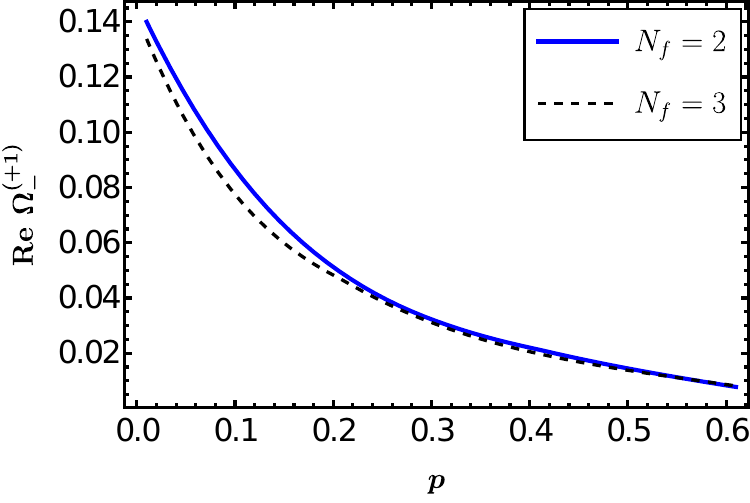}
\caption{Damping rate and quark energy dependence on soft momentum $p/m_{q}$ for plasmino mode scaled with a coefficient $g m_{q}$.}
\label{fig13}
\end{figure} 
\begin{figure}
\centering
\includegraphics[scale=0.5,keepaspectratio]{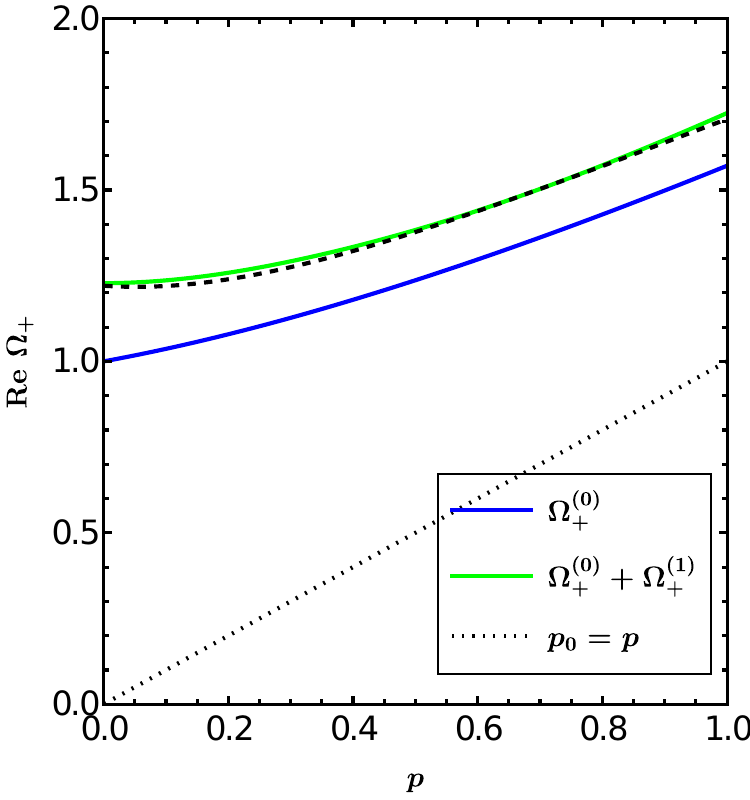} 
\qquad
\includegraphics[scale=0.5,keepaspectratio]{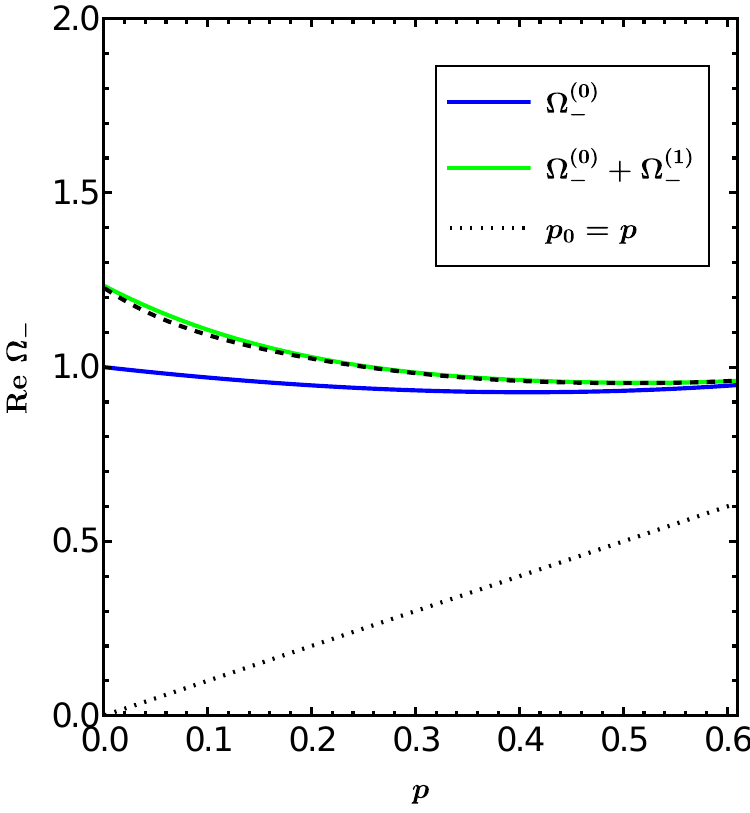}
\caption{Real part of dispersion relation variation with respect to soft momentum $p/m_{q}$ for both quark modes. The solid line is for $N_{f}=2$ case, while the dotted line shows the results for $N_{f}=3$ flavors.}
\label{fig_Omega_+1_-1}
\end{figure} 
\qquad
\begin{figure}
\centering
\includegraphics[scale=0.5,keepaspectratio]{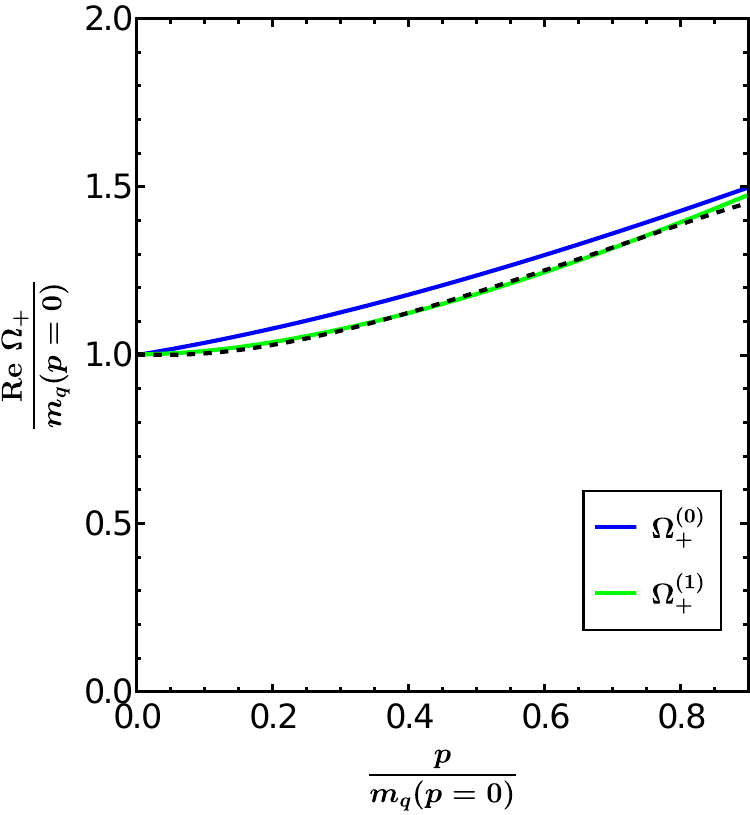} 
\qquad
\includegraphics[scale=0.5,keepaspectratio]{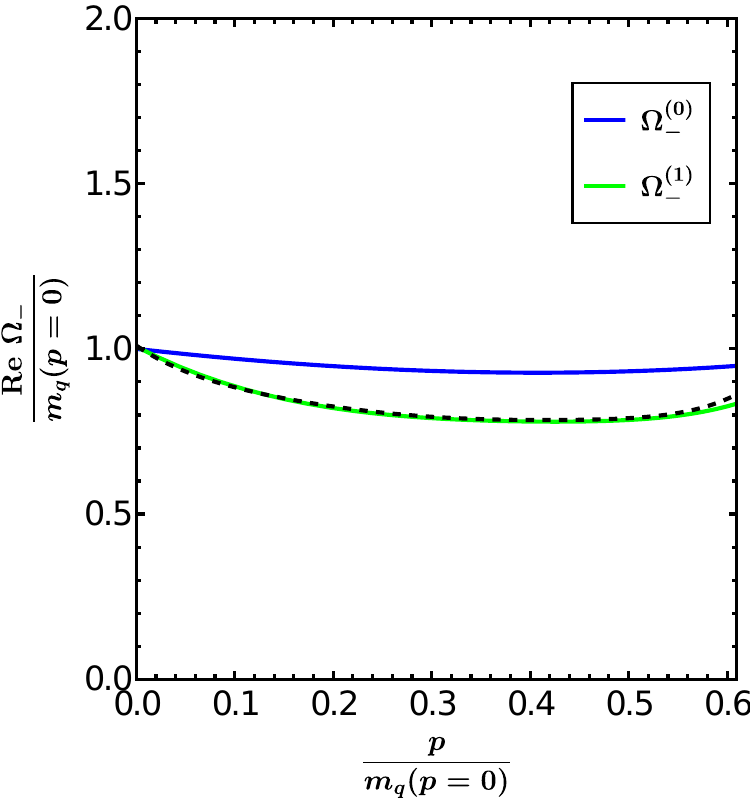}
\caption{Real part of dispersion relation variation with respect to soft momentum $p/m_{q}$ scaled with their corresponding thermal masses. The solid line shows the results for $N_{f}=2$ flavors, while the dotted line shows the results for $N_{f}=3$ flavors. The value of $\alpha_{s}=0.2$ is used.}
\label{fig_Omega_+1_-1_scaled}
\end{figure}  
We extract the numerical value of the dispersion relations, i.e., damping rate and mass, by taking the limit of $p \rightarrow 0$ in Figures~\ref{fig11} and~\ref{fig13} respectively. In order to compare our results with the existing results in the literature, one can work in the units of $m_{q}$. In the limit of zero momentum, we obtain the values $\gamma_{\pm}(0) \approx 0.159\, g^{2}T$ for $N_{f}=2$ and $\gamma_{\pm}(0) \approx 0.164 g^{2}T$ for $N_{f}=3$ which is, respectively $7\%$ and $9\%$ larger than the existing result obtained in Ref.~\cite{Braaten:1992gd}. Similarly, in the limit of zero momentum for the correction in the mass denoted by $\Delta m$, we obtain the values $\Delta m = 0.142 gm_{q}$ for $N_{f}=2$ which is $\sim 5 \, \%$ smaller as compared with Eq.~\eqref{m_q_nlo} and $\Delta m = 0.136 gm_{q}$ for $ N_{f}=3 $. The value of the coupling $\alpha_{s} =0.2$ viz. $g =1.58$ is used in the numerical evaluation of these results. The real part of the dispersion relation for both modes, i.e., mass, variation w.r.t. soft momentum, is shown in Figure~\ref{fig_Omega_+1_-1}. In Figure~\ref{fig_Omega_+1_-1_scaled}, we have shown the dependence of the real part of dispersion relation viz. mass on the soft momentum scaled with their corresponding thermal masses on both axes for both modes. Also, one can extract the velocity for both quark modes using the graphs in the Figure~\ref{fig_Omega_+1_-1}. Figure~\ref{fig_vel_+1_-1} shows the velocity variation w.r.t. soft momentum for real quark and plasmino modes. The result shows that for both quark modes, velocity is less than $c$ as expected. Also, these results show that velocity for both modes decreases in the limit of zero momentum. The decrease of the velocity in the limit of $ p\rightarrow 0 $ shows that quasi-particles become massive. As the momentum increases, medium effects gradually vanish, as seen in Figure~\ref{fig_vel_+1_-1}. Figure~\ref{fig_Mass_+1_-1} compares the real part of the dispersion relation with NLO correction and without NLO correction for soft momentum.    
\qquad
\begin{figure}
\centering
\includegraphics[scale=0.5,keepaspectratio]{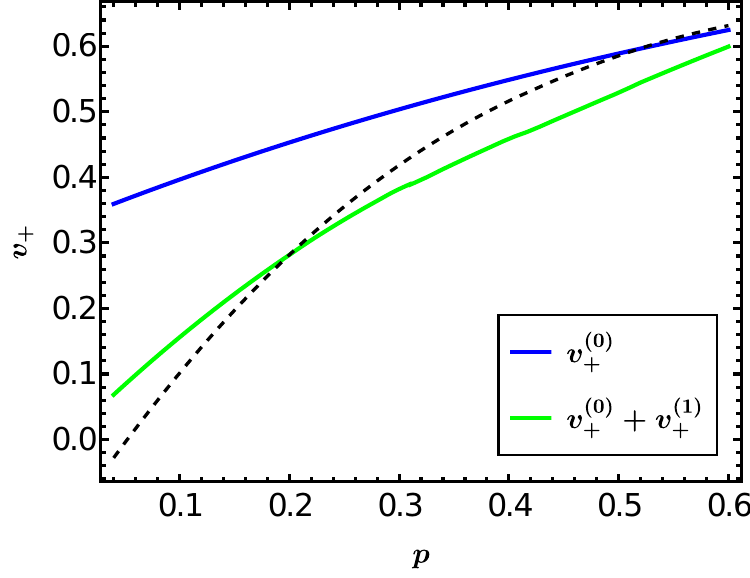} 
\qquad
\includegraphics[scale=0.5,keepaspectratio]{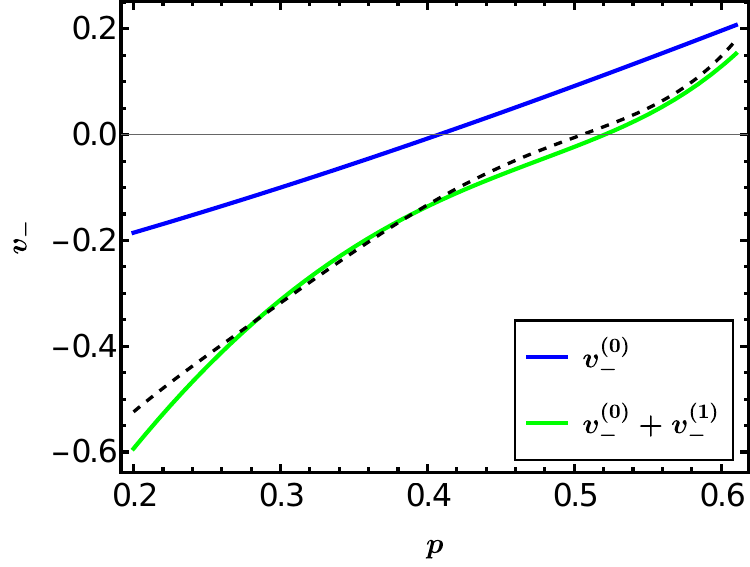}
\caption{Velocity variation with respect to soft momentum $p/m_{q}$ for quark mode and plasmino mode respectively. The solid and dotted line corresponds to $N_{f} =2,3$ flavors, respectively.}
\label{fig_vel_+1_-1}
\end{figure} 
\qquad
\begin{figure}
\centering
\includegraphics[scale=0.5,keepaspectratio]{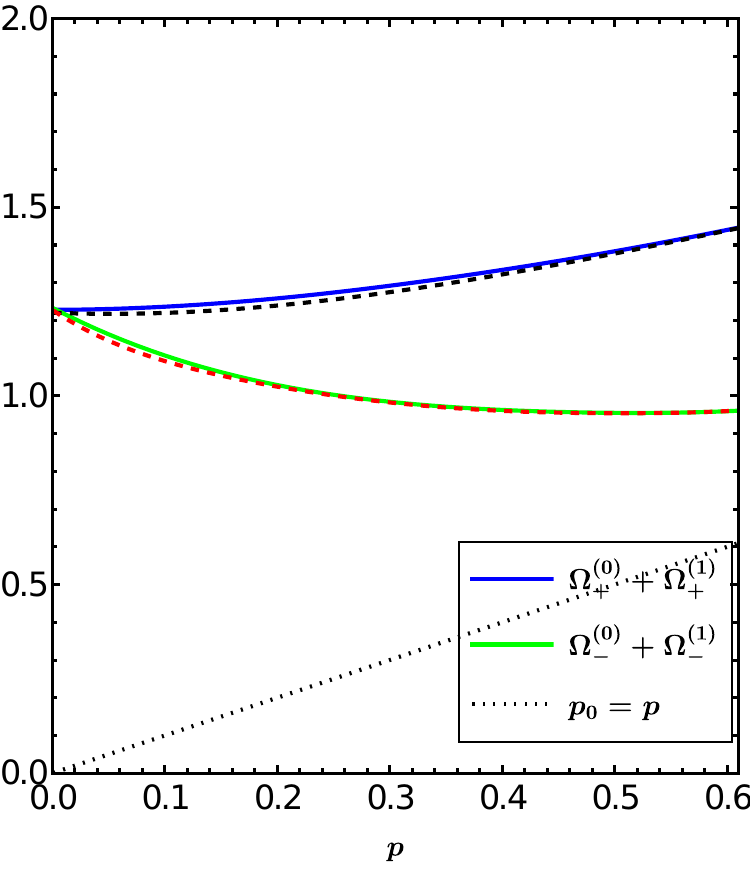} 
\qquad \qquad
\includegraphics[scale=0.5,keepaspectratio]{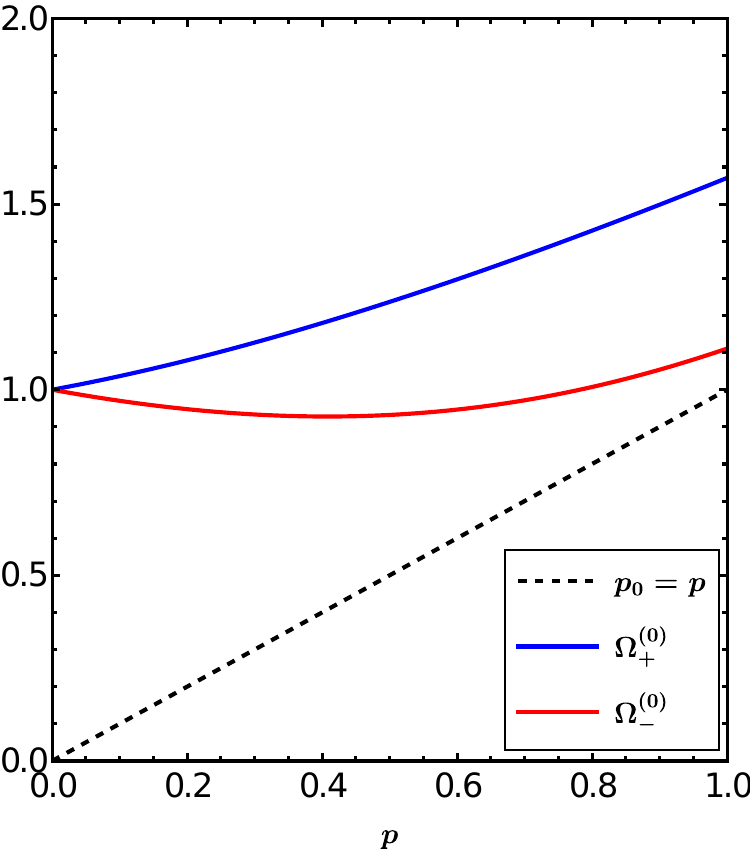}
\caption{\textit{Left Panel:} Comparison of the real part of dispersion relation variation with respect to soft momentum $p/m_{q}$ for both quark modes. The solid and dotted line corresponds to $N_{f} =2,3$ flavors, respectively.\ \textit{Right Panel:} LO dispersion relations are plotted for the two modes along with the massless free mode.}
\label{fig_Mass_+1_-1}
\end{figure} 
\section{Summary}\label{sec:summary}
In the present work, we have studied the NLO quark self-energy and their corresponding dispersion relations using the HTL resummation. To study the NLO quark self-energy, we have used the RTF of Keldysh indices, and the considered quarks are the slow-moving ones. The solution of the NLO dispersion laws gives us physical quantities like  NLO damping rate and NLO masses, and these observables come from the zeros of the HTL-dressed quark propagators. In the lowest order, the solution of the quark dispersion relation $\Omega_{\pm}(p)$ is real. To get the NLO contribution of the above-mentioned physical quantities, one needs to evaluate NLO quark self-energy (see Eq.~\eqref{disp_nlo}). 
In the current work, the NLO part of the quark self-energies $\Sigma_{\pm}^{(+1)}$ is given using the loop-four momenta integrals, which involves the effective HTL quark, gluon propagators, and three- and four-point vertex functions as done in Refs.~\cite{Carrington:2006gb, Abada:2014bma}. The effective three- and four-point vertex functions are derived separately using HTL approximation and expressed in terms of solid-angle integrals using the standard technique. The above ones are rewritten using the standard Feynman parameterization technique to evaluate these integrals. Further, we numerically evaluated the transverse and longitudinal parts of the expression mentioned in Eq.~\eqref{sigma_final}. The standard way of doing so is to use the spectral representation of the under-consideration dressed propagators, but we have tackled the integrals directly, which is non-trivial. One main difficulty in computing the integrals directly is the jumps the integrands encounter due to the divergences present in the propagators, more specifically when the fine-tuning parameter of the integrals $\varepsilon$ approaches zero. In evaluating the transverse and longitudinal parts of Eq.~\eqref{sigma_final}, we encountered the divergences arising from the integrands. This instability comes mainly from the transverse and longitudinal parts of the gluon propagator. To overcome these divergences, we have broken down the integration into some appropriate domains, and then integration has been done in each of the appropriate domains. The $\varepsilon$  dependence of all terms in NLO quark self-energy has been carried out independently, i.e., we have checked first the stability of all the terms involved in the NLO quark self-energy. After that, the usual integration has been done for that particular value of $\varepsilon$. The domain integration technique turns out to be very useful in handling those extensive integrations. In the end, we summed up all the contributions from the transverse and longitudinal terms, and we studied the dependence of the NLO quark self-energy on the variable $t = p/p_0$. Lastly, using the Eq.~\eqref{disp_nlo}, we plotted the NLO correction to dispersion relations.


\chapter{Mesonic Screening Masses using Gribov quantization}\label{Chapter_4}
\allowdisplaybreaks
\pagestyle{fancy}
\fancyhead[LE]{\emph{\leftmark}}
\fancyhead[LO]{\emph{\rightmark}}
\rhead{\thepage} 

In this chapter, we have studied and calculated the spatial correlation lengths of various mesonic observables using the non-perturbative Gribov resummation, both for quenched QCD and $(2+1)$ flavor QCD. This chapter is based on our work:
{\ \textit{QCD mesonic screening masses using Gribov quantization}}, {\bf Sumit}, Najmul Haque and Binoy Kishna Patra, {\bf \color{blue}Phys.Lett. B 845 (2023) 138143}~\cite{Sumit:2023hjj}.
\section{Introduction}\label{Int}	
Understanding experimental results at the relativistic heavy-ion colliders in high-energy nuclear and particle physics and the early cosmos' cosmic history depends heavily on QCD at finite temperatures. When the temperature rises, the theory switches from a confined phase with hadronic degrees of freedom, where chiral symmetry is spontaneously broken, to a deconfined phase of quarks and gluons, where chiral symmetry is restored.\\
The temporal and spatial directions are generally disconnected because the heat bath destroys the Lorentz symmetry at finite temperatures. The temporal correlation functions in the Minkowski space can be used to define spectral functions for various operators in Fourier space. The spectral functions then indicate the plasma's essential ``real-time" features, such as the particle production rates~\cite{McLerran:1984ay}. In investigations involving heavy ion collisions, these are then immediately observable. Nevertheless, the correlation functions in the spatial direction answer questions like At what length scales does thermal fluctuations correlation occur? What length scales are external charges screened? These static observables are also physical quantities eminently suited for lattice experiments' measurements.\\
The usual perturbation theory has a severe infrared problem at the higher orders in coupling~\cite{Linde:1980ts, Gross:1980br}, but the infrared problems are only connected with the static, soft gluons. It has led to the development of the dimensional reduction technique ~\cite{Ginsparg:1980ef, Appelquist:1981vg}, which reduces the non-perturbative infrared behavior to a more straightforward effective theory consisting of soft modes only, enabling the precise estimation of the lower order contributions to the perturbation theory. This way, the terms involving dynamical quarks can be handled perturbatively in four dimensions. At the same time, the computationally costly non-perturbative methods can be used for a more straightforward three-dimensional theory. Naturally, the QCD coupling is significant at moderate temperatures, and higher-order corrections cannot be disregarded.\\
The dimensional reduction method has been extensively used in QCD and electroweak theory to study thermodynamics ~\cite{Appelquist:1981vg, Kajantie:1995dw, Braaten:1995jr, Kajantie:2002wa, Blaizot:2003iq, Vuorinen:2003fs, Ipp:2006ij, Gynther:2005dj} and different gluonic correlators~\cite{Kajantie:1997tt, Laine:1997nq, Hart:2000ha, Hart:1999dj, Cucchieri:2001tw} in the bosonic sector. On the other hand, the fermionic modes, only affect the dimensionally reduced theory's parameters, which are generally integrated out from the effective theory. However, several intriguing observables built of quark fields are sensitive to IR physics and, as a result, require some resummations, elegantly arranged as successive effective theories. The initial attempts to include this class of correlators in theory with reduced dimensions go back almost 30 years~\cite{Hansson:1991kb, Koch:1992nx}. These works, however, did not consistently include every term in a fixed order. In ref.~\cite{Huang:1995tz}, it was demonstrated that the dimensionally reduced effective theory could be expressed in terms of nonrelativistic quarks, and the accurate power counting of various operators was done. The other motivation for studying the observables built of quark fields, i.e., ``mesonic and baryonic" observables, is related to their closer signs with experimental signatures. In recent years, significant work has been done on one class of these observables, namely the various quark number susceptibilities (see ref.~\cite{Gottlieb:1987ac, Blaizot:2001vr, Gavai:2002kq, Gavai:2002jt, Chakraborty:2001kx, Chakraborty:2003uw, Vuorinen:2002ue, Andersen:2012wr, Haque:2013qta, Haque:2018eph}). \\
The other class of observables, more closely related to IR physics and more sensitive toward it, is the correlation lengths of the mesonic operators, which are gauge-independent quark bilinears formed from the light quark flavors~\cite{Detar:1987kae}. The quark bilinears' two-point spatial correlation functions can determine the correlation lengths. Generally, these correlation functions are dominated by the screening masses at large distances, defined as the inverse of the correlation lengths. These screening masses give information about the QGP response when a meson is included in the system. A large number of studies have been done in the literature to study the meson correlation function using perturbative methods as well as via non-perturbative lattice computations. On the perturbative side, the mesonic correlation functions using HTL have been studied in refs.~\cite{Karsch:2000gi, Alberico:2004we}. Additionally, the meson correlation function at finite momentum in QCD plasma has been studied in ref.~\cite{Alberico:2006wc}. Using the pQCD, the LO result $(2\pi T)$ for the meson screening mass is obtained in ref.~\cite{Eletsky:1988an, Florkowski:1993bq}. The NLO correction for meson and gluon screening masses has been calculated using an effective theory at zero chemical potential in ref.~\cite{Laine:2003bd, Laine:2009dh}. The connection between thermal screening masses and real-time rates has been explored in ref.~\cite{Brandt:2014uda} in the spectral representation. The NLO correction of meson screening masses at finite chemical potential has also been extended ref.~\cite{Vepsalainen:2007ke}. \\
On the lattice side, the temporal and spatial hadronic correlation function in QCD plasma has been studied in ref.~\cite{Born:1991zz, Boyd:1994np,QCD-TARO:2000hup}. Also, screening masses in purely $SU(2)$ and $SU(3)$ gauge theories using LQCD is studied in ref.~\cite{Datta:1998eb, Datta:1999yu, Cheng:2010fe, Gavai:2001ie}. Recently, mesonic screening masses have been studied for the first time in a large range covering the temperature from $\sim 1$ GeV to $\sim 160$ GeV using LQCD in ref.~\cite{DallaBrida:2021ddx}. Also, the recent LQCD results for the meson screening masses in $(2+1)$ flavor QCD in the temperature range from $0.14 \hspace{1mm} \text{GeV} \leq T \leq 2.7 \hspace{1mm}\text{GeV} $ is presented in ref.~\cite{Bazavov:2019www} and for the recent study of mesonic screening mass at finite chemical potential using LQCD see ref.~\cite{Thakkar:2022frk, Thakkar:2023nvr, Pushkina:2004wa}. In the high-temperature limit, the perturbative result of meson screening masses calculated in ref.~\cite{Laine:2003bd} is comparable with the lattice results. However, no such analytic calculation in the literature can explain the lattice data for  $T \leq 2$ GeV in the low-temperature regime. In this article, we have tried to overcome this gap by using a non-perturbative scheme offered by Gribov quantization.\\
Since QCD's IR region is strongly coupled, the conventional resummed perturbative approach will not give appropriate results~\cite{Andersen:2004fp}. One of the effective ways to handle the IR region is to consider the GZ approach~\cite{Gribov:1977wm, Zwanziger:1989mf} (For recent reviews on the GZ approach, see~\cite{Dokshitzer:2004ie, Vandersickel:2012tz} and some recent works on extended Gribov approach follow~\cite{Dudal:2008sp, Capri:2016aqq, Dudal:2017kxb, Gotsman:2020ryd, Gotsman:2020mpg, Justo:2022vwa, Mena:2023mqj, Gracey:2010cg}). This approach improves Yang-Mills theory's infrared dynamics by removing the leftover gauge transformations after employing the Faddeev-Popov quantization. In recent years, this approach has been utilized to improve the thermodynamic quantities in QCD by evaluating free energy ~\cite{Fukushima:2013xsa}. Also, in kinetic theory, the transport coefficients have been explored using this scheme in ref.~\cite{Florkowski:2015dmm, Florkowski:2015rua,Jaiswal:2020qmj}. In this approach, the gluon propagator modifies its form, and a new (chromo)magnetic scale enters the theory through the mass parameter. Using the modified gluon propagator in Gribov action, quark dispersion relation~\cite{Su:2014rma}, the dilepton production rate and quark number susceptability~\cite{Bandyopadhyay:2015wua} have been calculated, which sheds light on some of the new physical insights of the said observables. Recently, this approach has been applied to heavy quark phenomenology to study the heavy quark potential~\cite{Wu:2022nbv, Debnath:2023dhs} and heavy quark diffusion coefficient~\cite{Madni:2022bea}.\\
The following is how this chapter is set up. We shall establish necessary notations and review some of the mesonic correlators' well-known characteristics at high temperatures in section~\ref{Section_2}. Section \ref{Section_3} covers how a dimensionally reduced effective field theory~\cite{Huang:1995tz} can describe high-temperature QCD in general while incorporating the mesonic correlators along the lines of work~\cite{Laine:2003bd}. Nonrelativistic QCD or three-dimensional (3d) $ \text{NRQCD}_{3} $ will refer to the quark portion of the lagrangian entering. In section \ref{Section_4}, we will give details about the Gribov quantization and do the required matching computation to fix the unknown parameter in the effective lagrangian by matching the dispersion relation of QCD and NRQCD$_{3}$, using Gribov formalism. Section \ref{Section_5} evaluates the dynamics of the effective theory with Gribov's gluon propagator inclusion. Results of the screening masses for the quenched QCD case and $N_f=3$ are compared with the recent lattice data in section \ref{Section_6}. We summarize the chapter in the last section, \ref{Section_7}. 
\section{ High-temperature static correlators in QCD}\label{Section_2}	
The Euclidean Lagrangian of the quarks at finite temperature QCD is given by
\begin{equation}\label{quark_lag.}
\mathcal{L}_{\mathrm{E}}^{\mathrm{Q}}=\bar{\Psi}\left(\gamma_\mu \mathcal{D}_\mu+M\right) \Psi,
\end{equation}
Here, the covariant derivative ($\mathcal{D}_{\mu}$) of the fermionic field is given by
\begin{equation}
\mathcal{D}_\mu \Psi \equiv \partial_\mu \Psi-i g \mathcal{A}_\mu^a T^a \Psi
\end{equation}
where $T^{a}$ are the generators defined in fundamental represenation of group $SU(N_c)$, which are hermitian in nature. The quark field $\Psi$ is an $N_F$-component vector in flavor space. For the sake of simplicity, we will assume M to be a diagonal and degenerate matrix, $\text{M} = \text{diag}(m,\cdots m),$ and we will take the assumption of $m=0$. \\
We may use quark fields to define bilinear objects with varying spin and flavor structures. We are interested in the operators having the form,
\begin{equation}\label{opera_form}
O^{a} = \bar{\Psi} F^{a}\Gamma\Psi,
\end{equation}
where $\Gamma$ can take the values $\{1,\gamma_{5},\gamma_{\mu},\gamma_{\mu}\gamma_{5}\}$ for the different channels, namely scalar, pseudoscalar, vector, and pseudo vector $S^{a}, P^{a}, V_{\mu}^{a}$ and $A_{\mu}^{a}$ respectively. The traceless matrices $F^{a}$ along with the identity matrix $F^{s}$ provides the flavor basis,
\begin{eqnarray}
F^{a}\equiv \{F^{s},F^{n}\} , \ F^{s} \equiv \mathbf{1}_{{N_F}*N_{F}}, \  \text{Tr}[F^{a} F^{b}] = \frac{1}{2}\delta^{ab} ,\qquad
\end{eqnarray}
with $a,b = 1,2,\cdots N^{2}_{F} - 1$. For the operators defined in Eq.~\eqref{opera_form}, we will focus on the correlators having the structure as
\begin{equation}\label{corre_def}
\mathcal{C}_{\mathbf{q}}\left[O^a, O^b\right] \equiv \int_0^{\frac{1}{T}} \mathrm{d} \tau \int \mathrm{d}^3 x \hspace{1mm} e^{i \mathbf{q} \cdot \mathbf{x}}\left\langle O^a(\tau, \mathbf{x}) O^b(0,0)\right\rangle,
\end{equation}
or, in position space,
\begin{equation}\label{Corr_in_position}
\mathcal{C}_{\mathbf{x}}\left[O^a, O^b\right] \equiv \int_0^{1 / T} \mathrm{~d} \tau\left\langle O^a(\tau, \mathbf{x}) O^b(0,0)\right\rangle
\end{equation}
The expectation values are considered in a region that is finite in the temporal direction with width $1/T$ but infinite in spatial directions. It is presumed that gauge fields will adhere to periodic and fermions anti-periodic boundary constraints around the temporal dimension.\\
Due to rotational invariance, the correlators' general structure in terms of the three momenta $\textbf{q}$ takes the form as
\begin{align}
\mathcal{C}_{\mathbf{q}}\left[S^a, S^b\right], \mathcal{C}_{\mathbf{q}}\left[V_0^a, V_0^b\right]  = & \, \sum_{\mathrm{fg}} F_{\mathrm{fg}}^a F_{\mathrm{gf}}^b \hspace{1mm} f\left(q^2\right) \nn
\mathcal{C}_{\mathbf{q}}\left[V_i^a, V_j^b\right] = & \, \sum_{\mathrm{fg}} F_{\mathrm{fg}}^a F_{\mathrm{gf}}^b\left[\left(\delta_{i j}-\frac{q_i q_j}{q^2}\right) t\left(q^2\right)+\frac{q_i q_j}{q^2} l\left(q^2\right)\right],
\end{align}
and  for $P^{a}$ and $A^{a}_{\mu}$, similar structure of correlation functions can be written. Here $q\equiv |\textbf{q}|$. In general, one may anticipate that these functions $f(q^{2}), t(q^{2})$ and $l(q^{2})$ are characterized by simple zeroes around the origin $q^{2}=0$, which corresponds to a collection of bound states in a $(2+1)$ D system derived after doing the analytic continuation.\\	
Using the rotational invariance, one can choose the correlation measurement direction as $z$ in Eq.~\eqref{corre_def}. With this choice, one can take the average over the $x_{1} x_{2} $ surface, giving a correlator that is easier to handle
\begin{eqnarray}\label{corre_in_z}
\mathcal{C}_z\left[O^a, O^b\right]&=&\int \mathrm{d}^2 \mathbf{x}_{\perp} \mathcal{C}_{\left(\mathbf{x}_{\perp}, z\right)}\left[O^a, O^b\right]
\end{eqnarray}
where $\mathcal{C}_{\left(\mathbf{x}_{\perp}, z\right)}$ is given in Eq.~\eqref{Corr_in_position}. The spatially separated two-point correlation function mentioned in Eq.~\eqref{corre_in_z} defines the screening masses as 
\begin{equation}\label{screeM_def}
m_{z}=-\lim _{z \rightarrow \infty} \frac{d}{dz} \ln \left[\mathcal{C}_z\left[O^a, O^b\right]\right]
\end{equation}
The above Eq.~\eqref{screeM_def} describes the correlation function's exponential falloff at large distances.
\begin{figure}
\centering
{\includegraphics[scale=1.3]{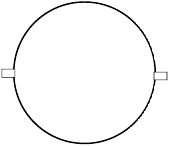}}
\quad \quad \quad
{\includegraphics[scale=1.3]{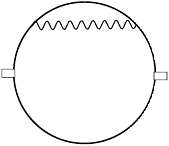}}
\quad \quad \quad
{\includegraphics[scale=1.3]{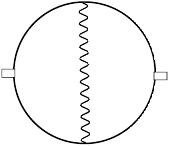}}
\caption{The diagrams which contribute to meson correlation function: (a) free theory correlator (b) quark self-energy graph (c) interaction of quark and antiquark through gluon exchange.}
\label{fig_correlation_diagrams}
\end{figure}
Let us focus on the correlation function's behavior at high temperatures. Due to asymptotic freedom, the correlators with free fermions depicted in Figure~\ref{fig_correlation_diagrams} (a) can be evaluated using perturbation theory. By using the dimensional regularisation technique, one would get the result for the diagram~\ref{fig_correlation_diagrams} (a) as
\begin{align}
\mathcal{C}_{\boldsymbol{q}}\left[O^a, O^b\right] = & \,\operatorname{Tr}\left[F^a F^b\right] N_c T \sum_{n=-\infty}^{\infty} \int \frac{\mathrm{d}^{3-2 \varepsilon} p}{(2 \pi)^{3-2 \varepsilon}}\nonumber \nn
\times & \,  \frac{1}{\left[p_n^2+p^2\right]\left[p_n^2+(p+q)^2\right]} \operatorname{Tr}\left[(\slashed{p}+\slashed{q}) \Gamma^a \slashed{p} \Gamma^b\right]
\label{corre_free}
\end{align}
Here, $N_c$ denotes number of colors, $\Gamma$ is the dirac matrix coming from the opeartor form $O^{a}$, $\slashed{p} \equiv \gamma_{\mu}p_{\mu}$ and $p_{n}$ refers to the fermionic Matsubara modes having the form, $p_{n} = 2\pi T (n+1/2)$. After doing the trivial Dirac algebra calculation, which gives some constant terms, the above correlation function in Eq.~\eqref{corre_free} contains the function
\begin{align} 
A_{3\mathrm{d}}\left(2 p_n\right) \equiv & \, \int \frac{\mathrm{d}^3 p}{(2 \pi)^3} \frac{1}{\left[p_n^2+p^2\right]\left[p_n^2+(p+q)^2\right]} = \, \frac{i}{8 \pi q} \ln \frac{2 p_n-i q}{2 p_n+i q}.
\label{A3d_func}
\end{align}
From the Eq.~\eqref{A3d_func}, it is clear that the singularity of the function $A_{3d}(2p_{0})$ appears at the point $2p_{n}$. Thus, the correlator dominates at large distances for the zeroth Matsubara frequencies defined as $\pm p_{0} = \pm \pi T$. Now, this frequency excitation occurs in the 3-dimensional theory. One can also consider this three-dimensional theory as $(2+1)$-dimensional and consider the time direction as the one in which the correlation is assessed. In this case, screening mass is equivalent to determining the screening states of a pair of on-shell heavy quarks, each with a mass of $``p_{0}"$. Also, if we consider the free theory correlator not around the pole $q_{0}= \pm 2i p_{0}$, for huge q, the correlation functions can be studied using the standard operator product expansion (OPE) approach.
\section{Effective Theory}\label{Section_3}
Using the non-perturbative effects, we want to determine the NLO correction to the mesons screening mass. To go beyond the leading order, we must calculate the diagrams shown in Figure~\ref{fig_correlation_diagrams} (b) and Figure~\ref{fig_correlation_diagrams} (c). Nevertheless, as we know, in finite temperature QCD, it will not be sufficient to calculate only these two diagrams. In principle, we can have a large number of diagrams that can behave in the same order as the two depicted in Figure~\ref{fig_correlation_diagrams} (b) and Figure~\ref{fig_correlation_diagrams} (c) do.\\
Since we evaluate corrections to a mesonic operator near the threshold for the formation of two free quarks, the quarks can almost be on-shell with $|1/\slashed{p}| \sim \mathcal{O}(1/g^{2}T) $. This includes the coupling factors from the vertices. To calculate the first-order correction to the free quarks propagator in a consistent way,  it is necessary to do the resummation of the diagrams with an arbitrary number of zero modes of the low-momentum exchanged gluon between quarks. The effective field theory approach provides a convenient way to perform such resummations. Only the diagram containing one-gluon exchange would contribute to order $g^{2}$ depicted in Figure~\ref{fig_correlation_diagrams} (c). As shown in Figure~\ref{fig_correlation_diagrams} (b) for non-zero gluon modes, the gluon exchange diagrams do not need to be computed because this diagram alone cannot alter the pole location or the screening mass. These diagrams would contribute to the overall normalizing factor of the correlator under consideration. \\
Conversely, in the bosonic section, the dynamics of the gluonic zero modes after the dimensional reduction technique describe the effective three-dimensional gauge theory~\cite{Appelquist:1981vg}, known as EQCD. The action takes the form as  
\begin{align}
S_{\mathrm{EQCD}} =& \, \frac{1}{g_{\mathrm{E}}^2} \int d^3 x\left\{\frac{1}{2} \operatorname{Tr}\left[F_{i j} F_{i j}\right]+\operatorname{Tr}\left[\left(\mathcal{D}_j \mathcal{A}_0\right)\left(\mathcal{D}_j \mathcal{A}_0\right)\right]\right. + \, \left.m_{\mathrm{E}}^2 \operatorname{Tr}\left[\mathcal{A}_0^2\right]\right\}+\ldots,
\label{EQCD_action}
\end{align}
where the dots refers to higher dimensional operators~\cite{Laine:2016hma} and $i =1,2,3$ , $F_{ij} = ig_{E}^{-1}[\mathcal{D}_{i},\mathcal{D}_{j}]$ , $\mathcal{D}_{i} = \partial_{i} - ig_{E}\mathcal{A}_{i}$ , $g_{E}^{2} = g^{2}T$. The temporal component of gauge field $\mathcal{A}_{0}$ acts as a scalar field with mass $m_{E}$. As we know, since at very high temperature $T$, the coupling $g$ is small, and therefore, three different energy level hierarchies occur as  
\begin{equation}\label{scale_hirar.}
\frac{g_{\mathrm{E}}^2}{\pi} \ll m_{\mathrm{E}} \ll \pi T .
\end{equation}
However, we are interested in the processes that occur at $\mathcal{O}(g_{E}^{2})$; then one can integrate out the scalar field from the theory in Eq.~\eqref{EQCD_action}. After integrating the temporal part, we are left with the effective theory known as MQCD, which is given by 
\begin{equation}\label{MQCD}
S_{\mathrm{MQCD}}=\frac{1}{g_{\mathrm{E}}^2} \int d^3 x\left\{\frac{1}{2} \operatorname{Tr}\left[F_{i j} F_{i j}\right]\right\}+\ldots
\end{equation}
This three-dimensional theory has non-perturbative dynamics; thus, it must be handled completely in a non-perturbative way~\cite{Linde:1980ts}. \\
As previously stated, one can interpret the inverse of screening masses in this instance as (2+1)-dimensional confined phases of heavy particles with mass $p_{0}$ much greater than the IR scales $gT$, $g^{2}T$ of gauge field dynamics at high temperatures~\cite{Ginsparg:1980ef}. Using the NRQCD, an effective theory, the masses of these heavy bound states may be calculated, similarly studied for quarkonia at $T = 0$~\cite{Caswell:1985ui}. Our situation results in some substantial departures from this framework, such as that these masses, $p_{0}$, are not actual masses in the traditional sense but directly scale invariant physical variables that retain chiral invariance. These NRQCD techniques in the context of a dimensional reduction approach in the fermionic sector were first used in ref.~\cite{Huang:1995tz}. \\	
\subsection{Tree-level NRQCD$_{3}$}\label{Section_3.1}
In the fermionic sector, we will consider only the lowest fermionic modes $p_{0} \equiv \pi T$ because the other modes will not contribute much to the correlator at large distances as seen in Eq.~\eqref{A3d_func}. The Euclidean quark lagrangian for the field $\Psi(x)$ having Matsubara mode $\omega_{n}$ reads as 
\begin{equation}
\mathcal{L}_{\mathrm{Q}}=\bar{\Psi}\left[i \gamma_0 \omega_n-i g \gamma_0 \mathcal{A}_0+\gamma_j \mathcal{D}_j+\gamma_3 \mathcal{D}_3\right] \Psi
\end{equation}
where  $j=1,2$ and $\mathcal{A}_{0}$ is the zero gluonic mode interacting with $\Psi(x)$. Utilizing the rotational invariance to measure the correlation function in the $z$ direction, the term $\mathcal{D}_{3}\Psi$ has been separated from the transverse components.\\	
In the nonrelativistic domain, one can treat particle and antiparticle features separately. We considered the Euclidean Dirac matrices to be of a different form than the standard one. In this representation, we have
\begin{equation}
\gamma_0=\left(\begin{array}{cc}
0 & 1_2 \\
1_2 & 0
\end{array}\right), \hspace{0.1cm} \gamma_i=\left(\begin{array}{cc}
\epsilon_{i j} \sigma_j & 0 \\
0 & -\epsilon_{i j} \sigma_j
\end{array}\right), \hspace{.1cm} \gamma_3=\left(\begin{array}{cc}
0 & -i \\
i & 0
\end{array}\right),
\end{equation}
so that
\begin{equation}
\gamma_0 \slashed{p} = \left(\begin{array}{cc}
\left(p_0+i p_3\right) 1_2 & -\epsilon_{i j} p_i \sigma_j \\
\epsilon_{i j} p_i \sigma_j & \left(p_0-i p_3\right) 1_2
\end{array}\right),
\end{equation}
where $\epsilon_{i j}$ is antisymmetric two rank tensor and $\epsilon_{12}=+1$. As the quarks are very heavy, the quark fields can be considered static fields, and the Dirac spinors can be rewritten as
\begin{equation}
\Psi=\left(\begin{array}{l}
X \\
\Phi
\end{array}\right)
\end{equation}
where $X$, $\Phi$ are the two-component spinor objects that lead to the Lagrangian 
\begin{align}
\mathcal{L}_{\mathrm{Q}} = & \, i X^{\dagger}\left(p_0-g \mathcal{A}_0+\mathcal{D}_3\right) X+i \Phi^{\dagger}\left(p_0-g \mathcal{A}_0-\mathcal{D}_3\right) \Phi  +  \, \Phi^{\dagger} \epsilon_{i j} \mathcal{D}_i \sigma_j X-X^{\dagger} \epsilon_{i j} \mathcal{D}_i \sigma_j \Phi
\end{align}
The quarks in the considered correlators are almost ``on-shell," and for the free theory, the on-shell point is given by $p_{0}^{2}+p^{2} = 0$, i.e., $p_{3} = \pm ip_{0}$. At the tree level, quarks with a fixed Matsubara frequency $\omega_{n}$ interact with zero gluonic modes only; thus, it is expected that quarks' off-shellness is related to the gluonic momentum scale as $|p_{3}\pm i p_{0}|\lesssim gT$. Now, after expanding around a state consisting of a free pair of quark-antiquark, one of the components is heavy, and another is light compared to the dynamical scale, viz. $gT$, $g^{2}T$, one can solve the equation of motion for the heavy component and the light mode. After doing an expansion in powers of $\frac{1}{p_{0}}$, the effective action for the fermionic mode can be written as 
\begin{align}
S_\Psi^{\text {eff}} = & \, \int d^3 x \{ i X^{\dagger}\left[p_{0}-g_{\mathrm{E}} \mathcal{A}_0+\mathcal{D}_3-\frac{1}{2 p_{0}}\left(\mathcal{D}_k^2\nonumber\right.\right.  \left.\left. \frac{g_{\mathrm{E}}}{4 i}\left[\sigma_k, \sigma_l\right] F_{k l}\right)\right]X\nn
+&  \, i \Phi^{\dagger}\left[p_{0}-g_{\mathrm{E}} \mathcal{A}_0-\mathcal{D}_3 \right. - \left.\left.\frac{1}{2 p_{0}}\left(\mathcal{D}_k^2+\frac{g_{\mathrm{E}}}{4 i}\left[\sigma_k, \sigma_l\right] F_{k l}\right)\right] \Phi\right\} \nn
+& \, \mathcal{O}\bigg(\frac{1}{p_{0}^{2}}\bigg)
\label{tree_level_action}
\end{align}	
\subsection{{Power Counting Arguments}}\label{Section_3.2}
We must consider the quantum corrections subjected to the parameters in Eq.~\eqref{tree_level_action} to ascertain the radiative corrections. One must take into account all additional operators that symmetries might permit. Setting up a power counting for the likely operators is crucial. \\
In the NRQCD$_{3}$ side, quarks have momentum scale as $|\textbf{p}_{\perp}| \lesssim gT$ and their off-shellness $\Delta p_{3} = p_{3}+ip_{0}$ is of the same order. By requiring the action mentioned in Eq.~\eqref{tree_level_action} to be of the order of unity, we have
\begin{align}
\int \mathrm{d} z \mathrm{~d}^2 \mathrm{x}_{\perp} \hspace{1mm} X^{\dagger} \partial_3 X \sim & \, 1, \Rightarrow X \sim 1 /\left|\mathrm{x}_{\perp}\right| \sim g T \nn
\int \mathrm{d} z \mathrm{~d}^2 \mathrm{x}_{\perp} \hspace{1mm} \mathcal{A} \partial_3^2 \mathcal{A} \sim & \, 1 \Rightarrow A \sim\left(z / \mathrm{x}_{\perp}^2\right)^{1 / 2} \sim g^{1 / 2} T^{1 / 2}
\end{align}
On-shell relativistic gluons have the same energy and momentum order viz., $p_{3} \sim \textbf{p}_{\perp}$. On the other hand, as the considered quarks are nonrelativistic,  their kinetic energy is directly proportional to momentum squared in $(2+1)$ dimensional theory. For a nearly on-shell quark with transverse momentum $|\textbf{p}_{\perp}| \lesssim gT$ in NRQCD$_{3}$ , the off-shellness in longitudinal momentum becomes
$\Delta p_3 \sim \mathbf{p}_{\perp}^2 / p_0 \sim g^2 T$ and this gives $\partial_{3}\sim g^2T$ that acts on quarks.\\
One can determine the number of possible operators from this power counting. For instance, any four-quark operator would be of the order $g^{4}T^{4}$, whereas the leading order behaves as $g^{2}T^{3}$. According to this argument, the coefficient would have to be of the order of $\mathcal{O}(g^{0})$ to produce corrections that are no greater than $\mathcal{O}(g^{2})$, but since this term is not present at the tree level, it may be ignored. \\
Similar power counting arguments demonstrate that $g_{E}\mathcal{A} \sim g^{3/2}T$, which means that this term will be of higher order compared to the derivative term $\partial_{i}$ present in the transversal covariant derivative. As $\partial_{\perp}^{2}$ is already of $\mathcal{O}(g^{2})$, one can left out the trasnverse gluons. Thus, the only parameter that needs to be matched beyond tree-level is the zero point energy $p_{0}\equiv M$. Thus, to find out the $\mathcal{O}(g^{2})$ corrections to meson correlation lengths, the following lagrangian is sufficient to use  
\begin{align}
\mathcal{L}_{\text {eff }}^{\mathrm{f}} = & \, i X^{\dagger}\left(M-g_{\mathrm{E}} \mathcal{A}_0+\mathcal{D}_3-\frac{\nabla_{\perp}^2}{2 p_0}\right) X + i \Phi^{\dagger}\left(M-g_{\mathrm{E}} \mathcal{A}_0-\mathcal{D}_3-\frac{\nabla_{\perp}^2}{2 p_0}\right) \Phi ,
\label{effective_lag.}
\end{align}
where $\mathcal{D}_{3}= \partial_{3} - ig_{E} \mathcal{A}_{3}$. Note that, here $\mathcal{A}_{0}$ and $\mathcal{A}_{3}$ play an important dynamical scale role. On the other hand, transverse gluons $\mathcal{A}_{1}$, $\mathcal{A}_{2}$ can be ignored as long as one is interested in the energy shift of $\mathcal{O}(g^{2}T)$. This means that gluons will not transfer transverse momentum to quarks in this order. To be consistent at $\mathcal{O}(g^{2}T)$, we should replace, energy i.e., $p_{0}$ of tree-level effective Lagrangian by a matching coefficient $M = p_{0} + \mathcal{O}(g^{2}T)$, which we will determine in the next section.   \\
\section{Matching Conditions from QCD to NRQCD$_{3}$, with Gribov}\label{Section_4}
We will determine the one-loop correction to the variable $M$ by matching the Green function, calculated on the QCD side and NRQCD$_{3}$ side, using Gribov formalism. Here, the matching will be done by finding the finite temperature Euclidean dispersion relation so that we do not need to worry about the overall normalization factors arising from the fields. This computation produces gauge-invariant results in both sectors, and we will do the matching order by order in $1/p_{0}$ on the NRQCD$_{3}$ side so that dimensional regularisation copes up with the power counting inputs as done in Ref.~\cite{Manohar:1997qy}. \\
In the conventional quantization of QCD, the gauge condition is not ideal, as proposed by Faddeev and Popov. Thus, there remains a residual gauge discrepancy in the infrared region, which is pointed out by Gribov and known as the Gribov copy problem~\cite{Gribov:1977wm}. Gribov proposed that the functional integral should be constrained to the first Gribov region, which collects gauge fields without zero modes for the FP operator. The YM partition function in Euclidean space-time in the Gribov quantization has the following representation:
\begin{equation}
Z=\int_{\mathit{\Omega}} D \mathcal{A}(x) \hspace{1mm} V(\mathit{\Omega}) \delta(\partial \cdot \mathcal{A}) \operatorname{det}[-\partial^{\mu} \mathcal{D}_{\mu}(\mathcal{A})] e^{-S_{\mathrm{YM}}}
\end{equation}
Here, $\mathit{\Omega}$ represents the Gribov region which is defined conventionally as
\begin{equation}\label{Omega_Gribov}
\mathit{\Omega} \equiv\{\mathcal{A}: \partial \cdot \mathcal{A}=0,-\partial^{\mu}  \mathcal{D}_{\mu}(\mathcal{A}) \geq 0\},
\end{equation}
One must include the function $V(\mathit{\Omega})$ in the partition function to limit the functional integration in the first Gribov region described in Eq.~\eqref{Omega_Gribov}, where
\begin{equation}
V(\mathit{\Omega})= \mathit{\Theta}(1-\sigma(0))=\int_{-i \infty+0^{+}}^{+i \infty+0^{+}} \frac{d \beta}{2 \pi i \beta} e^{\beta[1-\sigma(0)]}
\end{equation}
Here, $1-\sigma(P)$ is the inverse of the ghost dressing function $Z_{G}(P)$~\cite{Fukushima:2013xsa}. The integration variable $\beta$ is dubbed as the Gribov mass parameter. The expression for the Gluon propagator with Gribov quantization in the general covariant gauge is written as
\begin{equation}\label{Gribov_prop}
D_{\mu\nu}^{ab}(P)=\delta^{a b} \left(\delta_{\mu \nu}-(1-\xi)\frac{P_\mu P_\nu}{P^2}\right)\frac{P^2}{P^4+\gamma_{\mathrm{G}}^4}
\end{equation}
where $\xi = 0,1$ corresponds to Landau and Feynman gauges. On the QCD side, using the Gribov propagator mentioned in Eq.~\eqref{Gribov_prop}, the quark propagator's inverse, defined as $\Sigma(P)$, is given by,
\begin{align}
-i\Sigma(P) =& \, \slashed{P} -\left. g^{2} C_{F} \SumInt_{Q} \frac{\gamma_{\mu} (\slashed {P}-\slashed{Q}) \gamma_{\mu}}{(P-Q)_{f}^{2}}\left(\frac{Q^{2}}{Q^{4}+\gamma_{G}^{4}}\right)_{b} \right.\nn
+ & \, g^{2} C_{F}\left.\SumInt_{Q} \frac{\slashed{Q} (\slashed {P}-\slashed{Q}) \slashed{Q}}{Q^{2}(P-Q)_{f}^{2}}\left(\frac{Q^{2}}{Q^{4}+\gamma_{G}^{4}}-\frac{\xi\hspace{1mm} Q^{2}}{Q^{4}+\gamma_{G}^{4}}\right)_{b}\right.
\label{Disp_QCD}
\end{align}
where the gluonic four-momentum in Euclidean space-time reads $Q=\left(q_{0},q\right)$, with $q_0=2 n \pi T$, and $N_c$ represents the number of colors. This calculation uses a dimensional regularization technique with $\overline{\mathrm{MS}}$ renormalization scheme. In this technique, the sum-integral is given as
\begin{equation}
\SumInt_{Q} \equiv\left(\frac{e^{\gamma_E} \Lambda^2}{4 \pi}\right)^\varepsilon T \sum_{q_0=2 n \pi T} \int \frac{d^{d-1} q}{(2 \pi)^{d-1}},
\end{equation}
with $d=4-2 \varepsilon$ as the space-time dimensions. Also, $(\cdots)_{f} , (\cdots)_{b}$ refers to fermionic and bosonic Matsubara modes respectively, and $C_{F} = (N_{c}^{2}-1)/2N_{c}$. 
As we are calculating the $\mathcal{O}(g^{2})$ corrections to $\Sigma(P)$, one can utilize the free theory constraints namely $P^{2}=0$ and $\slashed{P} u(P) =0$ to handle the terms which are in proportion to $P^{2}$ and $\slashed{P}$. Thus, the longitudinal gauge-dependent part then vanishes, leaving only the transverse terms of Gribov's gluon propagator to contribute to the calculation. The outcome is irrespective of the gauge fixing parameter $\xi$ as expected.
\begin{figure}[t]
\centering		\includegraphics[width=8cm,height=8cm,keepaspectratio]{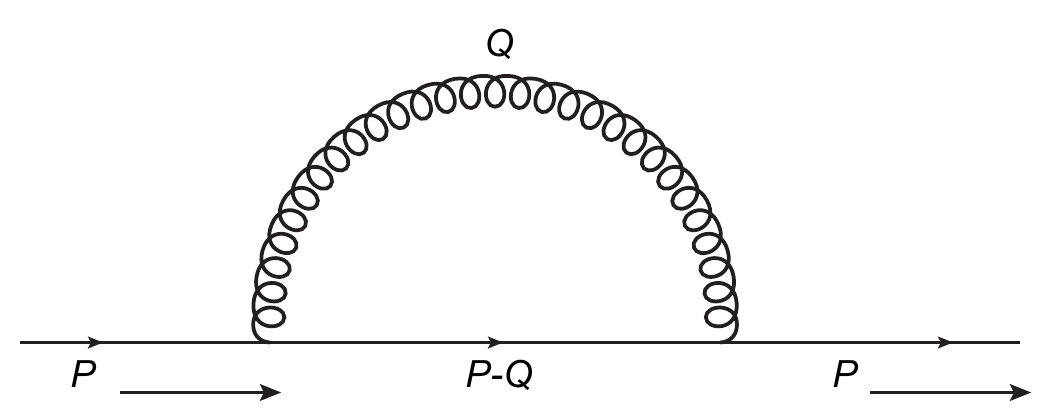}
\caption{Quark self-energy correction to one-loop}
\label{quark_self_energy}
\end{figure}
Because of the rotational invariance of the remaining terms in Eq.~\eqref{Disp_QCD}, we can do the matching of the Green's function at some particular momentum and set $p_{\perp} = 0$. After the multiplication of $\gamma_{0}$ from the left to Eq.~\eqref{Disp_QCD}, the expression becomes block-diagonal. Thus, we can focus on the particular term, say, $[\gamma_{0}\Sigma(P)]_{11}$, which is given by 
\begin{eqnarray}\label{gamma0_Sigma}
\hspace{-1cm} -i	\left[\gamma_0 \Sigma(P)\right]_{11} &=& p_0+i p_3 -g^2 C_F \SumInt_Q \left(\frac{Q^{2}}{Q^{4}+\gamma_{G}^{4}}\right)_{b} \times \frac{2(-p_{0}-ip_{3}+q_{0}+iq_{3})}{(P-Q)_f^2}
\end{eqnarray}
In Eq.~\eqref{gamma0_Sigma}, we have the contribution coming from the transverse gluons, i.e., $\mathcal{A}_{1}$ and $\mathcal{A}_{2}$ component only. In contrast, the gauge field components $\mathcal{A}_{0}$ and $\mathcal{A}_{3}$ vanish altogether at the free theory pole position $p_{0}=-ip_{3}$ as they occur with opposite signs. Thus, the only integral that we need to evaluate to find the $\mathcal{O}(g^{2})$ correction is 
\begin{align}
I =& \, \left.\SumInt_{Q} \frac{2\left(q_{0}+i q_{3}\right)}{(P-Q)_{f}^{2}}\left(\frac{Q^2}{Q^4+\gamma_{G}^4}\right)_{b}\right|_{p_{0}=-i p_{3}} \nn
=& \, \left.\frac{-1}{p_{0}}\SumInt_{Q} \left(\frac{Q^2}{Q^4+\gamma_{G}^4}\right)_{b}\right|_{p^{2}=0} + \left.\frac{1}{p_{0}}\SumInt_{Q} \frac{Q^2}{(P-Q)^{2}_{f}}\right. \left.\left(\frac{Q^2}{Q^4+\gamma_{G}^4}\right)_{b}\right|_{p^{2}=0} \nonumber\\
=& \, I_1 +  I_2.
\end{align}
Now, the first integral  $I_1$ becomes
\begin{eqnarray}\label{int_I1}
I_1 = - \hspace{1mm} \frac{1}{p_{0}} \int_{0}^{\infty} \frac{q^{2}dq}{(2\pi)^{2}}\bigg[\frac{n^+}{E_+} +	 \frac{n^-}{E_-} \bigg].
\end{eqnarray} 
Here,  $n^\pm$ represents the BE distribution function with $ E_\pm=\sqrt{q^2\pm i\gamma_{G}^2} $. Additionally, the second integral $I_{2}$ can also evaluated as 	
\begin{eqnarray}\label{int_I2}
I_2 &=& \left.\frac{1}{p_{0}}\SumInt_{Q} \frac{Q^2}{(P-Q)^{2}_{f}}\left(\frac{Q^2}{Q^4+\gamma_{G}^4}\right)_{b}\right|_{p^{2}=0}  =\frac{1}{p_{0}}\bigg[ \frac{-T^{2}}{24} + X^{\prime} \bigg],
\end{eqnarray}
where $X^{\prime}$ is given by
\begin{align}
X^{\prime} = & \, \frac{\gamma_{G}^{4}}{T^2} \int \frac{ d^3q}{(2\pi)^3}\frac{1}{8E E_{+} E_{-}}\bigg[\left(\frac{\tilde{n}+n^{-}}{i\pi-E+E_{-}} - \frac{\tilde{n}+n^{+}}{i\pi-E+E_{+}}\right) \nn
+ & \, \left(\frac{\tilde{n}+n^{+}}{i\pi+E-E_{+}} - \frac{\tilde{n}+n^{-}}{i\pi+E-E_{-}}\right)\bigg]\frac{1}{E_{+}-E_{-}}
\label{def._X} 
\end{align}
where $\tilde{n}$ is the FD distribution function with energy $E = q-p_3(1+\cos\theta)$. 
To get the $T^{2}$ dependence in the integrals $I_{1}$ and $I_{2}$, we need to do the trivial re-scaling of the parameters as $q\rightarrow q^{\prime} = q/T$. 
Thus, the Euclidean dispersion relation on the QCD side using the Gribov propagator becomes
\begin{equation}
p_{3} \approx i\bigg[p_{0}-g^{2}C_{F}(I_{1}+I_{2})\bigg]
\end{equation}
On the NRQCD$_{3}$ side, one can extract the Feynman rules from the lagrangian in Eq.~\eqref{effective_lag.}. For instance, free propagators are 
\begin{align}
\left\langle X^u(p) X^{v*}(q)\right\rangle = & \,\delta^{u v}(2 \pi)^3 \delta(p-q) \frac{-i}{M+i p_3+\mathbf{p}_{\perp}^2 / 2 p_0} \nn
\left\langle\Phi^u(p) \Phi^{v*}(q)\right\rangle = & \,\delta^{u v}(2 \pi)^3 \delta(p-q) \frac{-i}{M-i p_3+\mathbf{p}_{\perp}^2 / 2 p_0}
\end{align}	
or, in the position space, it reads as
\begin{align}
\left\langle X^u(x) X^{v*}(y)\right\rangle =& \,-i \delta^{u v} \Theta\left(x_3-y_3\right) \frac{p_0}{2 \pi\left(x_3-y_3\right)} 
\times	e^{-M\left(x_3-y_3\right)-\frac{p_0\left(\mathbf{x}_{\perp}-\mathbf{y}_{\perp}\right)}{2\left(x_3-y_3\right)}} \nonumber \nn
\left\langle\Phi^u(x) \Phi^{v*}(y)\right\rangle =& \,-i \delta^{u v} \Theta\left(y_3-x_3\right) \frac{p_0}{2 \pi\left(y_3-x_3\right)} 
\times  e^{-M\left(y_3-x_3\right)-\frac{p_0\left(\mathbf{x}_{\perp}-\mathbf{y}_{\perp}\right)}{2\left(y_3-x_3\right)}} .
\label{two_point_func}
\end{align}
The Eq.~\eqref{two_point_func} makes it apparent that $X$ field propagates forward in time direction $x_{3}$ and $\Phi$ field backward. The one-loop contribution to the quark self-energy vanishes on NRQCD$_{3}$ side because the two gauge field components, namely $\mathcal{A}_{0}$ and $\mathcal{A}_{3}$, are the same and come with opposite signs as is the case in QCD side. Thus, the inverse propagator is 
\begin{align}
-i \Sigma^{\prime}(p)=& \, M+i p_3-g_{\mathrm{E}}^2 C_F \int \frac{\mathrm{d}^{3-2 \epsilon} q}{(2 \pi)^{3-2 \epsilon}} \frac{1}{M+i p_3-i q_3}\nn
\times&	\left[\left(\frac{Q^{2}}{Q^{4}+\gamma_{G}^{4}}\right)_{A_{0}}-\left(\frac{Q^{2}}{Q^{4}+\gamma_{G}^{4}}\right)_{A_{3}}\right]
\end{align}
So, the pole location is simply $p_{3} = i M$ on NRQCD$_{3}$ side. Now, after doing the matching, we will get 
\begin{equation}\label{def._M}
M = p_{0}-g^{2}C_{F}(I_{1}+I_{2})
\end{equation}
The integrals $I_{1}$ and $I_{2}$ need to be evaluated numerically to find the matching parameter $M$. This matching accounts only for the hard gluons in the loop integral of Figure~\ref{fig_correlation_diagrams} (b).
\section{Dynamics of effective theory}\label{Section_5}
After considering the hard gluons for Figure~\ref{fig_correlation_diagrams} (b), we still need to take care of the soft gluons involved in Figure~\ref{fig_correlation_diagrams} (b) and as well as Figure~\ref{fig_correlation_diagrams} (c), i.e., we need to solve the dynamics of the effective theory at $\mathcal{O}(g^{2}T)$. The dynamics will be studied by computing the correlators, as done in section \ref{Section_2}, by using the effective theory mentioned in Eqs.~\eqref{MQCD},~\eqref{effective_lag.}. The different bilinears mentioned in Eq.~\eqref{opera_form} can be expressed in terms of $X$ and $\Phi$ fields, which consists of spin and flavor structure (not given explicitly here), shown in Eq.~\eqref{bilinears_chi+phi}. Nevertheless, since we are interested in the correlation functions which decay more slowly, the combinations of these operators are of the form $X^{\dagger} \Phi+\Phi^{\dagger} X$.
\begin{align}
S  = & \, X^{\dagger} \Phi+\Phi^{\dagger} X  \hspace{1cm}
P  =X^{\dagger} \Phi-\Phi^{\dagger}  X \nn
V_0  = & \, X^{\dagger} X+\Phi^{\dagger} \Phi \hspace{1cm}
V_k  =-\epsilon_{k l}\left(X^{\dagger} \Phi-\Phi^{\dagger}  X\right) \nn
A_0  = & \, \Phi^{\dagger}  \Phi-X^{\dagger}  X  \hspace{1cm}
A_k  =-i\left(X^{\dagger}  \Phi+\Phi^{\dagger}  X\right)  \nn
V_3  = & \, X^{\dagger} X-\Phi^{\dagger} \Phi \hspace{1cm}
A_3  =-i\left(X^{\dagger} X+\Phi^{\dagger} \Phi\right)
\label{bilinears_chi+phi}
\end{align}
The modified form of the correlation function of Eq.~\eqref{corre_in_z} is
\begin{equation}
\mathcal{C}_z\left[O^a, O^b\right] \sim \int \mathrm{d}^2 x_{\perp}\left\langle O^a\left(x_{\perp}, t\right) O^b\left(\mathbf{0}_{\perp}, 0\right)\right\rangle
\end{equation}
Although each of the bilinears in Eq.~\eqref{bilinears_chi+phi} consists of two terms, $\sim X^{\dagger} \Phi+\Phi^{\dagger} X$, only one of these terms affects the correlation function under consideration. In that particular term, both quarks will be moving in the forward direction, $\sim \hspace{1mm} \langle\Phi^{\dagger}(z)X(z)\Phi(0)X(0)\rangle$ (see Eq.~\eqref{two_point_func}). One may assume that the exchange of gluons between quarks and antiquarks occurs instantly in the nonrelativistic domain, and the static potential for the quark-antiquark pair may be determined by integrating the gauge fields. After obtaining the potential, we may use the usual Schr\"odinger equation to solve for the screening states and arrive at a solution. \\
Thus, we need to establish Green's function for the quantity of the form 
\begin{align}
\int \mathrm{d}^2 x_{\perp}\Phi_u^*\left(\mathbf{x_{\perp}}, z\right) \Gamma_{u v} X_v\left(\mathbf{x_{\perp}}, z\right)  \, ,
\end{align} 
where $\Gamma_{u v}$ is a constant term that consists of $2\times2$ matrix, which does not affect the overall calculation if we drop out that term. To find out the static potential, one can use the trick of including a point-splitting in the correlator, which means putting the quark pairs at a particular distance $\mathbf{r}$ apart, which is finite, and then determining the Schr\"odinger equation obeyed by the correlator and set $\mathbf{r} \rightarrow 0 $ afterward. Thus, the updated correlator is   
\begin{equation}
\mathcal{C}\left(\mathbf{r}, z\right) \equiv \int_{\mathbf{R}}\left\langle\Phi^{\dagger}\left(\mathbf{R}+\frac{\mathbf{r}}{2}, z\right)  X\left(\mathbf{R}-\frac{\mathbf{r}}{2}, z\right) X^{\dagger}(0) \Phi(0)\right\rangle,
\end{equation}
For the Gribov's gluon propagator $D_{\mu \nu}^{a b}(P)$, we have
\begin{align}
\left\langle \mathcal{A}_0^a(P) \mathcal{A}_0^b(Q)\right\rangle =& \,\delta^{a b}(2 \pi)^{4-2 \epsilon} \delta^{(4-2 \epsilon)}(P+Q) \frac{P^{2}}{P^{4}+\gamma_{G}^{4}} \nn
\left\langle \mathcal{A}_i^a(P) \mathcal{A}_j^b(Q)\right\rangle = & \,\delta^{a b}(2 \pi)^{4-2 \epsilon} \delta^{(4-2 \epsilon)}(P+Q)\left[\frac{P^{2}}{P^{4}+\gamma_{G}^{4}}\right. \nn
\times& \left(\delta_{i j}-\frac{P_i P_j}{P^2}\right)\left. +\frac{P_i P_j}{P^2} \frac{\xi P^{2}}{P^{4}+\gamma_{G}^{4}}\right] .
\end{align}
In general, the equation of motion (EOM) satisfied by Green's function, for large values of $z$, is of the form $(\partial_{z}-H)G(z)=c\hspace{0.5mm} \delta(z)$, where $H$ is hamiltonian of the system and $c$ denotes a constant value. Thus, the EOM obeyed by the correlation functions at the tree level reads as  
\begin{equation}\label{tree-level_EOM}
\left[\partial_z + 2 M-\frac{1}{p_0} \nabla_{\boldsymbol{r}}^2\right] \mathcal{C}^{(0)}(r, z) = 2 N_{c}\hspace{0.5mm} \delta(z) \hspace{0.5mm}\delta^{(2)}(\mathbf{r}),
\end{equation}
where the Hamiltonian can be read from the Eq.~\eqref{effective_lag.}. For the one-loop diagrams shown in Figure~\ref{fig_correlation_diagrams} (b) and Figure~\ref{fig_correlation_diagrams} (c), the equation of motion is given by
\begin{align}
\left[\partial_z+2 M-\frac{1}{p_0} \nabla_{\boldsymbol{r}}^2\right] \mathcal{C}^{(1)}(\boldsymbol{r}, z) = & \, -g_{\mathrm{E}}^2 C_F \mathcal{C}^{(0)}(\boldsymbol{r}, z)	\, \mathcal{K}\left(\frac{1}{z p_0}, \frac{\nabla_{\boldsymbol{r}}}{p_0}, \frac{\gamma_{G}^{4}}{p_0^4}, \boldsymbol{r} p_0\right) 
\label{one-loop_EOM}
\end{align}
where the kernel $\mathcal{K}$ is a dimensionless quantity that can be expanded to its first two arguments to find out the one-loop static potential, 
\begin{align}
V(r)  \equiv & \, g_{\mathrm{E}}^2 C_F  \mathcal{K}\left(0,0, \frac{\gamma_{G}^{4}}{p_0^4}, r p_0\right) = -2g_{\mathrm{E}}^2 C_F \int \frac{\mathrm{d}^{2} q}{(2 \pi)^{2}} \, \frac{Q^{2}}{Q^{4}+\gamma_{G}^{4}}\ e^{i \boldsymbol{q} \cdot \boldsymbol{r}}
\label{static_poten.}
\end{align}
Now the Eqs.~\eqref{tree-level_EOM},~\eqref{one-loop_EOM},~\eqref{static_poten.} can be clubbed together to get the final equation of motion up to one-loop order as
\begin{equation}
\left[\partial_z+2 M-\frac{1}{p_0} \nabla_{\boldsymbol{r}}^2+V(r)\right] \mathcal{C}(r, z) = 2 N_{c} \delta(z) \delta^{(2)}(\boldsymbol{r}) ,
\end{equation}
For the Gribov propagator, the expectation values of temporal gauge fields and spatial gauge fields are the same in the computation of the static potential. 
After solving the above integral in Eq.~\eqref{static_poten.}, we get the final form of one-loop static potential as
\begin{equation}
V(r)=g_{\mathrm{E}}^2 \frac{C_F}{2 \pi}\left[\ln \frac{{\gamma_{G}} r}{2}+\gamma_E-K_0\left(\gamma_{G} r\right)\right] .
\end{equation}
where $K_{0}$ is the modified Bessel function whose asymptotic behavior is given by $K_{0}(y) = -\ln\frac{y}{2}-\gamma_{E}+\mathrm{O}(y)$. Now, this potential will determine the correlation length $\zeta^{-1}= m $, through
\begin{equation}\label{sch_eq.}
\left[2 M-\frac{\nabla_{\boldsymbol{r}}^2}{p_0}+V(\mathbf{r})\right] \Psi_0 = m \Psi_0 ,
\end{equation}
where $\Psi_{0}$ represents the ground state wave function. We will numerically solve this Eq.~\eqref{sch_eq.} to find the screening mass. For the easiness of the problem, we can do the re-scaling as 
\begin{equation}\label{screening_mass_final}
r \equiv \frac{r^{\prime}}{\gamma_{G}}, \quad m -2 M \equiv g_{\mathrm{E}}^2 \frac{C_F}{2 \pi} E_0
\end{equation}
In the polar coordinates, the Schr\"odinger Eq.~\eqref{sch_eq.} modifies as
\begin{equation}\label{sch_eq_polar}
\hspace{-0.5cm}	\left[\left(\frac{\mathrm{d}^2}{\mathrm{d} r^{\prime 2}}+\frac{1}{r^{\prime}} \frac{\mathrm{d}}{\mathrm{d} r^{\prime}}\right)-\eta\left(\ln \frac{r^{\prime}}{2}+\gamma_E-K_0(r^{\prime})- E_0\right)\right] \Psi_0=0,
\end{equation}
where
$$\eta=\frac{p_0 g_{\mathrm{E}}^2 C_F}{2 \pi \gamma_{G}^{2}}$$
To find out the numerical solution of Eq.~\eqref{sch_eq_polar}, we need to find out the wave function $\Psi_{0}(r^{\prime})$ around the origin. Thus one finds that~\cite{Laine:2003bd}, 
\begin{equation}
\Psi_0(r^{\prime}) \approx \Psi_0(0)\left[1+\frac{1}{2} \eta r^{\prime 2}\left(\ln \frac{r^{\prime}}{2}+\gamma_E-1-\frac{1}{2} E_0\right)\right] .
\end{equation}
$\Psi_{0}(0)$ is some finite number to make the wave function bounded. To find out the $E_{0}$, one can integrate the Eq.~\eqref{sch_eq_polar} for a large value of $r^{\prime}$ and by requiring the square integrability condition. 
\section{Results and Discussion}\label{Section_6}
In this section, we will outline our findings for the screening masses and contrast them with the most recent lattice data obtained in Ref.~\cite{Bazavov:2019www}. The screening mass can be obtained from the Eqs.~\eqref{def._M} and~\eqref{screening_mass_final} as 
\begin{align}
m =& \, 2 M + g^{2} T \frac{C_{F}}{2\pi} E_{0} \nn
= & \, 2\pi T + g^{2}T \frac{C_{F}}{2\pi} \bigg(E_{0} - \frac{4\pi}{T}\big(I_{1}+I_{2}\big)\bigg)
\label{screening_mass_Final}
\end{align}
In the calculation of screening mass, we have used the lattice-fitted running coupling obtained in Ref.~\cite{Fukushima:2013xsa}, which is given by
\begin{equation}
\alpha_s\left(T / T_c\right) \equiv \frac{g^2\left(T / T_c\right)}{4 \pi}=\frac{6 \pi}{11 N_{\mathrm{c}} \ln \left[a\left(T / T_c\right)\right]},\label{alpha}
\end{equation}
where $a = 1.43$ for the IR case and $a=2.97$ for ultarviolet (UV) case respectively. To obtain the one loop coupling in Eq.~\eqref{alpha}, the authors fitted the lattice data of running coupling extracted from the IR and UV behavior of heavy-quark free energy from Ref.~\cite{Kaczmarek:2004gv}. In our calculation, since we are looking at the non-perturbative region, we use the infrared coupling case. 
\begin{figure}[]
\centering
\includegraphics[scale=.75]{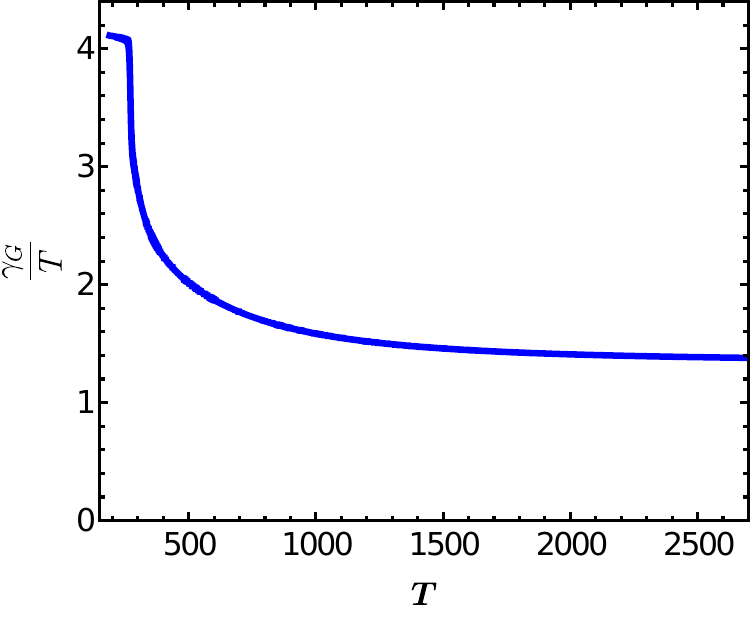}
\caption{Temperature variation of scaled Gribov mass parameter obtained using lattice (thermodynamics) data.}
\label{gammaG_vs_T}
\end{figure}
The integrals $I_{1}$ and $I_{2}$ consist of the Gribov parameter $\gamma_{G}$, which is often determined utilizing the one-loop or two-loop gap equation (for details, see~\cite{Gracey:2005cx}). Figure~\ref{gammaG_vs_T} shows the temperature variation of the scaled Gribov mass parameter $\gamma_{G}/T$ has been obtained after doing the matching with the lattice thermodynamics data, as obtained in Ref.~\cite{Jaiswal:2020qmj}. For the values of $\gamma_{G}/T$ shown in Figure~\ref{gammaG_vs_T}\, the integrals $I_{1}$ and $I_{2}$ have been evaluated numerically. The integral $I_{2}$ also contains the imaginary part, which eventually comes from the pole condition used to evaluate integral $X^{\prime}$ shown in Eq.~\eqref{def._X}. 
\begin{figure}[]
\centering
\includegraphics[scale=.75]{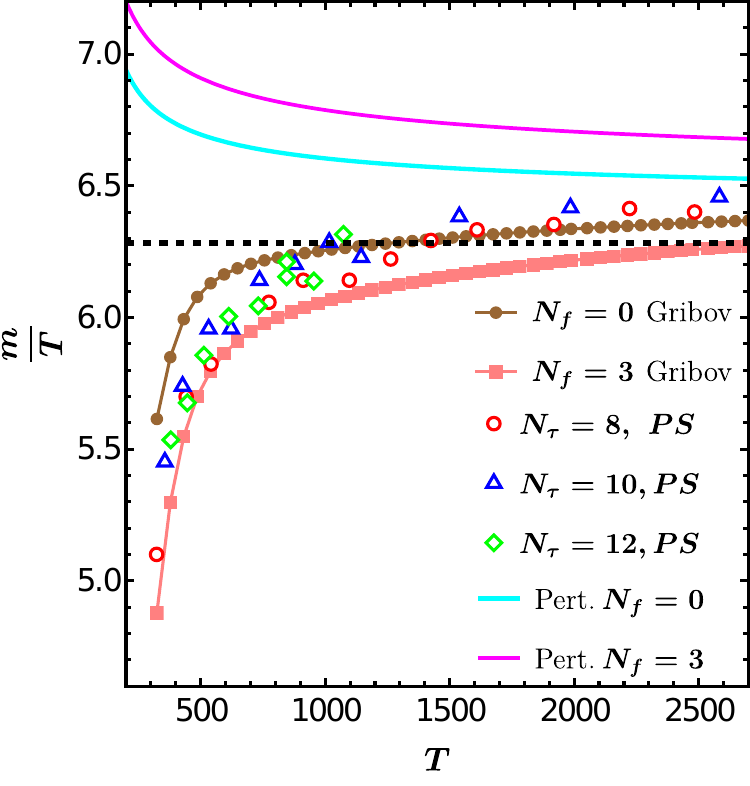}
\caption{The temperature dependence of the scaled screening mass. The dashed line represents the free theory result from $(m =2\pi T)$. We compare the Gribov results for quenched and $(2+1)$ flavor QCD case with the perturbative and lattice results for various $N_{\tau}$. Here, $PS$ represents the pseudo scalar channel.}
\label{screening_mass_vs_T}
\end{figure}
However, since screening mass depends on the real value of parameter $M$ defined in Eq.~\eqref{def._M}, by using Eq.~\eqref{screening_mass_Final}, we plot the scaled screening mass $m/T$ with temperature in Figure~\ref{screening_mass_vs_T} for quenched QCD $(N_{f}=0)$ case and for $N_{f} = 3$ case. We found a good agreement with the lattice data reported in ref.~\cite{Bazavov:2019www} and a good improvement over the perturbative results obtained in ref.~\cite{Laine:2003bd} in the low-temperature domain. The main outcome of Figure~\ref{screening_mass_vs_T} shows that screening mass significantly decreases from the free theory results in the low-temperature region. At the same time, in the high-temperature realm, it approaches the screening mass result obtained in ref.~\cite{Laine:2003bd}.
\section{Summary}\label{Section_7}
In this chapter, we used the non-perturbative resummation using the Gribov quantization approach to study the mesonic screening masses. We began by examining the fundamental characteristics of the static meson correlators at high temperatures and saw that the zeroth fermionic Matsubara mode dominates the large distance correlator. Then, we discussed the framework of the effective theory, namely NRQCD$_{3}$, in which we need to determine the parameters of the effective lagrangian to find out the non-perturbative NLO correction to mesonic screening mass. This matching is done by finding the Euclidean dispersion relation on the NRQCD$_{3}$ side and QCD side, using the Gribov approach. After this, we calculated the static quark-antiquark potential using Gribov's gluon propagator in the context of effective theory to understand the dynamics of the theory. This static potential determines the coefficient of exponential falloff by numerically solving the usual Schr\"odinger equation. We plotted the screening mass in the temperature ranges from$~300$ MeV to $2700$ MeV for $N_{f} = 0$ and $N_{f} = 3$, respectively. We compare our results with the recently obtained lattice data~\cite{Bazavov:2019www} and found a good agreement.  


\chapter{Heavy quark dynamics via Gribov-Zwanziger approach}\label{Chapter_5}
\allowdisplaybreaks
\pagestyle{fancy}
\fancyhead[LE]{\emph{\leftmark}}
\fancyhead[LO]{\emph{\rightmark}}
\rhead{\thepage} 

In the present chapter, we have investigated the heavy quark dynamics, namely momentum-dependent drag and diffusion coefficient, specific shear viscosity, and estimated energy loss of $\mathrm{HQs}$ moving in the QGP background using non-perturbative Gribov resummation approach. The present chapter is based on our work: {\em Heavy quark dynamics via Gribov-Zwanziger approach}, {\bf Sumit}, Arghya Mukherjee, Najmul Haque and Binoy Krishna Patra,
{\bf \color{blue}Phys.Rev.D 109 (2024) 11, 114043}~\cite{Sumit:2023oib}.
\section{Introduction}\label{Introduction}
The ultimate aim of the ongoing experiments, namely the RHIC at BNL and LHC at the CERN, is to create and study the new state of matter where bulk properties of this matter are governed by light quarks and gluons~\cite{Shuryak:2004cy, Jacak:2012dx}. It is now widely proven that this new state, which is the deconfined state of quarks and gluons dubbed as strongly interacting quark-gluon plasma (sQGP), is produced in these high energies nuclei collisions~\cite{Gyulassy:2004zy}. The models which successfully describe the space-time evolution of QGP fireball are governed by relativistic hydrodynamic models~\cite{Teaney:2000cw, Huovinen:2001cy, Nonaka:2006yn, Song:2007fn, Luzum:2008cw, Qiu:2011hf, Pang:2012uw, Gale:2012rq}, which gives information that the shear viscosity to entropy density ($\eta/s$) ratio of produced QGP is very small. Also, the experimental data analysis at RHIC suggests that $\eta/s \approx 0.1-0.2$~\cite{Bernhard:2019bmu, JETSCAPE:2020mzn} which is a strong indicator that the produced QGP in these collisions is strongly coupled because for a strongly coupled system $\eta/s$ is small. For a weakly coupled system, this ratio is large. One of the essential ways to characterize the properties of sQGP is by using hard probes, which are created in the initial stages of these highly energetic collisions, as their production requires a large momentum transfer. One of the promising hard probes is offered by $\mathrm{HQs}$, mainly charm and a bottom quark because the thermodynamic properties of QGP are governed mainly by the light quarks, namely up (u), down (d) and strange (s) quarks, and the force carrier among them via gluons. The other reason is that because of $\mathrm{HQs}$ large mass in comparison to the temperature scale generated in these URHIC~\cite{Rapp:2009my}. $\mathrm{HQs}$ travel in the expanding medium as generated after these collisions and interact with the light particles of the medium. However, their number is most likely to be conserved because of their considerable $M/T$ ratio where $M$ is the mass of the HQ and $T$ is the temperature of the medium. Thus, $\mathrm{HQs}$ can experience the complete evolution of the QGP. As they are produced in out-of-equilibrium, they are expected to retain their memory of interaction with plasma evolution~\cite{Rapp:2009my, Andronic:2015wma, Prino:2016cni, Aarts:2016hap, Greco:2017rro, Zhao:2020jqu}. Also, as $\mathrm{HQs}$ are so heavy that their masses are much larger than the QGP temperature generated at RHIC and LHC energies, $M_{\mathrm{HQ}}= 1-5$ GeV and $T \sim 0.3-0.5$ GeV. Thus, their thermal production and annihilation in the QGP can be safely ignored. In a perturbative QCD (pQCD) framework, the thermalization time of heavy quark ($\mathrm{HQ}$) has been estimated which is of the scale of $10-15$ fm/c for charm quark and the scale of $25-30$ fm/c for bottom quark~\cite{Rapp:2009my, Moore:2004tg,vanHees:2005wb, Cao:2011et} for the temperature scales required for QGP formed in RHIC and LHC experiments. Nevertheless, since the lifetime of QGP is around $4-5$ fm/c at RHIC and $10-12$ fm/c at LHC. Therefore, one should not expect the complete thermalization of $\mathrm{HQs}$ in uRHICs. For the small momentum exchange, the multiple scattering of $\mathrm{HQ}$ in a thermalized system can be dealt with as a Brownian motion, and Boltzmann equation in that approximation reduced to Fokker-Planck equation~\cite{Rapp:2009my, Moore:2004tg, Svetitsky:1987gq, GolamMustafa:1997id} which constitutes a simplified version of in-medium dynamics. This method has been widely used~\cite{Moore:2004tg,vanHees:2005wb, Cao:2011et, GolamMustafa:1997id,vanHees:2007me, Cao:2012au, Young:2011ug, Alberico:2011zy, Akamatsu:2008ge, He:2012df} to study the experimental observables such as nuclear modification factor ($R_{AA}$)~\cite{PHENIX:2006iih, STAR:2006btx, PHENIX:2005nhb, ALICE:2012ab} and elliptic flow ($v_{2}$)~\cite{PHENIX:2006iih} for nonphotonic electron spectra.   

$\mathrm{HQ}$ production has been explored in the perturbative QCD approach up to the NLO. In the perturbative realm, before the first experimental result, it was anticipated that their interaction with the medium particles could be described using a pQCD technique, which leads to the expectation of a small suppression of the final spectra and a small value of the elliptic flow. Nevertheless, experimental results come with a surprise in which the spectrum of nonphotonic electrons coming from the heavy quark decays has been observed in Au-Au collision at $\sqrt{s}=200$GeV at RHIC~\cite{PHENIX:2006iih, STAR:2006btx, PHENIX:2005nhb}. This result shows a relatively small $R_{AA}$ and a large value of elliptic flow $v_{2}$, which clearly indicates that there is a strong correlation between the $\mathrm{HQ}$ and medium constituents, which is beyond the pQCD explanations~\cite{Moore:2004tg,vanHees:2005wb, vanHees:2004gq}. This motivates one to go beyond pQCD to tackle the problem in a non-perturbative manner. One of the approaches is to consider the non-perturbative contribution~\cite{vanHees:2007me} from the quasi-hadronic bound states with subsequent hadronization from coalescence and fragmentation~\cite{Greco:2003vf, Greco:2003xt}. Another method consists of the hard thermal loop (HTL) in the pQCD framework to calculate the debye mass and running coupling~\cite{Gossiaux:2008jv, Alberico:2011zy}. This technique includes the non-perturbative contributions through the inclusion of thermal mass $\sim g(T) T$ where the running coupling has been fitted via lattice thermodynamics~\cite{Plumari:2011mk, Berrehrah:2014kba}. All these models are built upon the assumption that collisional energy loss serves as the predominant process in the low-momentum range of charm spectra~\cite{Cao:2011et, Cao:2012au, Cao:2016gvr}, $p_{T}\lesssim (3-5)$ $M_{\mathrm{HQ}}$. On the other hand, at high $p_{T}$, the radiative effects are dominating even collisional ones can not be disregarded although~\cite{Djordjevic:2011dd, Gossiaux:2010yx, Das:2010tj}. In the low transverse momentum region, the collisional energy loss process dominated because of the effect that the phase space for the in-medium induced gluon radiation is constrained because of $\mathrm{HQ}$ mass, i.e., ``dead-cone effect''~\cite{Dokshitzer:2001zm, Abir:2012pu}. However, now, at LHC experiments, heavy meson spectra can be observed around $30$ GeV. At such high $p_{T}$, even $\mathrm{HQs}$ become ultrarelativistic, and thus, radiative energy loss effects become important.

Energetic particles traversing the QCD medium suffer energy loss through the elastic process and gluon-bremsstrahlung. The drag and diffusion of $\mathrm{HQs}$ cause them to lose their energies in the medium. Much work has been done in the literature to study the energy loss of $\mathrm{HQs}$. $\mathrm{HQ}$ energy loss due to hard and soft collision processes has been studied in~\cite{Braaten:1991jj, Braaten:1991we} and for radiative processes in~\cite{Mustafa:1997pm, Mustafa:2004dr, Qin:2007rn, Abir:2012pu}. Recently, the soft contribution of the parton energy loss has been studied within a chiral imbalance in~\cite{Ghosh:2023ghi}. Many studies have been done recently in literature to understand the $\mathrm{HQ}$ dynamics like $\mathrm{HQ}$ potential~\cite{Thakur:2016cki, Thakur:2020ifi, Sebastian:2022sga}, spectral properties~\cite{Thakur:2021vbo}, transport coefficients without~\cite{Das:2012ck, Das:2010tj, Hong:2023cwl} and with bulk viscous medium~\cite{Shaikh:2021lka, Shaikh:2023qei}. Transport phenomenon has been studied for various other cases like Polyakov loop plasma~\cite{Singh:2018wps}, semi-QGP~\cite{Singh:2019cwi}, and memory effects in $\mathrm{HQ}$ dynamics~\cite{Ruggieri:2022kxv, Khowal:2021zoo}.   

Another method for studying the non-perturbative phenomenon in $\mathrm{HQ}$ dynamics can be made by using the Gribov-Zwanziger~\cite{Gribov:1977wm, Zwanziger:1989mf} technique. This method improves the infrared dynamics of QCD through a scale of the order of $g^{2} T $, which is known as the magnetic scale of the theory. This model deals with the non-perturbative resummation of the theory, having a mass parameter that captures the non-perturbative essence of the theory. For some good reviews, one can look at~\cite{Dokshitzer:2004ie, Vandersickel:2012tz}. This approach has been extended by including the impact of a local composite operator, which consists of a mass term of the order of electric scale $g(T) T$. For more details on this extended Gribov-Zwanziger method, see some of the recent works in~\cite{Dudal:2008sp, Capri:2014bsa, Capri:2015ixa, Capri:2015nzw, Capri:2016aqq, Dudal:2019ing, Dudal:2017kxb, Gotsman:2020ryd, Gotsman:2020mpg, Justo:2022vwa, Gracey:2010cg} and references therein. This extended approach with mass term inclusion in the propagator gives results that are very promising with lattice calculations in the infrared domain, as shown in~\cite{Dudal:2008sp}. Also, at zero temperature, it has been shown in~\cite{Tissier:2010ts, Mintz:2018hhx} that this mass term inclusion in gluon and ghost propagator of the usual Faddeev-Popov quantization is in excellent agreement with lattice results. For more details, look at the recent review~\cite{Pelaez:2021tpq}.  

Without any mass term, this scheme has been quite successful in describing the QCD thermodynamics when a comparison with lattice simulations has been made in~\cite{Fukushima:2013xsa}. Also, the other exciting studies which have been done in the recent literature explore quark dispersion relations~\cite{Su:2014rma}, connection between Gribov quantization and confinement-deconfinement transition~\cite{Kharzeev:2015xsa}, the transport coefficients~\cite{Florkowski:2015dmm, Florkowski:2015rua, Jaiswal:2020qmj}, the dilepton production rate has been calculated along with quark number susceptability~\cite{Bandyopadhyay:2015wua}, screening masses of mesons~\cite{Sumit:2023hjj} and electromagnetic debye mass~\cite{Bandyopadhyay:2023yjp}, which gives some interesting results of the said observables. In the context of $\mathrm{HQ}$ phenomenology, which we are interested in here, the heavy quarkonium potential has been calculated using this method in~\cite{Bandyopadhyay:2023yjp, Wu:2022nbv, Debnath:2023dhs}, the collisional energy loss of $\mathrm{HQs}$ has been estimated in~\cite{Debnath:2023zet} incorporating the formalism of Wong equations and $\mathrm{HQ}$ diffusion coefficient using Langevin dynamics has been studied in~\cite{Madni:2022bea}. 

In this work, we explored the finite momentum-dependent~\cite{Mazumder:2011nj} drag and diffusion coefficient of $\mathrm{HQs}$ using the Gribov gluon propagator. Earlier in the literature, carrying forward the calculation of drag and diffusion coefficient using the perturbative approaches, there was a need to set some infrared scale to tackle the infrared divergences that arise mainly in $t-$channel exchange diagrams. The significant advantage of this approach is that one does not need any infrared cut-off to put by hand in the matrix element calculation; instead, it comes automatically in the model calculations. Also, as discussed earlier, the ratio of shear viscosity to entropy density ratio, which is an essential observable to quantify the nature of QGP, has been studied earlier using the perturbative methods~\cite{Mazumder:2013oaa}. It was found that the inclusion of radiative effects in the calculation improves this ratio significantly. 

The chapter is organized as follows: Following this concise introduction in Section~\ref{Introduction}, we will delve into the conventional formalism for calculating the drag and diffusion coefficients of $\mathrm{HQs}$. This will be accomplished using the widely recognized Fokker-Planck method, as detailed in Section~\ref{sec._2}. In this section, we will discuss the scattering of $2\rightarrow 2$ collisional process as well as the $2\rightarrow3$ radiative process. We give the required matrix element calculation, which has been done using the Gribov propagator. Section~\ref{sec._3} focuses on our results for the drag and diffusion coefficient as estimated using the Gribov propagator. A critical observable $\eta/s$, which is required to understand the nature of the QGP, i.e., whether a medium behaves like a weakly coupled or strongly coupled system, has been plotted using the Gribov propagator. Also, the energy loss of $\mathrm{HQs}$ has been discussed within this model for the charm and bottom quarks while traversing the medium. In section ~\ref{sec.4}, we summarize the chapter. 


\section{Formalism: Drag and diffusion coefficients}\label{sec._2}
As discussed earlier, the motion of $\mathrm{HQs}$ in the QCD medium can be considered as a Brownian motion and is well described by the Fokker-Planck equation~\cite{Svetitsky:1987gq, GolamMustafa:1997id}
\begin{equation}\label{FP_equa.}
\frac{\partial \mathfrak{f}_{\mathrm{HQ}}}{\partial t}=\frac{\partial}{\partial p_i}\left[\mathcal{A}_i(\bm{p}) \mathfrak{f}_{\mathrm{HQ}}+\frac{\partial}{\partial p_j}\left[\mathcal{B}_{i j}(\bm{p}) \mathfrak{f}_{\mathrm{HQ}}\right]\right] \, ,
\end{equation}
where $\mathfrak{f}_{\mathrm{HQ}}$ represents the $\mathrm{HQ}$ momentum distribution in the medium. In this approach, the interaction of $\mathrm{HQ}$ with the medium constituent particles, which are light quarks, anti-quarks, and gluons, is encoded in the drag and diffusion tensors $\mathcal{A}_{i}$ and  $\mathcal{B}_{ij}$ respectively, which naturally arise from the momentum expansion of the collision integral of the Boltzmann transport equation (BTE)~\cite{Svetitsky:1987gq}. In the following, we briefly discuss the essential steps to obtain the drag and diffusion tensor of the $\mathrm{HQ}$. For clarity, the collisional and the radiative contributions are discussed in separate subsections.      

\subsection{Collisional Processes:}
Let us start with the two-body elastic scattering process:
$\mathrm{HQ}(P)+l(Q) \rightarrow$ $\mathrm{HQ}\left(P^{\prime}\right)+l\left(Q^{\prime}\right)$,  where $l$ denotes light particles viz. light quarks, anti-quarks, and gluons. Here, the four-momentum of the $\mathrm{HQ}$ and the constituent particle before the collision is represented by $P = (E_{p},\bm{p})$ and $Q = (E_{q},\bm{q})$  respectively. The corresponding four-momentum after the collision is denoted with primes.
Note that, in the case of the $\mathrm{HQ}$, the energy is given by $E_{p} = (|\bm{p}|^{2}+M_{\mathrm{HQ}}^{2})^{1/2}$ whereas the light particles are considered to be massless with $E_{q} = |\bm{q}|$. The  drag and the diffusion tensor that govern the dynamics  of the $\mathrm{HQ}$ in the QGP medium can be related to the $2 \rightarrow 2$ scattering  amplitude as~\cite{Svetitsky:1987gq}
\begin{eqnarray}
\mathcal{A}_i&= & \frac{1}{2 E_p} \int \frac{d^3 \bm{q}}{(2 \pi)^3 2 E_q} \int \frac{d^3 \bm{q}^{\prime}}{(2 \pi)^3 2 E_{q^{\prime}}} \int \frac{d^3 \bm{p}^{\prime}}{(2 \pi)^3 2 E_{p^{\prime}}} \frac{1}{\gamma_{\mathrm{HQ}}} \nonumber\\
& \times& \sum\left|\mathcal{M}_{2 \rightarrow 2}\right|^2(2 \pi)^4 \delta^4\left(P+Q-P^{\prime}-Q^{\prime}\right) \mathfrak{f}_k({E_{q}}) \nonumber\\
&\times &\left[1+a_k \mathfrak{f}_k\left({E_{q^{\prime}}}\right)\right]\left[\left(\bm{p}-\bm{p}^{\prime}\right)_i\right]=\llangle\left(\bm{p}-\bm{p}^{\prime}\right)_i\rrangle , \hspace*{-.0cm}  
\end{eqnarray}
\begin{equation}
\begin{aligned}
\mathcal{B}_{i j}= & \frac{1}{2}\llangle[\Big]\left(\bm{p}-\bm{p}^{\prime}\right)_i\left(\bm{p}-\bm{p}^{\prime}\right)_j
\rrangle[\Big] \, .
\end{aligned}\label{diffusion_definition}
\end{equation} 
The expressions above indicate that the drag force represents the thermal average of the momentum transfer $\left(\bm{p}-\bm{p}^{\prime}\right)$ resulting from interactions. On the other hand, momentum diffusion quantifies the average square of the momentum transfer. In these expressions, $\gamma_\mathrm{HQ}$ represents the statistical degeneracy factor of the $\mathrm{HQ}$, and the subscript $k$ denotes the particle species in the medium. The quantity $a_{k} = 1,-1$ represents respectively the near-equilibrium Bose-Einstein and the Fermi-Dirac distributions denoted in general as $\mathfrak{f}_{k}$. The delta function enforces the energy-momentum conservation. The computation of the matrix amplitude $\mathcal{M}_{2 \rightarrow 2}$  for the allowed $ 2 \rightarrow 2$ scattering processes will be discussed in the following subsection. It should be noted that the drag force depends only on $\mathrm{HQ}$ momentum. Thus, one can decompose it as
\begin{equation}
\mathcal{A}_{i}=p_i \mathcal{A} (p^2) \, , \quad \quad  \mathcal{A}=\llangle 1\rrangle-\frac{\llangle\bm{p} \cdot \bm{p}^{\prime}\rrangle}{p^2} \,.
\end{equation}
where $p^{2} = |\bm{p}|^{2}$ and $\mathcal{A}$ is the drag coefficient of $\mathrm{HQ}$. Similarly, one can decompose the diffusion tensor $\mathcal{B}_{i j}$ in terms of transverse and longitudinal components with respect to $\mathrm{HQ}$ momentum as
\begin{equation}
\mathcal{B}_{i j}=\left(\delta_{i j}-\frac{p_i p_j}{p^2}\right) \mathcal{B}_0\left(p^2\right)+\frac{p_i p_j}{p^2} \mathcal{B}_1\left(p^2\right)\, ,
\end{equation}
where the transverse diffusion coefficient $\mathcal{B}_0$ and longitudinal diffusion coefficient $\mathcal{B}_1$ take the following forms
\begin{equation}
\begin{aligned}
& \mathcal{B}_0=\frac{1}{4}\left[\llangle[\big] p^{\prime 2}\rrangle[\big]-\frac{\llangle[\big]\left(\bm{p}^{\prime} \cdot \bm{p}\right)^2\rrangle[\big]}{p^2}\right]\, , \\
\end{aligned}
\end{equation}
\begin{equation}
\begin{aligned}
& \mathcal{B}_1=\frac{1}{2}\left[\frac{\llangle[\big]\left(\bm{p}^{\prime} \cdot \bm{p}\right)^2\rrangle[\big]}{p^2} - 2\llangle[\big]\bm{p}^{\prime} \cdot \bm{p}\rrangle[\big]+p^2\llangle[\big] 1\rrangle[\big]\right] \,.
\end{aligned}
\end{equation}
One can study the kinematics of the $2 \rightarrow 2$ process in the center-of-momentum $\mathrm{(COM)}$ frame for simplification. The average of a generic function $\mathcal{F}(\bm{p})$ in the $\mathrm{COM}$ frame can be written as~\cite{Svetitsky:1987gq, Kumar:2021goi}
\ba
\llangle \mathcal{F}(\bm{p})\rrangle &= & \frac{1}{\left(512 \pi^4\right) E_p \gamma_{\mathrm{HQ}}} \int_0^{\infty} q\, d q\left(\frac{s-M_{\mathrm{HQ}}^2}{s}\right) \mathfrak{f}_k\left(E_q\right) \nn
&\times&  \int_0^\pi d \chi \sin \chi \int_0^\pi d \Theta_{\mathrm{cm}} \sin \Theta_{\mathrm{cm}} \sum\left|\mathcal{M}_{2 \rightarrow 2}\right|^2 \nn
&\times& \int_0^{2 \pi} d \Phi_\mathrm{cm}\left[1+a_k \mathfrak{f}_k\left(E_{q^{\prime}}\right)\right] \mathcal{F}(\bm{p}) \, .
\ea
where $\chi$ refers to the angle between the incident $\mathrm{HQ}$ and the medium constituent particles in the laboratory frame, while $\Theta_{\mathrm{cm}}$ and $\Phi_{\mathrm{cm}}$ are respectively the zenith and azimuthal angles in the $\mathrm{COM}$ frame. The Mandelstam variables $s, t$ and $u$ are defined as follows
\ba
s & =& (P+Q)^{2} = \left(E_p+E_q\right)^2-\left(p^2+q^2+2 p q \cos \chi\right) \, , \nn
t & =& (P^{\prime}-P)^{2} = 2 p_{\mathrm{cm}}^2\left(\cos \Theta_{\mathrm{cm}}-1\right) \, , \nn
u & =& (P^{\prime}-Q)^{2} = 2 M_{\mathrm{HQ}}^2-s-t \, .
\ea
Here $p_{\mathrm{cm}} = |\bm{p}_{\mathrm{cm}}|$ is the magnitude of the initial momentum of the $\mathrm{HQ}$ in the $\mathrm{COM}$ frame. The other quantity required in order to obtain the drag and diffusion coefficients is $\left(\bm{p}\cdot\bm{p}^{\prime}\right)$. In order to find this quantity, we need the Lorentz transformation that relates the laboratory frame and the $\mathrm{COM}$ frame via the relation $ \bm{p}^{\prime}=\gamma_{\mathrm{cm}}\left(\bm{p}_{\mathrm{cm}}^{\prime}+\bm{v}_{\mathrm{cm}} E_{\mathrm{cm}}^{\prime}\right),$ where  $\gamma_{\mathrm{cm}}=(E_p+E_q)/\sqrt{s}$ and the velocity in the $\mathrm{COM}$ is given by $\bm{v}_{\mathrm{cm}}=(\bm{p}+\bm{q})/(E_p+E_q)$. Now, the energy conservation dictates $p_{\mathrm{cm}}^{\prime 2}=p_{\mathrm{cm}}^2$. In the $\mathrm{COM}$ frame, $ \bm{p}_{\mathrm{cm}}^{\prime}$ can be decomposed as
$\bm{p}_{\mathrm{cm}}^{\prime}=  p_{\mathrm{cm}}\left(\cos \Theta_{\mathrm{cm}} \bm{x}_{\mathrm{cm}}+\sin \Theta_{\mathrm{cm}} \sin \Phi_{\mathrm{cm}} \bm{y}_{\mathrm{cm}}+\sin \Theta_{\mathrm{cm}} \cos \Phi_{\mathrm{cm}} \bm{z}_{\mathrm{cm}}\right),$
where $p_{\mathrm{cm}}=(s-M_{\mathrm{HQ}}^2)/(2 \sqrt{s})$ is the momentum and $E_{\mathrm{cm}}=$ $\left(p_{\mathrm{cm}}^2+M_{\mathrm{HQ}}^2\right)^{1 / 2}$ is the energy of the $\mathrm{HQ}$ in the $\mathrm{COM}$ frame. The axes $\bm{x}_{\mathrm{cm}}, \bm{y}_{\mathrm{cm}}$, and $\bm{z}_{\mathrm{cm}}$ are defined in~\cite{Svetitsky:1987gq}. Utilizing the above definitions, one can obtain
\begin{equation}
\begin{aligned}
\bm{p} \cdot \bm{p}^{\prime} & = E_p E_p^{\prime}-E_{\mathrm{cm}}^2+ p_{\mathrm{cm}}^2 \cos \Theta_{\mathrm{cm}} \, .
\end{aligned}
\end{equation}
\subsection{Matrix Elements for $2 \rightarrow 2$ processes}
\begin{figure}
		\centering
\begin{subfigure}		
		{\includegraphics[scale=1.5]{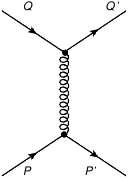}}\label{fig1(a)}
\end{subfigure}		
		\quad \quad \quad
\begin{subfigure}		
		{\includegraphics[scale=1.5]{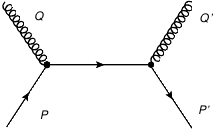}}\label{fig1(b)}
\end{subfigure}
		\quad \quad \quad
\begin{subfigure}		
		{\includegraphics[scale=1.5]{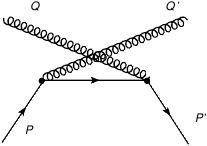}}\label{fig1(c)}
\end{subfigure}		
				\quad \quad \quad
\begin{subfigure}		
		{\includegraphics[scale=1.5]{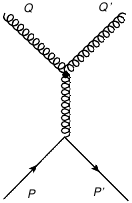}}\label{fig1(d)}
\end{subfigure}
		\caption{Feynman diagrams for $\mathrm{HQ}$ $2$ $\rightarrow$ $2$ processes with (a) gluon (t-channel), (b) gluon (s-channel), (c) gluon (u-channel), (d) light quark/anti-quark (t-channel).}%
		\label{fig:1}%
	\end{figure}

The leading order Feynman diagrams for $2 \rightarrow 2$ processes are shown in Fig.~\ref{fig:1}. There are three topologically distinct diagrams contributing to quark-gluon scattering and one diagram for quark-quark or quark-antiquark scattering shown in Fig.~\ref{fig:1}~\cite{Matsui:1985eu}. Note that two out of four diagrams shown in Fig.~\ref{fig:1} consist of a gluon propagator, which in the present work has been replaced with the Gribov-modified gluon propagator. The modified gluon propagator in the Landau gauge is written as~\cite{Fukushima:2013xsa}
\begin{equation}\label{Gribov_prop}
D_{\mu\nu}^{ab}(P)=\delta^{a b} \left(\delta_{\mu \nu}-\frac{P_\mu P_\nu}{P^2}\right)\frac{P^2}{P^4+\gamma_{\mathrm{G}}^4}\,.
\end{equation}
where $\gamma_{\mathrm{G}}$ is the Gribov mass parameter which is generally derived from the one-loop or two-loop gap equation~\cite{Gracey:2010cg}. The matrix elements for the diagrams shown in Fig.~\ref{fig:1}, using the Gribov propagator, are given by:
\begin{align}
 \mathcal{M}_{(\mathrm{a})} =& -g^2 \varepsilon_\mu(2) \varepsilon_\nu^{*}(4) f_{a b c}\left[g^{\mu \nu}\left(-Q-Q^{\prime}\right)^\rho+g^{\nu \rho}\left(2Q^{\prime}-Q\right)^\mu+g^{\rho \mu}\left(2 Q-Q^{\prime}\right)^\nu\right] \nn
\times & \frac{(P^{\prime}-P)^{2}}{(P^{\prime}-P)^{4}+\gamma_{G}^{4}} \bar{u}^{i}(3) \gamma_\rho \lambda_c u^{i}(1) \, , \nn
 \mathcal{M}_{(\mathrm{b})} =& \, -i g^2 \varepsilon_\mu(2) \varepsilon_\nu^{*}(4) \bar{u}^{i}(3) \gamma^\mu \lambda_a \frac{\slashed{P}+\slashed{Q}+M_{HQ}}{\left(P+Q\right)^2-M_{HQ}^2} \gamma^\nu \lambda_b u^{i}(1) \, , \nn
 \mathcal{M}_{(\mathrm{c})} =& \, -i g^2 \varepsilon_\mu(2) \varepsilon_\nu^{*}(4) \bar{u}^{i}(3) \gamma^\nu \lambda_b \frac{\slashed{P}^{\prime}-\slashed{Q}+M_{HQ}}{\left(P^{\prime}-Q\right)^2-M_{HQ}^2} \gamma^\mu \lambda_a u^{i}(1) \, , \nn
 \mathcal{M}_{(\mathrm{d})} =& \, i g^2 \bar{u}^{i}(3) \gamma^\mu \lambda_a u^{i}(1) \frac{(P^{\prime}-P)^{2}}{(P^{\prime}-P)^{4}+\gamma_{G}^{4}} \bar{u}^{i}(4) \gamma_\mu \lambda_a u^{i}(2) \, .
\end{align}
Here, the abbreviated notations used are $\varepsilon_{\mu}(1) = \varepsilon_{\mu}(P,\zeta_{P}),\, \varepsilon_{\mu}(2) = \varepsilon_{\mu}(Q,\zeta_{Q}),\, \varepsilon_{\mu}(3) = \varepsilon_{\mu}(P^{\prime},\zeta_{P^{\prime}})$ and $\varepsilon_{\mu}(4) = \varepsilon_{\mu}(Q,\zeta_{Q^{\prime}})$ for gluon polarization vectors, $i$ denotes different flavors and $u(1) = u(P,s_{P})$ for quark spinors. The symbols $\lambda_{a}$ represent $SU(3)$ matrices normalized by $\text{Tr}(\lambda_{a}\lambda_{b}) = \frac{1}{2}\delta_{ab}$, satisfying $[\lambda_{a},\lambda_{b}] = i f_{abc} \lambda_{c}$ and $f_{abc}$ are the structure constants. The summation of squared matrix elements over the initial and final quark spin states transforms the quark spinors into projection operators as per the relation:
\ba 
\sum_{s=1,2} u_\alpha^i(P, s) \bar{u}_\beta^i(P, s)&=& \left(\slashed{P}+M_{HQ}^{i}\right)_{\alpha \beta} \, . 
\ea
During summation over gluon polarizations $\zeta_{z}$ where $z=1,2,3,4$, in order to avoid the contributions from the unphysical states, one can remove the terms containing  $\varepsilon_{\mu}(P,\zeta)P^{\mu}$. Thus the amplitude $\mathcal{M}_{(\mathrm{a})}$ becomes:
\begin{align}
 \mathcal{M}_{(\mathrm{a})} =& \, -g^2 \varepsilon_\mu(2) \varepsilon_\nu^{*}(4) f_{a b c}\left[g^{\mu \nu}\left(-Q-Q^{\prime}\right)^\rho+g^{\nu \rho}\left(2Q^{\prime}\right)^\mu+g^{\rho \mu}\left(2 Q\right)^\nu\right] \nn
\times & \frac{(P^{\prime}-P)^{2}}{(P^{\prime}-P)^{4}+\gamma_{G}^{4}} u^{i}(1) \gamma_\rho \lambda_c \bar{u}^{i}(3) \, ,
\end{align}
Now, one can do the trace over the Lorentz indices utilizing the relation
\begin{equation}
\begin{aligned}
\sum_{\zeta = 1,2} \varepsilon_{\mu}^{*}(P,\zeta)\varepsilon_{\nu}(P,\zeta) = - g_{\mu\nu} \, .
\end{aligned}
\end{equation}
The squared matrix elements can be conveniently written in terms of Mandelstam variables, 
which satisfy the relation $s+t+u=2M_{\mathrm{HQ}}^{2}$. After doing the summation over the spins, polarizations, and color indices, one would get the final expressions as follows: (i) for the process $\mathrm{HQ}(P)+g(Q) \rightarrow$ $\mathrm{HQ}\left(P^{\prime}\right)+g\left(Q^{\prime}\right) $, one obtains 
\begin{align}
\left|\mathcal{M}_{(a)}\right|^2 = & \gamma_{\mathrm{HQ}} \gamma_g\left[32 \pi^2 \alpha^2 \frac{\left(s-M_{\mathrm{HQ}}^2\right)\left(M_{\mathrm{HQ}}^2-u\right)t^{2}}{\left(t^{2}+\gamma_{\mathrm{G}}^4\right)^2}\right] \, ,  \nn
\left|\mathcal{M}_{(b)}\right|^2 = & \gamma_{\mathrm{HQ}} \gamma_g\left[\frac{64 \pi^2 \alpha^2}{9} \frac{\left(s-M_{\mathrm{HQ}}^2\right)\left(M_{\mathrm{HQ}}^2-u\right)+2 M_{\mathrm{HQ}}^2\left(s+M_{\mathrm{HQ}}^2\right)}{\left(s-M_{\mathrm{HQ}}^2\right)^2}\right] \, , \nn
\left|\mathcal{M}_{(c)}\right|^2 = & \gamma_{\mathrm{HQ}} \gamma_g\left[\frac{64 \pi^2 \alpha^2}{9} \frac{\left(s-M_{\mathrm{HQ}}^2\right)\left(M_{\mathrm{HQ}}^2-u\right)+2 M_{\mathrm{HQ}}^2\left(M_{\mathrm{HQ}}^2+u\right)}{\left(M_{\mathrm{HQ}}^2-u\right)^2}\right] \, , \nn
\mathcal{M}_{(a)} \mathcal{M}_{(b)}^*  = & \mathcal{M}_{(b)}^* \mathcal{M}_{(a)} = \gamma_{\mathrm{HQ}}  \gamma_g\left[8 \pi^2 \alpha^2 \frac{\left(s-M_{\mathrm{HQ}}^2\right)\left(M_{\mathrm{HQ}}^2-u\right)+M_{\mathrm{HQ}}^2(s-u)}{\left(\frac{t^{2}+\gamma_{\mathrm{G}}^4}{t} \right)\left(s-M_{\mathrm{HQ}}^2\right)}\right] \, , \nn
\mathcal{M}_{(a)} \mathcal{M}_{(c)}^* = & \mathcal{M}_{(c)}^* \mathcal{M}_{(a)}= \gamma_{\mathrm{HQ}} \gamma_g\left[8 \pi^2 \alpha^2 \frac{\left(s-M_{\mathrm{HQ}}^2\right)\left(M_{\mathrm{HQ}}^2-u\right)-M_{\mathrm{HQ}}^2(s-u)}{\left(\frac{t^{2}+\gamma_{\mathrm{G}}^4}{t} \right)\left(M_{\mathrm{HQ}}^2-u\right)}\right] \, , \nn
\mathcal{M}_{(b)} \mathcal{M}_{(c)}^* = & \mathcal{M}_{(b)}^* \mathcal{M}_{(c)}= \gamma_{\mathrm{HQ}} \gamma_g\left[\frac{8 \pi^2 \alpha^2}{9} \frac{M_{\mathrm{HQ}}^2\left(4 M_{\mathrm{HQ}}^2-t\right)}{\left(s-M_{\mathrm{HQ}}^2\right)\left(M_{\mathrm{HQ}}^2-u\right)}\right] \, , \nn
\left|\mathcal{M}_{(i)}\right|^2 = & \left|\mathcal{M}_{(a)}\right|^2+\left|\mathcal{M}_{(b)}\right|^2+\left|\mathcal{M}_{(c)}\right|^2+2 \mathcal{R} e\left\{\mathcal{M}_{(a)} \mathcal{M}_{(b)}^*\right\}+2 \mathcal{R} e\left\{\mathcal{M}_{(b)} \mathcal{M}_{(c)}^*\right\} \nn
 + & 2 \mathcal{R} e\left\{\mathcal{M}_{(a)} \mathcal{M}_{(c)}^*\right\} \, ,
\end{align}
and (ii) for the process $\mathrm{HQ}(P)+lq(Q)/l\bar{q}(Q) \rightarrow$ $\mathrm{HQ}\left(P^{\prime}\right)+lq\left(Q^{\prime}\right)/l\bar{q}\left(Q^{\prime}\right)$, one obtains
\begin{equation}
\begin{aligned}
\left|\mathcal{M}_{(d)}\right|^2= \gamma_{\mathrm{HQ}} \gamma_{lq/l \bar{q}}\left[\frac{64 \pi^2 \alpha^2}{9} \frac{\left(\left(s-M_{\mathrm{HQ}}^2\right)^2+\left(M_{\mathrm{HQ}}^2-u\right)^2 + 2 M_{\mathrm{HQ}}^2 (\frac{t^{2}+\gamma_{\mathrm{G}}^4}{t} )\right)t^{2}}{\left(t^{2}+\gamma_{\mathrm{G}}^4\right)^2}\right] \, .
\end{aligned}
\end{equation}
Here, $\gamma_{\mathrm{HQ}} = N_{s}\times N_{c},\,  \gamma_{g} = N_{s} \times (N_{c}^{2}-1)$ and $\gamma_{lq/l \bar{q}} = N_{s}\times N_{c} \times N_{f}$ are the degeneracy factor for $\mathrm{HQ}$, gluon, and light quark respectively with $N_{s}=2, N_{f} =3$ and $N_{c} = 3$ have been used. 
\subsection{Radiative Process:}
In general, the transport coefficient can be written as follows~\cite{Mazumder:2013oaa}
\begin{equation}\label{tran_coeff._def.}
\text{X(p)} = \int \text{Phase space} \times \text{interaction}\times \text{transport part}
\end{equation}
The Eq.~\eqref{tran_coeff._def.} can be used in order to estimate the radiative contribution of the drag and diffusion coefficient by changing the two-body phase space and invariant amplitude with their three-body counterparts, keeping the transport part the same~\cite{Mazumder:2013oaa}. Let us consider the $2\rightarrow 3 $ inelastic process: $\mathrm{HQ}(P)+l(Q) \rightarrow$ $\mathrm{HQ}\left(P^{\prime}\right)+l\left(Q^{\prime}\right) + g(K^{\prime})$, where $K^{\prime} = (E_{k^{\prime}},\mathbf{k}_{\perp}^{\prime},k_{z}^{\prime})$ is the four-momenta of the emitted soft-gluon by HQ in the final state. The general expression for the thermal averaged $\llangle \mathcal{F}(p)\rrangle $ for $2$ $\rightarrow$ $3$ process is given by~\cite{Mazumder:2013oaa}:
\begin{figure}
		\centering
		{\includegraphics[scale=1.2]{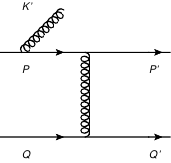}}
		\quad \quad \quad
		{\includegraphics[scale=1.2]{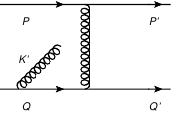}}
		\quad \quad \quad
		{\includegraphics[scale=1.2]{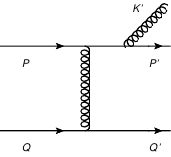}}
				\quad \quad \quad
		{\includegraphics[scale=1.2]{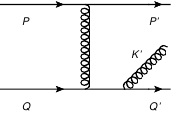}}
				\quad \quad \quad
		{\includegraphics[scale=1.2]{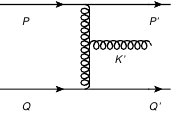}}
\captionof{figure}{Feynman diagrams for process $\mathrm{HQ}(P)+l(Q) \rightarrow$ $\mathrm{HQ}\left(P^{\prime}\right)+l\left(Q^{\prime}\right) + g(K^{\prime})$, showing an inelastic scattering of HQ with light quark and a soft gluon emission.}
\label{fig_2}
	\end{figure}	
\begin{align}
\llangle \mathcal{F}(\bm{p})\rrangle_{\mathrm{rad}}  =& \frac{1}{2 E_p \gamma_{\mathrm{HQ}}}\int\hspace{-0.1cm} \frac{d^3 \bm{q}}{(2 \pi)^3 E_q} \int\hspace{-0.1cm}\frac{d^3 \bm{q}^{\prime}}{(2 \pi)^3 E_{q^{\prime}}}\int\hspace{-0.1cm} \frac{d^3 \bm{p}^{\prime}}{(2 \pi)^3 E_{p^{\prime}}} \nn
\times &  \int\hspace{-0.1cm}\frac{d^3 \bm{k}^{\prime}}{(2 \pi)^3 E_{k^{\prime}}} \sum\left|\mathcal{M}_{2 \rightarrow 3}\right|^2  \mathfrak{f}_k\left(E_q\right)\left(1 \pm \mathfrak{f}_k\left(E_{q^{\prime}}\right)\right) \nn
\times &  \left(1+\mathfrak{f}_g\left(E_{k^{\prime}}\right)\right) \Theta_1\left(E_p-E_{k^{\prime}}\right) \Theta_2\left(\tau-\tau_F\right) \nn
\times & \mathcal{F}(\bm{p})(2 \pi)^4\delta^{(4)} \left(P+Q-P^{\prime}-Q^{\prime}-K^{\prime}\right) \, .
\label{rad._process}
\end{align}
where $\tau$ is the scattering time of $\mathrm{HQ}$ with the medium constitutents and $\tau_{F}$ is the formation time of gluons. The theta function $\Theta_{1}(E_{p}-E_{k^\prime})$ in Eq.~\eqref{rad._process} imposes the constraints on the process that the emitted gluon energy should be less than the initial energy of $\mathrm{HQ}$. Whereas, the second theta function $\Theta_{2}(\tau-\tau_{F})$ makes sure that the formation time of gluon should be lesser than the scattering time of $\mathrm{HQs}$ with medium constituents that accounts for the Landau-Pomerancguk-Migdal (LPM) effect~\cite{Wang:1994fx, Gyulassy:1993hr, Klein:1998du}. Also, $\mathfrak{f}_{g}(E_{k^\prime}) = 1/[\exp{(\beta E_{k^\prime})}-1]$ is the distribution of the emitted gluon where $\beta = 1/T$ i.e. the gluons in Figure~\ref{fig:1} and Figure~\ref{fig_2} are in thermally equilibrated state. The term $\left|\mathcal{M}_{2 \rightarrow 3}\right|^2$ denotes the matrix element squared for the $2 \rightarrow 3$ radiative process as depicted in Figure~\ref{fig_2}. It can be expressed in terms of the matrix element of the collision process multiplied by the probability for soft gluon emission~\cite{Abir:2011jb} as follows:
\begin{equation}
\left|\mathcal{M}_{2 \rightarrow 3}\right|^2=\left|\mathcal{M}_{2 \rightarrow 2}\right|^2 \times \frac{48 \alpha_{s}(T)}{{k}_{\perp}^{\prime 2}}\left(1+\frac{M_{H Q}^2}{s} e^{2\eta}\right)^{-2} \, ,
\end{equation}
where $\alpha_{s}(T)$ is the strong coupling constant defined at one-loop as 
\begin{equation}
\alpha_{s}(T) = \frac{g^{2}}{4\pi} = \frac{6\pi}{11N_{c}-2N_{f}}\frac{1}{\ln (2\pi T/\Lambda_{\overline{\text{MS}}})} 
\end{equation}
having scale $\Lambda_{\overline{\text{MS}}}=0.176$ GeV~\cite{Haque:2014rua} for $N_{f}=3$, $\eta$ is the rapidity of the emitted gluon and $\left(1+\frac{M_{\mathrm{HQ}}^2}{s} e^{2 \eta}\right)^{-2}$ is the suppression factor for the $\mathrm{HQ}$ due to the dead-cone factor~\cite{Dokshitzer:2001zm, Abir:2011jb}. From the Eq.~\eqref{rad._process} we have
\begin{equation}
\begin{aligned}
\llangle \mathcal{F}(\mathbf{p})\rrangle_{\mathrm{rad}}= & \llangle \mathcal{F}(\mathbf{p})\rrangle_{\mathrm{coll.}} \times \mathcal{I}(\mathbf{p}) \, ,
\end{aligned}
\end{equation}
where $\mathcal{I}(\mathbf{p})$ is given by
\begin{align}
\mathcal{I}\left(\mathbf{p}\right) = & \, \int \frac{d^3 k^{\prime}}{(2 \pi)^3 2 E_{k^{\prime}}} \frac{48 \alpha_{s}(T)}{k_{\perp}^{\prime 2}}\left(1+\frac{M_{H Q}^2}{s} e^{2 \eta}\right)^{-2} \nn
\times & \left(1+\mathfrak{f}_g\left(E_{k^{\prime}}\right)\right) \Theta_1\left(E_p-E_{k^{\prime}}\right) \Theta_2\left(\tau-\tau_F\right) \, .
\label{Int_I_k}
\end{align}
In the limit of soft gluon emission ($\Theta_{k^{\prime}}\rightarrow 0$), one will get $ \left(1+\frac{M_{H Q}^2}{s} e^{2 \eta}\right)^{-2} \approx\left(1+\frac{4 M_{H Q}^2}{s \Theta_{k^{\prime}}^2}\right)^{-2},$ where $\theta_{k^{\prime}}$ is the angle between the radiated soft gluon and the HQ which can be related to the rapidity parameter through the relation $\eta=-\ln \left[\tan \left(\Theta_{k^{\prime}} / 2\right)\right]$. 
In order to simplify the Eq.~\eqref{Int_I_k}, one can convert emitted gluon four-momentum in terms of the rapidity variable as
\begin{equation}
\begin{aligned}
E_{k^{\prime}} = k_{\perp}^{\prime} \cosh\eta \, , \quad \quad k_{z}^{\prime} = k_{\perp}^{\prime} \sinh\eta \, ,
\end{aligned}
\end{equation}
with $d^{3}k^{\prime} = d^{2}k_{\perp}^{\prime}dk_{z}^{\prime} = 2\pi k_{\perp}^{\prime 2}dk_{\perp}^{\prime}\cosh\eta \,d\eta$. The interaction time $\tau$ is related to the interaction rate $\Gamma = 2.26 \alpha_{s}T$~\cite{Shaikh:2021lka} and the $\Theta_{2}(\tau-\tau_{F})$ impose the constraint
\begin{equation}
\begin{aligned}
\tau = \Gamma^{-1} > \tau_{F} = \frac{\cosh\eta}{k_{\perp}^{\prime}} \, ,
\end{aligned}
\end{equation}
which shows that $k_{\perp}^{\prime} > \Gamma \cosh\eta = (k_{\perp}^{\prime})_{\text{min.}}$. Further, from the other theta function $\Theta_{1}(E_{p}-E_{k^{\prime}})$ we have, 
\begin{equation}
\begin{aligned}
E_{p} > E_{k^{\prime}} = k_{\perp}^{\prime} \cosh\eta \, , \quad \quad (k_{\perp}^{\prime})_{\text{max.}} = \frac{E_{p}}{\cosh\eta} \, .
\end{aligned}
\end{equation}
Also, the Bose enhancement factor for the emitted gluon in the limiting case ($E_{k^{\prime}} \ll T$) can be written as 
\begin{equation}
\begin{aligned}
1+\mathfrak{f}_g\left(E_{k^{\prime}}\right)
=\frac{T}{k_{\perp}^{\prime} \cosh\eta} \, .
\end{aligned}
\end{equation}
Thus the integral $\mathcal{I}(\mathbf{p})$ becomes
\begin{equation}
\begin{aligned}
\mathcal{I}\left(\mathbf{p}\right)= & \, \frac{6}{\pi} \alpha_{s} T \int_{\Gamma \cosh \eta}^{E_p / \cosh \eta} d k_{\perp}^{\prime} \int_{-\eta_{1}}^{\eta_{1}} d \eta  \times\left(1+\frac{M_{H Q}^2}{s} e^{2 \eta}\right)^{-2} \frac{1}{k_{\perp}^{\prime} \cosh \eta} \, .
\end{aligned}
\end{equation}
where rapidity integration limits are decided based on the pseudo-rapidity coverage of the detector accordingly. In the next section, we have used the value of $\eta_{1}=20$ for practical calculations.
\section{Results and Discussion}\label{sec._3}
\begin{figure}[H]
\centering
{\includegraphics[scale=0.7]{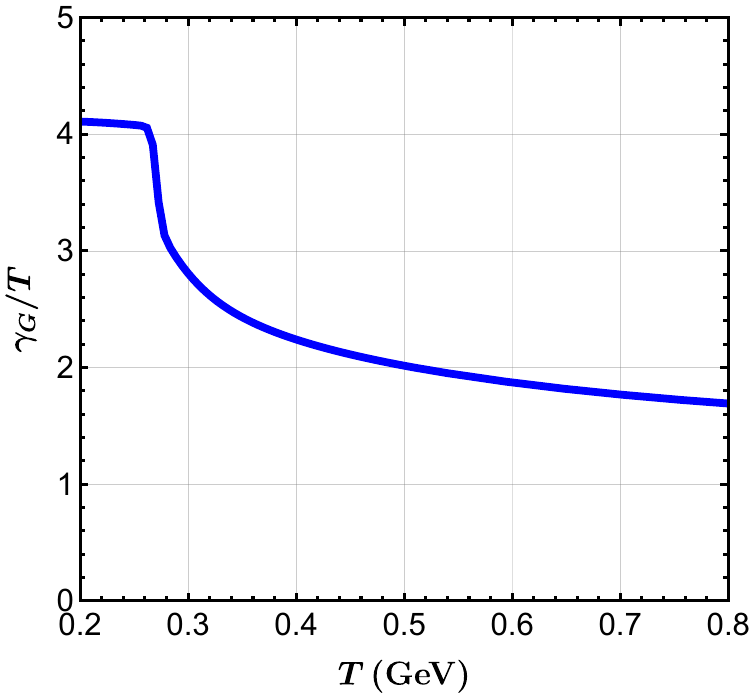}}
\caption{Temperature variation of the scaled Gribov mass parameter obtained by matching the thermodynamics of the quasi-particle approach with the pure gauge lattice  data~\cite{Borsanyi:2012ve}.}
\label{gammaG_variation}
\end{figure}

In order to do a numerical evaluation of drag and diffusion coefficients, firstly, we must fix the Gribov mass parameter $\gamma_{G}$ appearing in the Gribov propagator. To do so, the authors of Ref.~\cite{Jaiswal:2020qmj} have done the matching of temperature-dependent scaled trace anomaly results of lattice~\cite{Borsanyi:2012ve} with the equilibrium thermodynamic quantities. In Figure~\ref{gammaG_variation}, we showed the scaled Gribov mass parameter variation $\gamma_{G}/T$ with temperature $T$. This dependence of $\gamma_{G}$ will be used in the estimation of other quantities evaluated further.
Now, we will present our numerical results for the transport coefficient, namely drag and diffusion for elastic and inelastic processes, the specific shear viscosity of the QGP medium, and the estimation of collisional and radiative energy loss in the separate subsections.
\subsection{Drag and diffusion coefficient for collisional and radiative processes}
\begin{figure}[H]
\centering
{\includegraphics[scale=0.53]{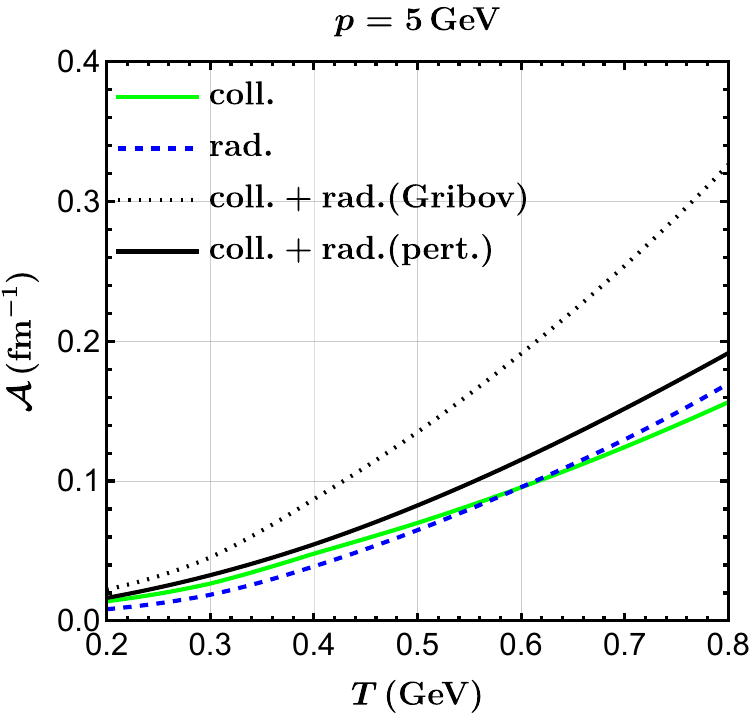} }
\quad \quad
{\includegraphics[scale=0.53]{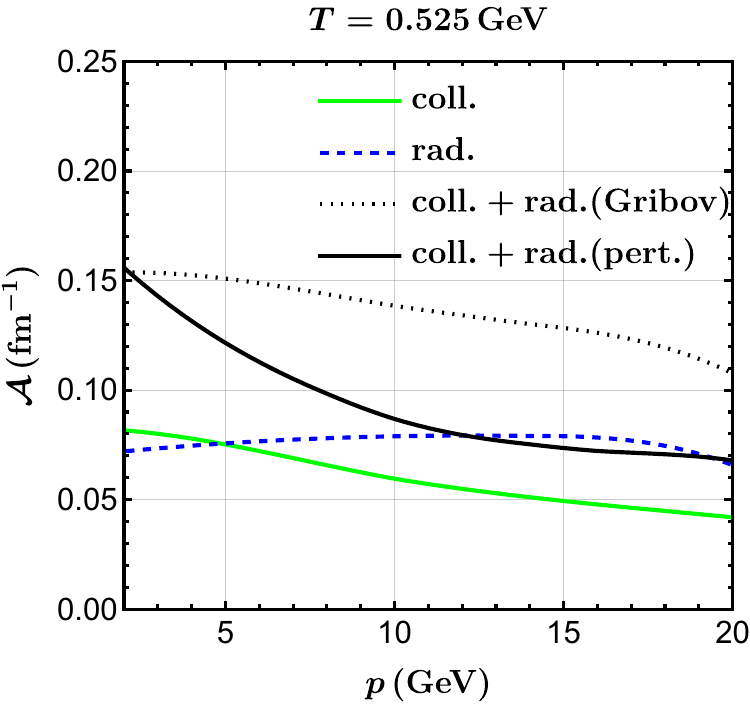} }
\vspace{.1cm}
\captionof{figure}{Variation of charm quark drag coefficient with temperature and momentum at (a) $p = 5$ GeV and (b) $T = 0.525$ GeV, respectively, where “coll.” stands for collisional processes and “rad.” stands for radiative processes. The same abbreviations have been used for the rest of the Figures.}
\label{drag_coeff_total}
\end{figure}

\begin{figure}[H]
\centering
{\includegraphics[scale=0.53]{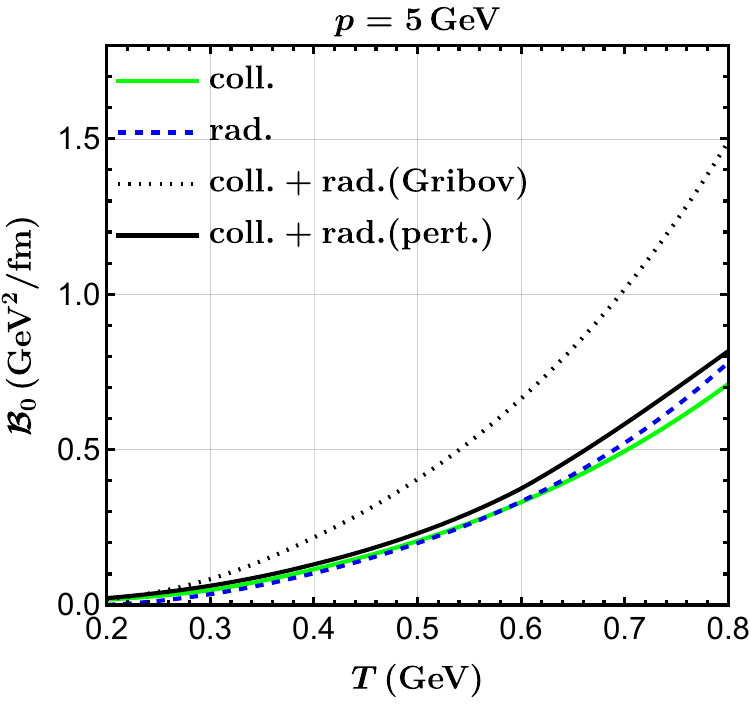} }
\quad \quad
{\includegraphics[scale=0.53]{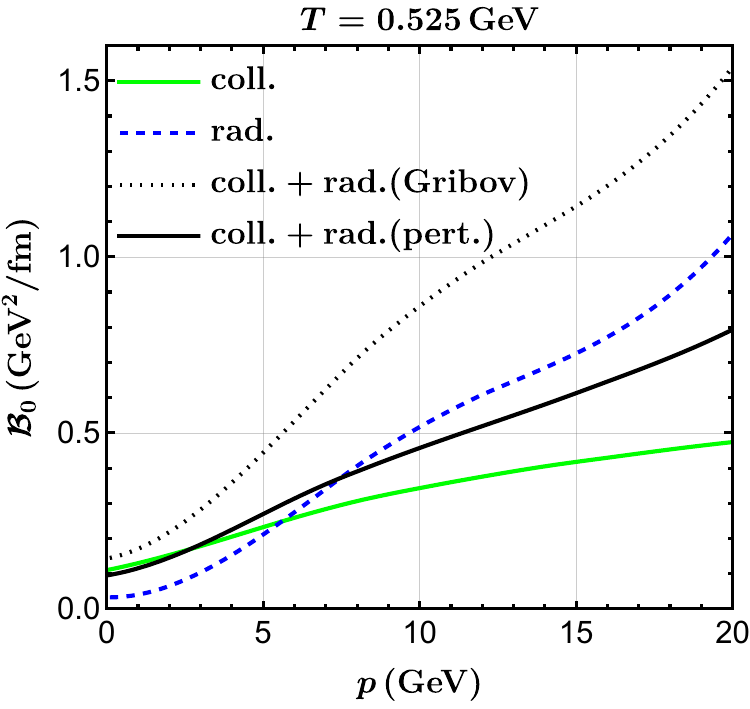} }
\vspace{.1cm}
\captionof{figure}{Variation of charm quark transverse diffusion coefficient with temperature and momentum at (a) $p = 5$ GeV and (b) $T = 0.525$ GeV respectively.}
\label{diff_tran}
\end{figure}

\begin{figure}[H]
\centering
{\includegraphics[scale=0.53]{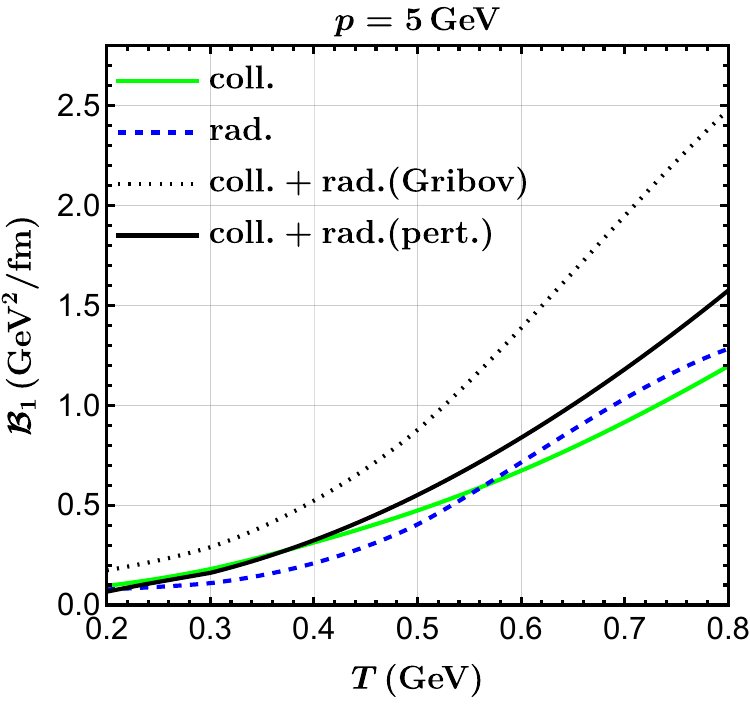} }
\quad \quad
{\includegraphics[scale=0.53]{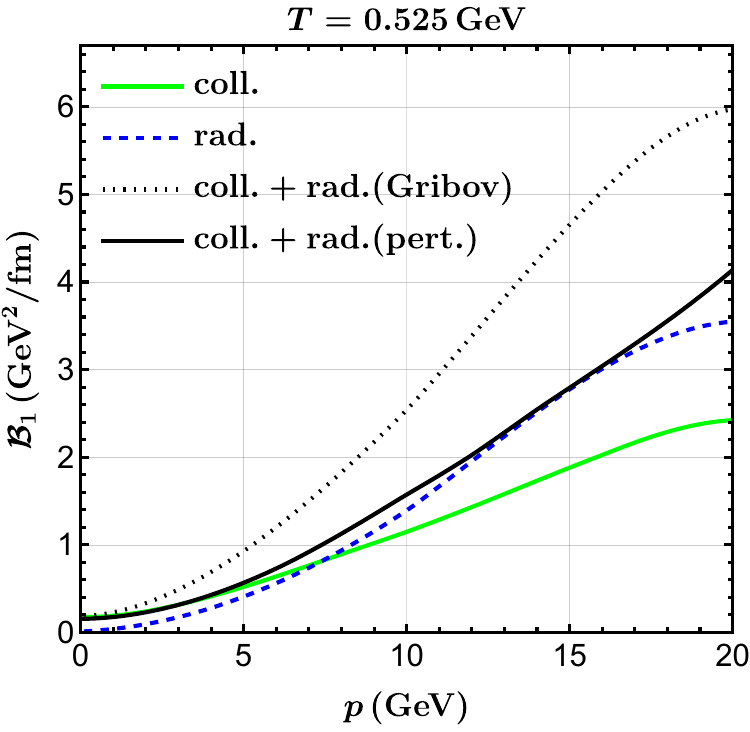} }
\vspace{.1cm}
\captionof{figure}{Variation of charm quark longitudinal diffusion coefficient with temperature and momentum at (a) $p = 5$ GeV and (b) $T = 0.525$ GeV respectively.}
\label{diff_long}
\end{figure}

In Figure~\ref{drag_coeff_total} (left panel), the temperature dependence of the drag coefficient has been shown at $p = 5$ GeV. Here, we have taken the charm quark mass $1.3$ GeV. The contributions from both processes have been shown as these processes occur independently in the thermal medium. Figure~\ref{drag_coeff_total} (left panel) shows that the collisional process contributes more at the low temperature than the radiative one. However, as the temperature increases, the radiative process starts dominating, indicating that inelastic processes are more important at LHC energies than RHIC energy within this model calculations. As the temperature increases, the total contribution to the drag coefficient increases compared to the elastic process. Qualitatively, the drag coefficient has a similar nature within this modeling compared to earlier perturbative results~\cite{Mazumder:2013oaa}. However, the overall magnitude of both processes is higher after a $T=0.4$ GeV and lower before $T=0.4$ GeV, which can be inferred from the non-perturbative nature of the Gribov propagator. As the system with the Gribov gluon propagator is strongly interacting, the $\mathrm{HQs}$ feel a strong drag force compared to the weakly interacting matter in the high-temperature domain while at lower temperature $\mathrm{HQ}$ drag coefficient is less compared to earlier perturbative estimation. In other words, one would expect a larger drag coefficient in Gribov plasma for large temperatures and a lower drag coefficient for lower temperatures. Thus, the overall magnitude of the drag coefficient for collisional and radiative processes is higher in the high-temperature domain via the Gribov-Zwanziger approach than it was with earlier perturbative results.
In Figure~\ref{drag_coeff_total} (right panel), the drag coefficient of $\mathrm{HQ}$ has been plotted with its momentum for a temperature $T = 0.525$ GeV. It has been observed that after a momentum of $5$ GeV, the radiative contribution dominates in the medium despite the dead cone effect. 
In Figure~\ref{diff_tran} (left panel) and Figure~\ref{diff_long} (left panel), the variation of the transverse and longitudinal diffusion coefficient of the charm quark is shown with respect to temperature. 
In Figure~\ref{diff_tran} (left panel) and Figure~\ref{diff_long} (left panel), the variation of the transverse and longitudinal diffusion coefficient of the charm quark is shown with respect to temperature. Similar to the drag coefficient, the radiative effects start dominating for the high-temperature range around $T = 0.6$ GeV. The transverse and the longitudinal diffusion coefficient have a smaller magnitude before $T\sim 0.45$ GeV and a larger magnitude after $T\sim 0.45$ GeV compared to earlier perturbative results~\cite{Mazumder:2013oaa}, pertaining to the more non-perturbative nature. Similarly, Figure~\ref{diff_tran} (right panel) and Figure~\ref{diff_long} (right panel) show the transverse and longitudinal diffusion coefficient variation with the charm momentum at $T = 0.525$ MeV. The variation with the momentum $p$ for the transverse diffusion is smaller than the longitudinal diffusion coefficient. Although the nature of the diffusion coefficient variation differs from the drag coefficient, the radiative effects dominations after $p = 5$ GeV are clearly evident, showing the importance of radiative effects at high momenta.    
\\
A similar analysis of drag and diffusion coefficients can be done for bottom quarks having a mass of approximately $4.2$ GeV easily. We report that for the bottom quarks, the drag coefficient magnitudes decrease compared to the charm quark drag coefficient magnitude because of the greater mass of the bottom quark compared to the charm quark. Similar behavior is also observed for the transverse and longitudinal diffusion coefficients as well.
\subsection{Shear viscosity to entropy density ratio ($\eta/s$) of QGP}
\begin{figure}[]
\centering
{\includegraphics[scale=0.6]{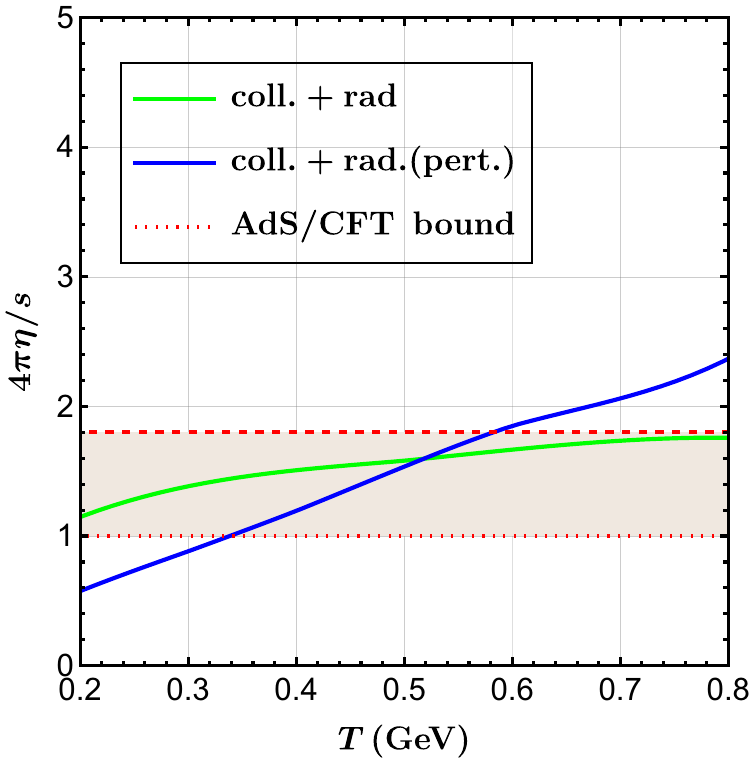} }
\vspace{.1cm}
\captionof{figure}{The value of $4\pi \,\eta /s$ for a charm quark with momentum $\langle p_{z}\rangle$ $= 5$ GeV propagating in QGP medium of temperature $T$.}
\label{shear_viscosity}
\end{figure}

As discussed earlier, the value of shear viscosity to entropy density ratio, $\eta/s$, is a vital quantity to study to understand the nature of the QGP, i.e., whether a medium behaves like a weakly coupled or strongly coupled system. In this subsection, we estimate this ratio by using the Gribov propagator, which enters the interaction part of the diffusion coefficient. The transverse momentum diffusion coefficient $\mathcal{B}_{0}$ can be written as
\begin{equation}
\mathcal{B}_{0}=\frac{1}{2}\left(\delta_{i j}-\frac{p_i p_j}{p^2}\right) \mathcal{B}_{i j}  \, ,
\end{equation}
By using Eq.~\eqref{diffusion_definition} and putting $(p^{\prime}-p)_{i} = k_{i}$,
\begin{equation}
\begin{aligned}
\mathcal{B}_{0} = \frac{1}{4} \llangle[\Bigg]\left({k}^{2} - \frac{(\mathbf{p}\cdot\mathbf{k})^{2}}{{p}^{2}}\right)\rrangle[\Bigg]\, ,
\end{aligned}
\end{equation}
If HQ momentum is considered in the $\hat{z}$ direction, then 
\begin{equation}
\begin{aligned}
\mathcal{B}_{0} = \frac{1}{4} \llangle[\big]k_{\perp}^{2} \rrangle[\big]\ = \frac{1}{4} \hat{q} \, .
\end{aligned}
\end{equation}
where $\hat{q}$ is the jet quenching parameter, which is also an important quantity for the characterization of QGP. Recently, the relation between these two parameters, namely specific shear viscosity $\eta/s$ and dimensionless quenching parameter $\hat{q}/T^{3}$ has been calculated up to NLO in terms of coupling constant using perturbative QCD approach in Ref.~\cite{Muller:2021wri}. Thus, we estimated $\eta/s$ of QGP using the following expression:
\begin{equation}
\frac{\eta}{s} = 1.63 \frac{T^3}{\hat{q}} \, ,
\end{equation}
Thus,
\begin{equation}
4 \pi \frac{\eta}{s} = 1.63 \pi \frac{T^3}{\mathcal{B}_{0}} \, .
\end{equation}
In Figure~\ref{shear_viscosity}, we plotted $4\pi \,\eta/s$ with respect to temperature $T$ within this model calculations. We compared it with the standard KSS bound having values of $4\pi \eta/s = 1.0 - 1.8$ as obtained in~\cite{Kovtun:2004de}, as well with the earlier perturbative result obtained in~\cite{Mazumder:2013oaa}. The obtained results show that the value of $4\pi \, \eta/s$ comes strictly within the AdS/CFT bound after the inclusion of radiative processes, which further improves the earlier perturbative results and shows a good agreement with the experimental values~\cite{Bernhard:2019bmu, JETSCAPE:2020mzn}. For the earlier perturbative results Debye mass $(m_{D})$ acts as a infrared regulator and the value of $m_{D}=\sqrt{6 \pi \alpha_{s}} T$ is used in Figures.~\ref{drag_coeff_total},~\ref{diff_tran}
,~\ref{diff_long} and \ref{shear_viscosity} for the perurbative result comparison with GZ estimations. Thus, one can infer that the Gribov-Zwanziger technique improves the perturbative results in the low-temperature domain as well as in the high-temperature domain, as can be observed in Figure~\ref{shear_viscosity}.  
%
\subsection{Collisional and radiative energy loss:}
\begin{figure}[H]
\centering
{\includegraphics[scale=0.53]{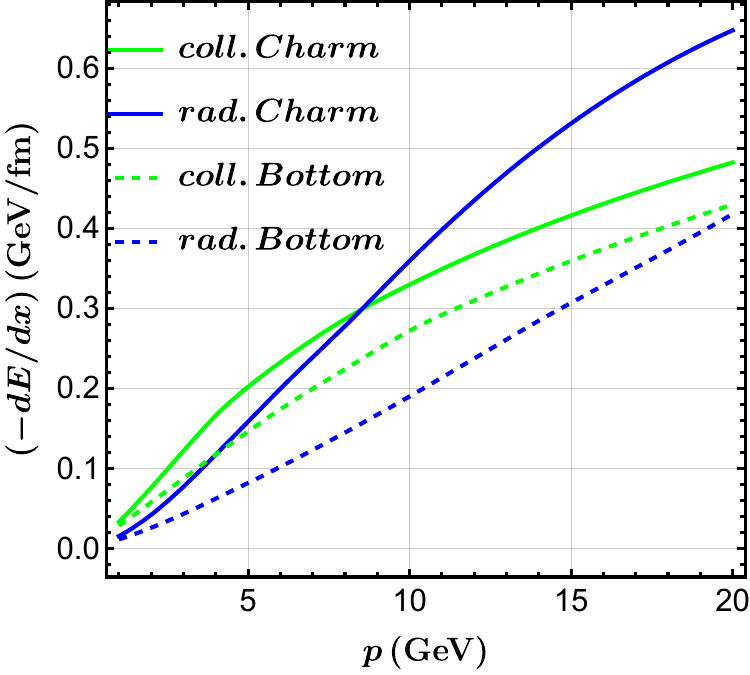} }
\quad \quad
{\includegraphics[scale=0.53]{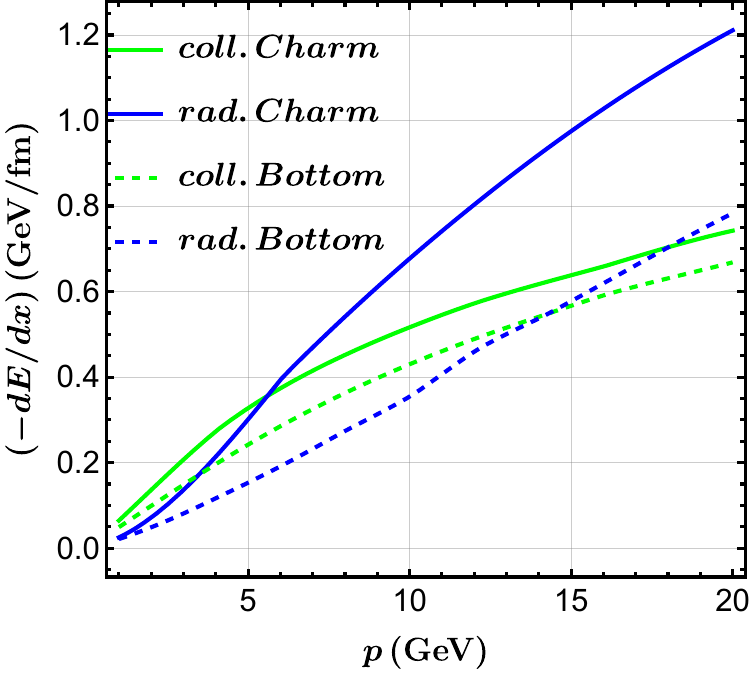} }
\vspace{.1cm}
\captionof{figure}{ Momentum variation of elastic and radiative energy loss of the heavy quark in the medium for the RHIC energy at $T = 360 $ MeV (left panel) and the LHC energy at $T = 480 $ MeV (right panel). Solid lines are for charm quarks, while dotted ones are for bottom quarks.}
\label{energy_loss}
\end{figure}

The differential energy loss of the $\mathrm{HQ}$ is related to the drag coefficient and can be expressed as \cite{GolamMustafa:1997id}
\begin{equation}
-\frac{dE}{dx} = \mathcal{A}(p^{2},T) \,p \,. 
\end{equation}
In Figure~\ref{energy_loss}, we have plotted the energy loss of $\mathrm{HQs}$ with respect to their momentum $p$, showing collisional and radiative contributions independently at RHIC and LHC energies. In Figure~\ref{energy_loss} (left panel) 
energy loss at RHIC energy ($T = 0.36$ GeV) for charm (solid lines) and bottom (dotted lines) quarks are shown. Similarly, Figure~\ref{energy_loss} (right panel) has been plotted for temperature $T = 0.48$ GeV, i.e., at LHC energy. As expected, energy loss for the bottom quark is less compared to the charm quark because of more drag offered to the bottom quark in the medium due to its large mass. The collisional processes dominate in the initial momentum range around $5$ GeV due to restricted phase space. However, after that, the radiative process dominates the collisional one for charm quark at both energies at LHC and RHIC. In the case of the bottom quark, the collisional process contribution dominates in the whole momentum range at RHIC. At the same time, at LHC energy, this nature continues till $\sim 15$ GeV, then the radiative process dominates. This suppression in the radiative energy loss in the case of the bottom quark in comparison to the charm quark can be accounted for because of the dead cone factor, which prohibits the HQ from radiating gluon at a small angle. Thus, for the higher mass, the dead-cone angle will be large, and the probability of energy loss due to radiation will be lesser.  
\section{Summary and Conclusion}\label{sec.4}
In this chapter, we have explored the momentum and temperature dependence of the drag and diffusion coefficient of $\mathrm{HQs}$ propagating in the QGP medium. These transport coefficients play a pivotal role in the HQ phenomenology as they essentially govern the dynamics of the HQs in the Fokker-Planck approach. In the present work, our primary focus has been to incorporate the non-perturbative effects in the estimation of drag and diffusion coefficients, especially in the temperature regime close to the crossover. For this purpose, we take recourse to the $\mathrm{GZ}$ approach. In this framework, the gluon propagators present in the scattering amplitudes have been replaced by the Gribov-modified propagators. It should be noted here that it has been a standard practice to include the Debye mass ($m_{D}\sim g(T)T$) as an infrared regulator in the t-channel matrix amplitude in order to circumvent the infrared divergence. However, in the present approach, the mass scale in the modified gluon propagator arises naturally within the model framework, resulting in a finite t-channel contribution. The temperature dependence of the mass scale has been extracted by matching the thermodynamics of the Gribov plasma with the pure gauge lattice results. Once the temperature dependence has been fixed, we incorporate this modified gluon propagator in the collisional and radiative contributions to obtain the momentum and temperature dependences of the drag and diffusion coefficient, which show a significant increment compared to earlier perturbative estimations with the exception of the momentum dependence of drag coefficient. Moreover, the estimation of the specific shear viscosity using the Gribov approach are closer to estimates based on AdS/CFT calculations. Finally, we have investigated the collisional and radiative energy loss of charm and bottom quark traversing through the medium. We find that the energy loss in both cases is higher in magnitude compared to the earlier perturbative estimations. \\
It should be mentioned here that the Gribov framework presented in this work is a simplistic approach to considering the non-perturbative effects pertinent near the phenomenologically accessible temperature regime. Nevertheless, the present study serves as an important first step toward estimating the impact of Gribov-like approaches on the HQ dynamics. 

\chapter{Summary and outlook}\label{Chapter_6}
\allowdisplaybreaks
\pagestyle{fancy}
\fancyhead[LE]{\emph{\leftmark}}
\fancyhead[LO]{\emph{\rightmark}}

In this chapter, we will provide the principal findings of the theoretical research done in this dissertation, followed by the prospective directions that can be pursued for future projects. The main scope of this thesis is to examine some of the features of the deconfined matter produced in URHIC at BNL and CERN laboratories, mainly at zero chemical potential. Other experiments like NICA and FAIR, which deal with finite baryon density, are currently underway and planned. The theoretical techniques that are utilized in order to study the different properties of this deconfined matter, i.e., QGP, are resummed perturbation theory and nonperturbative resummation, namely as Gribov quantization. 
\\
We started with the brief history of a theory, which describes strong interaction, namely QCD, its fundamental properties, and the QCD phase diagram in chapter~\ref{Chapter_1}. Then we discussed the HIC systems, which are experimental tools to study the matter at extreme temperatures and densities, the different signatures of QGP created in these collisions, and the different approaches, which are classified mainly into perturbative and nonperturbative ones, in order to study this deconfined matter in this chapter~\ref{Chapter_1}. After this, we discuss the methodology utilized to study the different features of deconfined matter. We review two main formalisms of thermal field theory, ITF and RTF, which are commonly used in theoretical techniques. We discussed the application of these formalism and effective theories, namely HTL resummation, in detail, as well as their outcomes. These effective theory techniques are utilized in the further calculation in chapter~\ref{Chapter_3}, and a brief overview of the Gribov approach to QCD is presented in chapter~\ref{Chapter_2}, which is required for the completeness of chapter~\ref{Chapter_4} and chapter~\ref{Chapter_5}.      
\\
In chapter \ref{Chapter_3}, we have studied the NLO quark self-energy and the NLO correction to the fermionic (QCD counterpart) dispersion relation for soft-moving quarks. This study was done in RTF and takes effective theory, namely HTL resummation, into account. When going to NLO for the soft-moving quarks, one needs to take into account the effects of resummation. We have replaced all the propagators and bare vertices with the effective HTL-ones in the usual self-energy diagram. Apart from the three-point vertex contribution, there is also a four-point vertex diagram, which also contributes to NLO quark self-energy. Both those diagrams have been calculated separately. Using the convenient summation structure, we express the integrals containing three and four-point vertex functions in terms of solid angle integrals. After computing the solid angle integrals, we calculate the momentum integrals in NLO quark self-energy and plot the results as a function of the ratio of momentum and energy. The NLO quark self-energy is required in order to obtain the NLO correction to dispersion relation, which gives NLO correction to quark mass and quark damping rate (Leading order quark damping rate is zero, contribution starts from NLO since the Imaginary part of quark self-energy is zero at leading order). Using the NLO quark self-energy, we plotted the correction to the dispersion relations within the numerical uncertainties present in the numerical computation.        
\\
In chapter~\ref{Chapter_4}, we have studied and calculated the mesonic correlation lengths of various mesonic observables using the nonperturbative Gribov resummation. In this study, we have used the effective theory, which has a similar structure as the NRQCD effective theory, which is quite helpful in describing the heavy quarkonia physics at zero temperature. We find that the nonperturbative Gribov approach improves the earlier perturbative results in the low-temperature domain and is well suited to the recent lattice measurements. Thus, one can say that the Gribov approach improves the infrared dynamics of the underlined theory. We have calculated the mesonic screening masses for both quenched QCD cases and $N_{f}=3$ flavor QCD. The primary outcome shows that the mesonic screening masses results approach the perturbative analysis results in the high-temperature regime and, at the same time, are in better agreement with the lattice results in the low-temperature domain.    
\\
Finally, we investigate the $\mathrm{HQ}$ dynamics using the $\mathrm{GZ}$ approach in chapter~\ref{Chapter_5}. In this chapter, we have studied the momentum-dependent drag and diffusion coefficients of $\mathrm{HQs}$ in the QGP background within the Fokker-Planck approach. The interaction of $\mathrm{HQs}$ with the medium constituents, which are encoded in the matrix elements, is computed at leading order in coupling using the $\mathrm{GZ}$ propagator. The intermediate gluons are considered as Gribov ones to incorporate the nonperturbative effects that are relevant to the phenomenologically accessible temperature regime. We have shown the temperature and momentum variation of drag and diffusion coefficients of $\mathrm{HQs}$. The obtained drag coefficient has been used to obtain the energy loss of $\mathrm{HQs}$ at RHIC and LHC energy. We reported a higher energy loss of the propagating $\mathrm{HQ}$ compared to the earlier perturbative results available in the literature. The other transport coefficients, mainly the transverse diffusion coefficient, are utilized to estimate the temperature estimation of specific shear viscosity $\eta/s$ of QGP medium. We observed that the $\eta/s$ ratio complies with the AdS/CFT estimation over a significantly more comprehensive temperature regime compared to the perturbative results. 
\\
For future directions, it would be interesting to utilize the results obtained for the NLO dispersion relation in chapter~\ref{Chapter_3} in order to study the transport phenomena more precisely. Recently, the mass correction to HTLs and mass effects to heavy fermion self-energy in QED and QCD has been investigated, showing the importance of finite quark mass in the calculations~\cite{Comadran:2021pkv, Comadran:2023vsr}. So, the calculations presented in chapter~\ref{Chapter_3} can be extended for the finite quark mass and finite chemical potential in order to study the quark number susceptability~\cite{Haque:2018eph}. It is expected that the NLO contribution will resolve the ambiguity observed in the HTL approximation results with the lattice data~\cite{Haque:2018eph}. Also, since the calculation provided in chapter~\ref{Chapter_4} does not respect the different channels like a scalar or vector, it would be interesting to study, by adopting some strategy, to include the different channels and different mesons in the framework provided in chapter~\ref{Chapter_4}. Another exciting study can be done by examining the finite chemical potential in theory for a more general case study of mesonic screening masses, which can give insights into the nature of the QCD phase diagram. Since, as discussed in chapter~\ref{Chapter_4}, Gribov approach improves the infrared dynamics of the theory, it encourages one to study further different experimental observables like nuclear modification factor $R_{AA}$~\cite{Debnath:2023zet}, elliptic flow $v_{2}$  as well as other transport properties of the medium~\cite{Florkowski:2015dmm, Florkowski:2015rua, Jaiswal:2020qmj, Madni:2024xyj, Madni:2024ubw} within this approach. Thus, the interesting future direction in this regard would be to consider the dissipative effects in the estimation of drag and diffusion coefficients of HQ in QGP background~\cite{Srivastava:2016igg, Prakash:2021lwt, Kumar:2021goi, Prakash:2023wbs, Singh:2023smw}. Also, the non-trivial backgrounds like the strong external magnetic field may have a significant impact~\cite{Fukushima:2015wck, Bandyopadhyay:2021zlm, Bandyopadhyay:2023hiv} on the transport properties of the Gribov modified plasma medium.    
\appendix
\chapter{}
\section{HTL dressed vertex integrals} \label{sec:appendix_A_1}
For the sake of completeness, this appendix summarises the derivation of 2-quark-1-gluon and 2-quark-2-gluon vertex functions within HTL approximation in the CTP formalism, which is utilized in section~\ref{section_3}. 
\subsection{Two quark and one gluon vertex integral}
\begin{figure}[tbh]
\centering
\includegraphics[scale=2.,keepaspectratio]{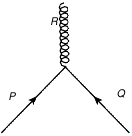} 
\qquad
\includegraphics[scale=1.1,keepaspectratio]{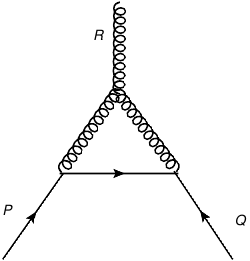}
\caption{$1$-loop Feynman diagrams for the two-quarks-one-gluon vertex function.}
\label{3PT.}
\end{figure}

The one-loop quark-gluon vertex function within HTL approximation is defined as 
\begin{equation}
\Gamma^{\mu} = \gamma^{\mu} + \delta\Gamma^{\mu}.
\end{equation}
where $\gamma^{\mu}$ is the bare vertex contribution and  $\delta\Gamma^{\mu} $ is the one-loop HTL correction. Within the $\{12\}$ basis of the Keldysh indices, the bare vertex function $\gamma^{\mu}$ is given by
\begin{align}
\gamma_{i j k}^{\mu} = & \left\{\begin{array}{l}
(-1)^{i-1} \gamma^{\mu} \quad \text { when } \quad i=j=k \\
0 \quad \text { otherwise }
\end{array}\right. ,
\end{align}
where $i,j,k$ can take the values $1,2$. The $1$-loop diagrams which contribute to the quark-gluon three vertex function are shown in Figure~\ref{3PT.}. The HTL contributions $\delta\Gamma^{\mu}$ to the three-point vertex can be obtained from the $1$-loop graphs shown in Figure~\ref{3PT.} as
\begin{equation}
\delta \Gamma_{i j k}^{\mu}(P, Q, R)=4 i g^{2} C_{F} \int \frac{d^{4} K}{(2 \pi)^{4}} K^{\mu} \slashed{K} \mathcal{V}_{i j k}^{\prime}-2 i g^{2} N_{c} \int \frac{d^{4} K}{(2 \pi)^{4}} K^{\mu} \slashed{K}\left(\mathcal{V}_{i j k}^{\prime}+\mathcal{V}_{i j k}\right),\label{Gamma_V}
\end{equation}
We can use HTL approximation to neglect the external momenta compared to the loop momentum $K$. The functions $\mathcal{V}_{i j k}$ and $\mathcal{V}_{i j k}'$ are
\ba
\mathcal{V}_{i j k} &=& - (-1)^{i+j+k} \bar{D}_{i j}(K) D_{j k}(K-Q) D_{k i}(K+P) ;\nn
\mathcal{V}_{i j k}^{\prime} &=& - (-1)^{i+j+k} D_{i j}(K) \bar{D}_{j k}(K-Q) \bar{D}_{k i}(K+P) .\label{Vijk_def} 
\ea
The functions $D_{ij}(K)$ are the bare bosonic propagator defined in the $\{12\}$ basis as
\begin{align}
D(K) =& \,\left(\begin{array}{ll}
D_{11}(K) & D_{12}(K) \\
D_{21}(K) & D_{22}(K)
\end{array}\right) \nn
=& \, \left(\begin{array}{cc}
\frac{1+n_{B}\left(k_{0}\right)}{K^{2}+i \varepsilon}-\frac{n_{B}\left(k_{0}\right)}{K^{2}-i \varepsilon} & \frac{\Theta\left(-k_{0}\right)+n_{B}\left(k_{0}\right)}{K^{2}+i \varepsilon}-\frac{\Theta\left(-k_{0}\right)+n_{B}\left(k_{0}\right)}{K^{2}-i \varepsilon} \\
\frac{\Theta\left(k_{0}\right)+n_{B}\left(k_{0}\right)}{K^{2}+i \varepsilon}-\frac{\Theta\left(k_{0}\right)+n_{B}\left(k_{0}\right)}{K^{2}-i \varepsilon} & \frac{n_{B}\left(k_{0}\right)}{K^{2}+i \varepsilon}-\frac{1+n_{B}\left(k_{0}\right)}{K^{2}-i \varepsilon}
\end{array}\right) .
\end{align}
The quantity $ \bar{D}_{ij}(K)$ is a fermionic propagator having the same expression as $D_{ij}(K)$ except BE distribution function $n_{B}(k_0)$ is replaced by the negative of the FD distribution function $-n_{F}(k_0)$. We can express the functions $D_{ij}(K)$ in terms of  symmetric $(F)$ , advanced $(A)$ and retarded $(R)$ propagators as
\begin{align}
D_{11} = & \, \frac{1}{2}(F+A+R) ; \quad D_{12}=\frac{1}{2}(F+A-R) ;\nn
D_{21} = & \,\frac{1}{2}(F-A+R) ; \quad D_{22}=\frac{1}{2}(F-A-R) ,\label{gluon_propagator}
\end{align}
with
\begin{align}
R(K) = & \, \frac{\Theta\left(k_{0}\right)}{K^{2}+i \varepsilon}+\frac{\Theta\left(-k_{0}\right)}{K^{2}-i \varepsilon}; \nn
A(K) = & \, \frac{\Theta\left(-k_{0}\right)}{K^{2}+i \varepsilon}+\frac{\Theta\left(k_{0}\right)}{K^{2}-i \varepsilon} ;\nn
F(K) = & \, \left(1 \pm 2 n_{B, F}\left(k_{0}\right)\right)\left(\frac{1}{K^{2}+i \varepsilon}-\frac{1}{K^{2}-i \varepsilon}\right).
\end{align}
The vertex functions in the $\{RA\}$ basis have linear correspondence with the functions mentioned in the $\{12\}$ basis. For instance, the $3$-point vertex function $\Gamma_{RAA}$ in $\{R.A.\}$ basis is given by the following relation
\begin{equation}
\Gamma_{RAA} = \Gamma_{111} + \Gamma_{112} +\Gamma_{121} +\Gamma_{122}.\label{RA_12}
\end{equation}
As the vertex functions are related with the functions $\mathcal{V}$ and $\mathcal{V}^{\prime}$ via the Eq.~\eqref{Gamma_V},  the relation in Eq.~\eqref{RA_12} also valid to the functions $\mathcal{V}$ and $\mathcal{V}^{\prime}$. Utilizing the expressions in Eq.~\eqref{Vijk_def} of the functions $\mathcal{V}$ and $\mathcal{V}^{\prime}$, and the relations in Eq.~\eqref{gluon_propagator}, we get 
\begin{equation}\label{V_func}
\begin{aligned}
\mathcal{V}_{RAA} = \frac{1}{2} \left(A_1 A_2 A_3 + R_1 R_2 R_3 + F_1 A_2 A_3 + R_1 F_2 A_3 + R_1 R_2 F_3  \right) ,
\end{aligned}
\end{equation}
where, for shorthand notation, $1$, $2$, and $3$ represents the arguments $K$, $K-Q$, and $K+P$, respectively. The loop integration $K$ can be done in two steps: firstly, over quark energy, i.e., $k_0$, performed utilizing the residue theorem in the complex $k_0$ plane resulting $A_1 A_2 A_3$ and $R_1 R_2 R_3 $ terms vanishes. The other three terms become
\begin{align}
F A A = & \,\frac{+i}{8 \pi^{2}} \int_{0}^{+\infty} k d k N_{B, F}(k) \int \frac{d \Omega_{s}}{4 \pi} \frac{S^{\mu} \slashed S}{(P S-i \varepsilon)(Q S+i \varepsilon)} ;\nn
R F A = & \,\frac{-i}{8 \pi^{2}} \int_{0}^{+\infty} k d k N_{B, F}(k) \int \frac{d \Omega_{s}}{4 \pi} \frac{S^{\mu} \slashed S}{(Q S+i \varepsilon)((P+Q) S-i \varepsilon)} ;\nn
R R F = & \,\frac{-i}{8 \pi^{2}} \int_{0}^{+\infty} k d k N_{B, F}(k) \int \frac{d \Omega_{s}}{4 \pi} \frac{S^{\mu} \slashed S}{(PS - i\varepsilon)((P+Q) S-i \varepsilon)} . 
\label{three_terms}
\end{align}
The functions $N_{B,F} (k)$ are mentioned in Eq.~\eqref{dist_func} and $S = (1,\hat{s})$ is a time-like unit four-vector where $\hat{s}$ is defined as $\vec{k}/k$. Since in HTL approximation, one can neglect the external momenta in front of loop momentum $K$, when we sum up all the contributions together as mentioned in Eq.~\eqref{three_terms}, we get the final expression as
\ba
\delta \Gamma_{R A A} &=&-m_{q}^{2} \int \frac{d \Omega}{4 \pi} \frac{S^{\mu} \slashed S}{(P S-i \varepsilon)(Q S+i \varepsilon)} , 
\ea
where thermal quark mass is defined as $m_q = \sqrt{C_F/8}gT$. Similarly, the other $3$-point HTL vertex integrals can be obtained in the $\{R.A. \}$ and $\{12\}$ bases analogously to Eq.~\eqref{RA_12} as shown in \cite{Abada:2014bma} as well. For each of those other vertex functions, one can perform the same steps as done for $\Gamma_{RAA}$ and eventually, one gets: 
\begin{align}
\delta \Gamma_{A R A}(P, Q, R) = & \, -m_{q}^{2} \int \frac{d \Omega_{s}}{4 \pi} \frac{S^{\mu} \slashed S}{(P S+i \varepsilon)(Q S-i \varepsilon)} ;  \nn
\delta \Gamma_{A A R}(P, Q, R)= &\, -m_{q}^{2} \int \frac{d \Omega_{s}}{4 \pi} \frac{S^{\mu} \slashed S}{(P S+i \varepsilon)(Q S+i \varepsilon)} ; \nn
\delta \Gamma_{R R R}(P, Q, R) = &\, -m_{q}^{2} \int \frac{d \Omega_{s}}{4 \pi} \frac{S^{\mu} \slashed S}{(P S-i \varepsilon)(Q S-i \varepsilon)} ; \nn
\delta \Gamma_{R R A}(P, Q, R) = & \, \delta \Gamma_{R A R}(P, Q, R)=\delta \Gamma_{A R R}(P, Q, R)=0 . 
\end{align}
\subsection{Two quark and two gluon vertex integral}
\begin{figure}
\centering
\includegraphics[scale=0.5,keepaspectratio]{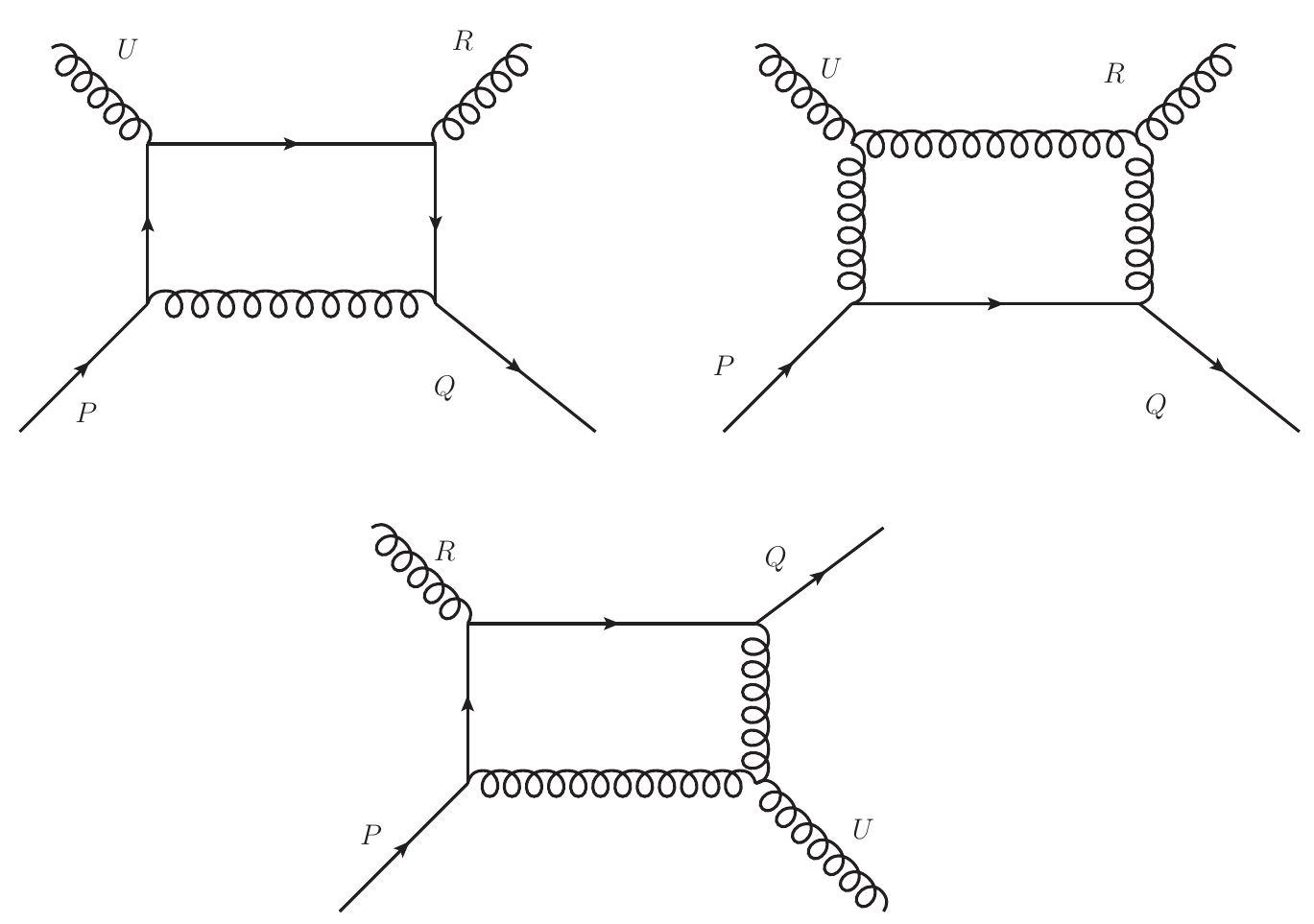} 
\caption{$1$-loop Feynman diagrams for the two-quarks-two-gluons vertex function.}
\label{two_quark_2_gluons}
\end{figure}

Figure~\ref{two_quark_2_gluons} shows the $1$-loop graphs contributing to the quark-gluon four vertex function. The contribution coming from $4$-point vertex functions is written in the $\{12\}$ basis as shown in Eq.~\eqref{Gamma_mu_nu}.
\begin{align}
\delta \Gamma_{i j k l}^{\mu \nu}(P, Q, R, U) = & \, -8 i g^{2}(-1)^{i+j+k+l} \int \frac{d^{4} K}{(2 \pi)^{4}} K^{\mu} K^{\nu} \slashed{K} \big[ C_{F} D_{i j}(K) \bar{D}_{j k}(K-Q) \nn 
\times & \, \bar{D}_{k l}(K+P+U) \bar{D}_{l i}(K+P) - N_{c} \bar{D}_{i j}(K) D_{j k}(K-Q)  \nn 
\times & D_{k l}(K+P+U) D_{l i}(K+P) + \frac{ N_{c}}{2} D_{l i}(K-P) \nn
\times & \, D_{j l}(K-P-U) \bar{D}_{k j}(K+R) \bar{D}_{i k}(K)\big].
\label{Gamma_mu_nu}
\end{align}
Let's consider the $\{RA\}$-component $\delta\Gamma_{RAAA}$, given by 
\begin{equation}
\delta\Gamma_{RAAA} = \delta\Gamma_{1111} + \delta\Gamma_{1112} + \delta\Gamma_{1121} + \delta\Gamma_{1211} + \delta\Gamma_{1122} + \delta\Gamma_{1212} + \delta\Gamma_{1221} + \delta\Gamma_{1222} 
\end{equation}
We denote the terms with the factor $C_F$ in eq.~\eqref{Gamma_mu_nu} as $\delta\Gamma^{1}_{RAAA}$. Using the decomposition as done in Eq.~\eqref{gluon_propagator}, one obtain, having a similar symbolic notation as in Eq.~\eqref{V_func}, the following relation
\begin{equation}\label{Gamma_RAAA}
\delta\Gamma^{1}_{RAAA} = \frac{1}{2}F_1 A_2 A_3 A_4 + \frac{1}{2}R_1 F_2 A_3 A_4 +\frac{1}{2}R_1 R_2 F_3 A_4 +\frac{1}{2}R_1 R_2 R_3 F_4  
\end{equation}
Here, we again used the shorthand notation $1,2,3,$ and $4$, which denotes the momenta $K, K-Q, K+P+U$, and $K+P$, respectively. The following two terms $ \frac{1}{2}A_1 A_2 A_3 A_4$ and $\frac{1}{2}R_1 R_2 R_3 R_4$ to $ \delta\Gamma^{1}_{RAAA} $ gives null each in the $k_0$-complex-plane integration and are not shown explicitly in Eq.~\eqref{Gamma_RAAA}. The other terms in Eq.~\eqref{Gamma_RAAA} give a contribution as follows
\begin{align}
F A A A = & \, \int_{0}^{\infty} \frac{k d k}{16 \pi^{2}} \int \frac{d \Omega_{s}}{4 \pi} \frac{i N_{B}(k) S^{\mu} S^{\nu} \slashed S}{((2 P+U) S+i \varepsilon)((P-R) S+i \varepsilon)(P S+i \varepsilon)}, \nn
R F A A =& \, \int_{0}^{\infty} \frac{k d k}{16 \pi^{2}} \int \frac{d \Omega_{s}}{4 \pi} \frac{-i N_{B}(k) K^{\mu} K^{\nu} \slashed S}{((2 P+U) S+i \varepsilon)((R+P+U) S-i \varepsilon)((P+U) S-i \varepsilon)}, \nn
R R F A =& \, \int_{0}^{\infty} \frac{k d k}{16 \pi^{2}} \int \frac{d \Omega_{s}}{4 \pi} \frac{i N_{F}(k) S^{\mu} S^{\nu} \slashed S}{((R-P) S-i \varepsilon)((R+P+U) S-i \varepsilon)(R S+i \varepsilon)}, \nn
R R R F =& \, \int_{0}^{\infty} \frac{k d k}{16 \pi^{2}} \int \frac{d \Omega_{s}}{4 \pi} \frac{i N_{F}(k) S^{\mu} S^{\nu} \slashed S}{(P S+i \varepsilon)((P+U) S-i \varepsilon)(R S+i \varepsilon)}.
\label{four_terms}
\end{align}
Summing all the contributions in Eq.~\eqref{four_terms} gives the following results
\begin{equation}
\delta \Gamma_{R A A A}^{1}=\frac{i}{8 \pi^{2}} \int_{0}^{+\infty} k d k \int \frac{d \Omega_{s}}{4 \pi} \frac{\left(n_{B}(k)+n_{F}(k)\right) S^{\mu} S^{\nu} \slashed S}{(P S-i \varepsilon)(Q S+i \varepsilon)((P+U) S-i \varepsilon)}.
\end{equation}
Now, three-momenta $k$ integrations can be done analytically. We calculate the terms with the factors $N_c$ and $N_{c}/2$ in Eq.~\eqref{Gamma_mu_nu} similarly, and they cancel each other. Finally, the expression for the HTL four-vertex function becomes
\begin{align}
\delta \Gamma_{R A A A}(P, Q, R, U) =& \, m_{q}^{2} \int \frac{d \Omega_{s}}{4 \pi} \frac{S^{\mu} S^{\nu} \slashed S}{(P S-i \varepsilon)(Q S+i \varepsilon)}
\left[\frac{1}{(P+U) S-i \varepsilon} \right.\nn
+ & \, \left.\frac{1}{(P+R) S-i \varepsilon}\right].
\end{align}
The other $\{RA\}$ four-vertex HTLs can be worked out similarly, and one finds that
\begin{align}
\delta \Gamma_{A R A A}(P, Q, R, U) = & \, m_{q}^{2} \int \frac{d \Omega_{s}}{4 \pi} \frac{S^{\mu} S^{\nu} \slashed S}{(P S+i \varepsilon)(Q S-i \varepsilon)} 
\times\left[\frac{1}{(P+U) S+i \varepsilon}+\frac{1}{(P+R) S+i \varepsilon}\right], \nn
\delta \Gamma_{A A R A}(P, Q, R, U) = & \, m_{q}^{2} \int \frac{d \Omega_{s}}{4 \pi} \frac{S^{\mu} S^{\nu} \slashed S}{(P S+i \varepsilon)(Q S+i \varepsilon)} 
\times\left[\frac{1}{(P+U) S+i \varepsilon}+\frac{1}{(P+R) S-i \varepsilon}\right], \nn
\delta \Gamma_{\text {AAAR }}(P, Q, R, U) = & \, m_{q}^{2} \int \frac{d \Omega_{s}}{4 \pi} \frac{S^{\mu} S^{\nu} \slashed S}{(P S+i \varepsilon)(Q S+i \varepsilon)} \times\left[\frac{1}{(P+U) S-i \varepsilon}+\frac{1}{(P+R) S+i \varepsilon}\right] , \nn
\delta \Gamma_{R R A A}(P, Q, R, U)= & \, \delta \Gamma_{R A R A}(P, Q, R, U)=\delta \Gamma_{R A A R}(P, Q, R, U)=0 .
\end{align}
The remaining eight components of four-vertex functions can be computed directly or can be obtained from the above results using the KMS conditions.
\subsection{Change of Notations}
The notations used in the main text can be connected with the notation used in this appendix via 
\begin{equation}
\begin{aligned}
\Gamma_{I_{1} I_{2} I_{3}}^{\mu}\left(P_{1}, P_{2}, P_{3}\right) &=\Gamma_{i_{1} i_{3} i_{2}}^{\mu}\left(P_{1},-P_{2}\right); \\
\Gamma_{I_{1} I_{2} I_{3} I_{4}}^{\mu \nu}\left(P_{1}, P_{2}, P_{3}, P_{4}\right) &=\Gamma_{i_{1} i_{4} i_{3} i_{2}}^{\mu \nu}\left(P_{1}, P_{4}, P_{3},-P_{2}\right),
\end{aligned}
\end{equation}
Here, \begin{equation*}
I_{j}=R(A) \leftrightarrow i_{j}=\mathrm{a}(\mathrm{r})
\end{equation*}
So, the $3$-point vertex functions are
\begin{align}
\Gamma_{\text {arr }}^{\mu}(P, Q) = & \, \Gamma_{R A A}^{\mu}(P,-Q, R)=\gamma^{\mu}+I_{--}^{\mu}(P, Q); \nn
\Gamma_{\text {rar }}^{\mu}(P, Q) = & \, \Gamma_{A A R}^{\mu}(P,-Q, R)=\gamma^{\mu}+I_{+-}^{\mu}(P, Q);\nn
\Gamma_{\mathrm{aar}}^{\mu}(P, Q) = & \, \Gamma_{R A R}^{\mu}(P,-Q, R)=0;\nn
\Gamma_{\mathrm{rra}}^{\mu}(P, Q) = & \, \Gamma_{A R A}^{\mu}(P,-Q, R)=\gamma^{\mu}+I_{++}^{\mu}(P, Q); \nn
\Gamma_{\mathrm{ara}}^{\mu}(P, Q) = & \, \Gamma_{R R A}^{\mu}(P,-Q, R)=0,
\end{align}
having following notation
\begin{equation}\label{I_eta1_eta2}
I_{\eta_{1} \eta_{2}}^{\mu}(P, Q)=m_{q}^{2} \int \frac{d \Omega_{s}}{4 \pi} \frac{S^{\mu}\slashed S}{\left(P S+i \eta_{1} \varepsilon\right)\left(Q S+i \eta_{2} \varepsilon\right)}.
\end{equation}
For the four-point vertex functions, we have
\begin{align}
\Gamma_{\text {arr }}^{\mu \nu}(P, K) \equiv & \, \Gamma_{\text {arrr }}^{\mu \nu}(P, K,-K, P)=\Gamma_{R A A A}^{\mu \nu}(P,-P,-Q, Q)=I_{--}^{\mu \nu}(P, K); \nn
\Gamma_{\text {aarr }}^{\mu \nu}(P, K) \equiv & \, \Gamma_{\text {aarr }}^{\mu \nu}(P, K,-K, P)=\Gamma_{R A A R}^{\mu \nu}(P,-P,-Q, Q)=0; \nn
\Gamma_{\text {arar }}^{\mu \nu}(P, K) \equiv & \,  \Gamma_{\text {arar }}^{\mu \nu}(P, K,-K, P)=\Gamma_{R A R A}^{\mu \nu}(P,-P,-Q, Q)=0,
\end{align}
with the notation
\begin{align}
I_{\eta_{1} \eta_{2}}^{\mu \nu}(P, K) = & \, m_{q}^{2} \int \frac{d \Omega_{s}}{4 \pi} \frac{-S^{\mu} S^{\nu} \slashed S}{\left[P S+i \eta_{1} \varepsilon\right]\left[P S+i \eta_{2} \varepsilon\right]} 
 \left[\frac{1}{(P+K) S+i \eta_{1} \varepsilon} \right.\nn 
+ & \left.\frac{1}{(P-K) S+i \eta_{2} \varepsilon}\right].
\label{I_mu_nu}
\end{align}
Eq.~\eqref{I_eta1_eta2} and Eq.~\eqref{I_mu_nu} have been utilised in the chapter \ref{Chapter_3}, see Eqs.~\eqref{solid_ang_int1},~\eqref{solid_ang_int2}.
\section{HTL vertex integrals using Feynman technique}\label{appendix_A_2}
In this appendix, we will calculate the solid-angle integrals present in eqs.~\eqref{solid_ang_int1} and~\eqref{solid_ang_int2}. This will be done by utilizing the Feynman parametrization technique. By using eqs.~\eqref{htl_terms_1},~\eqref{htl_terms_2}, and~\eqref{sigma2_htl}, we have only two types of solid-angle integrals which we need to evaluate, namely,
\begin{align}
J_{\eta_{1} \eta_{2}}^{\mu \alpha}(P, Q) = & \, \int \frac{d \Omega_{s}}{4 \pi} \frac{S^{\mu} S^{\alpha}}{\left[P S+i \eta_{1} \varepsilon\right]\left[Q S+i \eta_{2} \varepsilon\right]} ;\label{Jmualpha}\nn
I_{\eta_{1} \eta_{2}}^{\mu \nu \alpha}(P, K) = & \, \int \frac{d \Omega_{s}}{4 \pi} \frac{S^{\mu} S^{\nu} S^{\alpha}}{\left[P S+i \eta_{1} \varepsilon\right]\left[P S+i \eta_{2} \varepsilon\right]} \left[\frac{1}{(P+K) S+i \eta_{1} \varepsilon} \right.\nn
+& \, \left.\frac{1}{(P-K) S+i \eta_{2} \varepsilon}\right].
\end{align}
The `$00$' component of $J_{\eta_{1} \eta_{2}}^{\mu \alpha}$ is the simplest of all these integrals and becomes
\begin{equation}
J_{\eta_{1} \eta_{2}}^{00}(P, Q)=\int \frac{d \Omega_{s}}{4 \pi} \frac{1}{\left(P S+i \eta_{1} \varepsilon\right)\left(Q S+i \eta_{2} \varepsilon\right)}.\label{J00}
\end{equation}
Here, $S = (1,\hat{s})$ and the integration has to be done over the solid angle of the unit vector $\hat{s}$. We will drop the i$\varepsilon$ prescription for some time. Utilizing the Feynman technique, equation~\eqref{J00} becomes
\begin{equation}
J_{\eta_{1} \eta_{2}}^{00}(P, Q)=\int_{0}^{1} d u \int \frac{d \Omega_{s}}{4 \pi} \frac{1}{[(P-K u) S]^{2}}=\int_{0}^{1} \frac{d u}{(P-K u)^{2}},\label{J00_1}
\end{equation}
where $K = P-Q$. The integration over $u$ in eq.~\eqref{J00_1} can be done analytically to get
\begin{equation}\label{J00_2}
J_{\eta_{1} \eta_{2}}^{00}(P, Q)=\frac{1}{2 \sqrt{\Delta}} \ln \frac{\left(1-u_{1}\right) u_{2}}{\left(1-u_{2}\right) u_{1}},
\end{equation}
where we have used the notation $u_{1,2} = ((PK) \pm \sqrt{\Delta})/K^{2}$
and $\Delta = (PK)^{2} - P^{2}K^{2} $. Note that $ PS+ i\eta _{1} \varepsilon = p_{0} + i\eta_{1}\varepsilon - \vec{p}\cdot \hat{s}$, so i$\varepsilon$'s appearance in the final result~\eqref{J00_2} can be achieved by changing the variables $p_0 \rightarrow p_0 + i\eta_{1}\varepsilon$ and $q_{0} \rightarrow q_{0} + i\eta_{2}\varepsilon$. This will also apply to the next two terms. Now, the `$0i$' component of $J_{\eta_{1} \eta_{2}}^{\mu \alpha}$ is given by
\begin{equation}
J_{\eta_{1} \eta_{2}}^{0 i}(P, Q)=\int \frac{d \Omega_{s}}{4 \pi} \frac{\hat{s}^{i}}{\left(P S+i \eta_{1} \varepsilon\right)\left(Q S+i \eta_{2} \varepsilon\right)}.\label{J0i_def}
\end{equation}
Again, by using the same approach as done above and denoting $R = P -(P - Q)u$, eq.~\eqref{J0i_def} becomes
\begin{equation}\label{J0i}
\begin{aligned}
J_{\eta_{1} \eta_{2}}^{0 i}(P, Q) &=\int_{0}^{1} d u\left[\frac{r_{0}}{r_{0}^{2}-r^{2}}-\frac{1}{2 r} \ln \frac{r_{0}+r}{r_{0}-r}\right] \frac{r^{i}}{r^{2}}.
\end{aligned}
\end{equation}
Similarly the `$ij$' component of $J_{\eta_{1} \eta_{2}}^{\mu \alpha}$ is given by
\begin{equation}
J_{\eta_{1} \eta_{2}}^{i j}(P, Q)=\int \frac{d \Omega_{s}}{4 \pi} \frac{\hat{s}^{i} \hat{s}^{j}}{\left(P S+i \eta_{1} \varepsilon\right)\left(Q S+i \eta_{2} \varepsilon\right)}.\label{Jij_def}
\end{equation}
The spatial component `$ij$' can be computed in the similar way  $J_{\eta_{1} \eta_{2}}^{0 i}(P, Q)$ ws calculated in eq.~\eqref{J0i} and eq.~\eqref{Jij_def} becomes
\ba
J_{\eta_{1} \eta_{2}}^{i j}(P, Q) &=&\int_{0}^{1} d u\left(A_{\eta_{1} \eta_{2}} \delta^{i j}+B_{\eta_{1} \eta_{2}} \hat{r}^{i} \hat{r}^{j}\right), \label{Jij_int}
\ea
where
\begin{align}
A_{\eta_{1} \eta_{2}} = &-\frac{1}{r^{2}}\left(1-\frac{r_{0}}{2 r} \ln \frac{r_{0}+r}{r_{0}-r}\right) ; \nn
B_{\eta_{1} \eta_{2}} = & \, \frac{1}{r_{0}^{2}-r^{2}}+\frac{3}{r^{2}}\left(1-\frac{r_{0}}{2 r} \ln \frac{r_{0}+r}{r_{0}-r}\right) .
\end{align}
The integration over $u$ in eq.~\eqref{Jij_int} will be done numerically.
Considering, $P_{1,2} \equiv (p_{0} + i\eta_{1,2}\varepsilon,\vec{p})$, we rewrite
\begin{align}
I_{\eta_{1} \eta_{2}}^{\mu \nu \alpha}(P, K)  = & \, J_{\eta_{1} \eta_{2}}^{\mu \nu \alpha}(P, K)+J_{\eta_{2} \eta_{1}}^{\mu \nu \alpha}(P,-K) ;\nn
J_{\eta_{1} \eta_{2}}^{\mu \nu \alpha}(P, K)  = & \, \int \frac{d \Omega_{s}}{4 \pi} \frac{S^{\mu} S^{\nu} S^{\alpha}}{P_{1} S P_{2} S\left(P_{1}+K\right) S} .
\label{J_mu_nu_alpha}
\end{align}
We can denote the scalar products as
$ A=P_{2}S $, $B= P_{1}S$,  $C=(P_{1}+K)S$ , and the four-vector as 
\ba
T \equiv (t_0,\vec{t})= P_{2} + u_{1}(P_{1}-P_{2}) +  u_{1}u_{2}K , \label{T_def}
\ea
with
\ba
t_{0} &=& u_{1}u_{2}k_{0} + iu_{1}(\eta_{1}-\eta_{2})\varepsilon + p_{0} + i\eta_{2} \varepsilon \quad \quad
\vec{t} = u_{1}u_{2}\vec{k} + \vec{p}.
\ea
With the previous definitions, $J_{\eta_{1} \eta_{2}}^{\mu \nu \alpha}$ in eq.~\eqref{J_mu_nu_alpha} becomes
\begin{equation}
J_{\eta_{1} \eta_{2}}^{\mu \nu \alpha}(P, K)=2 \int_{0}^{1} d u_{1} u_{1} \int_{0}^{1} d u_{2} \int \frac{d \Omega_{s}}{4 \pi} \frac{S^{\mu} S^{\nu} S^{\alpha}}{\left(t_{0}-\vec{t} . \hat{s}\right)^{3}}.
\end{equation}
The dependence of $ \eta_{1} \eta_{2}$ in r.h.s. comes through the zeroth component of $T$, defined in eq.~\eqref{T_def}. As one can see, the quantity $J_{\eta_{1} \eta_{2}}^{\mu \nu \alpha}(P, K)$ has symmetry in its Lorentz indices, so there will be only four independent components to work out. Thus, we will get
\begin{equation}
\begin{aligned}
J_{\eta_{1} \eta_{2}}^{000}(P, K) &=2 \int_{0}^{1} d u_{1} u_{1} \int_{0}^{1} d u_{2} \frac{t_{0}}{\left(t_{0}^{2}-t^{2}\right)^{2}}.
\end{aligned}\label{J000}
\end{equation}
\begin{equation}
\begin{aligned}
J_{\eta_{1} \eta_{2}}^{00 i}(P, K) &=2 \int_{0}^{1} d u_{1} u_{1} \int_{0}^{1} d u_{2} \frac{t^{i}}{\left(t_{0}^{2}-t^{2}\right)^{2}}.
\end{aligned}\label{J00i}
\end{equation}
The third solid-angle integral consists of a symmetric tensor of rank two as
\ba
J_{\eta_{1} \eta_{2}}^{0 i j}(P, K) &=&2 \int_{0}^{1} d u_{1} u_{1} \int_{0}^{1} d u_{2}\left(C_{\eta_{1} \eta_{2}} \delta^{i j}+D_{\eta_{1} \eta_{2}} \hat{t}^{i} \hat{t}^{j}\right) ,\label{J0ij} \ea
where
\begin{align}
C_{\eta_{1} \eta_{2}} = & \, \frac{t_{0}}{2 t^{2}\left(t_{0}^{2}-t^{2}\right)}-\frac{1}{4 t^{3}} \ln \frac{t_{0}+t}{t_{0}-t} ; \nn
D_{\eta_{1} \eta_{2}} = & \, \frac{t_{0}\left(5 t^{2}-3 t_{0}^{2}\right)}{2 t^{2}\left(t_{0}^{2}-t^{2}\right)^{2}}+\frac{3}{4 t^{3}} \ln \frac{t_{0}+t}{t_{0}-t} .\label{C_D_def}
\end{align}
Similarly, the fourth solid-angle integral consists of a completely symmetric tensor of rank three.
\begin{align}
J_{\eta_{1} \eta_{2}}^{i j k}(P, K) =& \, 2 \int_{0}^{1} d u_{1} u_{1} \int_{0}^{1} d u_{2}\left[E_{\eta_{1} \eta_{2}}\left(\hat{t}^{i} \delta^{j k}+\hat{t}^{j} \delta^{k i}+\hat{t}^{k} \delta^{i j}\right)\right. \nn 
+ & \, \left. F_{\eta_{1} \eta_{2}} \hat{t}^{i} \hat{t}^{j} \hat{t}^{k}\right],
\label{Jijk}
\end{align}
with
\begin{align}
E_{\eta_{1} \eta_{2}} = & \, \frac{1}{2 t^3}\left(2+\frac{t_{0}^{2}}{t_{0}^{2}-t^{2}}-\frac{3 t_{0}}{2 t} \ln \frac{t_{0}+t}{t_{0}-t}\right) ;\nn
F_{\eta_{1} \eta_{2}} = & \, \frac{t}{\left(t_{0}^{2}-t^{2}\right)^{2}}-\frac{5}{2 t^{3}}\left(2+\frac{t_{0}^{2}}{t_{0}^{2}-t^{2}}-\frac{3 t_{0}}{2 t} \ln \frac{t_{0}+t}{t_{0}-t}\right).
\label{E_F_def}
\end{align} 
\section{HTL dressed propagators}\label{sec:appendix_A_3}
This appendix summarizes the derivation of HTL-dressed transverse and longitudinal gluon and quark propagators.
\subsection{Transverse gluon propagator}
The transverse gluon HTL-dressed propagator ${D}_T (K)$ is  
\begin{equation}
D_{T}^{-1}(K)=K^{2}-m_{g}^{2}\left[1+\frac{K^{2}}{k^{2}}\left(1-\frac{k_{0}}{2 k} \ln \frac{k_{0}+k}{k_{0}-k}\right)\right]
\end{equation}
For retarded gluon propagator 
\begin{equation}
D_{\mu \nu}^{\mathrm{R}}(K) \equiv {D}_{\mu \nu}^{\mathrm{ra}}(K)= {D}_{\mu \nu}\left(k_{0}+i \varepsilon, \vec{k}\right)
\end{equation}
Thus, 
\ba
{D}_{T}^{R{(-1)}}(k,k_0,\varepsilon) &=& k_{0}^2 -\varepsilon^2 + 2k_{0} i \varepsilon -k^2 -m_{g}^{2} X
\label{glu_ret}
\ea
where 
\begin{align}
X  = & \, \left(\frac{k_{0}^{2}}{k^{2}}-\frac{\varepsilon^2}{k^2}+\frac{2k_0i\varepsilon}{k^2} \right)+ \left(\frac{k_{0}+i\varepsilon}{2 k}\right)\ln A + \left(\frac{-3 i \varepsilon k_0^2}{2 k^3} \ln A -\frac{3k_{0} \varepsilon^2}{2k^3} \ln A \right) \nn
-&\, \left(\frac{k_{0}^3 - i \varepsilon^3}{2k^3} \right)\label{X_def}
\end{align}
and
\begin{equation}
\ln A = \ln \frac{k_0+k+i\varepsilon}{k_0-k+i\varepsilon}
\end{equation}
As for $N_c=3$ and $N_f=2$, $ m_{g}^2 = 4 $, so Eq.~\eqref{glu_ret} becomes
\begin{align}
{D}_{T}^{R{(-1)}}(k,k_0,\varepsilon)
=& \, -\left[\left(k^{2} -k_{0}^{2} + \frac{4k_{0}^2}{k^{2}} - 2k_{0} i\varepsilon + \frac{8k_0 i \varepsilon}{k^2} \right) + \left(\frac{2k_{0}}{k} - \frac{2k_{0}^{3}}{k^3} \right)\right. \ln A \nn
+ & \, \left. \left(\frac{2 i \varepsilon}{k} - \frac{2 i \varepsilon k_0^2}{k^3} \right) \ln A \right]
\label{DelTRinv}
\end{align}
Now, the $\ln A $ term can explicitly be written as
\ba
- \ln A 
&=& \frac{1}{2} \ln \left[\frac{\left(k_0 - k\right)^2 + \varepsilon^2}{\left(k_0+k\right)^2 + \varepsilon^2} \right] + i \left[\tan^{-1} \frac{\varepsilon}{k_0 - k} -  \tan^{-1} \frac{\varepsilon}{k_0 + k} \right]\label{lnA}
\ea
Now, using the expression of $\ln A$ from Eq.~\eqref{lnA}, Eq.~\eqref{DelTRinv} becomes
\begin{align}
{D}_{T}^{R{(-1)}}(k,k_0,\varepsilon) = & \, -\frac{4k_0^{2}}{k^{2}} - \left(k^2 - k_0^{2} \right) \left[1 - \frac{k_0}{k^3}  \ln \frac{\left(k_0 - k\right)^2 + \varepsilon^2}{\left(k_0+k\right)^2 + \varepsilon^2} \right]\nn
-& \,  i \left[\frac{2k_{0}}{k^3} \left( k_0^{2} - k^2 \right) \left(\tan^{-1} \frac{\varepsilon}{k_0 - k} -  \tan^{-1} \frac{\varepsilon}{k_0 + k} \right) - \varepsilon \Theta(k_0) \right]
\end{align}
\subsection{Longitudinal gluon propagator}
The HTL gluon propagator longitudinal part ${D}_L (K)$ is given by 
\begin{equation}
{D}_{L}^{-1}(K)=\left[K^{2}+2 m_{g}^{2} \frac{K^{2}}{k^{2}}\left(1-\frac{k_{0}}{2 k} \ln \frac{k_{0}+k}{k_{0}-k}\right)\right].
\end{equation}
As the scaled longitudinal part of the HTL gluon propagator $\tilde{D}_L(K)$ is defined as
\begin{equation}
\tilde{D}_L(K) = \frac{D_L(K)}{K^2},
\end{equation}
So, the retarded part of the inverse of the longitudinal gluon propagator becomes
\ba
\tilde{D}_{L}^{R (-1)}(K) 
&=& \left(k_{0}^{2} -k^{2} \right)^2 \left[1+\frac{8}{k^2} - \frac{4k_0}{k^3} \ln A \right] +\mathcal{O}(\varepsilon)^3.\label{DelLR}
\ea
Using the expression of $\ln A$ from Eq.~\eqref{lnA}, Eq.~\eqref{DelLR} becomes
\begin{align}
\tilde{D}_{L}^{R (-1)}(k,k_0,\varepsilon) = & \,\left(k_{0}^{2} -k^{2} \right)^2 \left[1+ \frac{8}{k^2} + \frac{2k_0}{k^3} \ln \frac{\left(k_0 - k\right)^2 + \varepsilon^2}{\left(k_0+k\right)^2 + \varepsilon^2}\right. \nn 
+& \, i \left.\left\{\frac{4k_0}{k^3} \left(\tan^{-1} \frac{\varepsilon}{k_0 - k} -  \tan^{-1} \frac{\varepsilon}{k_0 + k} \right) + \varepsilon \Theta(k_0) \right\} \right].
\end{align}
\subsection{Quark propagator}
The HTL-dressed quark propagator $\Delta_\pm (Q)$ is given by 
\begin{equation}
\Delta_{\pm}^{-1}(Q)=q_{0} \pm q-\frac{m_{q}^{2}}{q}\left[\mp 1+\frac{1}{2 q} m_{q}^{2}\left(q \pm q_{0}\right) \ln \frac{q_{0}+q}{q_{0}-q}\right] .
\end{equation}
The inverse of the retarded quark propagator for plasmino mode can be written as
\begin{align}
\hspace{-.2cm}\Delta_{-}^{R(-1)}(q,q_0,\varepsilon) 
= &  q_{0} + i \varepsilon - q - \frac{1}{q} + \frac{q_{0} - q}{2q^2}  \ln \frac{q_{0}+q+i \varepsilon}{q_{0}-q+ i \varepsilon}  + \frac{i \varepsilon}{2 q^2}  \ln \frac{q_{0}+q+i \varepsilon}{q_{0}-q+ i \varepsilon}.
\label{DelQ_inv}
\end{align}
Using Eq.~\eqref{lnA}, Eq.~\eqref{DelQ_inv} becomes
\begin{align}
\Delta_{-}^{R(-1)}(q,q_0,\varepsilon) = & \, -\left[\frac{1}{q} + q - q_0 - \frac{q_0-q}{4q^2} \ln \frac{(q_{0}+q)^2+ \varepsilon^2}{(q_{0}-q)^2+ \varepsilon^2}+\frac{\varepsilon}{2 q^2}\left\{\tan^{-1} \left(\frac{\varepsilon}{q_0 + q }\right) \right\}\right.\nn
-& \, \tan^{-1} \left(\frac{\varepsilon}{q_0-q}\right)- i \left[\varepsilon+  \frac{\varepsilon}{4q^2} \ln  \frac{(q_{0}+q)^2+ \varepsilon^2}{(q_{0}-q)^2+ \varepsilon^2} \right. \nn
+ & \left.\left. \frac{q_0-q}{2 q^2} \left\{\tan^{-1} \left(\frac{\varepsilon}{q_0 + q }\right) - \tan^{-1} \left(\frac{\varepsilon}{q_0-q}\right)\right\}\right]\right].
\end{align}
Similarly, the retarded quark propagator for the real quark mode comes out to be 
\begin{align}
\Delta_{+}^{R(-1)}(q,q_0,\varepsilon) = & \, q_0 + q + \frac{1}{q} - \frac{q_0+q}{4q^2} \ln \frac{(q_{0}+q)^2+ \varepsilon^2}{(q_{0}-q)^2+ \varepsilon^2}+\frac{\varepsilon}{2 q^2} \left\{\tan^{-1} \left(\frac{\varepsilon}{q_0 + q }\right)\right. \nn
-& \,\left. \tan^{-1} \left(\frac{\varepsilon}{q_0-q}\right)\right\} - i \left[-\varepsilon+  \frac{\varepsilon}{4q^2} \ln  \frac{(q_{0}+q)^2+ \varepsilon^2}{(q_{0}-q)^2+ \varepsilon^2}\right. \nn
+ & \, \left.\left. \frac{q_0+q}{2 q^2} \left\{\tan^{-1} \left( \frac{\varepsilon}{q_0 + q }\right) - \tan^{-1} \left(\frac{\varepsilon}{q_0-q}\right)\right\}\right]\right. .
\end{align}

\chapter{}

\section{Derivation of Eq.~\eqref{int_I1},~\eqref{int_I2}} \label{sec:appendix_B_1}
In this appendix, we will present the derivation of the integrals $I_{1}$ and $I_{2}$, which are used in evaluating the $M$ parameter in section \ref{Section_4}. 
\subsection{Derivation of Eq.~\eqref{int_I1}}
The integral $I_{1}$ is defined in section \ref{Section_4} as 
\begin{equation}\label{I1_def}
I_1\left.=\frac{-1}{p_{0}}\SumInt_{Q} \left(\frac{Q^2}{Q^4+\gamma_{G}^4}\right)_{b}\right|_{p^{2}=0} = \frac{-1}{p_{0}}\SumInt_{Q} \frac{q_{0}^2+q^{2}}{(q_{0}^2+q^2+i \gamma_{G}^{2})(q_{0}^2+q^2-i \gamma_{G}^{2})} 
\end{equation} 
We will use $\hspace{1mm} \gamma_{G} = \gamma \hspace{1mm}$ here for convenience. Here, the dimensionally regularized sum integral is given by
\begin{equation}
\SumInt_{Q} \equiv\left(\frac{e^\gamma \Lambda^2}{4 \pi}\right)^\epsilon T \sum_{q_0=2 n \pi T} \int \frac{d^{3-2 \epsilon} q}{(2 \pi)^{3-2 \epsilon}}
\end{equation}
By using the partial fraction technique, the above Eq.~\eqref{I1_def} can be re-written as  
\begin{align} \label{A3}
I_1 = & \, \frac{-1}{p_{0}}\SumInt_{Q}\quad \frac{Q^2}{Q^4+\gamma^4} =  \frac{i}{4\sqrt{q^2+i\gamma^2}}\bigg[\frac{1}{(q_{0}+i\sqrt{q^2+i\gamma^{2}})} - \frac{1}{(q_{0}-i\sqrt{q^2+i\gamma^{2}})}\bigg]\nn
+ & \, \frac{i}{4\sqrt{q^2-i\gamma^2}}\bigg[\frac{1}{(q_{0}+i\sqrt{q^2-i\gamma^{2}})} - \frac{1}{(q_{0}-i\sqrt{q^2-i\gamma^{2}})}\bigg]
\end{align} 
In general, for a meromorphic function $f(q_{0})$ which comes from the loop diagrams, can be written as (For details see~\cite{Bellac:2011kqa, Kapusta:2006pm, Mustafa:2022got})
\begin{equation}
T \sum_{q_0= 2n \pi T} f\left(q_0\right)=T \oint_C \frac{d q_0}{2 \pi i} f\left(q_0\right) \frac{\beta}{2} \coth \frac{\beta\left(i q_{0}\right)}{2}=-\frac{T}{2 \pi } \times \frac{\beta}{2} \times(2 \pi i) \sum \text { Residues, }
\end{equation} 
As poles of integral $I_{1}$ are $q_{0} = i\sqrt{q^2+i\gamma^2} ,\hspace{1mm} -i\sqrt{q^2+i\gamma^2},\hspace{1mm} i\sqrt{q^2-i\gamma^2},\hspace{1mm} -i\sqrt{q^2-i\gamma^2}$. So, we must find the residues at these poles to solve Eq.~\eqref{A3}. Let's consider the residue at $q_{0}= i\sqrt{q^{2}+i\gamma^{2}}$ of the integrand in Eq.~\eqref{A3} 
\begin{align}
\left.\frac{-i}{4\sqrt{q^2+i\gamma^{2}}}\bigg[\frac{1}{1-e^{-i\beta q_{0}}} - \frac{1}{1-e^{i\beta q_{0}}} \bigg]\right|_{q_{0} =  i\sqrt{q^{2}+i\gamma^{2}}}
=& \, \frac{-i}{4\sqrt{q^2+i\gamma^{2}}}\bigg[\frac{1}{1-e^{\beta \sqrt{q^2+i\gamma^2}}} \nn 
- & \, \frac{1}{1-e^{-\beta \sqrt{q^2+i\gamma^2}}} \bigg]
\end{align}
Residues at other poles can be found similarly, and after adding all four contributions, we will get 
\begin{align}
= & \, \frac{i}{\sqrt{q^2+i\gamma^{2}}} n_{B}(q,\gamma) + \frac{i}{\sqrt{q^2-i\gamma^{2}}} n_{B}^{\prime}(q,\gamma)
\end{align} 
where $\hspace{1mm}$ $n_{B}(q,\gamma) = \bigg[\frac{1}{e^{\beta \sqrt{q^2+i\gamma^2}}-1}\bigg]$ , $\quad$$ $$n_{B}^{\prime}(q,\gamma) = \bigg[\frac{1}{e^{\beta \sqrt{q^2-i\gamma^2}}-1}\bigg] $. Thus the integral $I_{1}$ takes the form as 
\begin{equation}
I_1 = \frac{-1}{p_{0}}\SumInt_{Q} \left(\frac{Q^2}{Q^4+\gamma^4}\right) = \frac{-1}{p_{0}}\hspace{1mm}  \int_{0}^{\infty} \frac{q^{2}dq}{(4\pi^{2})} \bigg[\frac{n_{B}(q,\gamma)}{\sqrt{q^2+i\gamma^{2}}}+ \frac{n_{B}^{\prime}(q,\gamma)}{\sqrt{q^2-i\gamma^{2}}} \bigg]
\end{equation}  
Now in order to get the $T^{2}$ behavior we can do the trivial re-scaling as $q^{\prime}\rightarrow \frac{q}{T}$ $ \quad dq^{\prime} = \frac{dq}{T}$. Thus we get
\begin{align}
I_1 = & \, \frac{-1}{p_{0}}\SumInt_{Q} \left(\frac{Q^2}{Q^4+\gamma_{G}^4}\right) = \frac{-T^{2}}{p_{0}}\hspace{1mm} \int_{0}^{\infty} \frac{q^{\prime2}dq^{\prime}}{(4\pi^{2})} \bigg[\frac{1}{\sqrt{q^{\prime2}+i(\frac{\gamma}{T})^{2}}}\frac{1}{e^{\sqrt{q^{\prime2}+i(\frac{\gamma}{T})^{2}}}-1} \nn
+ & \, \frac{1}{\sqrt{q^{\prime2}-i(\frac{\gamma}{T})^{2}}}\frac{1}{e^{\sqrt{q^{\prime2}-i(\frac{\gamma}{T})^{2}}}-1}\bigg]
\end{align} 
\subsection{Derivation of Eq.~\eqref{int_I2}} 
The integral $I_2$ is defined earlier as 
\begin{align}
I_2 = & \, \left.\frac{1}{p_{0}}\SumInt_{Q} \frac{Q^2}{(P-Q)^{2}_{f}}\left(\frac{Q^2}{Q^4+\gamma_{G}^4}\right)_{b}\right|_{p^{2}=0} \nn
= & \, \frac{1}{p_{0}}\SumInt_{Q} \frac{1}{(P-Q)^{2}_{f}} - \frac{1}{p_{0}}\SumInt_{Q} \frac{1}{(P-Q)^{2}_{f}}\left(\frac{\gamma^{4}}{Q^4+\gamma_{G}^4}\right) = \frac{1}{p_{0}} \bigg[ \frac{-T^{2}}{24} + X \bigg]
\end{align} 
where we have used the standard value of one-loop sum-integral, 
\begin{equation}
\SumInt_{Q} \frac{1}{(P-Q)^{2}_{f}} = \SumInt_{Q} \frac{1}{(Q)^{2}_{f}} = \frac{-T^{2}}{24}
\end{equation}
and $X$ is given by the integral 
\begin{equation}
X = \frac{-1}{p_{0}}\SumInt_{Q} \frac{1}{(P-Q)^{2}_{f}}\left(\frac{\gamma^{4}}{Q^4+\gamma^4}\right) = \frac{-1}{p_{0}} \SumInt_{Q} \frac{\gamma^{4}}{(P-Q)^{2}_{f}\left(Q^2+i\gamma^2)(Q^2-i\gamma^2)\right)} 
\end{equation}
which can further be written as 
\begin{equation}
X = -\gamma^{4} T \sum_{n} \int \frac{d^3 q}{(2\pi)^3} \hspace{1mm} \tilde{\Delta}(i\omega_{n},E) \Delta(i(\omega_{1}- \omega_{n}),E_1) \Delta(i(\omega_{2}- \omega_{n}),E_2)
\end{equation}
Let us consider the term $X_1$ as 
\begin{equation}
X_1 = T \sum_{n} \tilde{\Delta}(i\omega_{n},E) \Delta(i(\omega_{1}- \omega_{n}),E_1) \Delta(i(\omega_{2}- \omega_{n}),E_2)
\end{equation}
By re-writing the propagators,
\begin{equation}
X_1 = \frac{-T}{8 E E_{1} E_{2}} \sum_{n,s,s_{1},s_{2}} \frac{s s_1 s_2}{(i\omega_{n}- s E)(i(\omega_{1}-\omega_{n})-s_1 E_1)(i(\omega_2 -\omega_{n})-s_2 E_2)}
\end{equation} 
Again, using the partial fraction technique, we get 
\begin{align}
X_1 = & \, T\sum_{n,s,s_1,s_2} \frac{1}{i(\omega_1-\omega_2) -s_1 E_1 +s_2 E_2}\bigg[\frac{s_2}{2E_2}\tilde{\Delta_{s}}(i\omega_{n},E) \Delta_{s_1}(i(\omega_{1}-\omega_{n}),E_1) \nn
- & \, \frac{s_1}{2E_1}\tilde{\Delta_{s}}(i\omega_{n},E) \Delta_{s_2}(i(\omega_{2}-\omega_{n}),E_2) \bigg] 
\label{X1_form}
\end{align}
Since the standard frequency sum for two propagator cases is given by~\cite{Bellac:2011kqa} 
\begin{equation}
T\sum_{n} \Delta_{s_1}(i\omega_{n},E_1) \Delta_{s_2}(i(\omega-\omega_{n}),E_2) = \frac{-s_1 s_2}{4E_1 E_2}\hspace{1mm}\frac{1+f(s_1 E_1) + f(s_2 E_2)}{i\omega-s_1 E_1 -s_2 E_2}
\end{equation}
Thus Eq.~\eqref{X1_form} becomes
\begin{align}
X_1 = & \, -\sum_{s,s_1,s_2} \frac{ss_1s_2}{8E E_1 E_2} \frac{1}{i(\omega_1-\omega_2) -s_1 E_1 +s_2 E_2} \bigg[\frac{1-\tilde{f}(s E) + f(s_1 E_1)}{i\omega_1-s E -s_1 E_1} \nn
- & \, \frac{1-\tilde{f}(s E) + f(s_1 E_1)}{i\omega_1-s E -s_1 E_1}\bigg]
\end{align}
where the different notations are given as follow as $s,s_1,s_2=\pm 1 ,\quad f(E) = \frac{1}{e^{\beta E}-1} ,\quad \tilde{f}(E) = \frac{1}{e^{\beta E}+1} \quad \quad \quad$ \\
$f(-E) = -(1+f(E)) ,\quad \tilde{f}(-E) = (1-\tilde{f}(E)) ,  \quad \quad E = q-p_3(1+\cos\theta) ,\quad E_1 = \sqrt{q^2+i\gamma^2} ,\quad E_2 = \sqrt{q^2-i\gamma^2} $\\ \\
Here, $\theta$ is the angle between momentum $\vec{p}$ and $\vec{q}$. In order to get the $T^{2}$ behavior, we need $ s=-s1=-s2 $. Thus, after some simplification, we get
\begin{align}
X_1 = & \, \frac{-1}{8E E_1 E_2} \bigg[\frac{1}{E_1-E_2} \bigg\{\frac{\tilde{n}(E)+n(E_2)}{ip_0-E+E_2} - \frac{\tilde{n}(E)+n(E_1)}{ip_0-E+E_1}\bigg\} \nn
+ & \, \frac{1}{E_1-E_2} \bigg\{\frac{\tilde{n}(E)+n(E_1)}{ip_0+E-E_1} - \frac{\tilde{n}(E)+n(E_2)}{ip_0+E-E_2}\bigg\} \bigg]
\end{align}
Now, 
\begin{equation}
X = -\gamma^{4} \int \frac{q^2 dq}{(2\pi)^{3}} \sin\theta d\theta d\phi \hspace{1mm}(X_1)
\end{equation}
which take the following form
\begin{align}
X = & \, \frac{\gamma^{4}}{8E E_1 E_2} \int \frac{q^2 dq}{(2\pi)^2}\bigg[\frac{1}{E_1-E_2}\bigg\{\frac{\tilde{n}(E)+n(E_2)}{p_3-E+E_2} - \frac{\tilde{n}(E)+n(E_1)}{p_3-E+E_1}\bigg\} \nn
+ & \, \frac{1}{E_1-E_2} \bigg\{\frac{\tilde{n}(E)+n(E_1)}{p_3+E-E_1} - \frac{\tilde{n}(E)+n(E_2)}{p_3+E-E_2}\bigg\} \bigg]
\end{align}
In order to get the overall $T^{2}$ contribution one can again do the trivial re-scaling $q \rightarrow q^{\prime} = \frac{q}{T} , \quad  dq^{\prime} = \frac{dq}{T}$  as done earlier
\begin{align}
X = & \, \frac{\gamma^{4}}{T^2} \int \frac{q^{\prime 2} dq^{\prime}}{(2\pi)^2}\frac{\sin\theta d\theta}{8E^{\prime} E_{1}^{\prime} E_{2}^{\prime}}\bigg[\frac{1}{E_{1}^{\prime}-E_{2}^{\prime}}\bigg\{\frac{\tilde{n}(E^{\prime})+n(E_{2}^{\prime})}{\frac{p_3}{T}-E^{\prime}+E_{2}^{\prime}} - \frac{\tilde{n}(E^{\prime})+n(E_{1}^{\prime})}{\frac{p_3}{T}-E^{\prime}+E_{1}^{\prime}}\bigg\}\nn
+ & \, \frac{1}{E_{1}^{\prime}-E_{2}^{\prime}} \bigg\{\frac{\tilde{n}(E^{\prime})+n(E_{1}^{\prime})}{\frac{p_3}{T}+E^{\prime}-E_{1}^{\prime}} - \frac{\tilde{n}(E^{\prime})+n(E_{2}^{\prime})}{\frac{p_3}{T}+E^{\prime}-E_{2}^{\prime}}\bigg\} \bigg]
\end{align}
Here, $E^{\prime} = q -\frac{p_{3}}{T}(1+\cos\theta) , \quad E_{1}^{\prime} = \sqrt{q^{\prime 2}+ i (\frac{\gamma}{T})^2},  \quad E_{2}^{\prime} = \sqrt{q^{\prime 2}- i (\frac{\gamma}{T})^2}$ $\quad$ and $\quad$ $\tilde{n}(E^{\prime}) = \frac{1}{e^{q^{\prime}}+1}, \quad \\ n(E_{1}^{\prime}) = \frac{1}{e^{\sqrt{q^{\prime2} + i(\frac{\gamma}{T})^{2}}}} ,
\quad n(E_{2}^{\prime}) = \frac{1}{e^{\sqrt{q^{\prime2} - i(\frac{\gamma}{T})^{2}}}}$\\ \\
Now calling, $q^{\prime}$ as $q$, $E^{\prime}$ as $E$, $E_{1}^{\prime}$ as $E_1$ , $E_{2}^{\prime}$ as $E_2$. Also, the pole condition reads as $p_3 = i\pi T$. Thus, we get 
\begin{align}
X = & \, \frac{\gamma^{4}}{T^2} \int \frac{q^{2} dq}{(2\pi)^2}\frac{\sin\theta d\theta}{8E E_{1} E_{2}}\bigg[\frac{1}{E_{1}-E_{2}}\bigg\{\frac{\tilde{n}(E)+n(E_{2})}{i\pi-E+E_{2}} - \frac{\tilde{n}(E)+n(E_{1})}{i\pi-E+E_{1}}\bigg\} \nn
+ & \, \frac{1}{E_{1}-E_{2}} \bigg\{\frac{\tilde{n}(E)+n(E_{1})}{i\pi+E-E_{1}} - \frac{\tilde{n}(E)+n(E_{2})}{i\pi+E-E_{2}}\bigg\} \bigg]
\label{def._X_old}
\end{align}
\section{Derivation of Static $Q\bar{Q}$ potential in Effective theory using Gribov approach}\label{sec:appendix_B_2}
This appendix will give the detailed derivation of the static $Q\bar{Q}$ potential in the context of NRQCD$_{3}$ theory using Gribov's gluon propagator. 
Let us start with the modified correlator defined earlier as 
\begin{equation}
\mathcal{C}(\mathbf{r}, z) \equiv \int_\mathbf{R}\left\langle \Phi_{\mathrm{i}}^{\dagger}(\mathbf{R}, z) X_{\mathbf{j}}(\mathbf{R}, z) X_{\mathbf{j}}^{\dagger}\left(\frac{\mathbf{r}}{  2},0\right) \Phi_{\mathrm{i}}\left(-\frac{\mathbf{r}}{  2},0\right)\right\rangle
\end{equation}
In the momentum configuration, after doing the straightforward integration over $\mathbf{R}$ variable, which gives a $\delta(\mathbf{q}_{\perp}-\mathbf{p}_{\perp})$ function and then doing the $\mathbf{q}_{\perp} \equiv \mathbf{q}$ integration leads to  
\begin{align}
\mathcal{C}(\mathbf{r}, z) = & \, \int \frac{\mathrm{d}^3 p}{(2 \pi)^3} \frac{\mathrm{d}^3 p^{\prime}}{(2 \pi)^3} \frac{\mathrm{d}^3 q^{\prime}}{(2 \pi)^3} \frac{\mathrm{d} q_3}{2 \pi}e^{i\left(q_3-p_3\right) z-i\left(\mathbf{p}^{\prime}+\mathbf{q}^{\prime}\right) \cdot \mathbf{r} / 2} \nn
\times & \, \left\langle\Phi_{\mathrm{i}}^{\dagger}(p) X_{\mathbf{j}}\left(\mathbf{p}, q_3\right) X_{\mathrm{j}}^{\dagger}\left(q^{\prime}\right) \Phi_{\mathrm{i}}\left(p^{\prime}\right)\right\rangle .
\end{align}
For the tree order diagram shown in Figure.~\ref{fig_correlation_diagrams} (a), the correlation function for a pair of free quark propagators is
\begin{align}
\mathcal{C}^{(0)}(\mathbf{r}, z)  =& 2 N_c \int \frac{\mathrm{d}^3 p}{(2 \pi)^3} \frac{\mathrm{d} q_3}{2 \pi} \frac{-i}{M-i p_3+\frac{\mathrm{p}^2}{2 p_{0 \mathrm{i}}}}  
\frac{-i}{M+i q_3+\frac{\mathrm{p}^2}{2 p_{0 \mathrm{j}}}}
e^{i\left(q_3-p_3\right) z-i \mathbf{p} \cdot \mathbf{r}}
\end{align}
After doing the $p$ and $q_{3}$ integrals , we get
\begin{eqnarray}\label{leading_order_scr_mass}
\hspace{-.6cm}		\mathcal{C}^{(0)}(\mathbf{r}, z)
& =\theta(z) N_c \frac{p_{0}}{2\pi z} \exp \left[-2M z-\frac{p_{0}}{4 z} \mathbf{r}^2\right],
\end{eqnarray}
From the above Eq.~\eqref{leading_order_scr_mass}, it can be inferred that the leading order screening mass is $\zeta^{-1} =  2\pi T$. Similar computation can be done for the graphs of Figure~\ref{fig_correlation_diagrams} (b) and Figure~\ref{fig_correlation_diagrams} (c), being the non-trivial pole structure in $p_{3}$ and $q_{3}$ integrations. The contribution from the Figure~\ref{fig_correlation_diagrams} (b) cancels out as the temporal and spatial gauge fields contribute equally and come with the opposite sign, while Figure~\ref{fig_correlation_diagrams} (c) contribution to the NLO term in the correlation function  becomes
\begin{align}
\mathcal{C}^{ 1(\mathrm{c})}(\mathbf{r}, z) = & \, 2g_{\mathrm{E}}^2 C_{\mathrm{F}} \int \frac{\mathrm{d}^3 q}{(2 \pi)^3} \frac{q^{2}}{q^4+\gamma_{G}^4} \frac{e^{i \mathrm{q} \cdot \mathbf{r}}}{q_3^2} \left(1-e^{i q_3 z}\right) 
\left(1-e^{-i q_3 z}\right) \mathcal{C}^{(0)}(\mathbf{r}, z)
\end{align}
Thus, the NLO correction in the correlation function comes out to be  
\begin{align}
\mathcal{C}^{(1)}(\mathbf{r}, z) = & \, \hspace{1mm} 2 g_{\mathrm{E}}^2 C_{\mathrm{F}} \, \mathcal{C}^{(0)}(\mathbf{r}, z) \int \frac{\mathrm{d}^3 q}{(2 \pi)^3}  \frac{q^{2}}{q^4+\gamma_{G}^4} \frac{e^{i \mathrm{q} \cdot \mathbf{r}}}{q_3^2} \left(2-e^{i q_3 z}-e^{-i q_3 z}\right) 
\end{align}
For the leading order correlation function, the EOM satisfied is given by 
\begin{equation}
\left[\partial_z + 2 M-\frac{1}{p_0} \nabla_{\boldsymbol{r}}^2\right] \mathcal{C}^{(0)}(\mathbf{r}, z) = 2 N_{c}\hspace{0.5mm} \delta(z) \hspace{0.5mm}\delta^{(2)}(\mathbf{r})
\end{equation}
On the other hand, a similar kind of equation can be written for first order, as mentioned in section \ref{Section_5} as well
\begin{align}
\left[\partial_z+2 M-\frac{1}{p_0} \nabla_{\boldsymbol{r}}^2\right] \mathcal{C}^{(1)}(\boldsymbol{r}, z) = & \, -g_{\mathrm{E}}^2 C_F  \, \mathcal{C}^{(0)}(\boldsymbol{r}, z) \, \mathcal{K}\left(\frac{1}{z p_0}, \frac{\nabla_{\boldsymbol{r}}}{p_0}, \frac{\gamma_{G}^{4}}{p_0^4}, \boldsymbol{r} p_0\right)
\label{One-loop_EOM}
\end{align}
This kernel $\mathcal{K}$ is a dimensionless quantity that can be expanded into its first two arguments so that the terms remain of the order of unity inside the kernel, as we have $g^{2}$ coefficient already there in the potential calculation. Thus the kernel $\mathcal{K}$ is obtained as   
\begin{align}
\mathcal{K}\left(0, 0, \frac{\gamma_{G}^{4}}{p_0^4}, \boldsymbol{r} p_0\right) = & \, -\lim _{z \rightarrow \infty} \partial_z \int \frac{\mathrm{d}^3 q}{(2 \pi)^3} \frac{1}{q_3^2}\left[\frac{q^{2}}{q^4+\gamma_{G}^4}\right] \left(2-e^{i q_3 z}-e^{-i q_3 z}\right)\left(2 e^{i \mathbf{q} \cdot \mathbf{r}}\right) \nn
 =& \, \frac{1}{2 \pi}\left[\ln \frac{{\gamma_{G}} r}{2}+\gamma_E-K_0\left(\gamma_{G} r\right)\right] 
\end{align}
Thus, the static one-loop potential is obtained as
\begin{align}
V(\mathbf{r}) = & \, g_{E}^{2} C_{F} \mathcal{K}\left(0, 0, \frac{\gamma_{G}^{4}}{p_0^4}, \boldsymbol{r} p_0\right) = -2 g_{E}^{2} C_{F} \bigg[\frac{1}{2\pi}\int_{0}^{\infty} \frac{q^{3}}{q^{4}+\gamma_{G}^{4}} \hspace{1mm}\mathrm{J}_{0}(q r) \hspace{1mm}\mathrm{d}q\bigg] \nn
=& \, g_{\mathrm{E}}^2 \frac{C_F}{2 \pi}\left[\ln \frac{{\gamma_{G}} r}{2}+\gamma_E-K_0\left(\gamma_{G} r\right)\right] .
\end{align}
\clearpage
\phantomsection 
\addcontentsline{toc}{chapter}{\textbf{Bibliography}}
\label{Bibliography}
\lhead{\emph{Bibliography}}




\end{document}